\documentclass[sn-basic]{sn-jnl}

\makeatletter
\input{aas_macros.sty}
\def\ref@jnl#1{{\jnl@style#1\ }}
\makeatother

\usepackage{graphicx}%
\usepackage{hyperref}
\usepackage{multirow}%
\usepackage{amsmath,amssymb,amsfonts}%
\usepackage{amsthm}%
\usepackage{mathrsfs}%
\usepackage[title]{appendix}%
\usepackage{xcolor}%
\usepackage{textcomp}%
\usepackage{manyfoot}%
\usepackage{booktabs}%
\usepackage{algorithm}%
\usepackage{algorithmicx}%
\usepackage{algpseudocode}%
\usepackage{listings}%
\usepackage{pdflscape} 
\usepackage{rotating}
\usepackage{longtable}
\usepackage{tabularx}

\begin{document}

\newgeometry{left=3cm,bottom=2.5cm,top=2.5cm}

\title[Intrinsic alignments]{A rising tide: Intrinsic alignments since the turn of the millennium}


\author*[1,2]{\fnm{Nora Elisa} \sur{Chisari}}\email{n.e.chisari@uu.nl}

\affil*[1]{\orgdiv{Institute for Theoretical Physics, Department of Physics}, \orgname{Utrecht University}, \orgaddress{\street{Princetonplein 5}, \city{Utrecht}, \postcode{3584 CC}, \country{The Netherlands}}}

\affil*[2]{\orgdiv{Leiden Observatory}, \orgname{Leiden University}, \orgaddress{\street{Niels Bohrweg 2}, \city{Leiden}, \postcode{NL-2333 CA}, \country{The Netherlands}}}

\abstract{The alignments of galaxies across the large-scale structure of the Universe are known to be a source of contamination for gravitational lensing, but they can also probe cosmology and the physics of galaxy evolution in many ways. In this review, I cover developments in our understanding of intrinsic alignments over the past 25 years on: (1) different approaches to model intrinsic alignments across a range of scales, (2) existing observational constraints, (3) predictions from cosmological numerical $N$-body and hydrodynamical simulations, (4) mitigation strategies to account for their contamination to lensing observables and (5) cosmological and astrophysical applications. While the review focuses mostly on two-point statistics of intrinsic alignments, I also give a summary of other statistics beyond two-point. Finally, I point out some of the open problems hindering the understanding or application of intrinsic alignments and how they might be overcome in the future.}




\maketitle

\tableofcontents

\section{Introduction}\label{sec:intro}

The preferential orientation of galaxies (or other cosmic structures) with respect to one another goes by the name of \emph{intrinsic alignments}. In a  seminal paper, \citet{Binggeli82} detected preferential orientations of galaxies across the large-scale structure. His study showed that Abell clusters point towards each other, and that the brightest galaxy was preferentially aligned with the shape of the cluster, implying they also point towards each other. Since then, many observational studies have confirmed the presence of intrinsic alignments. 

\begin{figure*}
    \centering
    \includegraphics[width=\textwidth]{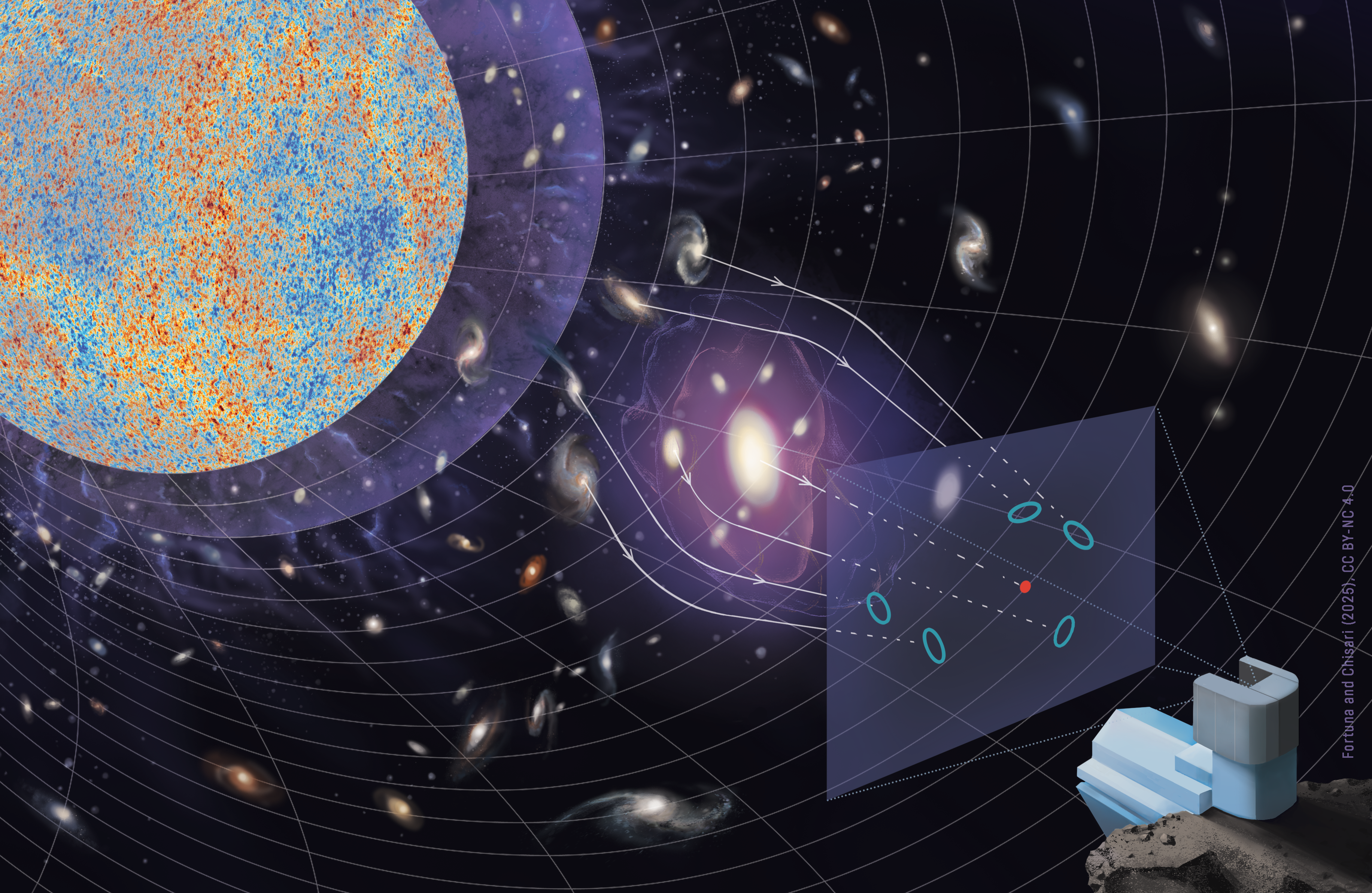}
    \caption{An artistic rendering of the distorted path of photons through the large-scale structure. This phenomenon, which distorts observed galaxy shapes in the tangential direction around matter overdensities, is known as gravitational lensing. Credit: Fortuna \& Chisari (2025), CC-BY-NC 4.0, adapted from Fortuna \& Chisari (2022), CC-BY-NC 4.0.}
    \label{fig:lensing}
\end{figure*}

Around the turn of the millennium, weak gravitational lensing was emerging as a new cosmological probe \citep{Kaiser92}. The per cent level distortion of the path of photons as they travel through the matter distribution in the Universe is reflected in coherent changes in the perceived shapes of galaxies (Fig.~\ref{fig:lensing}). These allow one to infer the distribution of matter in the Universe across cosmic time, probing the growth of structure and the expansion rate of the Universe through the distance-redshift relation. Enabled by a Stage III era of wide-fast-deep astronomical surveys, which includes the Kilo-Degree Survey \cite[KiDS,][]{deJong13}, the Hyper Suprime Cam Subaru Strategic Project \cite[HSC,][]{Aihara18}, and the Dark Energy Survey \cite[DES,][]{DES}, weak gravitational lensing has become one of the key probes of cosmic acceleration \citep{Weinberg13}. 

The primary observable is the auto-correlation of lensing distortions of galaxy shapes (`cosmic shear') but in recent years, shear has also been combined and/or cross-correlated with other tracers of the large-scale structure to deliver cosmological constraints. In particular, the combination of galaxy clustering with cosmic shear and galaxy-galaxy lensing (the cross-correlation of shear and density tracers) is now commonly adopted as a cosmological probe and referred to as a $3\times2$pt correlation. The main goal of weak lensing surveys is to constrain the $S_8=\sigma_8\sqrt{\Omega_{\rm m}/0.3}$ cosmological parameter, a combination of the variance of the matter overdensity field in spheres of $8\,h^{-1}$ Mpc radius, $\sigma_8$, and the fractional energy density in matter today, $\Omega_{\rm m}$. The next generation of Stage IV surveys is expected to also put stringent constraints on the equation of state of dark energy, parametrised as $w=w_0+w_a(1-a)$.

\begin{figure*}
    \centering
    \includegraphics[width=\textwidth]{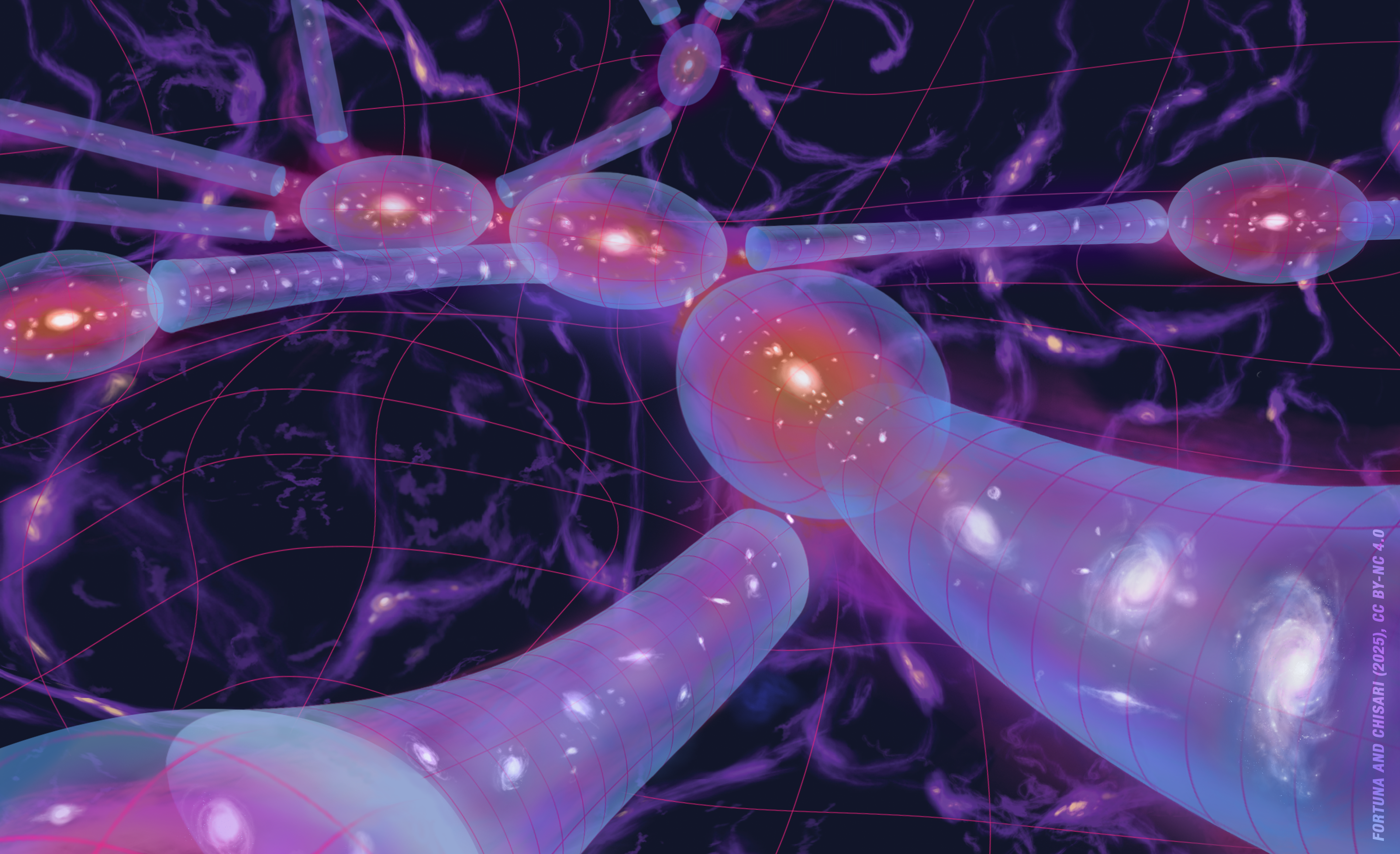}
    \caption{An artistic rendering of the intrinsic alignments of galaxies embedded in the large-scale structure of the Universe. Different signals contribute to the overlap alignment of shapes relative to the density field: the alignment of central galaxies with their haloes, the alignment of central galaxies relative to one another, the alignment of satellites within the halo and with the central shape, and the potential alignment of galaxies with filaments. This also leads one to hypothesize about the presence of an alignment around voids. Credit: Fortuna \& Chisari (2025), CC-BY-NC 4.0.}
    \label{fig:sketch}
\end{figure*}

Intrinsic alignments distort galaxy shapes not because photons alter their path, but because the actual distribution of stars in a galaxy is elongated with a preferential alignment relative to structures that can be even hundreds of Mpc away (Fig.~\ref{fig:sketch}). When observing the shape of an individual galaxy, in principle we do not know to what level its ellipticity and orientation is determined by random processes, by gravitational lensing or by physical mechanisms which correlate it with the large-scale structure. A priori, all of them might be present. For this reason, intrinsic alignments are considered an \emph{astrophysical systematic} to gravitational lensing. To extract unbiased cosmological information from weak gravitational lensing observables, intrinsic alignments need to be accounted for.

The need for a better understanding of intrinsic alignments for weak lensing applications has triggered interest in the topic in the last few decades. In parallel to the development and assessment of mitigation strategies, a number of studies have looked into how, when, which and by how much galaxies align in the Universe, and at possible mechanisms which can originate such alignments. Deeper, faster, wider surveys are enabling the measurement of hundreds of millions of galaxy shapes, which could also open opportunities to use intrinsic alignments of galaxies as a cosmological or astrophysical probe.

This review covers the developments on our knowledge of intrinsic alignments in the last $25$ years. For detailed accounts of the field prior to the turn of the millennium, we refer the reader to the exhaustive work of \citet{Joachimi15} (historical perspective), \citet{Kiessling15} (simulations), \citet{Kirk15} (observations) and \citet{Troxel15} (observations and mitigation). In addition, we will only gloss over the literature on the connection between the intrinsic alignments of galaxy shapes and that of angular momenta, which is covered by \citet{Schafer09}. For a very practical guide of intrinsic alignments, we refer the reader to \citet{IAGuide}. Extensive weak lensing reviews are available in \citet{Bartelmann,Kilbinger,Mandelbaum}, for example.

In this review, we first discuss the origin of the alignment signal from a theoretical perspective in Sect.~\ref{sec:model}. There, we give an overview of the different modelling options available in the literature. Observational evidence for intrinsic alignments, focusing on constraints coming from two-point statistics of galaxies and clusters in Sect.~\ref{sec:obs}. Our knowledge of intrinsic alignments from cosmological simulations ($N$-body and hydrodynamical) are presented separately in Sect.~\ref{sec:sims}. In sections \ref{sec:astro} and \ref{sec:cosmo}, we review for the first time different applications of intrinsic alignments to cosmology and galaxy evolution. Section~\ref{sec:astro} includes a short discussion on the connection to alignments of galaxy angular momenta. Section~\ref{sec:miti} covers the different mitigation strategies adopted by the community to account for the impact of intrinsic alignments mainly on weak gravitational lensing statistics as well as indirect constraints derived from these surveys on intrinsic alignment models. Section~\ref{sec:beyond} remarks on the usefulness of beyond two-point statistics in constraining and mitigating intrinsic alignments. Section~\ref{sec:future} gives some expectations in terms of future data sets. We conclude by further discussing the open problems in Sect.~\ref{sec:discuss} and presenting a summary and outlook in Sect.~\ref{sec:conclu}.

\section{Modelling}\label{sec:model}

This section looks into different options for modelling intrinsic alignments: from linear to perturbative to fully non-linear models. It also discusses the possibility of obtaining intrinsic alignment priors from theory. 

\subsection{Separation of scales}

In complete analogy to galaxy clustering, one can take different approaches to modelling intrinsic alignments depending on the scales that separate the galaxies. Because alignments are measured from intrinsic galaxy shapes, which are spin-2 tensors, the lowest order approximation relates them linearly to the tidal field of the large-scale structure. Therefore, sufficiently large scales ($\gtrsim 10-20\,h^{-1}$ Mpc) can be modelled linearly on the tidal field. Intermediate scales (above a few Mpc) are quasi-linear and requires more operators and higher order expansions, including terms that depend quadratically on the tidal field and are often associated with angular momenta alignments of galaxies (``tidal torquing''). The fully non-linear regime (below a few Mpc) is intractable with perturbation theory and requires a halo model approach (or numerical simulations, see Sect.~\ref{sec:sims}). In what follows, we present the models that are available for intrinsic alignments in these different regimes. Numerical implementations of: the linear and non-linear alignment models, the tidal alignment-tidal torquing model (TATT), the effective field theory (EFT) model, and the halo model, can be found publicly available in the {\tt PyCCL} library\footnote{\url{https://github.com/LSSTDESC/CCL/} and \url{https://github.com/LSSTDESC/CCLX/}} \citep{Chisari19}, FAST-PT\footnote{\url{https://github.com/JoeMcEwen/FAST-PT}} \citep{McEwen16,Fang17} and {\sc spinosaurus}\footnote{\url{https://github.com/sfschen/spinosaurus}} \citep{Chen24}.

\subsection{Linear alignment model}
\label{sec:linear}

The simplest model of intrinsic alignments posits that the shape of a galaxy is linearly related to the tidal field of the large-scale structure. This ``linear alignment'' (LA) model was put forward by \citet{Catelan01}, who directly postulated that the linear relation was satisfied between the \emph{projected} shape of a galaxy and the \emph{projected} tidal field of the matter distribution. More generally, one can postulate a linear relation between the three-dimensional shape of a large-scale structure tracer and the three-dimensional tidal field, which can then be projected onto the sky \citep{Vlah21}. At the linear level, these are equivalent. It is understood \citep{Hirata04,Blazek19,Vlah20} that this linear relation is the only one that can be constructed at the lowest order satisfying rotational invariance and General Relativity \citep{Vlah20}, as we will see in Sect.~\ref{sec:eft}.   

In three-dimensions, the shape of an object is described by a three-dimensional symmetric trace-free inertia tensor at a certain position and redshift, $\mathcal{I}_{ij}({\bf x},z)$. To linear order, this intrinsic shape is related to the three-dimensional tidal field of the large-scale structure, $K_{ij}$, in Fourier space as
\begin{equation}
    \overset{\sim}{\mathcal{I}}_{ij}({\bf k},z) = b_{1,I} \tilde K_{ij}({\bf k},z),
    \label{eq:IpropK}
\end{equation}
where $\overset{\sim}{\mathcal{I}}_{ij}$ and $\tilde K_{ij}$ denote the Fourier transform of the fields, $b_{1,I}$ is a free bias parameter that quantifies the linear response of a galaxy shape to the large-scale tidal field, i.e. the alignment strength of the sample and for a specific shape measurement choice, and 
\begin{equation}
    \tilde K_{ij}({\bf k},z) = \frac{k_ik_j}{k^2}\tilde{\delta}(\mathbf{k},z)-\frac{1}{3}\delta_{ij}
    \tilde{\delta}(\mathbf{k},z),
    \label{eq:Kij}
\end{equation}
where $\tilde\delta$ is the Fourier transform of the matter overdensity field and $\delta_{ij}$ the Kronecker delta function.

What we actually observe is the two-dimensional projection of the shape, $I_{ij}({\bf x},z)$. Hence, we need to project the tensor and the relation in Eq.~(\ref{eq:IpropK}) onto the sky. This projection operation is independent of the model choice \citep{Vlah21} and applies to any tensorial quantity whose observable we are trying to project onto two-dimensions. The projection operation is defined as
\begin{equation}
    I_{ij}({\bf x},z)={\rm TF}(\mathcal{P}^{ik}(\hat{\bf n})\mathcal{P}^{jl}(\hat{\bf n})\mathcal{I}_{kl}({\bf x},z))
    \label{eq:ProjectionOperator}
\end{equation}
where $\hat{\bf n}$ is the direction over which we project, TF stands for trace-free part and $\mathcal{P}^{ij}\equiv\delta^{ij}-\hat{\bf n}_i\hat{\bf n}_j$. Projection reduces the degrees of freedom of $\mathcal{I}_{ij}$ from five to two. 
Eq.~(\ref{eq:ProjectionOperator}) misses the effects of redshift-space distortions (RSDs, discussed in Sect.~\ref{sec:RSD}) and other General Relativistic corrections that could be incorporated in the future \citep{FJ12}. At linear order, the contribution of gravitational lensing (`shear') can be directly added, but in reality the relation between intrinsic shape, shear, and observed shear is non-linear \cite{BJ02}. There might also be additional corrections depending on how the galaxy shapes are measured.

Let us define the two remaining degrees of freedom in $I_{ij}({\bf x},z)$ as
\begin{eqnarray}
    \gamma_1\equiv I_{11}&\propto&\mathcal{I}_{11}-\mathcal{I}_{22},\label{eq:g1}\\
    \gamma_2\equiv {I}_{12}&\propto&2\mathcal{I}_{12}\label{eq:g2},
\end{eqnarray}
Here, the line-of-sight is taken along the $3$-axis and Eqs. (\ref{eq:g1}) and (\ref{eq:g2}) share the same normalization factor. The normalization is chosen such that it transforms as a scalar under rotation in three-dimensions and it can be expanded in perturbation theory \citep{Bakx23}. 
If measured with respect to the direction towards another galaxy, these components are re-labelled $\gamma_1\rightarrow \gamma_+$ and $\gamma_2\rightarrow\gamma_
\times$. $\gamma_+$ corresponds then to the ellipticity component aligned either radially or tangentially with the separation vector, while $\gamma_\times$ is rotated by $45\deg$ with respect to this direction. 

\begin{figure}
    \centering
    \includegraphics[width=0.7\textwidth]{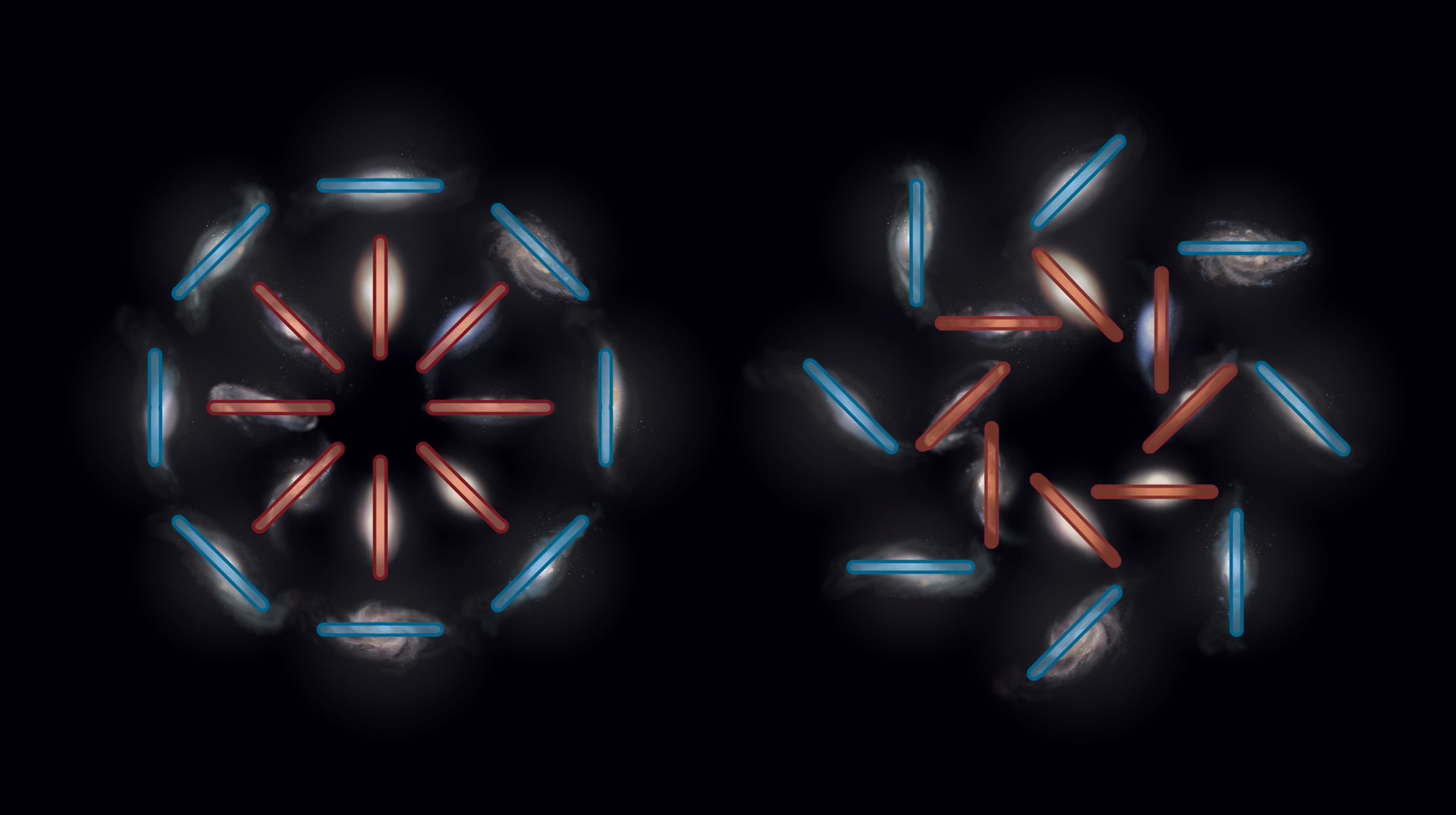}
    \caption{$E$ (left) and $B$-modes (right) of galaxy shapes. Credit: Fortuna and Chisari (2022), CC-BY-NC 4.0.}
    \label{fig:EB}
\end{figure}

Shape components can be transformed into ``electric'', $\tilde\gamma_E$, and ``magnetic'', $\tilde\gamma_B$, modes (whose characteristic patterns are shown in Fig.~\ref{fig:EB}) in Fourier space through:
\begin{eqnarray}
    \tilde\gamma_E&=&\cos(2\phi_k)\tilde\gamma_{1}+\sin(2\phi_k)\tilde\gamma_{2}\label{eq:gamma_E}\\
    \tilde\gamma_B&=&-\sin(2\phi_k)\tilde\gamma_{1}+\cos(2\phi_k)\tilde\gamma_{2}\label{eq:gamma_B},
\end{eqnarray}
where $\phi_k$ is the angle of the three-dimensional {\bf k} wavevector on the projected plane. This decomposition is fully analogous to the one adopted in the cosmic microwave background (CMB) polarization literature \citep{CMBPol}. We will refer to the Fourier-space shape components of Eqs. (\ref{eq:gamma_E}) and (\ref{eq:gamma_B}) as $E$- and $B$-modes, respectively, in what follows.

From Eqs. (\ref{eq:IpropK}) and (\ref{eq:Kij}), it is clear that there is a correlation between the intrinsic shape and the density field:
\begin{equation}
    \langle \tilde\delta({\bf k},z)\tilde\gamma_E({\bf k}',z)\rangle =(2\pi)^3P_{\delta E}(k,\mu,z)\delta^{(D)}({\bf k}+{\bf k}')
\end{equation}
It should also be then possible to predict a similar correlation for the shape components, and thereafter the auto-correlations of those shapes and their cross-correlation with any biased tracer, $g$:
\begin{eqnarray}
    \langle \tilde\delta_g({\bf k},z)\tilde\gamma_E({\bf k}',z)\rangle &=&(2\pi)^3P_{g E}(k,\mu,z)\delta^{(D)}({\bf k}+{\bf k}')\nonumber\\
    \langle \tilde\gamma_E({\bf k},z)\tilde\gamma_E({\bf k}',z)\rangle &=&(2\pi)^3P_{EE}(k,\mu,z)\delta^{(D)}({\bf k}+{\bf k}')\nonumber\\
    \langle \tilde\gamma_B({\bf k},z)\tilde\gamma_B({\bf k}',z)\rangle &=&(2\pi)^3P_{BB}(k,\mu,z)\delta^{(D)}({\bf k}+{\bf k}')
\end{eqnarray}
We will skip over the details of how these can be obtained (see \citealt{Vlah20,Vlah21,Bakx23}) and present directly the relevant power spectra for $E$- and $B$-modes at linear order ($L$):
\begin{eqnarray}
P_{\delta E}^L(k,\mu,z)&=&\frac{b_{1,I}}{2}(1-\mu^2)P_L(k,z);\label{eq:LA_deltaE}\\
P_{gE}^L(k,\mu,z)&=&\frac{b_{1,g}b_{1,I}}{2}(1-\mu^2)P_L(k,z);\label{eq:LA_gE}\\
P_{EE}^L(k,\mu,z)&=&\frac{b_{1,I}^2}{4}(1-\mu^2)^2 P_L(k,z);\label{eq:LA_EE}\\
P_{BB}^L(k,\mu,z)&=&0;\,P_{g B}^L(k,\mu,z)=0;\,P_{EB}^L(k,\mu,z)=0,
\end{eqnarray}
where $\mu=\hat{\bf n}\cdot{\hat {\bf k}}$,  $P_L(k,z)$ is the linear matter power spectrum, and  $b_{1,g}$ is the linear galaxy bias \citep{Desjacques18}. The first two lines represent non-zero cross-correlations between shapes and the matter overdensity and the galaxy density fields, respectively. The second line corresponds to non-zero auto-correlations of the shapes. $B$-modes are not generated at linear level, and both $gB$ and $EB$ cross-correlations are null at any order as long as parity is not violated (see Sect.~\ref{sec:cosmo}).

In the early formulation of the linear alignment model by \citet{Hirata04}, the projection is performed over a fixed axis ${\bf z}$. With this simplification, Eq.~(\ref{eq:LA_deltaE}) would reduce to 
\begin{equation}
    P_{\delta E}^L({\bf k},z)=\frac{b_{1,I}}{2}\frac{k_x^2-k_y^2}{k^2}P_L(k,z),\label{eq:LA_deltaE_zdirection}
\end{equation}
correctly reproducing the ${\bf k}$-dependence found in that work. \citet{Hirata04} related the projected shape of an object to the projected tidal field of the \emph{gravitational potential}. Because this is not the same as $\tilde K_{ij}$ (where the derivatives act on the density field), the proportionality constants that relate the shape to the operator differ. They are related by
\begin{equation}
    \frac{b_{1,I}}{2}  = -\frac{C_1\bar{\rho}(a)}{\bar D(a)}a^2 \label{eq:iabias_norm_LA}
\end{equation}
where $\bar{\rho}(a)=\Omega_{\rm m}\rho_{\rm crit}a^{-3}$, $\rho_{\rm crit}$ is the critical density today, $\Omega_{\rm m}$ is the fractional energy density in matter today,
${\bar D}(a)=D(a)/a$, and $D(a)$ is the linear growth factor normalised to unity today\footnote{Note that the choice of how to normalise $D$ varies in the literature.}. This particular redshift dependence is valid if alignments are a function of the \emph{primordial} gravitational potential (at the time the galaxy was formed) and not the instantaneous one. If the instantaneous tidal field is assumed, then the redshift evolution of the right-hand side of Eq.~(\ref{eq:iabias_norm_LA}) is different (see \citealt{Chisari13}). 

Instead of specifying a redshift-dependence of the alignment bias, some works attempt to constrain a power-law dependence with redshift of the form $[(1+z)/(1+z_0)]^\beta$ on the right-hand side of Eq.~(\ref{eq:iabias_norm_LA}), where $z_0$ is a pivot redshift of choice and $\beta$ is a free parameter to be constrained from the data. This power law makes us partially agnostic to the redshift scaling, but it is not motivated from theory. For a more detailed discussion on the redshift-dependence of intrinsic alignments and how it might also affect the scale-dependence of the signal \citep{Blazek15}.

Following the first significant detection of intrinsic alignments in a cosmic shear survey \citep{Brown02}, most works normalize the linear alignment model amplitude to the observed value in that work: $(C_1{\rho_{\rm crit}})|_{\rm fid}=0.0134$, such that they actually quote the measured alignment amplitude in terms of 
\begin{equation}
    A_{\rm IA}\equiv \frac{C_1\rho_{\rm crit}}{(C_1{\rho_{\rm crit}})|_{\rm fid}}.
\end{equation}
Other works (e.g., \citealt{Chisari19}) use the convention that $C_1=5\times 10^{-14} {\rm M_\odot}^{-1} h^{-2} {\rm Mpc}^3$. While the sign is also defined by convention, note that most alignment measurements suggest that galaxies or haloes align their major axis with the minor axis of the tidal field, which also corresponds to the direction in which matter is accreted. This results in a radial alignment of the projected shape towards overdensities -- opposite in sign to the gravitational lensing effect.

\subsubsection{Distinguishing between galaxy populations} 
\label{sec:diffgal}

Galaxies do not all align equally. A property that seems to be relevant to determine the alignment amplitude is colour, although we will see in following sections that colour often acts as a proxy for morphology and formation history. Some works therefore make predictions for intrinsic alignment observables splitting by the galaxy population. For example, \citet{FortunaHM} constructs the LA prediction on large scales from a combination of red and blue galaxies, weighted by their relative fractions:
\begin{eqnarray}
    P_{\delta E}^{L,\rm tot}(k,\mu,z)&=&[f^{\rm red}(z)b_{1,I}^{\rm red}(z)+f^{\rm blue}(z)b_{1,I}^{\rm blue}(z)]\frac{(1-\mu^2)}{2}P_L(k,z),\\
    P_{EE}^{L,\rm tot}(k,\mu,z)&=&[f^{\rm red}(z)b_{1,I}^{\rm red}(z)+f^{\rm blue}(z)b_{1,I}^{\rm blue}(z)]^2\frac{(1-\mu^2)^2}{4} P_L(k,z),
\end{eqnarray}
where the superscript ``${\rm tot}$'' refers to the total population. If blue galaxies do not align at the linear level, then the equations simplify to depend only on the fraction of red galaxies and their alignment bias.

\subsection{NLA model} 

Initial attempts to extend the linear alignment model to small scales effectively replaced $P_L(k,z)$ in Eqs. (\ref{eq:LA_deltaE}) and (\ref{eq:LA_EE}) by the non-linear one, $P_{NL}(k,z)$. This was first suggested by \citet{Hirata04} and formally adopted since \citet{Bridle07}. This option was dubbed the ``non-linear alignment model'' (NLA) despite it relying on linear theory. This minor theoretical inconsistency was a compromise to empirically extend the agreement between model and data to smaller scales than possible with LA. It is still widely used in the context of mitigation (Sect.~\ref{sec:miti}), as models with more free parameters to describe quasi- and non-linear scales are effectively not needed in the context of Stage III surveys. Most of the observational constraints we present in Sect.~\ref{sec:obs} are on this model and they will be given in terms of the alignment amplitude, $A_{\rm IA}$. 

\subsection{Effective Field Theory}
\label{sec:eft}

The effective field theory (EFT, \citealt{Mcdonald09,BaumannEFT,Carrasco12}) approach to intrinsic alignments postulates that the shapes of biased tracers in the Universe, averaged over scales larger than some smoothing scale $\Lambda$ can be expanded in a set of basis operators as
\begin{equation}
    \mathcal{I}_{ij}({\bf x},z)=\sum_\mathcal{O}b_\mathcal{O}(z) \mathcal{O}_{ij}({\bf x},z).\label{eq:EFT_expand_IA}
\end{equation}
Each coefficient $b_\mathcal{O}(z)$ multiplying the operators is a free parameter which needs to be found from the data and might be different for different populations of tracers. For shapes, this expansion is presented in \citet{Vlah20,Bakx23}. In the case of number counts, the tracer and the operators are scalar quantities, and a similar expansion is known. 

The expansion in Eq.~(\ref{eq:EFT_expand_IA}) is still missing two types of terms predicted by EFT. First, it should also include spatial derivatives of the operators $\mathcal{O}_{ij}$. Such ``higher derivative" terms are needed because tracer shapes are not perfectly local functions of the operators \citep{Matsubara99,Coles07}. For each derivative, there is a weighing by a power of $R_\star$, the typical spatial extent of the kernel (i.e., the halo). Second, one should also add the relevant stochastic contributions, among which is the typical ``shape noise'' (a combination of the dispersion in intrinsic ellipticities and the Poisson noise in the number of galaxies) considered in galaxy surveys. The functional form of those contributions can also be predicted within the framework of the theory, but not the free amplitude coefficients.

Each operator acts on the density field, which itself can be expanded into different orders as $\delta=\delta^{(1)}+\delta^{(2)}+\delta^{(3)}+...$. The lowest order term that contributes to Eq.~(\ref{eq:EFT_expand_IA}) is the tidal field $K_{ij}$ acting on $\delta^{(1)}$, consistently with Eq.~(\ref{eq:IpropK}). Instead of giving the full expansion of the shape field, which can be found in \citet{Bakx23}, we will only present the power spectra of alignments up to third order as predicted by the EFT and projected onto the sky following the procedure outlined in the previous section. The non-zero two-point power spectra are given by \citep{Vlah21,Kurita22}
\begin{eqnarray}
P_{\delta E}(k,\mu,z)&=&\frac{1}{2}\sqrt{\frac{3}{2}}(1-\mu^2)P_{02}^{(0)}(k,z),\label{eq:PDE}\\
P_{EE}(k,\mu,z)&=&\frac{3}{8}(1-\mu^2)^2P_{22}^{(0)}(k,z)+\frac{\mu^2}{2}(1-\mu^2)P_{22}^{(1)}(k,z)+\nonumber\label{eq:PEE}\\
&&+\frac{1}{8}(1+\mu^2)^2P_{22}^{(2)}(k,z),\\
P_{BB}(k,\mu,z)&=&\frac{1}{2}(1-\mu^2)P_{22}^{(1)}(k,z)+\frac{1}{2}\mu^2P_{22}^{(2)}(k,z)\label{eq:PBB},
\end{eqnarray}
where the expressions for the helicity power spectra up to third order can be found in \citet{Bakx23} and they involve integrals over perturbation theory kernels. Up to that order, these combinations of power spectra have $6$ free bias parameters. Two more are needed to describe higher order derivative terms and stochastic terms (which can in practice deviate from shape noise).

\subsection{Tidal alignment-Tidal torquing (TATT) model}
\label{sec:tatt}

The TATT model \citep{Blazek19} is a precursor to the EFT of IA based on standard perturbation theory (SPT) that incorporated linear, quadratic and third-order terms contributing to the intrinsic alignment signal. The early works of \citet{Catelan01} and \citet{Hirata04} had already identified two potential quadratic terms that could play a role in galaxy alignments, namely: the torquing of the angular momentum of galaxy by the same tidal field in which it was generated (``tidal torquing'') and the weighting of the linear tidal stretching term by the density field at the location where galaxies are measured (``density-weighting''). 

Consider an incipient proto-galaxy or proto-halo in the large-scale structure. As the object collapses, linear variations in the displacements of fluid elements from the centre of the object in the Euler-frame will result in a rotational motion in the comoving Lagrange-frame. Because displacements are generated by gravity, two nearby collapsing objects will have correlated angular momenta. For discs, this means in practice that their orientation is correlated. On the other hand, density-weighting arises because we only observe shapes at the location of biased tracers. Both of these, plus additional terms, are identified to contribute up to second order in \citet{Blazek15,Blazek19,Schmitz18}. 

There is an extensive literature on the tidal torquing model per se, which is reviewed by \citet{Schafer09} and we will not discuss in detail here. Some of the earliest works on the estimation of intrinsic alignment contamination to weak lensing were performed using this model \citep{Heavens88,Crittenden00}. Nowadays, elliptical galaxies are known to dominate the contamination signal in the redshift range of Stage III surveys. However, it is worth highlighting that if different types of galaxies have more or less sensitivity to this term, there might be advantages in predicting their alignment separately, as done in \citet{Tugendhat18}. Still, theory \citep{Hui08} and numerical simulations \citep{Zjupa22} suggests that spiral galaxies do receive a linear contribution, i.e., their alignment is not purely quadratic.

We can also think of the TATT model as a subset of the EFT of IA. Compared to the EFT of IA, TATT does not include third-order operators in the shape expansion. There are also some practical differences with regards to how this model has been implemented in the literature. First, the EFT of IA reduces to the TATT implementation only if two of the EFT bias parameters are equated  \cite[$b_{2,2}^s=b_{2,3}^s$ in the notation of][]{Bakx23}, and second, the TATT model implemented in observational analyses often relies on replacing the instances of $P_L(k,z)$ appearing in the linear terms with the fully non-linear matter power spectra. This is analogous to the difference between NLA and LA models. The first assumption is possibly problematic for haloes, since it is inconsistent with their Lagrangian prior, i.e., the idea that the LA model holds in Lagrangian coordinates. However, whether this is a problem for galaxies is still unclear.\footnote{Setting $b_{2,2}^s=b_{2,3}^s$ is equivalent to excluding the velocity-shear term, $t_{ij}$. This term has since been included in the TATT expansion.}

As implemented in recent cosmic shear analyses \citep{Secco22}, the Limber-approximated TATT model reads:
\begin{eqnarray}
    P_{\delta E}^{\rm TATT}(k,z)&=&A_1P_{NL}(k,z)+A_{1\delta}P_{0|0E}(k,z)+A_2P_{0|E2}(k,z),\\
    P_{EE}^{\rm TATT}(k,z)&=&A_1^2P_{NL}(k,z)+2A_1A_{1\delta}P_{0|0E}(k,z)+A_{1\delta}^2P_{0E|0E}(k,z)\\
    &&+A_2^2P_{E2|E2}(k,z)+2A_1A_2P_{0|E2}(k,z)+2A_{1\delta}A_2P_{0E|E2}(k,z),\\
    P_{BB}^{\rm TATT}(k,z)&=&A_{1\delta}^2P_{0B|0B}(k,z)+A_2^2P_{B2|B2}(k,z)+2A_{1\delta}A_2P_{0B|B2}(k,z),
\end{eqnarray}
where the $k$-dependent terms involve integrations over perturbation theory kernels and they can be found in \citet{Blazek19} and \citet{Schmitz18}. These can be evaluated using {\tt FAST-PT} \citep{Fang17,McEwen16}, for example. The density-weighting term arises because shapes are measured at the location of biased tracers. Effectively, this means that intrinsic ellipticities are multiplied by a factor $1+b_{\rm TA}\delta$. Generically, $b_{\rm TA}$ can be left free, but in recent TATT implementations, the density-weighting term is assumed to be proportional to $A_{1\delta}=b_{1,g}A_1$, and the linear galaxy bias, $b_{1,g}$, is either fixed or left to vary freely independent of the source clustering signal. Other works, such as \citet{HD22}, leave this parameter to be independent of galaxy bias, i.e. $b_{\rm TA}\neq b_{1,g}$. As exemplified in \citet{Secco22}, the alignment bias parameters $A_1$ and $A_2$ are often parametrized in terms of a fiducial amplitude and power-law redshift dependence. 

\subsection{Lagrangian Effective Field Theory}
\label{sec:lpt}

\citet{Chen24} proposed an alternative expansion of shapes within the framework of Lagrangian perturbation theory (LPT), similar to previous efforts to model galaxy clustering statistics \citep{Desjacques18}. In LPT, the observed shape field evolves from Lagrangian to Eulerian space by having its amplitude rescale through local changes in volume, but without shearing. One therefore assumes that:
\begin{equation}
    \mathcal{I}_{ij}({\bf x},\tau)d^3{\bf x}  = \mathcal{I}_{ij}({\bf q})d^3{\bf q},   
\end{equation}
where ${\bf x}={\bf q}+\psi({\bf q},\tau)$ and $\psi({\bf q},\tau)$ is the trajectory of a fluid element up to time $\tau$. The Lagrangian shape field then results from advecting the shape field in Eulerian space through
\begin{equation}
    \mathcal{I}_{ij}({\bf x},\tau)=\int d^3{\bf q}\delta^{(D)}({\bf x}-{\bf q}-\psi({\bf q},\tau))\mathcal{I}_{ij}({\bf q}).
\end{equation}
Notice that \citet{Chen24} included additional factors in this expression (``active'' advection, their Eq. 2.3) which eventually can be absorbed in the final expansion we will present here. In addition, they defined the shape field with a density-weighting factor as in \citet{Hirata04}, which accounts for the fact that shapes can only be measured at the location of biased tracers. This factor also results in terms that are degenerate with the expansion, though for consistency there are some advantages to making it appear explicitly \citep{Maion24}. 

The LPT expansion is performed in terms of the Lagrangian shear tensor $L_{ij}=\partial_i\psi_j$. At linear level, this is simply related to the density and tidal field ($K_{ij}$) as
\begin{equation}
    L_{ij}^{(1)}({\bf q})=-\frac{1}{3}\delta({\bf q})\delta_{ij}- K_{ij}.
\end{equation}
To one-loop order in perturbation theory,
\begin{equation}
    \mathcal{I}_{ij}[L_{ij}({\bf q})]=A_1 K_{ij}+A_{1\delta}\delta K_{ij}+A_t t_{ij}+A_2 {\rm TF}\{K^2\}_{ij}+A_{\delta t}\delta t_{ij}+A_3{\rm TF}\{L^{(3)}\}_{ij}+\alpha_s\nabla^2K_{ij}
    \label{eq:lptshape}
\end{equation}
plus stochastic terms and where only two cubic-order terms are included (the rest being degenerate). Effective theory considerations allow the authors to ensure the expansion is complete at a given order and to account for counterterms, derivative bias terms, etc. All $A_i$ factors are free bias parameters. We immediately notice the presence of the linear alignment model term, $A_1K_{ij}$. The term with prefactor $A_{1\delta}$ is the density-weighted tidal field and 
\begin{equation}
    t_{ij}({\bf q})=\frac{4}{3}{\rm TF}\{L^{(2)}\}_{ij}
\end{equation}
represents, at leading order, the difference between the second-order matter overdensity and the velocity divergence in Eulerian perturbation theory \citep{Schmitz18}. Terms with $A_1,A_{1\delta},A_2$ are included in the TATT expansion, but note that due to the displacement field including non-linearities, the LPT model cannot be immediately reduced to either TATT or LA by specifying the bias coefficients. The relevant power spectra needed to construct $P_{\delta E}, P_{EE}, P_{BB}$ can be obtained from the cross-spectra between advected operators. We will give a concrete example below.

\subsection{Hybrid effective field theory model (HYMALAIA)}
\label{ss:hymalaia}

The HYMALAIA model proposed by \citet{Maion24} suggests that the impact of advection can be predicted using the fully non-linear displacement field obtained from $N$-body simulations. Compared to \citet{Chen24,Chen24b}, HYMALAIA relies on a subset of the terms present in Eq.~(\ref{eq:lptshape}), 
\begin{equation}
    \mathcal{I}_{ij}({\bf q})\simeq (A_1+A_{1\delta}\delta)K_{ij}({\bf q})+A_2(K\otimes K)_{ij}({\bf q})+\alpha_s\nabla^2K_{ij}({\bf q}),
\end{equation}
plus a stochastic term. The authors chose not to consider the velocity-shear operator $t_{ij}$ because it did not improve the performance of the model. Concretely, the numerical implementation is as follows:
\begin{enumerate}
\item The linearly evolved Lagrangian density field, $\delta_L({\bf q},z)$ is smoothed over a scale of $\Lambda\,h^{-1}\,{\rm Mpc}$ (consistently with the EFT formalism of Sec. \ref{sec:eft}).
\item Using the smoothed density field, the Lagrangian fields $\{K_{ij},(K\otimes K)_{ij},\nabla^2K_{ij}\}$ are computed.
\item The $N$-body simulation is used to obtain the displacement field at a given redshift: $\psi={\bf x}(z)-{\bf q}$.
\item The advected operators are obtained by summing over the values of the fields at the Lagrangian positions within a region in Lagrangian space, $S_p$, which corresponds to the particles that end up at position ${\bf x}$: $\mathcal O_{ij}({\bf x})=\sum_{{\bf q}\epsilon S_{p}}\mathcal{O}_{ij}({\bf q})$.
\end{enumerate}

LPT has also been used to model missing power from large-scale modes in a finite survey volume. Many works have investigated the impact of such ``super-sample'' modes on large-scale structure observables. For intrinsic alignments, useful references that capture this effect using LPT are the works of \citet{SuperSample2} and \citet{SuperSample}.

\subsection{Halo model}

To model intrinsic alignments in fully non-linear scales, the only analytical option is to rely on the ``halo model'' \citep{Cooray02,Asgari23}. (We will discuss numerical options in Sect.~\ref{sec:sims}.) The halo model assumes all matter in the Universe is distributed in spherically symmetric haloes. In its simplest version, this model assumes dark matter is the only matter component present. If we want to derive intrinsic alignment observables from the halo model, we have to have a model for how to populate the dark matter haloes with galaxies. This is known as a halo occupation distribution (HOD) recipe. 

\begin{figure}[ht]
    \centering
\includegraphics[width=0.75\linewidth]{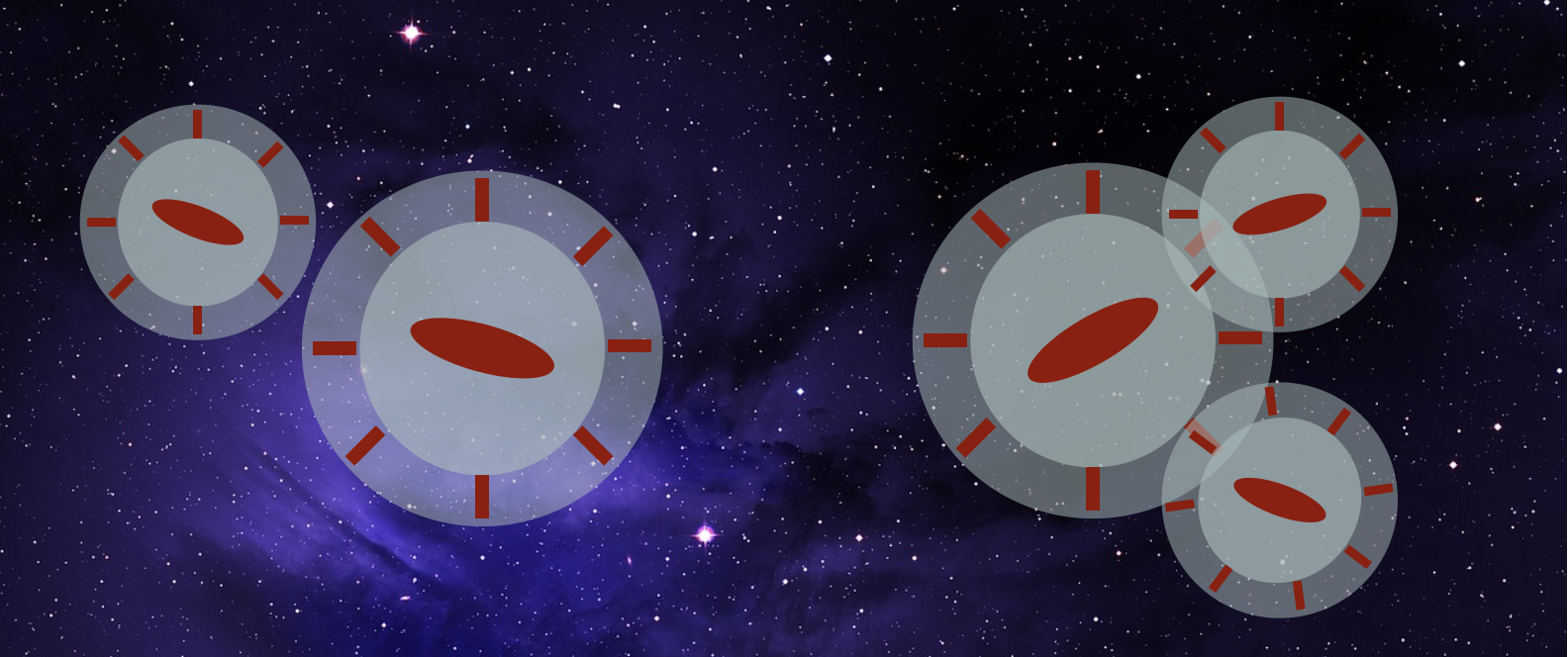}
    \caption{A cartoon representation of the assumptions behind the halo model \citep{Schneider10}. According to the halo model, matter in the Universe is distributed in a collection of spherical haloes. Each halo has a central galaxy whose shape and orientation are aligned pointing towards other haloes, while satellites are sticks pointing radially towards the centre of the halo. Credit: Chisari (2025), CC-BY-NC 4.0. Background image: \url{https://www.needpix.com/photo/1186728/}.}
    \label{fig:cartoonHM}
\end{figure}

For intrinsic alignment predictions, the galaxies that belong to each halo also have to be aligned in a particular way (see Fig.~\ref{fig:cartoonHM}). The first version of such a model was presented by \citet{Schneider10}. Here, central galaxies are aligned towards each other matching the alignment amplitude observed in the linear regime. Within a halo, satellite galaxies are modelled as sticks that point radially towards the centre of the halo, following numerical predictions \citep{Pereira08,Pereira10}. \citet{FortunaHM} revisited the halo model to include some modifications: (i) a possible scale-dependence of the satellite alignment signal within the halo, motivated by observations \citep{Georgiou19}, (ii) a distinction between red and blue galaxy alignments for both centrals and satellites (as in Sect.~\ref{sec:diffgal}), and (iii) a luminosity-dependence of the alignment amplitude of red centrals. We will present this particular formulation of the halo model here. Because the central galaxies are assumed to follow LA or NLA, we are only interested in the formulation of the alignments of satellite galaxies in the one- and two-halo regime.

If $\bar\gamma$ is the length of the stick and $\{r,\theta,\phi\}$ are the spherical coordinates that describe the location of the satellite in the halo, the projected intrinsic shape is given by 
\begin{equation}
   \gamma(r,M,c,z)=\bar\gamma(r,M,c,z)\sin\theta e^{2i\phi}
\end{equation}
where the factor $\sin\theta$ projects the stick along the line of sight and $\bar\gamma(r,M,c,z)$ needs to be informed by simulations or observations. The dependence with concentration, $c$, can be dropped if there is a deterministic relation between mass and concentration. In practice, satellite orientations should be randomized to a certain level, but this is degenerate with the amplitude of the signal. 

The density-weighted intrinsic ellipticity is then
\begin{equation} \gamma_\delta(r,M,z)=\bar\gamma(r,M,z)\sin\theta e^{2i\phi}N_g\,u(r|M,z),
\end{equation}
where $N_g$ is the number of galaxies in the halo and 
$u(r|M,z)$ is the halo density profile, $\rho_{\rm halo}(r|M,z)$, normalized by mass $M$:
\begin{equation}
   u(r|M,z)=\rho_{\rm halo}(r|M,z)/M.
\end{equation}

\begin{figure*}
    \centering
    \includegraphics[width=0.5\linewidth]{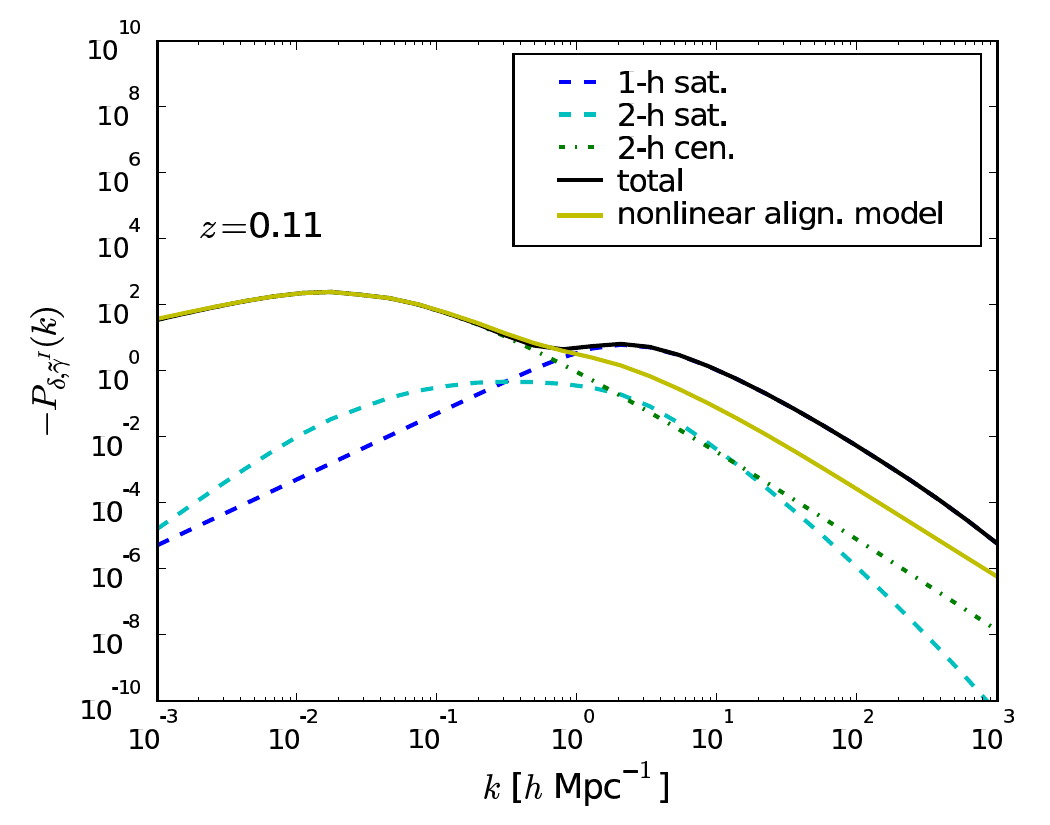}\includegraphics[width=0.5\linewidth]{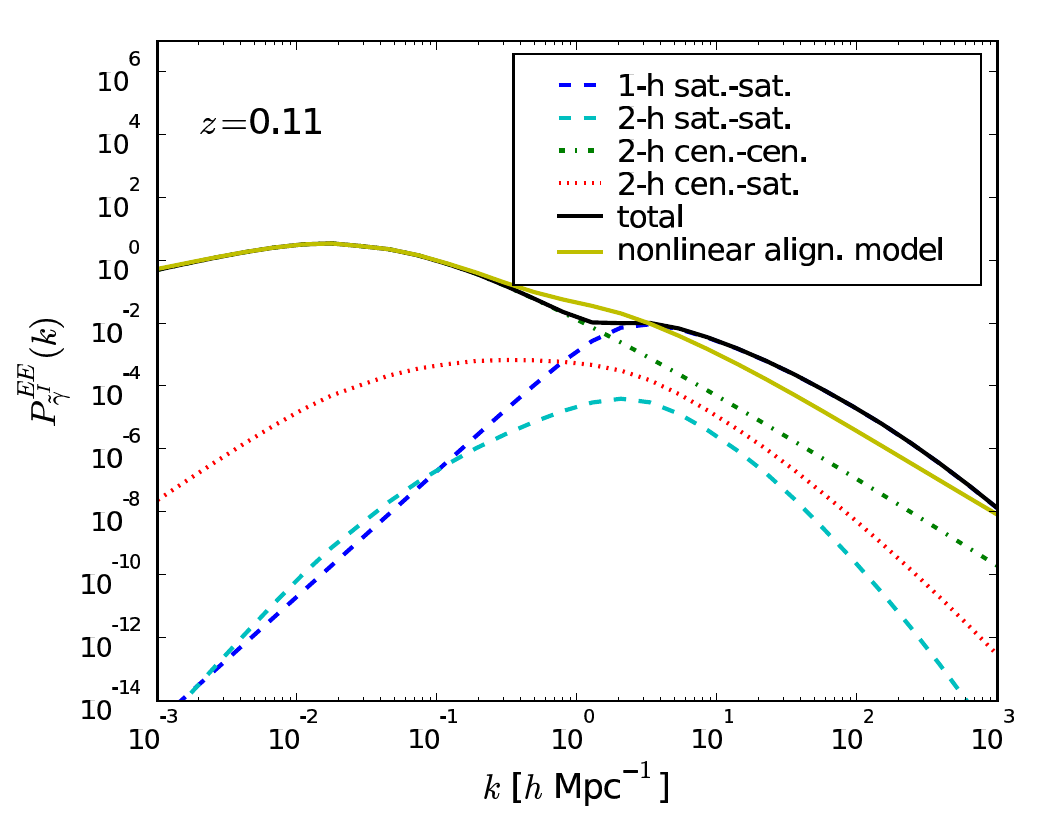}
    \caption{Different contributions to the matter-intrinsic shape power spectrum (left) and intrinsic shape auto-spectrum (right) at $z=0.11$ as predicted by the halo model of intrinsic alignments in the version of \citet{Schneider10}. The yellow line corresponds to NLA for comparison. Deviations from NLA are evidenced at large $k$ (small scales), where the one-halo satellite-satellite term (dark blue dashed) is seen to dominate the signal. Credit: Figure 4 of \citet{Schneider10}. Image reproduced with permission from \citet{Schneider10}, copyright by the author(s).}
    \label{fig:schneider}
\end{figure*}

A continuous density-weighted shape field can be constructed from adding the shapes of galaxies that populate each halo $i$ as:
\begin{equation}
    \gamma_\delta({\bf r},z)=\frac{1}{\bar n_g(z)}\sum_{i}N_{g,i}\,\gamma({\bf r}-{\bf r}_i,M_i,z)\,u({\bf r}-{\bf r}_i|M_i,z),
\end{equation}
where $\bar n_g(z)$ is the number density of galaxies at redshift $z$. The real and complex part of $\gamma_\delta$ are $\gamma_{\delta,1}({\bf r},z)$ and $\gamma_{\delta,2}({\bf r},z)$, respectively, and they can be Fourier-transformed and combined via Eqs. (\ref{eq:gamma_E}) and (\ref{eq:gamma_B}) to yield the $E$ and $B-$modes of the intrinsic ellipticity. 
This gives rise to the following $E-$mode power spectra for alignments of satellites with the density field and with each other in the one-halo regime:
\begin{eqnarray}
    P_{\delta E,\rm 1h}^s&=&\int dM n(M)\frac{M}{\bar\rho}f_s(z)\frac{\langle N_s|M\rangle}{\bar{n_s}(z)}|\tilde\gamma_E^{s}({\bf k}|M)|u(k,M)\\
    P_{EE,\rm 1h}^{ss}&=&\int dM n(M)f_s^2(z)\frac{\langle N_s(N_s-1)|M\rangle}{\bar{n_s}^2(z)}|\tilde\gamma_E^{s}({\bf k}|M)|^2
\end{eqnarray}
where $n(M)$ is the halo mass function; $f_s(z)$ is the fraction of satellites at a given redshift; $\langle N_s|M\rangle$ is the halo occupation distribution of the satellites; and
\begin{equation}
    \tilde\gamma_E^{s}({\bf k}|M)=\int d^3{\bf r}\, \gamma({\bf r},M)\,u(r,M) e^{i{\bf k\cdot r}}.
\end{equation}
Satellites dominate the contribution to both alignment power spectra in the one-halo regime (see Fig.~\ref{fig:schneider}) over a still present LA contribution from centrals. Similarly, satellites still contribute in the two-halo regime, but in \citet{FortunaHM}, those contributions are neglected. Full expressions for the two-halo contributions from central-satellite correlations and satellite-satellite correlations can be found in \citet{Schneider10}.

The halo model formalism relies on describing the matter field as contained in spherically symmetric haloes. Therefore, the portion of the alignment signal coming from the preferential orientation of either central or satellites with the anisotropic distribution of satellites inside an ellipsoidal halo would not be captured. There is evidence from numerical simulations \cite[e.g.,][]{Faltenbacher07,Shao16,Welker18} that those terms impact intrinsic alignment observables by up to 40\% \citep{Samuroff20} and could bias $S_8-\Omega_{\rm m}$ constraints by $1.5\sigma$ in Stage IV surveys. However, generalizing the halo model to predict alignments in ellipsoids is non-trivial \citep{Smith05}.

\subsection{From power spectra to observables on the sky}

We present here alternative observables of intrinsic alignments. Multipole moments of the alignment power spectra are defined as \cite[e.g.,][]{Kurita21}
\begin{equation}
    P_{XY}^{(\ell)}(k) \equiv \frac{2\ell+1}{2}\int_{-1}^1d\mu \mathcal{L}_{\ell}(\mu) P_{XY}(k,\mu) 
    \label{eq:Fourier_multipoles}
\end{equation}
where $X,Y$ are possible fields to be cross-correlated and $\mathcal{L}_{\ell}$ are Legendre polynomials. \citet{Okumura20b} and \citet{Vlah21} point out that due to the projection of the shapes being independent from the alignment model, the multipoles are expected to satisfy certain ratios. For example: $P_{\delta E}^{(2)}/P_{\delta E}^{(0)}=-1$. 

Angular power spectra are obtained by integrating the intrinsic alignment power spectra over the line of sight with the appropriate kernels \citep{Vlah21}:
\begin{eqnarray}
    C_{gE}(\ell)&=&\int d\chi\frac{W_{g}(\chi)W_{\gamma}(\chi)}{\chi^2}P_{gE}(k=\ell/\chi,\mu=0,z=z(\chi)), \\
    C_{EE}(\ell)&=&\int d\chi\frac{W_{\gamma}^2(\chi)}{\chi^2}P_{EE}(k=\ell/\chi,\mu=0,z=z(\chi)),\\
    C_{BB}(\ell)&=&\int d\chi\frac{W_{\gamma}^2(\chi)}{\chi^2}P_{BB}(k=\ell/\chi,\mu=0,z=z(\chi)),
\end{eqnarray}
where $W_{g}=dN_g/d\chi$ and $W_{\gamma}=dN_\gamma/d\chi$ are the normalized distributions of comoving distances of the galaxies used as position and shape tracers, respectively. For example, in the EFT of IA, the power spectra in the integrands would be replaced by Eqs. (\ref{eq:PEE}) and (\ref{eq:PBB}) in the second and third equations, and by $P_{gE}(k,\mu,z)$ in the first one. These expressions are only valid in the flat-sky and Limber approximation - for more generality, see Appendix B of \citet{Vlah21}.

More common are statistics of intrinsic alignments in real space. The conversion from Fourier power spectra to real space correlation functions, $\xi_{XY}({\bf r})$, is given by
\begin{eqnarray}
    \xi_{XY}({\bf r})=\int d^3{\bf k} \,e^{i{\bf k \cdot r}}\,P_{XY}(k,\mu),
\end{eqnarray}
where the radial coordinate is often expressed in cylindrical coordinates ${\bf r}=(r_p,\theta,\Pi)$ and often the angular dependence on $\theta$ is integrated over. 

Projecting over the line of sight increases signal-to-noise ratio and thus the most common real-space statistic adopted is the projected correlation function. For example, for the $+$ component,
\begin{equation}
    w_{g+}(r_p)=\int_{-\Pi_{\max}}^{\Pi_{\max}} d\Pi\, \xi_{g+}(r_p,\Pi)
    \label{eq:wg+}
\end{equation}
where $\Pi_{\max}$ is the maximum projection baseline in front and behind the position tracer.  

\citet{Kurita22} recently advocated for replacing the line-of-sight projection by real-space multipoles. In the notation of \citet{SinghMultipoles}, these are defined as
\begin{equation}
    \xi_{ab}^{l,s_{ab}}(r)=\frac{2l+1}{2}\frac{(l-s_{ab})!}{(l+s_{ab})!}\int d\mu_r L^{l,s_{ab}}(\mu_r)\xi_{ab}(r,\mu_r)
    \label{eq:multipoles}
\end{equation}
where $a,b$ refers to the different tracers (relevant to our problem: positions, galaxy shapes $+$ or $\times$), $s_{ab}$ is the spin of the tracer ($0$ for positions, $2$ for shapes), $L^{l,s_{ab}}$ are associated Legendre polynomials and $\mu_r=\Pi/r$ is the cosine of the angle between the separation vector the line-of-sight direction. Multipoles capture the geometrical effects of the projection more optimally than projected correlations and can therefore yield higher signal-to-noise ratio than projected correlations. There is also a direct transformation between multipoles and the projected correlation function \citep{Baldauf10}:
\begin{equation}
    w_{ab}(r_p)=\sum_l 2\int_0^{\Pi_{\max}}d\Pi \,\xi_{ab}^{l,s_{ab}}(r)L^{l,s_{ab}}\left(\frac{\Pi}{r}\right),
\end{equation}
where the loss in signal-to-noise ratio results from cropping the multipoles at the same $\Pi_{\max}$ regardless of the value of $r=(r_p^2+\Pi^2)^{1/2}$. 

\subsection{RSD and peculiar velocities}
\label{sec:RSD}

Observations give us access to the angular coordinates of galaxies on the sky and their redshift. Redshift does not directly translate into the coordinate along the line of sight: the peculiar velocity of a galaxy leaves an imprint on the redshift via the Doppler effect. In other words, the redshift-space coordinates of a galaxy are distorted by
\begin{equation}
{\bf s} = {\bf x}+\frac{v_z({\bf x})}{aH(a)}\hat{\bf z},
\end{equation}
where $\hat{\bf z}$ is the direction of the line of sight in the plane-parallel approximation. This phenomenon is dubbed ``redshift space distortions'' (RSDs).

On linear scales, RSDs are known as the ``Kaiser effect'' \citep{Kaiser87,Hamilton92}. The impact of RSD on the galaxy density field is well-understood. Galaxy shapes are not affected by RSD at linear order. Their shape in redshift space is the same as in configuration space: $\gamma_{+/\times}({\bf s})=\gamma_{+/\times}({\bf x})$ \citep{SinghBOSS}. This implies that intrinsic shape auto-correlations will not be affected by RSD at leading order, but cross-correlations with density tracers are modified according to the Kaiser factor:
\begin{equation}
P_{gE}^{(s)}({\bf k})=\left(1+\frac{f}{b_{1,g}}\mu^2\right)P_{gE}^{(r)}({\bf k})
\end{equation}
where $f\equiv d\ln D/d\ln a$ is the logarithmic growth rate. Thus, we can conclude that at linear order, only position-shape correlations are modified when going from configuration to redshift space. Interestingly, this suggests that $w_{g+}$, na\"ively projected along the line of sight, is not an optimal estimator for position-shape alignments. \citet{Lamman25} proposed to let $\Pi_{\max}$ vary with transverse separation instead. They also suggested that weighting based on the geometry of shape projection and RSD yields further improvements in signal-to-noise ratio. 

Following this rationale, RSD will only affect intrinsic alignments at higher orders. A simple example is the impact of RSD via density-weighting \citep{Okumura20b,Kurita21}. \citet{Taruya24} analysed the impact of the Kaiser effect in the modelling of the position-intrinsic alignments power spectrum to next-to-leading order. In addition, they provide a prescription for treating ``Fingers-of-God'' \citep{Scoccimarro04}, the RSD effect arising from the random peculiar motions of galaxies inside the deep potential wells of bound structures, such as galaxy clusters. Such motions are known to contribute significantly even at scales of hundreds of Mpc. \citet{Okumura24} expanded on this work by providing expressions for position-shape and shape-shape correlation function multipoles in configuration space with RSDs, expanded on a basis of standard and associated Legendre polynomials, but staying at linear order. A complete treatment of perturbative alignment models with RSDs is not yet available but can be constructed from these references. 

If we go beyond the flat-sky approximation, RSD have an imprint on the dipole of the position-intrinsic shape \citet{Saga23}. In such a case, the $g+$ dipole (Eq. \ref{eq:multipoles}) also picks up a correction from RSD that can be summarized as a transformation: $b_{1,g}A_{\rm IA}\rightarrow (b_{1,g}+f)A_{\rm IA}$.

\subsection{Theoretical priors on the alignment amplitude}
\label{sec:priors}

There have been a couple of (beautifully done) attempts at estimating the alignment amplitude directly from theory. In \citet{Camelio15}, the authors modelled a galaxy using the phase-space distribution function of its stars. They applied an instantaneous tidal field to this distribution function and found the response of the quadrupolar moment of the stellar matter distribution to the tidal field. This gives a direct estimate of $A_{\rm IA}$ (see their Eq. 34). They concluded that if the external tidal field is generated by a nearby filament or dark matter halo, the resulting $A_{\rm IA}$ is too low compared to observations. 

\citet{Ghosh24} performed a similar calculation for both elliptical and spiral galaxies. Here, they again estimated the quadrupolar distortion resulting from the application of an external tidal field. Their results can be phrased in terms of the time that it takes an elliptical or spiral galaxy to respond to a tidal field. The strength of this tidal field can also be associated to a typical timescale, and the ellipticity distortion induced is the (square) ratio between the timescale of stellar orbits and that of the tidal field. The authors confirmed their results with numerical simulations of the orbits of stars in an external tidal field. They also provided predictions for how the alignment amplitudes of ellipticals and spirals differ and depend on their properties. The authors do not give an estimate of $A_{\rm IA}$, though they remark that shape distortions would be small based on their scaling arguments. Compared to \citet{Camelio15}, the step of accounting for a realistic tidal field is missing here.

The calculations of \citet{Camelio15} essentially rule out the hypothesis that an instantaneous response to the tidal field is responsible for galaxy alignments. Instead, alignments must either be of primordial origin or build up significantly over time, whether by the cumulative effect of the tidal field on galaxies, or because accretion of material is along preferential directions that correlate with the tidal field as well. Based on these works, it is not possible to distinguish between these options. Still, the analyses of \citet{Camelio15} and \citet{Ghosh24} could be extended to account for the cumulative effect of tidal stretching over time. The impact of accretion and mergers would still be missing in such a scenario. 

\section{Observations}\label{sec:obs}

We will review here direct constraints on intrinsic alignments coming from observations. Indirect constraints derived in the cosmological analyses of cosmic shear surveys will be presented in Sect.~\ref{sec:miti}.

Intrinsic alignments for a given sample of objects can be measured directly from survey data by determining statistics related to the preferential orientation of one object with respect to another. In practice, certain conditions must be met for a robust detection:
\begin{enumerate}
\item There should be a sufficient number of pairs in the volume considered. In other words, the signal should be strong enough to overcome the noise imprinted by the intrinsic dispersion of shapes (``shape noise'') and cosmic variance. 
\item The pairs should not be separated by too large a distance along the line of sight. For example, \citet{Mandelbaum06} expects that integrating $w_{g+}$ over $30h^{-1}$~ Mpc$<\Pi<90 h^{-1}$ Mpc should give a null signal, and adopts this as a consistency check for systematics. The specific range depends on the covariance of the data.
\item Objects should have precise redshifts. Measurements with photometric redshifts are possible, but only for specific samples where the redshift error is small -- typically below a few per cent \citep{MegaZ,Fortuna24,Johnston21}.
\end{enumerate}

In addition, the signal might be contaminated by gravitational lensing. For position-shape correlations, contamination will be present if the redshift is high. In this case, position-shape is contaminated by a magnification-shear correlation \citep{Joachimi10}. Magnification is a change in the flux of objects due to gravitational lensing that effectively changes the number of objects observed at above some flux limit threshold. Similarly to shear, magnification is sourced and correlated with the intervening large-scale structure. If the distance along the line of sight over which the position-shape correlation is projected is large, one might obtain a position-shear contamination, or a magnification-intrinsic shape contamination. For shape-shape correlations, contamination from cosmic shear will arise at high redshift, or if the projection baseline is large. In such cases, lensing should be included in the modelling.

\subsection{Shape measurements}

The shape measurements used in intrinsic alignment studies are often inherited from weak lensing shape measurements and carry the relevant corrections for telescope optics and smearing of the point-spread function (PSF). We will see some clear examples of this below when discussing measurements of intrinsic alignments from the Sloan Digital Sky Survey (SDSS). The advantage of using weak lensing shape measurements is that any prior on intrinsic alignment can be directly translated to estimates of the contamination to weak lensing. The disadvantage is that weak lensing shape estimators might not give the best signal-to-noise ratio in the alignment signal. If the goal is to extract information about what causes galaxies to align, it might as well be that other shape measurement methods are better suited. Examples include shapes derived from profile fitting or isophotes \citep{SinghShapes} or moment-based methods \citep{Georgiou19,Georgiou25}.

\subsection{Estimators}
\label{sec:estimators}

The alignments of galaxies and other cosmic structures is most often estimated from fits to two-point functions, either in real or Fourier space, of shape-position correlations. Shape-shape correlations are not so frequently used due to their lower signal-to-noise ratio \citep{Blazek11} or because it is easier for them to be contaminated by systematics in shape auto-correlations, both in observations and in simulations\footnote{Systematics associated with the PSF are more likely to drop out in cross-correlation with galaxy positions.}. Still, with the increasing signal-to-noise ratio from Stage IV surveys, shape-shape correlations might start playing a role and we will therefore introduce them here as well. For statistics beyond two-point, see Sect.~\ref{sec:beyond}.

To describe the projected shape of an object, we need two numbers: an estimate of the ellipticity and the orientation angle of the major axis with respect to some reference system. If $f({\bf x})$ is the observed flux of an image in two-dimensional Cartesian coordinates, the weighted quadrupole moments of the image are
\begin{equation}
    q_{ij}=\frac{1}{F_0}\int d^2{\bf x}\,x_ix_j \,W({\bf x})f({\bf x})
\end{equation}
where $W({\bf x})$ is some weight function, $F_0=\int d^2{\bf x}\, \,W({\bf x})f({\bf x})$ is the total weighted flux and ${\bf x}$ is defined with respect to the center of the object. The ``polarisation'' of the image is given by $e$, whose components are \citep{Blandford91}:
\begin{equation}
    e_1=\frac{q_{11}-q_{22}}{q_{11}+q_{22}}; e_2=\frac{2q_{12}}{q_{11}+q_{22}};
\end{equation}
or in terms of the polar orientation angle of the major axis, $\beta$,
\begin{equation}
(e_1,e_2)=e(\cos2\beta,\sin2\beta),
\label{eq:e1e2}
\end{equation}
where $e=\sqrt{e_1^2+e_2^2}$. This is equivalent to defining $e\equiv(1-q^2)/(1+q^2)$, where $q=b/a<1$ is the axis ratio of the galaxy. There is no single way of estimating an ellipticity. For example, another commonly used estimator is $\epsilon\equiv(1-q)/(1+q)$.

\begin{figure}[ht]
    \centering
    \includegraphics[width=0.5\linewidth]{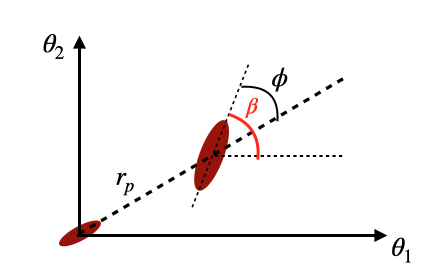}
    \caption{Two galaxies on the sky whose orientation is measured with respect to the angular system of coordinates on the sky ($\beta$) or the separation vector ($\phi$). Credit: Chisari (2025), CC-BY-NC 4.0.}
    \label{fig:rotation}
\end{figure}

In measuring the alignment with respect to another object, the coordinate system is rotated so that the angle of the major axis is measured relative to the separation vector to the other object, see Fig.~\ref{fig:rotation}. We can then work with the ellipticity components of the object, 
\begin{equation}
(e_+,e_\times)=\hat e(\cos2\phi,\sin2\phi),
\end{equation}
where $\hat e$ is the ellipticity estimator of choice. These components can be directly connected to theory predictions. 

If we are interested in obtaining priors on alignment amplitude for lensing contamination, the shape measurement method should match the one applied in weak lensing studies. Depending on which algorithm is used, one sometimes needs to correct the ellipticities with a ``responsivity'' factor to obtain an unbiased estimate of the shear \citep{BJ02}. While $\epsilon$ is an unbiased estimator of the shear, the components of the polarisation $e$ are related to those of the shear by \citet{Mandelbaum14}
\begin{equation}
    (\gamma_+,\gamma_\times)=(e_+,e_{\times})/(2\mathcal R),
\end{equation}
where $\mathcal R=1-\langle \sigma_e^2\rangle$, $\langle \sigma_e^2\rangle$ is the ellipticity dispersion per component. (Equivalent expressions hold for $\gamma_1$ and $\gamma_2$ in analogy to $e_1$ and $e_2$.)
In \citet{Georgiou19,SinghShapes}, the impact of different shape estimators on intrinsic alignment measurements are compared. We will come back to this in Sect.~\ref{sec:othermeasures}. 

The correlation function of position-shape alignments is constructed similar to the Landy-Szalay estimator for galaxy clustering \citep{LS93,Mandelbaum06}. It is measured as a function of comoving projected separation on the sky, $r_p$, and comoving line-of-sight distance, $\Pi$, through:
\begin{equation}
    \xi_{gS}(r_p,\Pi) = \frac{SD-SR_D}{R_DR_S}
    \label{eq:xig+_LS}
\end{equation}
where $S$ represents the objects with measured shapes (either $+$ or $\times$); $D$ are the density tracers, i.e. the position sample and $R_X$ are sets of random points. $\xi_{g\times}$ is expected to be null but it is often used for cross-checking the contribution of systematics is low. Random catalogues are often over-sampled compared to the $S$ and $D$ data sets (by factors of $10\times -20\times$), and normalization corrections are therefore necessary. For example, the specific combinations appearing in Eq.~(\ref{eq:xig+_LS}) used to construct $\xi_g+$ are:
\begin{eqnarray}
    S_+D&=&\sum_{i\in S,j\in D}\gamma_+(i|j),\\
    S_+R_D&=&\sum_{i\in S,j\in R_D}\gamma_+(i|j),\\
    R_SR_D&=&\sum_{i\in R_S,j\in R_D}1.
\end{eqnarray}
where $\gamma_+(i|j)$ represents the intrinsic $+$ shape of galaxy $i$ relative to the separation vector to galaxy $j$. While the term $S_+R_D$ should average to zero, including it effectively reduces the size of the error bars \citep{Singh17}. Whether the sign of $\xi_{g+}$ corresponds to radial or tangential alignment is fixed by convention.

One can analogously estimate shape-shape auto-correlations:
\begin{equation}
    \xi_{++}(r_p,\Pi) = \frac{S_+S_+}{R_SR_S};
 \,\,\,\,\xi_{+\times}(r_p,\Pi) = \frac{S_+S_\times}{R_SR_S}; \,\,\,\,\xi_{\times\times}(r_p,\Pi) = \frac{S_\times S_\times}{R_SR_S}.
    \label{eq:xiss}
\end{equation}
$\xi_{+\times}$ is expected to vanish due to parity \citep{Blazek11}.

These three-dimensional correlation functions can be measured directly, as in \citet{SinghShapes} or projected over the line of sight to yield an estimator of $w_{g+}$ (Eq. \ref{eq:wg+}), $w_{++}$ or $w_{\times\times}$. $\Pi_{\max}$ is usually between $60$ $h^{-1}$ Mpc and $200$ $h^{-1}$ Mpc to capture most of the signal. In practice, the integral is performed by direct summation over the $\Pi$ bins in which $\xi_{g+}$ is estimated. Increasing the integration range beyond $200$ $h^{-1}$ Mpc can be counterproductive because pairs separated by such long line-of-sight distances are less aligned with each other. This is however not true when the redshifts of the objects are poorly known. We discuss this case in Sect.~\ref{sec:photo}. Similarly to the $w_{g+}$ case, the integral in Eq.~(\ref{eq:multipoles}) must be performed numerically relying on the $(r,\mu)$ bins available. Optionally, to remove non-linear scales from the measurement, \citet{SinghMultipoles} applies a scale cut before performing the integration.

$w_{g\times}$ and $w_{+\times}$ are expected to be null unless the Universe breaks parity (see Sect.~\ref{sec:cosmo}) and they are used to check for the robustness of the measurements against observational systematics. 
In addition, galaxies separated by large $\Pi$ values are not expected to show a significant alignment signal. Therefore, another consistency check is to verify that integrating $w_{g+}$ only over large $\Pi$ gives a null result \cite[e.g.,][]{Mandelbaum06}. 

\citet{Kurita22} devised a Fourier space estimator for intrinsic alignments that was applied in \citet{Kurita23}. Galaxies and random points are placed on a grid using cloud-in-cell interpolation \citep{CIC} and weighted optimally using FKP weights \cite[$w_{\rm FKP,g}$ and $w_{\rm FKP,\gamma}$,][]{FKP}:
\begin{eqnarray}
F_{\rm g}({\bf x})&=& w_{\rm FKP,g}({\bf x})[n_{\rm g} ({\bf x}) - \alpha_{\rm g} n_{\rm r,g} ({\bf x})] \\
F_\gamma({\bf x})&=& w_{\rm FKP,\gamma}({\bf x})n_{\gamma} ({\bf x}) \gamma ({\bf x})
\end{eqnarray}
where $\gamma=\gamma_1+i\gamma_2$, $\gamma_1$ and $\gamma_2$ are the shear estimates derived from the ellipticity measurement, $n_{\rm g}$ is the number density of galaxies, $n_{\rm r,g}$ is the number density of random points, $n_{\gamma}$ is the number density of galaxies with measured shapes, and $\alpha_{\rm g}$ is the ratio of the number count of galaxies to number of random points. Note that the construction of $\{n_{\rm g},n_{\gamma}\}$ might involve additional weights depending on the selection effects of the survey (e.g., to account for fibre collisions). The estimator for the position-shape power spectrum multipoles is then given by
\begin{equation}
    \hat P_{g\gamma}^{(L)}(k_b)=\frac{2L+1}{I_{g\gamma}}\frac{(L-2)!}{(L+2)!}\int_{\bf \hat k_b}\hat F_g^{(L)}(-{\bf k})\hat F_\gamma^{(L)}({\bf k})
\end{equation}
where $I_{g\gamma}$ is a normalization constant that depends on the window functions of the survey(s), 
\begin{equation}
    I_{g\gamma}=\int d^3{\bf x} w_\gamma({\bf x})\bar n_\gamma({\bf x})
    w_g({\bf x})\bar n_g({\bf x}),
\end{equation}
where $w_\gamma$ and $w_{\rm g}$ are masks that account for incompleteness of the survey or weights while $\bar n_\gamma({\bf x})$ and $\bar n_{\rm g}({\bf x})$ are the mean number densities of shape and clustering tracers, respectively. The multipole of the shape field is defined as
\begin{equation}
    \hat F_\gamma^{(L)}({\bf k})\equiv \int d^3{\bf x}\, F_\gamma({\bf x})e^{-2i\phi} e^{-i{\bf k\cdot x}}\mathcal L_L^{m=2}({\bf \hat k\cdot\hat x}),
\end{equation}
$\phi$ is the angle that rotates the shape field on the plane perpendicular to the line-of-sight direction (i.e., it results in $E$- and $B$-modes as in Eqs. \ref{eq:gamma_E} and \ref{eq:gamma_B}), and $\mathcal L_L^{m=2}(\mu)$ is the associated Legendre polynomial. This is different from the clustering kernel case, where:
\begin{equation}
    F_g^{(\ell)}({\bf k})\equiv \int d^3{\bf x} \,F_g({\bf x})e^{-i{\bf k\cdot x}}\mathcal L_\ell({\bf \hat k\cdot\hat x}),
\end{equation}
and $\mathcal L_\ell$ are Legendre polynomials instead of associated ones.
The associated Legendre polynomial is needed to project the tensorial shape field onto the line of sight. In addition to the estimator for the power spectrum multipole measurement, \citet{Kurita22} provide estimators for its covariance.

\subsection{Constraints from spectroscopic samples}
\label{sec:spec}

When spectroscopic redshifts are available, it becomes easier to isolate objects that are close to each other. This improves the signal-to-noise ratio of the intrinsic alignment measurement, and it mitigates the impact of lensing. The first modern spectroscopic measurements of galaxy alignments were performed on luminous red galaxy (LRG) samples.

\citet{Mandelbaum06} measured the shape-shape and position-shape alignments of $\simeq 265,000$ galaxies with spectroscopic redshifts in the SDSS \citep{York00} with the goal of establishing the degree of contamination to cosmic shear. Previous studies attempting to establish the impact of alignments for cosmic shear contamination had been performed using the SuperCOSMOS \citep{Brown02} and COMBO-17 \citep{Heymans04} data sets, i.e. photometric samples, which we will discuss in Sect.~\ref{sec:photo}. The work by \citet{Mandelbaum06} instead allowed for cleanly separating the intrinsic alignments of galaxies from their lensing by isolating close pairs, thus deriving more stringent constraints on alignment contamination. The measurements spanned the redshift range $z\sim 0.05-0.2$ and an absolute $r-$band magnitude range $-19< M_r+5\log h\leq -23$, using weak lensing {\sc RegLens} shapes from \citet{Hirata03,Mandelbaum05a}.

To fit the $w_{g+}$ measurements, \citet{Mandelbaum06} used a power-law as well as empirical fits from mocks based on the angular momentum alignments between galaxy and halo (dubbed HRH$^*$ model, based on \citet{Heavens00}, see Sect.~\ref{sec:mocks}). When modelling $w_{g+}$ as a power-law, they found alignments to be significant for their overall sample and for the two most luminous bins they considered, for which $L>L_*$, where $L_*$ is the characteristic luminosity near the knee of the Schechter function \citep{Schechter76}. In terms of the alignment amplitude defined by HRH$^*$, they found the $w_{++}$ measurements consistent with null, in agreement with and marginally more constraining than previous works \citep{Brown02,Heymans04}. Mandelbaum et al. concluded that matter-shape correlations are likely to be the dominant contaminant to cosmic shear, as opposed to shape-shape. They hypothesized that this contamination arises from the correlation of the alignments of brightest cluster galaxies (BCGs) with clusters \citep{Binggeli82}.

The hypothesis that BCGs are the objects that dominate the matter-intrinsic shape correlation led \citet{Hirata07} to study a broader class of LRGs and compare their alignments to those of blue galaxies. They used a sample of $36,278$ LRGs from the SDSS spectroscopic sample between $0.15<z<0.35$, $7,758$ LRGs from the 2SLAQ survey at $0.4<z<0.8$ \citep{Cannon06}, and the SDSS Main sample from \citet{Mandelbaum06} ($266,000$ galaxies, split by colour) to probe the redshift, colour and luminosity-dependence of the position-shape correlation. For the SDSS Main sample, they found no significant signal of blue galaxy alignments around the galaxies in the full sample. Interestingly, red and blue alignment measurements were consistent with each other in the highest luminosity bin, which they attributed to the fact that their luminous blue sample may actually be on the edge of being red. For the SDSS LRG samples, they found a significant alignment out to $60\,h^{-1}$ Mpc. An attempt to separate brightest (central) galaxies in the sample yielded consistent results and is probably driven by the fact that $83\%$ of the galaxies are classified as centrals. The redshift evolution was also not significant for this volume-limited sample and there was a tendency for more luminous samples to align more strongly. The 2SLAQ LRG sample
is fainter and bluer than the fiducial spectroscopic LRG
sample from SDSS, and while the signal is of low significance, this still helps in constrain the redshift evolution. Joint power-law fits to the alignment signal performed by \citet{Hirata07} on the LRG samples suggested the following power-law indices with projected radial separation, luminosity and redshift, respectively: $\alpha \sim -0.9$, $\beta \sim 1.5$, $\gamma \sim -1$. $\alpha$ was consistent within error bars with the expected power-law index from the NLA model. 

\begin{figure}[ht]
    \centering
    \includegraphics[width=0.45\linewidth]{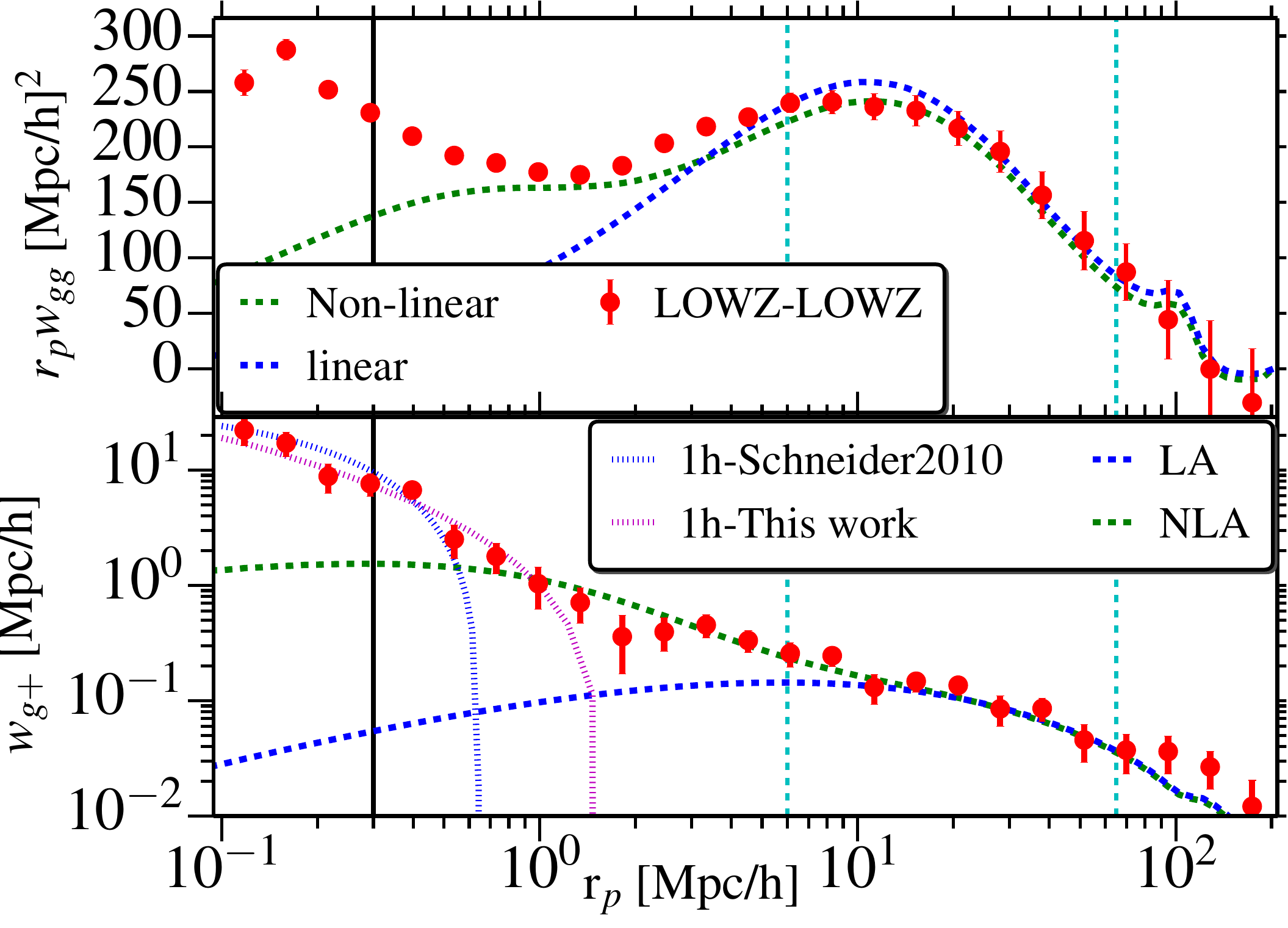}
    \includegraphics[width=0.45\linewidth]{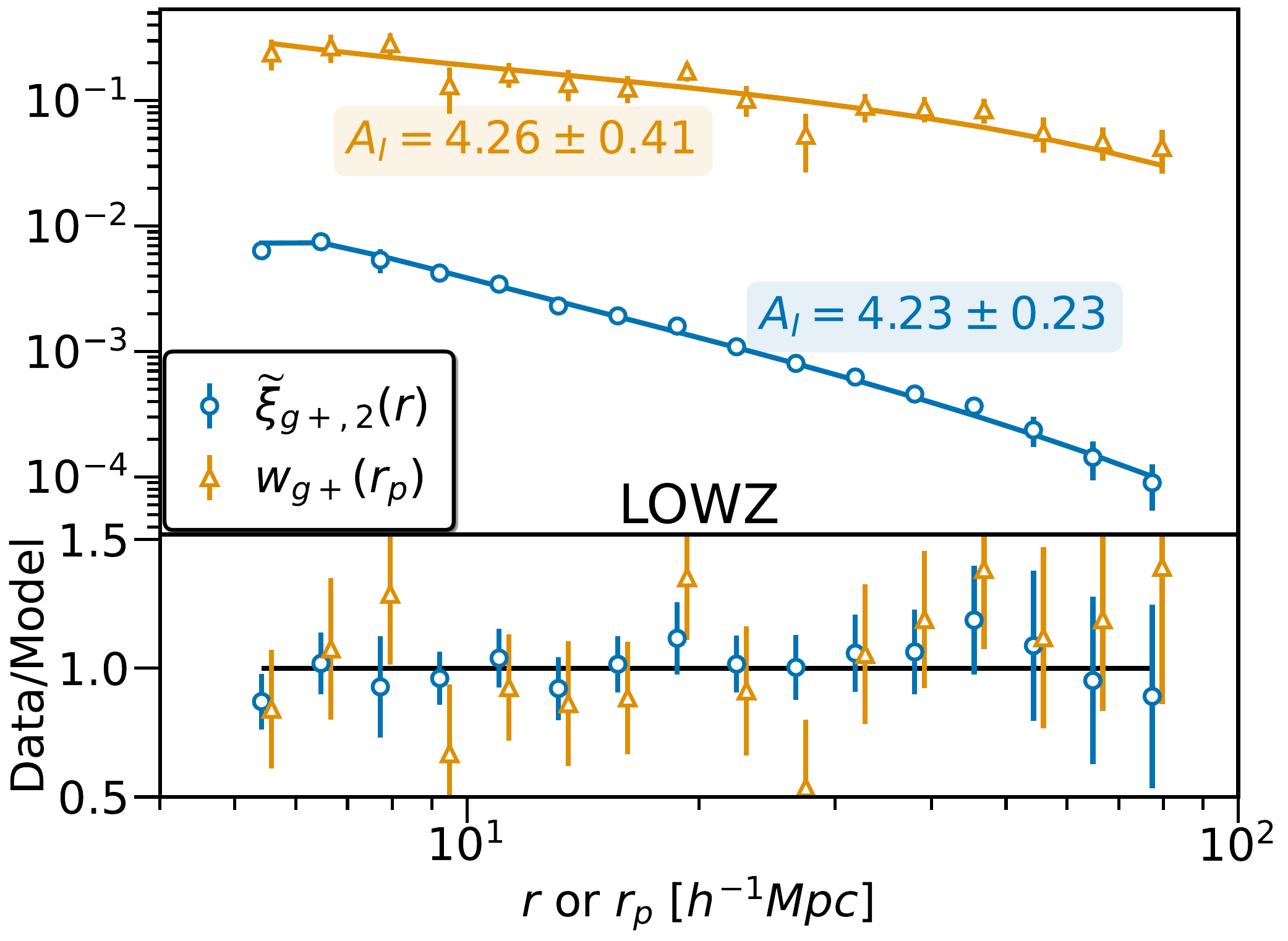}
    \caption{Intrinsic alignments of the BOSS LOWZ sample. Left: Measurement of $w_{gg}$ and $w_{g+}$ for BOSS LOWZ galaxies as a function of their comoving projected separation \citep{SinghBOSS}. The adopted convention means a positive $w_{g+}$ corresponds to a radial alignment. The LA model fit is shown in blue; the NLA fit is shown in green and the halo model fit, in pink. Progressively, more model complexity is needed to describe the small scales. Right: The projected correlation function of LOWZ alignments and the quadrupole of the same sample \citep{SinghMultipoles}. The NLA model fits have consistent $A_{\rm IA}$ amplitudes. Left panel image reproduced with permission from \citet{SinghBOSS}, copyright by the author(s). Right panel image reproduced with permission from \citet{SinghMultipoles}, copyright by the author(s)}
    \label{fig:singh}
\end{figure}

\citet {SinghBOSS} used the BOSS LOWZ sample, in the redshift range of $0.16<z<0.36$, to constrain alignments of LRGs at lower luminosities than previous works. Once again, they found a very significant alignment signal (Fig.~\ref{fig:singh}), corresponding to $A_{\rm IA} \sim 4$ and a power-law dependence of the strength of alignment on luminosity. Their measurements were complemented by galaxy-galaxy lensing measurements \citep{Hoekstra05,Mandelbaum05a}, which allowed them to constrain the mass of the host halo and to show that the alignment amplitude of luminous red galaxies has a power-law dependence with halo mass. 

\citet{Hirata07} had also suggested that while LRGs align strongly, blue galaxies do not. These observations were revisited in 
\citet{Mandelbaum11} using the WiggleZ Dark Energy Survey, a spectroscopic survey targeting $300,000$ UV-selected (blue) galaxies at $z\lesssim 1$ over $1,000$ sq. deg. At the time of the analysis, three overlapping regions with SDSS were used to obtain constraints on intrinsic alignments. The total number of WiggleZ galaxies with SDSS shapes was $76,084$. No intrinsic alignment signal was detected, but the work placed stringent constraints on blue galaxy alignments at intermediate redshift ($z\sim 0.6$). With blue galaxies constituting the majority of the sources of a cosmic shear sample, these constraints are particularly useful for putting a prior on intrinsic alignment contamination. Predictions made by the authors suggested that alignments of blue galaxies alone could not cause a shift in $\sigma_8$ larger than the statistical errors of Stage III surveys. Nevertheless, this statement is model dependent, and it is unclear how representative the target selected sample was, compared to a cosmic shear sample. 

To mitigate this, \citet{JohnstonKiDSGAMA} performed a study of blue galaxy alignment in KiDS+GAMA, a flux-limited sample. KiDS is a wide-field galaxy survey optimized for weak lensing in the $r$-band, and carried out at the VST telescope of the European Southern Observatory \citep{deJong13}. It overlaps with the Galaxy and Mass Assembly survey \cite[GAMA,][]{Liske15}, a spectroscopic survey flux-limited to $r<19.8$ (at $98\%$ completeness), over three patches of $60\,\deg^2$ each. \citet{JohnstonKiDSGAMA} complemented this dataset with SDSS spectroscopic data from the Main sample \citep{York00,Strauss02} to deliver tight constraints on blue galaxy alignments. In terms of the NLA model amplitude, this yielded $A_{\rm IA}^{\rm blue}=0.21^{+0.37}_{-0.36}$. Hence, similarly to WiggleZ, there was no detection of a blue galaxy alignment, but the authors demonstrated that the tight priors they were able to place on the alignment amplitude helped reduce the error budget on $S_8$ by $60\%$ for a Stage III survey like KiDS. 

\citet{JohnstonKiDSGAMA} emphasized the importance of implementing such priors in a colour-split cosmic shear analysis, some of which are starting to emerge \citep{Li21,McCullough24}. To date, the uncertainties in blue galaxy alignments remain still sufficiently large that their contamination to lensing cannot be entirely disregarded.
It is also interesting to note that for a flux-limited red sample,  \citet{JohnstonKiDSGAMA} found only a $9\sigma$ detection of red galaxy alignments, consistent with $A_{\rm IA}^{\rm red}=3.18^{+0.47}_{-0.46}$. The amplitude is thus smaller than for more luminous LRGs. 

The redshift evolution of the alignment signal is difficult to probe and interpret. Spectroscopic samples have different target selections at low and high redshifts, which represent different galaxy populations. Flux-limited spectroscopic samples on the other hand are highly complete, but they are typically small and restricted to low redshifts, which limits the possibility of splitting them to probe redshift evolution. In addition, it is difficult to distinguish redshift evolution from  an evolving red or satellite fraction, or from a varying mean halo mass as a function of redshift. What do we then know about intrinsic alignments at higher redshifts? 

The BOSS survey used colour and colour-magnitude cuts to identify and take spectra of massive galaxies in the redshift range of $0.4<z<0.7$. Note that these CMASS galaxies are not specifically red, as the colour-cuts initially imposed in the LOWZ sample analysed in \citet{SinghBOSS} were removed \citep{Reid16}. The alignments of CMASS galaxies were measured by \citet{SamuroffEBOSS} and by \citet{HervasPeters24}. \citet{SamuroffEBOSS} restricted the sample to $0.4<z<0.6$ to avoid overlap with eBOSS LRGs, yielding a total of $\sim 50,000$ galaxies with a mean redshift of $\langle z \rangle=0.52$. Their constraints on the alignment amplitude, obtained using shapes from the DES Year 3 (Y3) catalogue \citep{Gatti21}, were $A_{\rm IA}=2.72\pm0.47$ and are shown in the left panel of Fig.~\ref{fig:G25_AIA}. Constraints were also provided split by luminosity (in three bins) and redshift (in two luminosity bins). Notably, the CMASS sample lies below the power-law luminosity relation evidenced for luminous red galaxy samples in the literature, possibly due to their bluer colour. No significant redshift evolution is evidenced.

\citet{HervasPeters24} combined spectroscopic CMASS data with shapes from the Ultraviolet NearInfrared Optical Northern Survey (UNIONS) data set, which allowed them to measured the CMASS alignments to $13\sigma$ thanks to the increased area of overlap between shape measurement and spectroscopic data (yielding $\sim 200,000$ galaxies). Their measurements returned an alignment amplitude of $A_{\rm IA}=4.02\pm 0.31$ at $68\%$ confidence level. Their own reanalysis of CMASS galaxies with DES Y3 shapes yielded $A_{\rm IA}=3.18\pm 0.64$ at 68\% confidence level. This is slightly higher, though in agreement with \citet{SamuroffEBOSS} at $1.2\sigma$. Note as well that the CMASS-UNIONS sample extends to higher redshifts (up to $z=0.7$). Both CMASS-UNIONS and CMASS-DESY3 samples, split intro three luminosity bins, give evidence of a strong luminosity scaling.

\citet{Kurita23} used CMASS in combination with LOWZ to constrain primordial non-Gaussianity from the alignment signal (see Sect.~\ref{sec:cosmo}), which as a by-product produced a constraint on the alignment amplitude of the CMASS galaxies. The constraints were $A_{\rm IA}=3.66^{+0.92}_{-0.83}$ and $A_{\rm IA}=4.02^{+0.31}_{-0.36}$ for CMASS galaxies in the South and North Galactic caps, respectively, at 68\% confidence level. The result for the Southern Galactic cap sample is consistent with that of \citet{SamuroffEBOSS} and thoseof \citet{HervasPeters24}. It is also slightly smaller than the alignment amplitude for LOWZ galaxies determined by the same authors, which was quoted to be $A_{\rm IA}=4.74^{+0.72}_{-0.67}$ and $A_{\rm IA}=4.97^{+0.30}_{-0.30}$ for Southern and Northern samples, respectively, at the same confidence level. 

At even higher redshifts, constraints on galaxy alignments come from spectroscopic samples of emission line galaxies (ELGs) in eBOSS \citep{Raichoor17}. These are typically strongly star-forming galaxies with significant [OII] emission. Samuroff et al. \citet{SamuroffEBOSS} obtained constraints on the intrinsic alignments of $\sim 93,000$ eBOSS ELGs at $z\sim 0.8$, finding $A_{\rm IA}=0.07^{+0.32}_{-0.42}$ at 68\% confidence level for this sample in $r_p>2\,h^{-1}$ Mpc. The shapes of the galaxies were obtained from DES, which overlaps with eBOSS over $600\deg^2$ in the Southern Galactic cap. Similarly, \citet{HervasPeters24} used eBOSS in combination with UNIONS to constrain ELG alignments. Their results, compatible with null at $1\sigma$ suggested constraints on $A_{\rm IA}=3.0^{+3.3}_{-2.9}$ at 68\% confidence level. 

The first measurement of galaxy alignments in a spectroscopic sample at $z>1$ was presented in \citet{Tonegawa21} using a sample of H$\alpha$ emitters from the FastSound survey, a near-infrared spectroscopic survey of star-forming galaxies at $1.19<z<1.55$ carried out with the FMOS spectrograph on the Subaru Telescope. The total amount of area was limited to $17\,\deg^2$ overlapping with data from the Canada-France-Hawaii Telescope Lensing Survey (CFHTLenS, \citealt{HeymansCFHT}), which provides the shape measurements that complement the spectroscopic sample. One of the challenges of this measurement is the need to cross-match the two samples, which results in a significant level of contamination from redshift interlopers and significantly reduces the final sample to only $1,158$ galaxies. In addition, this high-redshift sample is poorly resolved by CFHTLenS, degrading the quality of the shapes. Overall, this is a challenging measurement but it has a large pay-off, as constraints on intrinsic alignments at such high redshift do not exist. The constraints on the alignment amplitude obtained by the authors were $A_{\rm IA}=27.48^{+11.53}_{-11.54}$ at 95\% confidence level. 

A much tighter constraint on the intrinsic alignment amplitude of blue star-forming galaxies at $z\sim 1.5$ was recently reported from the HSC narrow-band survey \citep{Tonegawa24}. The galaxies were selected based on the detection of the [OII] emission lines. Measuring $w_{g+}$ and $w_{++}$ correlations, the authors obtained constraints on the alignment amplitude of $A_{\rm IA}=1.38\pm2.32$ at $z=1.19$ and $A_{\rm IA}=0.45\pm2.09$ at $z=1.47$ at $95\%$ confidence level. The authors also suggested that a tentative signal for galaxy angular momentum-cosmic filament alignment is present. We will discuss the relation between shape and angular momentum alignment briefly in Sect.~\ref{sec:spin}.

So far, intrinsic alignments observations have been underused to study the quasilinear or non-linear behaviour of alignments. Some constraints have been derived from cosmic shear or other cross-correlations of photometric shear samples -- these will be described in Sect.~\ref{sec:miti}. 
As for direct constraints from spectroscopic observations, \citet{SinghBOSS} constrained the one-halo contribution to intrinsic alignments of LOWZ galaxies at small scales. They found that the approximate functional form of the one-halo power spectrum proposed by \citet{Schneider10} through a Monte Carlo experiment provided a good description of the observations. 

Some works have pursued a direct identification of satellites to determine their level of alignment. Following this approach, \citet{Agustsson16} found a radial alignment signal for the first time in SDSS and \citet{Georgiou19} measured $\langle e_+\rangle$ at small scales in KiDS-GAMA groups to obtain a functional form for the radial dependence of the signal. This was subsequently re-analysed by \citet{FortunaHM} to include a luminosity dependence of the alignment amplitude for both red and blue galaxies, motivated by \citet{Huang18}:
\begin{equation}
    \langle e_+\rangle=a_{\rm 1h}\left(\frac{L}{L_0}\right)^\zeta\left(\frac{r_{\rm sat}}{r_{\rm vir}}\right)^{-2}
\end{equation}
where $L_0$ is a pivot luminosity corresponding to $M_r=-22$, $r_{\rm vir}$ is the virial radius of the halo, $r_{\rm sat}$ is the radial coordinate of the satellite and $a_{\rm 1h}$ is a constant. 

\subsection{Constraints from photometric samples}
\label{sec:photo}

Purely photometric samples can also deliver constraints on intrinsic alignments if the photometric redshift scatter is small enough ($\sim$per cent level). This is the case of bright or red/early-type galaxy samples. Given the cost of obtaining spectroscopic redshifts and the lack of priors for intrinsic alignments for some populations, it is important to complement the observational efforts described in Sect.~\ref{sec:spec} with this method. Examples of works that have used photometric samples for delivering measurements of intrinsic alignments are: MegaZ \citep{MegaZ}, KiDS LRGs \citep{Fortuna21}, the Physics of the Accelerating Universe Survey \citep[PAUS][]{Johnston21}, redMaGiC \citep{SamuroffEBOSS} and KiDS Bright \citep{Georgiou25}. 

The first photometric sample used to measure intrinsic alignments was the MegaZ sample of 800,000 LRGs and it was instrumental to extend the magnitude ($4$ dex) and redshift range (up to $z<0.7$) of existing constraints. This reduced the uncertainty on the impact of alignments on weak lensing significantly. The KiDS survey has provided intrinsic alignment measurements of a photometric LRG sample in \citet{Fortuna21} based on the selection performed in \citet{Vakili19} over the redshift range $0.2<z<0.8$. The resulting constraints are similar to those from MegaZ. The selection of red galaxies enabled a very secure photometric redshift estimate, with a dispersion close to $\sigma_z\sim 0.015$. \citet{Fortuna21} suggested that the relation between alignment amplitude and luminosity of LRGs cannot be described by a single power-law. Instead, a broken power-law with two different indices at low and high luminosity was preferred. 

In a follow-up study, complementing their results with galaxy-galaxy lensing measurements of the same sample of galaxies, \citet{Fortuna24} showed that this broken power-law dependence was consistent with a single power-law dependence on halo mass. They hypothesized that the driver of a broken power-law relation between $A_{\rm IA}$ and luminosity is the shape of the stellar-to-halo mass relation \cite[e.g.,][]{Behroozi13}.

The KiDS Bright sample \citep{Georgiou25} is limited to $r<20$ and for each galaxy, a machine-learning photometric redshift has been produced, resulting in a low scatter $\sim 0.018(1+z)$ \citep{Bilicki21}. Compared to KiDS-GAMA \citep{JohnstonKiDSGAMA}, which features spectroscopic redshifts, the sample is larger by a factor $\sim 6$, but yields similar constraining power. This is due to the degradation of the signal coming from the photometric redshift scatter. The sample has other advantages such as the availability of different shape measurements and morphological classification. In fact, \citet{Georgiou25} also reported a tentative detection of residual sensitivity of $A_{\rm IA}$ of red galaxies to morphology, in addition to luminosity. 

\begin{figure}[ht]
    \centering
    \includegraphics[width=0.53\linewidth]{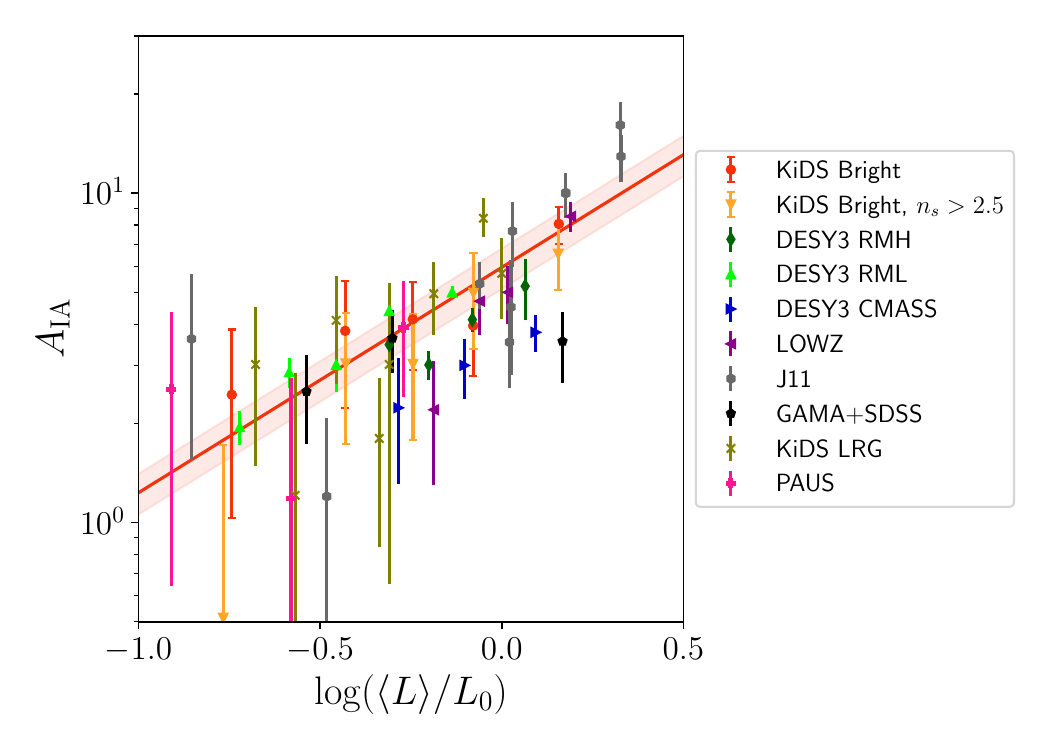}
    \includegraphics[width=0.4\linewidth]{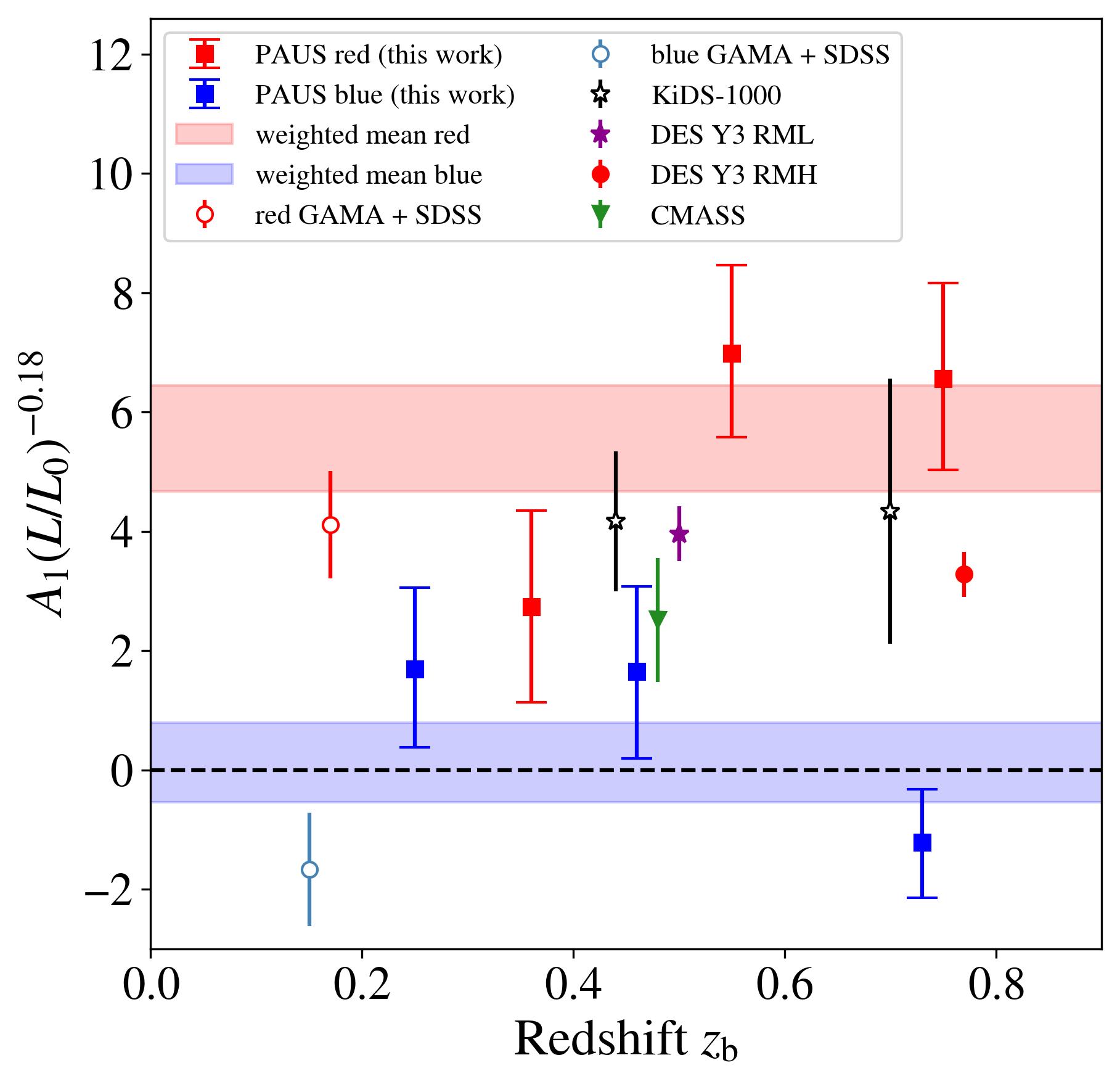}
    \caption{Left: The intrinsic alignment amplitude constrained by different surveys as a function of the mean $r$-band luminosity of different galaxy samples, adapted from \citet{Georgiou25}. Note the figure mixes samples at different redshifts and the assumed redshift dependence is that of the NLA model. The top two symbols in the legend correspond to measurements of the KiDS Bright sample alignments, performed by \citet{Georgiou25}. The DESY3 measurements are taken from \citet{SamuroffEBOSS}. The rest of the measurements are for the LOWZ \citep{SinghBOSS}, J11 \citep{MegaZ}, GAMA+SDSS \citep{JohnstonKiDSGAMA} and KiDS LRGs \citep{Fortuna21}. 
    Right: The intrinsic alignment amplitude of low luminosity galaxies as a function of redshift from \citet{Navarro24}. The $y$-axis has been re-scaled to remove the luminosity dependence. Images reproduced with permission from the authors.}
    \label{fig:G25_AIA}
\end{figure}

The largest photometric LRG samples available in the literature are two low- and high-redshift redMaGiC samples analysed by \citet{SamuroffEBOSS}, both selected from DES photometry \citep{Rozo16}. The large number of galaxies, thanks to the DES area, allowed the authors to place tight constraints on the luminosity-dependence of $A_{\rm IA}$, confirming the need for broken power-law to describe the low- and high-luminosity regimes \citep{Fortuna21}. They also remarked that the colour-space of their samples is slightly different than previous LRG samples, and suggested that an additional dependence of $A_{\rm IA}$ on colour might be present. \citet{Georgiou25} presents the most complete compilation of $A_{\rm IA}$ to date, reproduced here in the left panel of Fig.~\ref{fig:G25_AIA}.

PAUS breaks new ground with the strategy to obtain photometric redshifts for large samples of galaxies using $40$ narrow-band filters in the optical range, combined with broad-band data from CFHTLenS and KiDS. PAUS extends the magnitude limit of KiDS-GAMA by $2$ dex and is expected to provide around $2$ million redshifts with a photometric redshift scatter of $\sim 0.01(1+z)$ \citep{Navarro24}. 
\citet{Johnston21} provided upper limits on intrinsic alignments from PAUS over $19\,\deg^2$. These have been superseded by measurements presented in \citet{Navarro25} over $51\,\deg^2$. Blue galaxy alignments were constrained to have an amplitude between $A_{\rm IA}=0.68^{+0.53}_{-0.51}$. Red galaxy alignments were consistent with an amplitude of $A_{\rm IA}=2.78^{+0.83}_{-0.82}$. \citet{Navarro25} confirmed the trend of more luminous and massive galaxies showing more alignment and extended existing measurements to lower luminosities. By splitting the sample into three redshift bins, the authors found that the redshift evolution observed is simply compatible with changes of the luminosity distribution of galaxies (see right panel of Fig.~\ref{fig:G25_AIA}). 

\subsection{Other measurements}
\label{sec:othermeasures}

\begin{itemize}
\item {\it Alignments of Brightest Cluster Galaxies.} BCGs hold a special status in the alignment literature. Since the 1960s, their shapes and orientations have been explored in connection to that of their host clusters \citep{Sastry68,Carter80,Binggeli82} to determine whether they are affected by accretion along filaments \citep{Dubinski98}, or tidal interactions. In recent years, it has become clear that BCGs align with the shape of their host cluster \citep{Niederste10}. The connection seems to go further even to a correlation with the $X$-ray emitting gas in clusters \citep{Hashimoto08,Okabe18}. By probing the redshift evolution of the alignment, one can in principle distinguish between a primordial alignment scenario and the impact of the cluster's own tidal field \citep{Ciotti94}.
\item {\it Cross-correlations with other biased tracers.} The intrinsic shapes of galaxies are a biased tracer of the density field, and as such it should correlate with any other biased tracer that shares the same underlying large-scale structure. For example, one would expect for galaxies to be aligned in the direction of clusters, or for clusters to be aligned in the direction of galaxies. Similarly, galaxies should be aligned tangentially around cosmic voids. Most works have focused on establishing whether the direction of the angular momentum of spiral galaxies is oriented preferentially with respect to the centre of a void, with a variety or results \citep{Trujillo06,Slosar09,Varela12}. While we briefly discuss the angular momentum alignments in Sect.~\ref{sec:spin}, we concentrate here on shape alignments around voids. \citet{dAssignies21} hypothesized based on the linear alignment model that on large-scales, the alignment of galaxies around positively-biased voids should be radial, while it should become tangential on the rim of the void. Their attempt to find such alignment trends using CMASS galaxies and voids in SDSS resulted in tentative detections for alignments around small voids (linear scales only) and intermediate-size voids (small scales only). If confirmed, galaxy alignments around voids could give complementary constraints on $A_{\rm IA}$ and help constrain void bias. 
\item {\it Dependence on shape measurement method.} How one measures the shape of a galaxy has been shown to affect intrinsic alignment correlations and can have applications to how we do cosmology from galaxy shapes (Sect.~\ref{sec:cosmo}). This is expected, as the outer or inner region of a galaxy responds differently to the processes responsible for aligning it. For example, the inner region is more gravitationally bound and might respond less effectively to tidal alignment mechanisms. The consequence might be a change in the ellipticity of the galaxy, the orientation angle, or both. 

Evidence for this behaviour has been found in \citet{SinghShapes}, where the authors obtained the intrinsic alignment correlations of LOWZ galaxies using three different shape estimation methods. The methods applied were based on an algorithm to retrieve shear estimates (re-Gaussianization; \citealt{Hirata03}), isophotal measurements and elliptical profile fits. The first two methods were shown to be robust to PSF anisotropies and they differed in $A_{\rm IA}$ by up to $40\%$, independent of the galaxy luminosity or other properties. Consistently with these results, a similar analysis carried out by \citet{Huang18} specifically for satellite galaxies in galaxy clusters in SDSS found a stronger satellite alignment signal in isophotal shapes, followed by shapes derived from profile fits. For re-Gaussianization, the signal was only significant at high luminosity. 

\citet{Georgiou19} used the DEIMOS algorithm \citep{Melchior11} to measure the shape of galaxies at different radial scales in the KiDS-GAMA data set. Although they found no significance variation in the large-scale alignment signal of central galaxies, their results remain statistically compatible with \citet{SinghShapes} due to the difference in mass of the sample. Instead, \citet{Georgiou19} found a significant difference in the amplitude of alignment at small scales for red satellite galaxies, which explains the null detections in previous work by \citet{Schneider13,Chisari14,Sifon15}. \citet{Georgiou25} recently confirmed this result using the KiDS-1000 bright sample. 

In \citet{Fortuna21}, the authors presented a comparison between the intrinsic alignments of LRGs from DEIMOS and a model-fitting algorithm for determining shear estimates \cite[\emph{lens}fit,][]{lensfit}. The signals were similar for both measurements but the signal-to-noise ratio of the alignment signal recovered by shear-estimation code was lower due to the lack of galaxies at the bright end ($m_r<20$). When restricting the galaxies in common between both samples, the average difference in $w_{g+}$ was negligible. Therefore, whether $w_{g+}$ measurements of LRGs are sensitive to the shape measurement algorithm remains to be confirmed.

\item {\it Dependence on wavelength}. Intrinsic alignments are often measured from the photometric data with highest seeing quality. For Stage III weak lensing measurements, this is often the $r$-band. However, the intrinsic shape of a galaxy and its sensitivity to alignment processes might change in different bands. \citet{Georgiou19b} explored this hypothesis comparing $g$, $r$ and $i$-band data in KiDS+GAMA. Their results suggested that red satellite galaxies display a different alignment amplitude across bands, with the difference in $w_{g+}$ being positive for $g-r$ and negative for $r-i$. This trend is not monotonic with colour, which complicates the interpretation and it further suggests alignment priors used in weak lensing mitigation should be filter-dependent. 

\item {\it Multipoles.} As we argued in Sect.~\ref{sec:estimators}, multipoles are a promising way to measure and model intrinsic alignments with gains in the signal-to-noise ratio compared to projected correlation functions. Multipoles of the LOWZ galaxies alignments were presented for the first time in \citet{SinghMultipoles}. Their results demonstrated that multipoles yield consistent $A_{\rm IA}$ values compared to $w_{g+}$ and higher constraining power (roughly double). Similar gains have been evidenced in simulations (see Sect.~\ref{sec:sims}). 
\item {\it Intrinsic alignments of cluster shapes.} An optically-identified cluster of galaxies is comprised by a set of member galaxies. Using their spatial distribution (typically an elliptical NFW profile), one can assign an intrinsic shape to each cluster and examine alignments between them. The shape is often obtained from constructing a two-dimensional inertia tensor of the cluster where each galaxy is a weighted by some probability of belonging to the cluster. \citet{Smargon12} were the first to confirm that clusters align towards each other. Compared to previous attempts to measure cluster alignments, they used a larger statistical sample of clusters selected from more uniform photometry, which enabled them to obtain a significant alignment signal even across scales as large as 100 $h^{-1}$ Mpc. The first measurement of the alignment amplitude of clusters was performed by \citet{vanUitert17}, who studied the relative alignments of $\sim 26,000$ redMaPPer clusters in the SDSS data \citep{redmapper}. Classifying the clusters into nine bins of redshift and richness, $\lambda$, they found that cluster alignments in the linear regime satisfy: 
\begin{equation}
A_{\rm IA} = A_{\rm IA}^{\rm gen} \left(\frac{1+z}{1+z_0}\right)^\eta \left(\frac{\lambda}{\lambda_0}\right)^\beta
\end{equation}
where $z_0=0.3$ and $\lambda_0=30$. The best-fit alignment amplitude obtained from fitting the projected correlation of cluster shapes and cluster positions was $A_{\rm IA}^{\rm gen}=12.6^{+1.5}_{-1.2}$, and the power-law indices: $\eta=-3.2^{+1.31}_{-1.40}$ and $\beta=0.60^{+0.20}_{-0.27}$. Richness is a proxy for the mass of the cluster often constructed from summing over the probabilities of galaxies being members of the cluster. The mass-richness relation derived from weak lensing observations can be used to convert this into a dependency with halo mass \citep{Simet17}. Intriguingly, the alignment amplitude of clusters seems to lie on the same power-law mass relation between the $A_{\rm IA}$ of galaxies and halo mass (see also \citealt{FortunaMass}), despite their shape being measured very differently. While clusters are fewer than galaxies and their intrinsic alignments suffer from increased shape noise, their alignment amplitude is also higher, making them possible candidates for cosmological applications of alignments or a test-bed of alignment models \citep{Vedder21,Georgiou24}. When intrinsic alignment are measured between clusters and galaxies, this results in a higher signal-to-noise ratio \citep{Shi24}. 
\item {\it Intrinsic alignments of group shapes.} The alignments of lower mass haloes have also been studied using group catalogues in several surveys. The shape of the group is obtained from the (possibly weighted) spatial distribution of the group members and this shape is correlated with other position tracers to recover an alignment estimate. We will showcase here a few studies, but this is by no means a complete list. \citet{Wang09} used $\sim 300,000$ groups distributed over $4,500$ sq. deg. in SDSS to investigate the relative alignments of satellite and central galaxies in the groups with respect to their nearest-neighbour group. They found that both the distribution of satellites and the central galaxy shape are preferentially oriented towards their nearest-neighbour group, and more strongly so with increasing mass. A slightly different approach was taken in \citet{Paz11}, where the authors measured the orientation of $\sim 900,000$ groups in a later SDSS release, and estimated their level of alignment by measuring the clustering of galaxies in the direction parallel or perpendicular to the group's orientation. Their results show a significant anisotropy in the clustering correlation function (for groups above $6\times 10^{13}\,h^{-1}{\rm M}_\odot$) which increases with halo mass. This measurement seems analogous to that of multiplet alignments in Sect.~\ref{sec:beyond} and it might be possible to relate them to each other. 
\end{itemize}

\section{Numerical simulations}
\label{sec:sims}

Numerical simulations are used in the intrinsic alignment literature for many purposes. In this section, we will discuss their applications with regards to testing models and obtaining priors on the alignment amplitude of different samples. After presenting the methodology in Sect.~\ref{sec:methodnum}, we will cover the insights obtained from the alignments of haloes in $N$-body simulations in Sect.~\ref{sec:nbody} and in cosmological hydrodynamical simulations in Sect.~\ref{sec:hydro}. 

\subsection{Methodology in numerical simulations}
\label{sec:methodnum}

In simulations, galaxies and haloes are identified through some criteria to locate overdensities of particles, with or without taking into account whether they are bound (e.g., by using their velocities). There is usually some minimum threshold to call an object a halo or a galaxy --the simplest being that it contains a certain number of particles. These particles are then used to determine the shape of the object. 

The three-dimensional shapes of either galaxies or haloes in simulations are characterized by their inertia tensor (i.e., the distribution of mass): 
\begin{equation}
\mathcal{I}_{ij}=M^{-1} \sum_q^N m_q w_q x_{q,i} x_{q,j}
\label{eq:Iij3d}
\end{equation}
where $M=\sum_N m_q$ is the total mass of the object, $N$ is the number of particles, $m_q$ is the mass of each particle, and $x_{q,i},x_{q,j}$ are the Cartesian coordinates of the particle in the box relative to the centre of the object.

There are three types of inertia tensors: simple, reduced and luminosity-weighted. The simple inertia tensor (SIT) weights each particle equally. The reduced inertia tensor (RIT) weights particles inversely by their square distance to the centre of the object, $w_q=r_q^{-2}$.
The luminosity-weighted tensor, applied only in the case of galaxies, weights each stellar particle by the light it emits instead of the mass. In practice, luminosity-weighting has disappeared from the literature since \citet{Tenneti14} showed it gives equivalent results to the RIT.

To project shapes over one axis of the simulation box and obtain $I_{ij}$, one restricts the $i,j$ indices in Eq.~(\ref{eq:Iij3d}) to run over only two coordinates (most often, $x,y$). There is certain ambiguity in the literature as to whether the radius used in the weighting of the projected RIT is also the two-dimensional one (vs. three-dimensional). In practice, the effect is the same: up-weighing the inner, more luminous, region of the galaxy in the determination of the shape. This ambiguity is not too relevant when one considers that to match the quality of observations by a specific experiment it would also be necessary to mimic the path of light through the atmosphere, the telescope optics, the relevant filter(s) and the electronics of the camera \citep{Mandelbaum14}. Such realistic treatment of galaxy alignments has so far not been attempted in the literature. However, one step in this direction was taken by \citet{Hilbert17}, who estimated the responsivity directly by shearing the simulated galaxies by a small amount. The authors concluded that this smooths the shape distribution, resulting in significantly lower alignments. A realistic experiment-dependent prediction of observed galaxy shapes is likely to have a bigger impact on any derived priors for weak lensing than differences in shape definitions. Intrinsic alignment priors obtained from numerical simulations should be used with this caveat in mind.

An inertia tensor defines an ellipsoid in three dimensions or an ellipse in two. When diagonalized, the eigenvectors of the inertia tensor correspond to the axes of the ellipsoid. The eigenvalues correspond to the square of the axes lengths. I.e. the largest eigenvalue is $\lambda_a=a^2$. (Note the eigenvalue is assumed to be always positive, since the direction the eigenvector points carries no information.)

Iterative versions of the shape tensors defined above also exist \citep{Katz91}. After a first standard determination of the inertia tensor is performed, particles that lie outside the volume of the ellipsoid are removed and a new shape is obtained. The enclosed volume is kept constant in each iteration until the shape determination has converged (e.g., to within $1\%$) \citep{Schneider12,Tenneti15}. 

\citet{Tenneti15} measured the alignments of galaxies in the MassiveBlack-II (MB-II) cosmological hydrodynamical simulation using SIT, RIT and luminosity-weighted inertia tensors, including their iterative versions. They adopted a threshold of 1000 stellar or dark matter particles to accurately determine the shape of an object, which is conservative compared to most other works. Through random subsampling stellar particles in galaxies with $>$1000 particles, \citet{Chisari15} argued that 300 particles was enough for the uncertainty in projected shapes to be at least one order of magnitude smaller than the shape noise. 
Exceptionally, \citet{Hilbert17} adopted a particle cut of 100 to measure galaxy shapes with several different estimators in the Illustris simulation. The motivation to lower the cut was to be able to reach an $i$-band limiting magnitude of 24.5, therefore mimicking Stage III surveys. The risk is that the shapes of low mass galaxies might not be well-converged at such number of particles. The $w_{\delta +}$ signal in Illustris is dominated by satellites, which reinforces this concern. 

\citet{Zjupa22} argued that weighting each particle by its luminosity reduces ambiguity in the shape definition and is more motivated by observations. However, \citet{Tenneti15} found that including luminosity-weighting in the inertia tensor produces different shape distributions, but similar alignment correlations compared to mass-weighting. \citet{Tenneti15} also argued against using the non-iterative reduced tensor for shapes due to it producing too round objects. Nevertheless, this is still often adopted in the later literature. \citet{Hilbert17} concluded that the use of different shape estimators simply re-scales $w_{\delta+}$ by the variance of the ellipticity, but we will see in later sections that this does not necessarily generalize to all simulations.

Convergence tests on $N$-body simulations report that resolution can affect the alignment amplitude by a factor of $2$ in massive haloes \citep{Schneider12} or the misalignment angle by $2\deg$ for central galaxies in $M_h>10^{12}\,{\rm M}_\odot$ haloes \cite[][]{Herle25}. \citet{Herle25} suggested that the impact of resolution bias on the halo shape can be mostly accounted for by sampling noise and therefore corrected. Further work is needed to understand the impact of resolution on galaxy shapes in hydrodynamical simulations and a degeneracy with structure finding algorithms can be expected \citep{Chisari16,Herle25}. 

The majority of the intrinsic alignment studies in numerical simulations project the galaxy shape over one axis of the box only. One might be able to gain further information from accessing other projections. Because the galaxy population and the region of the Universe being simulated are the same, other projections are not independent. 

Once the shape is established, one can proceed to measure the alignment statistics\footnote{Redshift uncertainties are typically not modelled, though this could in principle be implemented for photometric samples.}: $w_{g+}$ (Eq. \ref{eq:wg+}), multipoles in real (Eq. \ref{eq:multipoles}) or Fourier space (Eq. \ref{eq:Fourier_multipoles}), as we referred to in Sect.~\ref{sec:obs}. Because the density field is directly accessible in simulations, $w_{\delta +}$ can also be measured, yielding direct insights into the modelling of contamination to gravitational lensing.

\citet{Hilbert17} take a slightly different approach. They first project the matter, intrinsic shape and galaxy position fields in $50$ logarithmically-space bins along one of the axis of the box, and they correlate them using fast Fourier transforms. Because of the limited size of the simulation box ($75\,h^{-1}$ Mpc), the authors measured correlations between 10 $h^{-1}$ kpc and 10 $h^{-1}$ Mpc only. From the point of view of observations, the regime $<100 h^{-1}$~kpc is difficult to access due to systematics such as blending \cite[e.g.,][]{Sanchez21}.

In simulations, one can also measure the alignment of shape tracers in three-dimensions. By convention, the three-dimensional position-shape correlation is measured by defining the average of the cosine of the alignment angle, $\theta_e$, between the major (or minor) axis of the shape tracer and the direction of the separation vector of the pair. This is known as the ellipticity-direction correlation (ED) and it is defined:
\begin{equation}
\eta_e(r) = \langle \cos^2\theta_e \rangle-1/3
\label{eq:ED}
\end{equation}
where $1/3$ is the expectation value if alignments are random. The three dimensional shape auto-correlations are similarly obtained considering the angle $\theta_{ee}$ between two major (or minor) axes, 
\begin{equation}
\eta_{ee}(r) = \langle \cos^2\theta_{ee} \rangle-1/3.
\label{eq:EE}
\end{equation}
Note that for grid-based simulation methods, two-point auto-correlations of shapes might be contaminated by numerical effects (`grid-locking'). Checking for this is necessary before using them to constrain any alignment model \citep{Codis15,Chisari15}.
The advantage of using these three-dimensional statistics is that there is no loss of signal-to-noise ratio because shapes are not projected. On the other hand, it has the disadvantage that it does not take into account the axes ratios of the ellipsoid. This issue has recently been overcome by \citet{Samuroff21}, who proposed a method to fit alignment models presented in Sect.~\ref{sec:model} directly to the three-dimensional shapes of objects in simulations, dubbed ``Direct Alignment Field Fitting'' (DAFF). This is a promising avenue for the interpretation of three-dimensional shapes and their correlations in numerical simulations in the future.

Because the majority of the works with simulations are done in real space and probe the non-linear regime, covariance matrices are difficult to obtain. The most common technique to attach an error bar to these statistics is the jackknife method \cite[e.g.,][]{HirataJK}. This captures both the shape noise as well as cosmic variance. Given the separations over which the alignment signal is measured, the number of bins and the size of the box, one must pay attention to how many jackknife regions are needed to robustly determine the inverse covariance, and the Hartlap correction normally applies \citep{Hartlap07}. The number of jackknife regions must generally be larger than $N_{\rm bin}^{3/2}$ (with $N_{\rm bin}$ the number of radial bins) and their area should be larger than the scales for measuring the signal (see appendix D in \citealt{HirataJK}).

Compared to the jackknife technique, bootstrapping over the galaxy sample only accounts for shape noise and thus not only underestimates the error bars, but it also fails to account for off-diagonal components in the covariance. Splitting a simulation box in sub-boxes to estimate the covariance of the signal also fails in this regard, and it further restricts the range of separations that can be accurately probed. Ideally, to reduce the computational cost one would use a theoretical covariance \citep{Chisari13,Samuroff21} with the free parameters of the model iteratively determined from fitting the simulation data itself.

The intrinsic shapes of galaxies are poorly resolved in hydrodynamical simulations of cosmological volumes, and their numbers are too few in otherwise smaller simulations. A simulation (or suite thereof) targeted to find priors on intrinsic alignments with sufficient statistical power and at the right redshifts for weak lensing applications would be an asset, but none has been specifically crafted for this purpose. Instead, one resorts to samples of galaxies with $\gtrsim 300$ stellar particles in simulations of $\sim 100$ Mpc on a side, which only allows us to constrain alignments of galaxies with $\gtrsim 10^9\,{\rm M_\odot}$ and up to separations of $20$ Mpc at $z\lesssim3$. This is still far from the ideal magnitude-limited sample needed for weak lensing cosmology in the Stage IV era, but it is also a regime that is otherwise not accessible via observations at the moment.

\subsection{Alignments of haloes in N-body simulations}
\label{sec:nbody}

An extensive literature uses $N$-body simulations to measure the intrinsic alignments of dark matter haloes starting from the year 2000 \citep{Heavens00,Crittenden00,Jing02}. Understanding how dark matter haloes align allows one to test intrinsic alignment models and to also decide how to populate $N$-body simulations with aligned galaxies in order to produce suitable mocks for predicting intrinsic alignment contamination to lensing. Here, we review some of the intrinsic alignment studies of dark matter haloes from the first perspective, while mock building will be the topic of Sect.~\ref{sec:mocks}. 
The alignments of dark matter haloes can also be directly related to those of galaxy groups or clusters \cite[e.g.,][]{Faltenbacher02,Hopkins05}, whose observations we discussed in Sect.~\ref{sec:othermeasures} and which might play a role in extracting cosmological information from intrinsic alignments (Sect.~\ref{sec:cosmo}). Note that some model comparisons are performed in real space, others in Fourier space, and there is not a straightforward correspondence between them. Care must be taken when translating one into another.

A compelling test of the LA model in simulations was performed in \citet{Schneider12} in Millennium and Millennium-II, two $N$-body simulations of different box sizes ($500\,h^{-1}$Gpc and $100\,h^{-1}$Mpc, respectively) and different dark matter particle mass (resp. $8.6\times 10^8\,h^{-1}{\rm M}_\odot$ and $6.89\times 10^6\,h^{-1}{\rm M}_\odot$). The test presented in \citet{Schneider12} was done using the anisotropic correlation function (ACF), a quantity which describes how many haloes cluster in the direction of the major axis of another halo \citep{Paz11} and a model for which is provided in \citet{Blazek11}. This suggested the LA model was successful in describing the measurement (see also \citealt{Faltenbacher09} and \citealt{Xia17}). 

\citet{Schneider12} found some of the first evidence that intrinsic halo alignments depend strongly on mass, and that this dependence is self-similar, leading to a potentially simple re-scaling of the alignment signal with cosmology. Explicitly, the dependence scales with $M/M_\star(z)$, where $M_\star(z)$ is defined as the mass where the r.m.s. density in top-hat spheres, $\sigma(M)$, is equal to the overdensity for spherical collapse, $\delta_{\rm sc}(z)$, at a given redshift. 
Conceptually, they attributed this dependence to higher-mass haloes being younger and more biased than lower-mass haloes. 

Performing a fit of the LA model to two-point correlation function of halo shapes, i.e. Eq.~(\ref{eq:xiss}) (dubbed $c_{11}$ in that work), \citet{Xia17} confirmed its validity on large scales and of its redshift dependence over $z=\{0,1\}$ for haloes above a $10^{12}\,h^{-1}{\rm M}_\odot$. To successfully reproduce the simulations, the authors needed to include the density-weighting of shapes at the location of haloes in their model. In doing this, they showed that the dependence of the strength of halo alignment with mass could be explained by the dependence of halo bias with mass. 

\begin{figure}[ht]
    \centering
    \includegraphics[width=0.7\linewidth]{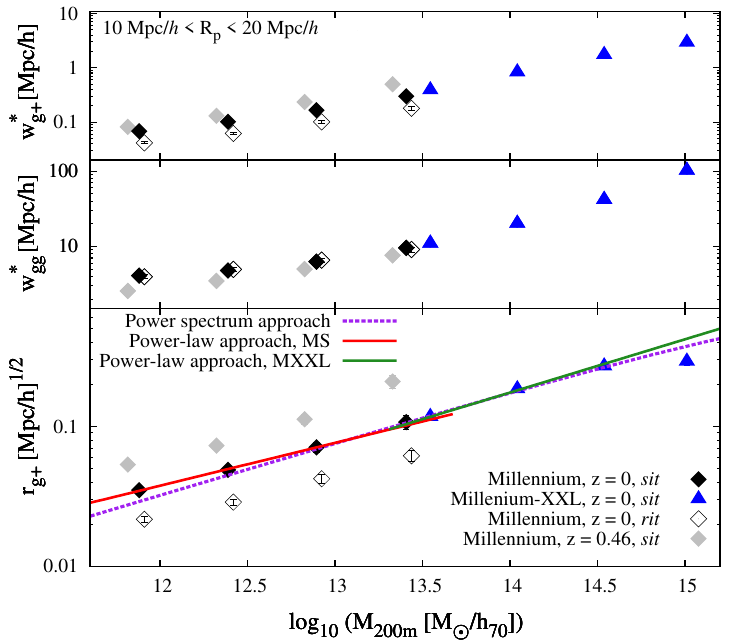}
    \caption{To assess the mass-dependence of the intrinsic alignment amplitude, \citet{Piras18} average the halo alignment and halo clustering projected statistics (top two panels) on the range $10 h^{-1}{\rm Mpc}<r_p<20\,h^{-1}{\rm Mpc}$ and take a ratio, $r_{g+}$ (bottom panel) that cancels the mass-dependence coming from the halo bias. Their results suggest that haloes in the Millennium and Millennium XXL simulations have a power-law dependence of $A_{\rm IA}$ with haloes mass. Credit: \citet{Piras18}. Image reproduced with permission from \citet{Piras18}, copyright by the author(s).}
    \label{fig:piras}
\end{figure}

A roughly simultaneous work by \citet{Piras18} using the higher volume Millennium-XXL simulation presented a similar argument to that of \citet{Camelio15} (Sect.~\ref{sec:priors}) to derive the power-law dependence on mass of the intrinsic alignments of haloes. They obtained excellent agreement with the simulated alignment correlations in the range between $10^{13}-10^{15}\,h^{-1}{\rm M}_\odot$, suggesting they follow a power-law scaling with mass with index $\sim 0.36$ (see Fig.~\ref{fig:piras}), in agreement with the findings of \citet{Kurita21}. It remains to be understood whether the mass dependence of the alignments of haloes results from density-weighting or from the sensitivity of the shapes to the tidal stretching mechanism, and whether this is compatible with the cosmological dependence of the alignment signal proposed by \citet{Schneider12}.

In recent years, the literature has also expanded in tests of the scale-dependence of intrinsic alignments. High volume $N$-body simulations allow for more stringent tests than possible with hydrodynamical boxes. \citet{Okumura20} tested the LA and NLA models in a $(2\,h^{-1}$Gpc$)^3$ box for subhaloes of mass $>10^{13}{\rm M}_\odot$ at $z=0.306$ and found that LA fails to provide an accurate fit to a variety of real-space alignment cross-correlations already at the baryon acoustic oscillation (BAO) scale, while NLA allows the fit to extend down to $50\,h^{-1}$Mpc, capturing also the smearing of the BAO feature. \citet{Kurita21} performed similar tests in a $(1\,h^{-1}$Gpc$)^3$ box using Fourier-space multipoles of halo alignments for haloes of mass $\geq 10^{12}h^{-1}{\rm M}_\odot$ at $0<z\lesssim1.5$. The LA model and the measurements were found to agree only at $k\leq 0.1\,h\,{\rm Mpc}^{-1}$, evidencing the need for more sophisticated models given the error bars of the simulations.

\begin{figure}[ht]
    \centering
    \includegraphics[width=0.5\linewidth]{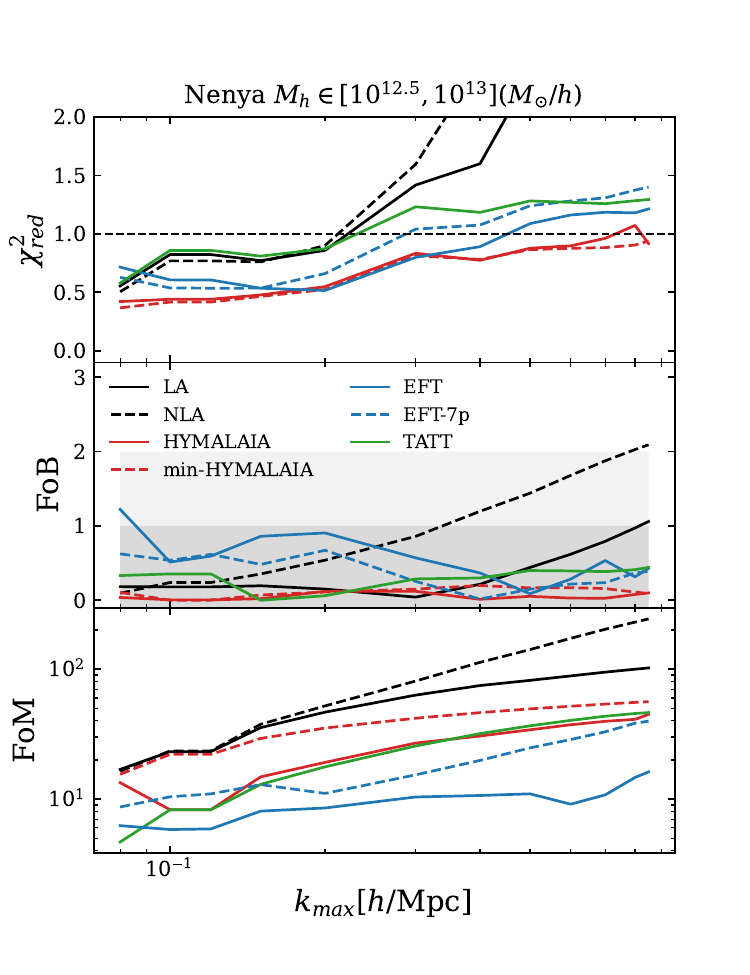}
    \caption{The goodness-of-fit (top) of several different intrinsic alignment models to the alignments of haloes in an $N$-body simulation. The middle panel shows the ``figure of bias'' (FoB), which quantifies how well the linear alignment bias is recovered. The bottom panel quantifies the ``figure of merit'' (FoM) of the model, i.e. an analogue of the signal-to-noise ratio. HYMALAIA outperforms all other available quasilinear models. Credit: \citet{Maion24}. Image reproduced with permission from \citet{Maion24}, copyright by the author(s).}
    \label{fig:maion}
\end{figure}

Tests of higher-order models by \citet{Bakx23} showed a good agreement between the EFT model and $N$-body simulations. The criteria used were, similarly to Fig.~\ref{fig:maion}, a combination of goodness of recovery of the LA amplitude and goodness of fit of the model per se. The EFT model surpassed LA, NLA and TATT in the accuracy and precision with which $b_{1,I}$ was recovered. LA, NLA and TATT were validated up to $k=0.05\,h\,{\rm Mpc}^{-1}$ only, while the EFT extended in applicability range up to $k\simeq 0.3\,h\,{\rm Mpc}^{-1}$. The number of free parameters could also be approximated from $8$ to $6$ if some of the terms contributing to the helicity power spectra were dropped due to their similar scale-dependence (these are third-order terms). 

\citet{Maion24} allowed for substantial extensions into the non-linear range by demonstrating that in its original form, HYMALAIA can provide successful modelling of alignment power spectra of dark matter haloes in an $N$-body simulation up to $k\simeq 0.85\,h\,{\rm Mpc}^{-1}$ (see Fig.~\ref{fig:maion}). A minimal version of the model (min-HYMALAIA) which only accounts for the terms corresponding to $A_1$ and $\tilde 
A_{1\delta}=A_{1\delta}+b_{1,g}A_1$ performs almost equally well, only losing $\sim 20\%$ constraining power. These results are showcased in Fig.~\ref{fig:maion}.

While these maximum scales might sound restrictive compared to the regime where intrinsic alignments need to be modelled for lensing mitigation (Sect.~\ref{sec:miti}), lensing observations are expected to be more forgiving\footnote{The works listed in this section have more direct applicability to spectroscopic measurements of intrinsic alignments.}. Works using realistic mock samples for lensing contamination \citep{Hoffmann22,FortunaHM} have found NLA and TATT to yield reasonable fits up to much smaller scales (a few $h^{-1} {\rm Mpc}$). In practice, the choice of model will depend on the tolerance set by the error bars of the observable.

In parallel to efforts to model the scale-dependence of the alignment observables, new approaches have emerged to quantify shape bias parameters including and beyond the linear alignment amplitude. \citet{Schmidt18} developed a methodology to include the impact of an external tidal field on haloes in an $N$-body simulation through an effective anisotropic ``separate universe'' \citep{Wagner15} expansion applied to a particle-mesh code. This was used by \citet{Stucker21} to measure $A_{\rm IA}$ to high significance through finite difference of simulation realizations with different external tidal fields. Resolving the shapes of haloes was possible because in this case the method was extended to work with a TreePM simulation. Consistently with other works, they found that the response remains linear (i.e., LA is valid) only up to $k\lesssim 0.1\,h\,{\rm Mpc}^{-1}$ at $z=0$, that haloes are preferentially aligned with the direction of maximum tidal stretch and that $A_{\rm IA}$ depends on halo mass. 

While \citet{Stucker21} were able to show that the response of halo alignments is suppressed on non-linear scales compared to linear ones, the work was not able to extract higher-order biases. \citet{Akitsu23} reported a detection of quadratic shape bias parameters using the technique proposed in \citet{Schmittfull15}, which projects measured alignment power spectra onto the specific operator fields constructed from the initial conditions of the simulation. They also established a universal relation between $A_{\rm IA}$ and $b_{1,g}$. 

\citet{Maion24b} proposed a new method to determine shape bias parameters which is conceptually based on the separate universe technique but has a much lower computational cost, as it can be computed within a single simulation. The method also provides an estimate of biases on an object-by-object basis, which allows one to easily average over different ensembles of interest. Using this method, \citet{Maion24b} were able to establish other universal relations for shape biases with the height of the density peak inhabited by a halo and to falsify the scenario that alignments are established after their formation. These new insights into alignment bias values and their origin could bear an analogue in the case of galaxies worth exploring.

\subsection{Cosmological hydrodynamical simulations}
\label{sec:hydro}

To aid in their mitigation, it is important to determine priors on how strong alignments are. Unfortunately, the existing observations, described in detail in Sect.~\ref{sec:obs}, do not cover the regime needed by weak lensing studies. Spectroscopic samples are targeted to specific populations of galaxies (LRGs or ELGs) or they are only complete to relatively bright magnitudes compared to the magnitude-limit of weak lensing surveys. Photometric samples allow us to extend constraints further in redshift, but they still target either relatively bright objects, or red ones. Cosmological hydrodynamical simulations are a useful tool to understand priors on intrinsic alignments models of dimmer samples. In this section, we will give an overview of the methodology used to extract alignment priors from cosmological hydrodynamical simulations and report on the constraints available from them.

Cosmological hydrodynamical simulations used for measuring intrinsic alignments vary in numerical schemes adopted, modelling of unresolved physical ``sub-grid'' processes like star formation or feedback, resolution, volume and degree of calibration to different observables \cite[e.g.][]{Schaye15}. All of these features can have an impact on the resulting intrinsic alignment signal, but exploring them systematically is computationally costly and has only been achieved in a few works \citep{Tenneti17,Soussana20}. In sub-sections \ref{sec:hydroiacorr} and \ref{sec:gxyhalo}, we describe the main results of different hydrodynamical simulations and comment on how these might depend on the underlying assumptions of these methods. The simulations that have been used for intrinsic alignment studies are summarized in Table~\ref{tab:hydrosims}.

\begin{landscape}
\begin{table*}[htbp]
\caption{Cosmological hydrodynamical simulations that have been used to measure intrinsic alignment correlations. The references correspond to the manuscripts where intrinsic alignment results were presented. $L_{\rm box}$ is the size of the box, $m_b$ is the mass of the baryonic particle (stars when possible, otherwise gas mass is reported), $m_{\rm DM}$ is the mass of the dark matter particle. The sixth column indicates the name of the numerical software tool used to run the simulation.}
\label{tab:hydrosims}
\small
\begin{tabular}{lcccccc}
     \toprule
     Cosmological & Number of & $L_{\rm box}$   &$m_b$ & $m_{\rm DM}$ &  Code  & Refs. \\
     simulation & particles & ($h^{-1}$Mpc) &($h^{-1}{\rm M_\odot}$) & ($h^{-1}{\rm M_\odot}$) &    &  \\
     \midrule
     MB-II & $2\times 1792^3$ & 100 & $1.1\times 10^6$ & $1.1\times 10^7$ & {\sc GADGET-2} & \cite{Tenneti14,Tenneti15,Tenneti16,Samuroff21}\\
     EAGLE Recal & $2\times 752^3$& 25 $h$ & $1.5\times 10^5$&   $8.2\times 10^5 $ & {\sc GADGET-3}  & \cite{Velliscig15,Velliscig15b} \\
     EAGLE Ref & $2\times 1504^3$&  100 $h$  & $1.2 \times 10^6$&   $6.6 \times 10^6$ & {\sc GADGET-3} & \cite{Velliscig15,Velliscig15b} \\
     cosmo-OWLS & $2\times 1024^3$ & 200 &  $8.7 \times 10^7$&  $4.1 \times 10^8$ & {\sc GADGET-3} & \cite{Velliscig15,Velliscig15b} \\
     cosmo-OWLS & $2\times 1024^3$ & 400 & $7.5 \times 10^7$&  $3.7 \times 10^9$&   {\sc GADGET-3} & \cite{Velliscig15,Velliscig15b} \\
     Horizon-AGN & $2\times 1024^3$& 100 & $2\times 10^6\,h$ & $8\times 10^7$& {\sc RAMSES} & \cite{Codis15,Chisari15,Chisari16}\\
     SIMBA & $2\times 1024^3$ & 100 & $1.82 \times 10^7\, h$ & $9.6 \times 10^7\,h$ & {\sc Gizmo} & \cite{Kraljic20} \\
     Magneticum & $2\times 576^3$& $48$ & $7.3\times 10^6$ & $3.6\times 10^7$ & {\sc GADGET-3} & \cite{Teklu15} \\
     Illustris-1 & $2\times 1820^3$& $75$ & $1.26 \times 10^6$ & $6.26 \times 10^6$ & {\sc AREPO} & \cite{Tenneti15,Hilbert17} \\
     TNG100 & $2\times 1820^3$ &75  & $1.4\times 10^6\,h$ & $7.5\times 10^6\,h$ & {\sc AREPO} & \cite{Samuroff21}\\
     TNG300 & $2\times 2500^3$ & 205 & $1.1\times 10^7\,h$ & $5.9\times 10^7\,h$ & {\sc AREPO} & \cite{Samuroff21} \\
     MTNG & $2\times 4320^3$ & 500 & $3.1\times 10^7$ & $1.7 \times 10^8$ & {\sc AREPO}  & \cite{Delgado23}\\
     \bottomrule
\end{tabular}
\end{table*}
\end{landscape}

\subsubsection{Galaxy shape correlations}
\label{sec:hydroiacorr} 

Generally speaking, all simulations find a correlation between the amplitude of the alignment signal and stellar or sub-halo mass (see Sect.~\ref{sec:gxyhalo}). Higher mass galaxies typically live in higher mass haloes and display a larger alignment signal, in line with observational results \citep{SinghBOSS,FortunaMass}. The first statistical measurements of galaxy alignments in hydrodynamical simulations that illustrated this trend were performed in the MB-II simulation \citep{Khandai15} by \citet{Tenneti15}. The alignment measurements consisted of $\eta_e$, $w_{\delta +}$ and $w_{g+}$ statistics. The former in particular are generally used to determine the level of agreement with observations at the high luminosity end \citep{SinghBOSS}. Most simulations are successful in reproducing those observations \citep{Velliscig15,Tenneti15,Chisari15}, sometimes within a factor of a few \citep{Hilbert17}. 

Alignment statistics from cosmological hydrodynamical simulations are used to test a number of hypotheses behind alignment models. One of them is the assumption that red ellipticals and blue spirals are subject to different alignment mechanisms. Separating galaxies by median $u-r$ rest-frame colour, \citet{Tenneti15} did not find any colour-dependence in the alignment signal: both red and blue samples aligned with similar strength in MB-II. This could be a consequence of the simulation not displaying any bimodality in galaxy colours. 

MB-II predicts a range of different observables successfully (e.g., galaxy clustering and halo occupation distribution statistics), but massive galaxies show signs of not being sufficiently quenched. The sub-grid modelling in MB-II was the same as previously adopted for smaller volumes and lower resolution \citep{DeGraf12}, and it was not actively re-calibrated to account for these changes. This can impact the relative abundance of red and blue galaxies, and the alignment signal.

For more realistic colours, one might turn to the Illustris simulation \citep{Vogelsberger14,Vogelsberger14b,Genel14,Sijacki15} where the $g-r$ colour distribution is bimodal, correlates well with galaxy spectral type \citep{Benitez00} and allows for a clear distinction between blue and red galaxies (whose abundance is $<9\%$). In Illustris, galaxy type fractions are roughly in agreement with observations at the bright end, but late-type galaxies are too abundant at the faint end. In addition, faint galaxies are larger than expected from observations. Possible drivers of these discrepancies are lack of resolution, issues with the identification of particles belonging to a galaxy or with feedback and star formation recipes. 

Despite reproducing the observed bimodality of galaxy types, intrinsic alignments are also radial in the Illustris simulation. A detailed comparison between Illustris and MB-II was presented by \citet{Tenneti15} with the goal of comparing the predictions for disc galaxies specifically. Their findings will be discussed more thoroughly in Sect.~\ref{sec:compare}. 

\citet{Velliscig15} used four simulation runs taken from the EAGLE and cosmo-OWLS suites, ranging in box sizes between $25\,h^{-1}$ Mpc and $400\,h^{-1}$ Mpc, to measure orientation-direction correlations (Eq. \ref{eq:ED}) of galaxies with $>300$ particles. In EAGLE and cosmo-OWLS \citep{Velliscig15}, alignments are also always radial. They confirmed a trend for alignments to increase with halo mass. Their study restricted to the SIT shape measurement but they made a distinction between shapes measured from all stellar particles and those within the radius that encloses half of the mass (``half-mass''). Alignments were found to be stronger when stars in the outskirts were included. Although that work made no distinction between red and blue, or early- and late-type galaxies, shapes measured from star-forming gas in EAGLE have been confirmed to show radial (albeit weaker) alignments \citep{Hill22}. 

\begin{figure}[ht]
    \centering
    \includegraphics[width=0.5\linewidth]{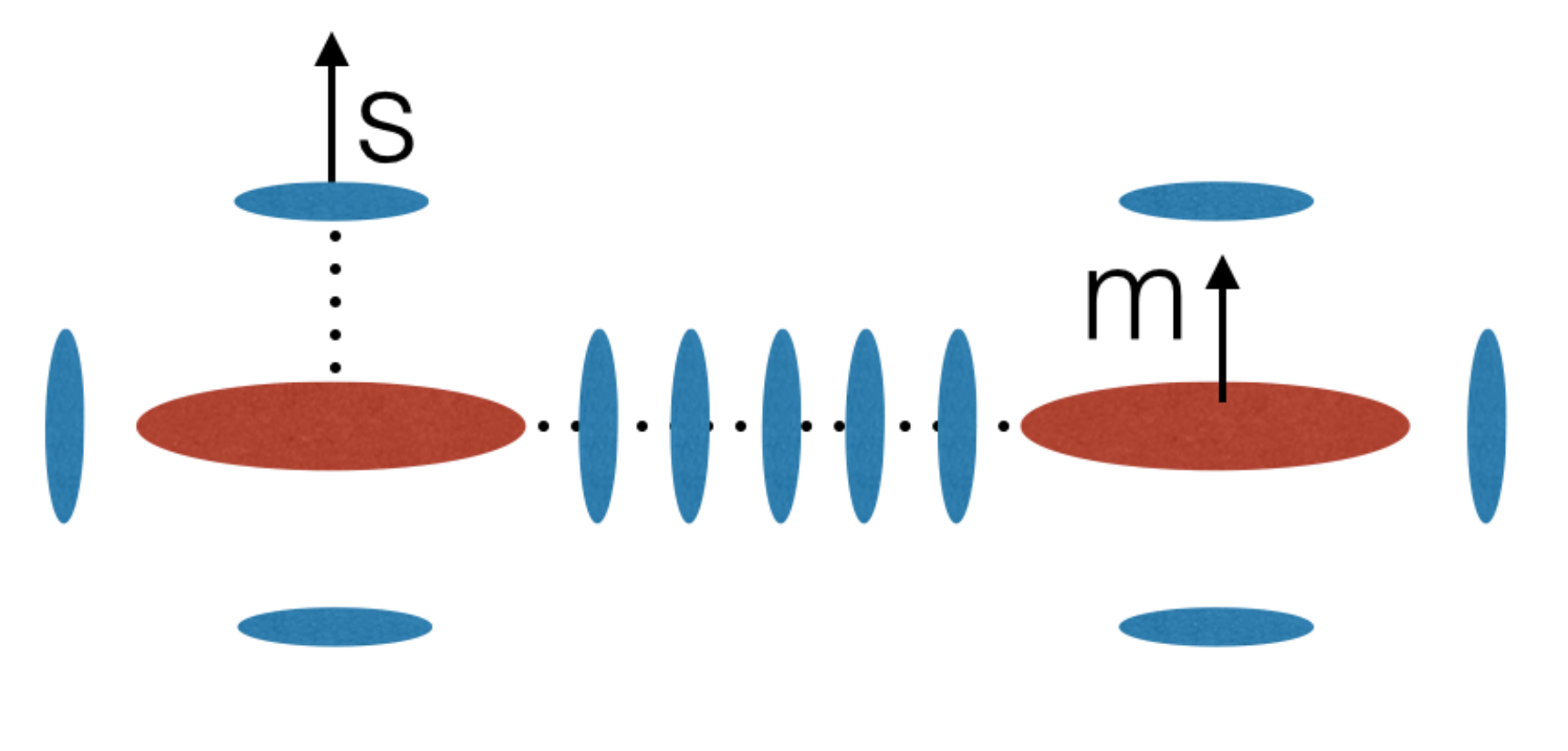}
    \caption{A cartoon of the two modes of galaxy alignment evidenced for elliptical and spirals in Horizon-AGN \citep{Chisari15}. Elliptical galaxies (red) point their minor axes perpendicular to the direction towards another galaxies, while spiral (blue) galaxies align their spin radially around them. Credit: \citet{Chisari15}. Image reproduced with permission from \citet{Chisari15}, copyright by the author(s).}
    \label{fig:cartoonHAGN}
\end{figure}

Shape measurement could also be a factor in determining the sign of alignment. \citet{Chisari15} selected elliptical and disc galaxies on basis of their dynamical properties (specifically, with a threshold of tangential velocity to velocity dispersion of $V/\sigma\sim 0.6$) and measured their alignments with galaxy position tracers and the density field at $z\sim 0.5$ in the Horizon-AGN simulation \citep{Dubois14,Kaviraj17}. Elliptical galaxies were seen to align radially around overdensities with an amplitude roughly consistent with observational values. They also found that disc alignments were null when projected shapes were considered, but their alignment in three dimensions showed evidence of a tangential signal around ellipticals, especially when considering the RIT shape. We reproduce a cartoon of this effect in Fig.~\ref{fig:cartoonHAGN}.

More evidence for a tangential alignment signal of disc galaxies comes from the SIMBA simulation \citep{Dave19}. \citet{Kraljic20} do not measure shapes but the direction of angular momenta of galaxies (often dubbed `spin'). According to their analysis, low mass galaxies in the redshift range $0\leq z\leq 2$ align their spin in the direction of the filament. Extrapolating from the Horizon-AGN results, it would be expected then that low mass galaxy RIT shapes in SIMBA would be aligned tangentially around overdensities. These results are also compatible with those found by \citet{Codis15} when estimating intrinsic alignment contamination to lensing based on the spins of galaxies in Horizon-AGN at $z=1.2$, given that the RIT and the spin trace each other closely for these galaxies.

\citet{Chisari16} anticipated that such level of disc alignment, if present, would still be a significant source of contamination for Stage IV surveys. Even if we consider the other extreme scenario, with discs aligned radially, \citet{Tenneti15} would predict that the alignment amplitude expected for Stage IV surveys would be a factor $5-18$ lower than for luminous ellipticals. If we allow ourselves to be pessimistic, this would correspond to $A_{\rm IA}\sim 1$, which is already near the realm of detection of Stage III surveys (see Sect.~\ref{sec:margin}).

Note that the distinction between discs and ellipticals in Horizon-AGN was based on a dynamical criterion, instead of colour or spectral type. One might expect this to make a difference in separating alignment mechanisms, but in practice \citet{Tenneti16} find that selecting galaxies based on a dynamical criterion in Illustris and MB-II leads to the same alignment sign as when separating by colour. 
Instead,  \citet{Shi21} found that the choice of shape can yield a redshift-dependent effect on the alignment measurements in the TNG300 simulation (see also \citealt{Moon23}). Choosing one or another shape estimator at $z=2$ in particular makes a significant difference, leading to even a change of sign. In fact, this is the only TNG analysis that reports a change of sign of the spin alignment of low vs. high mass galaxies. Still, the amplitude is very small: the authors forecast no expected detection of intrinsic alignments for ELGs selected by their star formation rate in ongoing or upcoming spectroscopic and imaging surveys.  

The redshift evolution of intrinsic alignments is of interest to determine the level of contamination to Stage IV surveys. The abundance of discs and ellipticals changes with redshift, with ellipticals typically more prominent at low redshift. \citet{Tenneti15} found no evidence of redshift evolution of the alignment signal at fixed mass in MB-II. At low redshift, despite their low abundance, ellipticals dominate the contribution to $w_{\delta+}$ in Illustris. This is similar to the case of MB-II and Horizon-AGN, as is the stellar mass dependence of the alignment amplitude. \citet{Chisari16} studied the redshift evolution of intrinsic alignments in Horizon-AGN from $z=0$ to $z=3$ selecting galaxies by their $r$-band luminosity. Elliptical galaxies had a stronger radial alignment amplitude which increased with redshift, but the overall alignment of the total population of galaxies decreased. This is because the disc alignments in Horizon-AGN are null or tangential, and they dominate the signal by number, particularly at high redshifts.

Many of the studies discussed in this section focus on fitting the NLA model to the data and often find good descriptions of the projected alignment statistics above a few $h^{-1}$ Mpc. For example, \citet{Tenneti15} reported that NLA is a good fit to $w_{g+}$ from $6$ to $25\,h^{-1}$ Mpc, but a one-halo term is required at smaller scales and dominated by the contribution of satellites.  
\citet{Velliscig15} added that alignments in the one-halo regime can be split into two contributions: the alignment of a central galaxy in the direction of the satellites that populate the halo anisotropically; and the alignment of satellite galaxies towards the centre of the halo, which quickly diminishes with increasing radius and seems to be independent of central subhalo mass. \citet{Chisari15} also remarked on a significant difference in power at small scales when $w_{\delta+}$ or $w_{g+}$ was considered. Similar conclusions were reached by \citet{Delgado23}, who measured the intrinsic alignments of galaxies in $0<z<1$ in the much larger Millennium TNG (MTNG) simulation. They found no significant redshift evolution in $w_{g+}$ and a very different behaviour for $w_{\delta+}$, suggesting that the evolution of galaxy bias plays a big role over this redshift range. 

\citet{Hilbert17} confirmed that in Illustris, $w_{\delta +}$ is heavily dominated by satellites, even at scales of a few $h^{-1}$ Mpc and regardless of how the shape measurements are made. These predictions might however be affected by the details of sub-halo finding \citep{Chisari16} and could be at odds with observational constraints where the satellite alignment signal has traditionally been elusive \citep{Chisari14b,Sifon15,Georgiou19}. Nevertheless, these findings still caution
against using the same template model for both statistics in non-linear scales. 

What evidence can be gathered from hydrodynamical simulations regarding the validity of different modelling options? \citet{Samuroff21} and \citet{Zjupa22}, employed the TNG100 and TNG300 hydrodynamical simulations, respectively, to directly test the validity of  alignment models in different samples. While \citet{Zjupa22} examined the quadratic model for spiral galaxies and linear model for elliptical galaxies, \citet{Samuroff21} opted to fit TATT (for the first time) and NLA to several different samples. With the exception of MB-II, both works generally found their results to be consistent with the LA or NLA model, including for spirals/blue galaxies, without the need for higher order terms. In MB-II, \citet{Samuroff21} found a slight preference for $A_2<0$ at $2-3\sigma$, which increases if the minimum scale of the fit is brought down from $6\,h^{-1}$ Mpc to $1\,h^{-1}$ Mpc. 

Notably, both works implemented a methodology similar to the field-level approach undertaken by \citet{Tsaprazi22} in the case of observations (see Sect.~\ref{sec:othermeasures}) and proposed in \citet{Desjacques18} to obtain constraints on alignment model parameters. \citet{Samuroff21} took advantage of the shapes and model in full three-dimensions through their DAFF approach. Both works emphasized the sensitivity of the method to the choice of smoothing scales, the need for incorporating higher-order terms to mitigate it, and the difficulty of interpretation this entails. In Sect.~\ref{sec:nbody}, we presented a few other strategies to test alignment models in $N$-body simulations.

Additional gains in model testing can be drawn from larger boxes. In MTNG, \citet{Delgado23} found that the larger size of the box compared to previously available simulations allowed for a significant detection of $w_{++}$ over a significant separation. If precisely measured, shape-shape correlations could provide ways to distinguish between alignment models and more direct predictions of contamination to cosmic shear.

\subsubsection{Galaxy-halo connection}
\label{sec:gxyhalo}

The alignment of haloes with the tidal field is stronger than with galaxies. In the context of the LA model, this is evidenced in their larger $A_{\rm IA}$ value. For a central galaxy in a halo to have a smaller alignment amplitude than the halo itself, its projected shape should be misaligned with respect to the projected shape of the halo and its axis ratio ($q$) might be different. \citet{OkumuraLRGs} distinguished between these two effects by populating dark matter haloes with LRGs and matching their misalignment an elongation to reproduce observations. They found that misalignment is the main driver of the discrepancy in alignment amplitudes, while axis ratio differences play a smaller role. 

For the purpose of creating realistic mock catalogues with aligned galaxies, we need to connect the shape and orientation of a galaxy to that of a halo. We will see in Sect.~\ref{sec:mocks} that early mocks were crafted assuming the shape and orientation of a central galaxy followed that of halo. However, according to hydrodynamical simulations \citep{Tenneti14,Velliscig15b,Chisari17}, the axis ratios and misalignment angles can vary across galaxies depending on their type, mass, or redshift. Generally speaking, the halo shape is a poor proxy of the galaxy shape, particularly if estimated using SIT. The distribution of stellar particles is generally less spherical than dark matter, and it becomes more prolate with increasing distance from the centre. Misalignment angles are a strong function of halo mass, with higher mass haloes hosting more aligned galaxies. Discrepancies exist in the redshift evolution and level of misalignment of galaxies of different types between these hydrodynamical simulations, which is perhaps not surprising given that $w_{g+}$ predictions are also discrepant. Certain implementations of sub-grid baryonic physics might also play a role in determining misalignment angles \citep{Velliscig15b,Herle25}.

\subsubsection{Comparison between simulations}
\label{sec:compare}

The discussion of whether blue/disc galaxies align and to what level is still open. Comparing the predictions from different numerical simulations requires applying uniform selection criteria and constructing observables in similar ways. Even then, the galaxy populations present in the simulations might be different due to resolution effects, the choice of galaxy and halo finders, and sub-grid modelling. An option is to weight galaxies in one simulation to better match the host halo mass distribution \citep{Samuroff21}. However, this does not fully solve the problem of the intrinsic properties of galaxies being different and the findings summarized in this sub-section should be considered in that light. 

\begin{figure}[ht]
    \centering
    \includegraphics[width=0.5\linewidth]{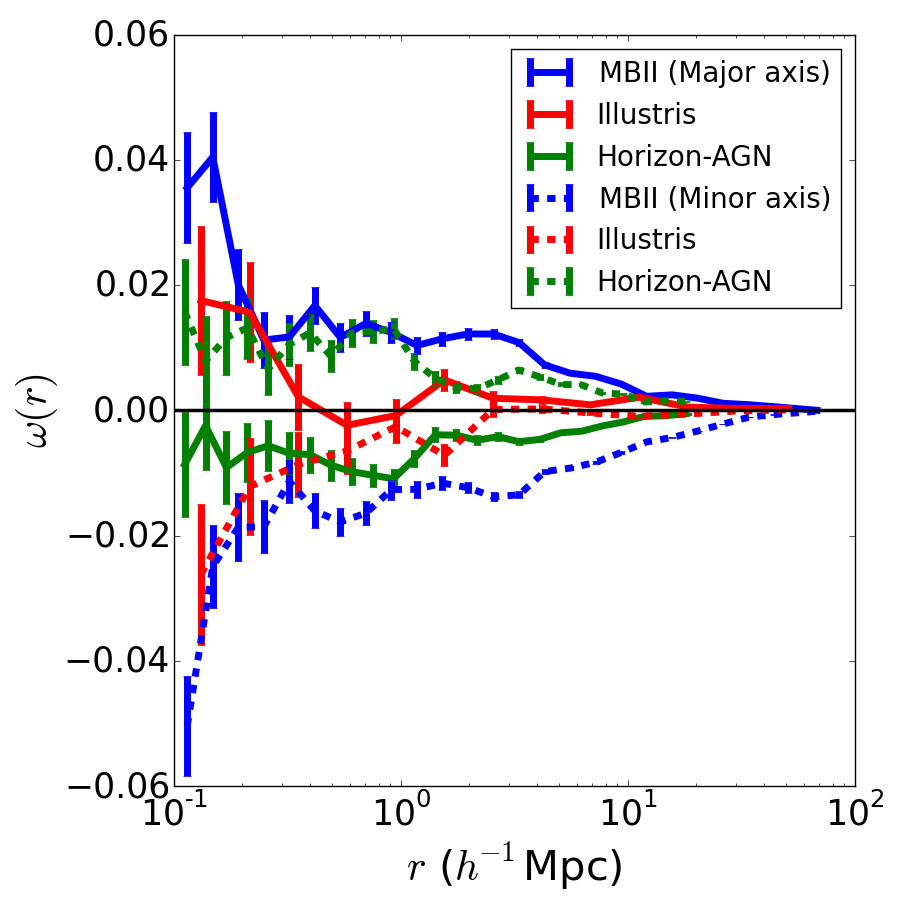}
    \caption{The ED correlation function, $\eta_e$ (here $\omega(r)$) of disc galaxies around the location of ellipticals in three different hydrodynamical simulations: Horizon-AGN, MB-II and Illustris. The solid lines quantifies orientations via the major axis. The dashed line uses the minor axis of the galaxy ellipsoid. The statistics are measured at high signal-to-noise ratio thanks to the availability of three-dimensional orientations. Horizon-AGN displays a signal which has the opposite sign to the other simulations. Credit: \citet{Tenneti16}. Image reproduced with permission from \citet{Tenneti16}, copyright by the author(s).}
    \label{fig:firstcomp}
\end{figure}

\citet{Tenneti16} presented the first comparison of alignment predictions across numerical simulations. Using iterative RIT shape measurements, they measured misalignment angle distributions, $\eta_e(r)$ and $w_{\delta+}$ correlations in Illustris and MB-II. They found that galaxies are typically rounder in Illustris and misalignments with the host halo more pronounced, leading to suppressed alignment signals in general. They split the galaxy population into discs and ellipticals based on a dynamical ``bulge-to-total ratio'' (BTR, \citealt{Scannapieco09}). With this criterion, more galaxies were classified as discs in Illustris than in MB-II at fixed stellar mass. Still, the alignments of discs around ellipticals was more prominent in MB-II and of opposite sign to Horizon-AGN (see Fig.~\ref{fig:firstcomp}). In addition, the disc alignment signal was only present at small scales and the $w_{\delta+}$ amplitude was boosted because of the elongated shapes of discs.

\begin{figure}[ht]
    \centering
    \includegraphics[width=0.9\linewidth]{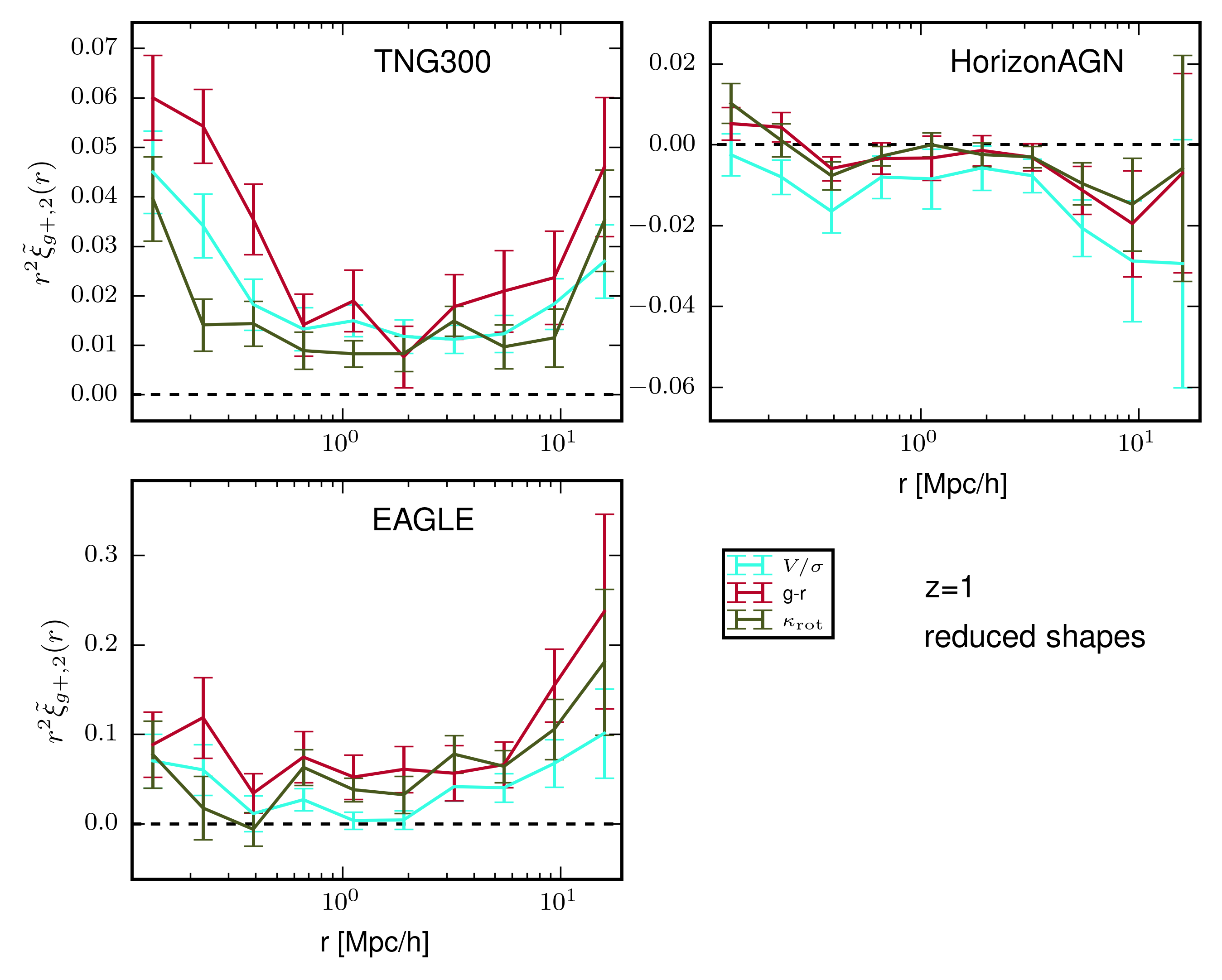}
    \caption{The quadrupole of the alignment signal of discs around ellipticals in three cosmological hydrodynamical simulations at $z=1$ for galaxies selected by colour, $V/\sigma$ and fractional rotational energy. Shapes are measured using the RIT. Horizon-AGN displays a negative signal when the selection is performed on $V/\sigma$, while disc alignment signals in the two other simulations are positive. Credit: adapted from van Heukelum et al. (in prep.), reproduced with permission of the author(s).}
    \label{fig:vH}
\end{figure}

A further comparison between intrinsic alignment predictions from hydrodynamical simulations was performed by \citet{Samuroff21} using TNG300, MB-II and Illustris. They selected well-resolved galaxies ($N>300$) with a lower mass threshold of $1.6\times 10^9\,h^{-1}{\rm M}_\odot$ and at least $1000$ particles in the dark matter halo. The shapes of galaxies in Illustris, measured using the SIT, were rounder than in other simulations or observations. Nevertheless, they found generally consistent $A_{\rm IA}$ values for the fiducial sample, with slightly higher values in MB-II. A slightly negative $A_2$ amplitude found for MB-II does not invalidate the fact that MB-II galaxies still show a radial alignment signal, compared to the case of Horizon-AGN or SIMBA, where the alignment is of opposite sign. While a tangential alignment signal has been found in \citet{Shi21b} for the spin of low mass discs in TNG300, \citet{Samuroff21} reported none based on a colour separation of galaxy populations in the same simulation.

van Heukelum et al. (in prep.) compared the measured alignment signal in three different cosmological hydrodynamical simulations (TNG300, EAGLE and Horizon-AGN) for galaxies selected in the same way. When necessary, outputs were re-processed to ensure that the selection was performed uniformly across all simulations. The analysis was performed for both SIT and RIT at $z=0$ and $z=1$. Galaxies were put into different bins according to their $g-r$ rest-frame colour or dynamical variables such as $V/\sigma$ or $\kappa_{\rm rot}$, the fraction of kinetic energy in coherent rotational motion. Only $10\%$ of the galaxies were labelled as ``ellipticals'' due to some of the simulations not clearly showing bimodal distributions (i.e., a red sequence). 

van Heukelum et al. found that TNG300 and EAGLE show consistent results in terms of disc and elliptical galaxies displaying both significant alignment signals of the same sign. All galaxies in these simulations are aligned radially towards overdensities regardless of redshift or shape measurement choice. On the other hand, Horizon-AGN displayed a disc alignment signal compatible with null when the correlation was performed around all galaxies in the simulation. A tangential alignment signal was found for Horizon-AGN discs around elliptical galaxies at $z=1$ when the RIT was adopted (in line with \citealt{Chisari16}). As shown in Fig.~\ref{fig:vH}, the signal has the opposite sign in TNG300 and EAGLE. 

This work further confirms that disc alignments in cosmological hydrodynamical simulations are not yet robustly predicted. Fortunately, whether positive or negative, the signal is small. Still, its amplitude is sufficient to be a source of contamination for cosmic shear surveys in the Stage IV era \citep{Chisari16}.

Finally, the only alignment measurement available from the Magneticum Pathfinder simulation\footnote{\url{http://magneticum.org/}} \citep{Hirschmann14} reports on the connection between the spin of a galaxy and a halo (see also Sect.~\ref{sec:spin}). \citet{Teklu15} used a 48 $h^{-1}$ Mpc on-a-side high resolution box to extract $\sim 500$ haloes for which the angular momentum of the DM halo, the stellar component and the gas could be measured. In addition, they classified the galaxies hosted by the halo into discs and ellipticals depending on their rotational dynamics \citep{Abadi03}. In summary, they found that gas and stars have angular momenta that are aligned with each other at high-redshift. For discs, this trend remains but for ellipticals, the alignment decreases as time passes. Moreover, the alignment between the stars and the dark matter halo’s angular momentum increases with time. These findings are in line with the results presented in \citet{Chisari17} for the case of Horizon-AGN.

\subsection{Mocks}
\label{sec:mocks}

High-resolution hydrodynamical simulations with reliable shapes have too small volumes compared to Stage III and IV surveys. To test mitigation schemes (Sec. \ref{sec:miti}), build covariance matrices for weak lensing applications or for cosmological applications, mock data over a sufficient volume is often needed. The primary strategy to create mock galaxy catalogues that include the intrinsic alignment signal is to populate haloes found in large-volume $N$-body simulations with galaxies oriented in specific ways. Early examples such as \citet{Croft00} directly used the shapes and alignments of dark matter haloes as a proxy for galaxy alignments. We know now that this methodology overestimates the alignment amplitude and yields a different redshift evolution (see Sect.~\ref{sec:gxyhalo}). 

\citet{Heavens00} included an additional recipe for the alignment of spiral galaxies which followed the orientation of the angular momentum of the haloes but continued to model elliptical alignment directly from the parent halo. 
Further measurements \citep{Mandelbaum06} implied that such modelling approach predicted a higher alignment than observed \citep{Heymans06}, but could not be completely disregarded. This called for including a misalignment angle between elliptical galaxies and their parent haloes in the production of mock catalogues \citep{OkumuraLRGs}, unless the central region of the halo is considered as a proxy for the central galaxy shape \citep{Faltenbacher09}. \citet{Heymans06} also suggested that the contamination to weak lensing would in fact be dominated by ellipticals, since their predicted spiral alignment signal, calibrated to reproduce hydrodynamical simulations \citep{vandenBosch02}, was consistent with null. 

The availability of the Hubble Space Telescope COSMOS survey data \citep{Scoville07} and the higher mass resolution of the Millennium $N$-body cosmological simulation \citep{Springel05} enabled two companion works \citep{Joachimi13,Joachimi13b} to improve upon the realism of mock aligned galaxies. To this goal, the authors implemented a comprehensive semi-analytical approach to modelling galaxy ellipticities and orientations informed by deep COSMOS data. They found that the intrinsic dispersion of galaxy ellipticities was different for early- and late-type samples, with weak dependence on magnitude and no evidence of redshift evolution (see also \citealt{Leauthaud07}) except for luminous late-type galaxies. This suggested that care needs to be taken when translating alignment priors from one survey to another because the fractional contribution of each population might differ. After testing different models for the galaxy ellipticities, they found that they are best reproduced when elliptical shapes mimic those of their parent haloes and spirals have a ratio of disc thickness to disc diameter of $0.1$. 

The question is then how to orient these galaxies in their haloes to produce a realistic alignment signal. \citet{Joachimi13b} implemented different prescriptions for central and satellite early- and late-type galaxy alignment (we reproduce a summary of Table 1 of that work in Table \ref{tab:galaxymodels}). They reported that early-type galaxies show a strong alignment that increases with redshift. Introducing a misalignment angle for centrals was not immediately necessary to reproduce the observational data, but a comparison to the work of \citet{Mandelbaum06} indicated that low-luminosity objects might not be as strongly aligned as assumed, whether centrals or satellites. Late-type galaxies only showed a small alignment signal at highly non-linear scales ($<5\,h^{-1}{\rm Mpc}$), without any significant dependence on luminosity or redshift, albeit with large error bars. They also concluded that models that use the RIT as proxy of the halo shape (which probe scales closer to the galaxy) and include misalignment of satellite galaxies yield better agreement with the data. 

\begin{table*}
\centering
\caption{Models for galaxy shapes adopted in \citet{Joachimi13b}. The distinction between central and satellite galaxies, and early- and late-type galaxies is based on the properties predicted by the semi-analytic model. Low-mass, poorly resolved, haloes are assigned galaxies with random orientations. This is an adapted version of Table 1 of \citet{Joachimi13b}, reproduced with permission of the author(s).}
\begin{tabular}[t]{llll}
\hline\hline
halo & galaxy & shape & alignment model \\
\hline
\hline
 & early & shape of halo; SIT & perfectly aligned\\
 & " & shape of halo; RIT & perfectly aligned \\
 & " & shape of halo; SIT & \cite{OkumuraLRGs} misalignment \\
central & late & disc with $r_{\rm edge-on}=0.25$ & disc $\perp {\bf L}$  \\
 & " & disc with $r_{\rm edge-on}=0.25$ & \cite{Bett12} misalignment  \\
 & " & disc with $r_{\rm edge-on}=0.1$ & \cite{Bett12} misalignment\\
\hline
 & early &  SIT from halo distribution & major axis $\rightarrow$ halo centre \\
 & " &  RIT from halo distribution & major axis $\rightarrow$ halo centre\\
satellite & " & \cite{Knebe08} modifications & \cite{Knebe08} misalignment \\
 & late-type & disc with $r_{\rm edge-on}=0.25$ & thick disc $\rightarrow$  halo centre\\
 & " & disc with $r_{\rm edge-on}=0.25$ & \citet{Bett12} misalignment\\
 & " & disc with $r_{\rm edge-on}=0.1$ & \citet{Bett12} misalignment \\
\hline
\end{tabular}
\label{tab:galaxymodels}
\end{table*}

\citet{Wei18} adopted the semi-analytic model of \citet{Joachimi13b} to incorporate intrinsic alignments in ray-tracing simulation to directly gauge their impact on cosmic shear observables. They similarly came to the conclusion that radially aligned satellite galaxies yield too much power on small scales. Intriguingly, they found that spiral galaxies contribute a positive $\delta E$ term which can be up to 15 per cent of the total shear correlation, but this is possibly a consequence of no misalignment being introduced for this sample. 

How much alignment is realistic? \citet{Hoffmann22} took advantage of low redshift observations of LRG alignments from BOSS \citep{SinghBOSS} to calibrate a semi-analytical model in the MICE $N$-body simulation. This allowed them to produce a realistic mock over $5000\,\deg^2$ up to redshift $z\sim 1.4$. Their goal was to establish whether adopting NLA vs. TATT in the cosmic shear analysis of the DES Y3 analysis would incur in biases in the cosmological parameters (see Sect.~\ref{sec:miti}). While TATT outperformed NLA at small scales in fitting $w_{\delta+}$, both models were acceptable for a DES Y3 analysis. This conclusion is likely due to the fact that intrinsic alignments predicted from these mocks were sub-dominant with respect to the measurement uncertainty. 

A completely different approach to generate intrinsic alignment mocks was presented in \citet{HD22}. The authors used the projection of the density field of $N$-body simulations onto the sky \citep{SLICS,cosmoSLICS} to estimate the projected tidal field. They used the tidal field at each location of the simulated map to establish the intrinsic axis ratio and orientation of galaxy, following the NLA model. Optionally, they also included density-weighting by multiplying intrinsic ellipticities by a factor $1+b_{\rm TA}\delta$ (see Sect.~\ref{sec:tatt}). This simple approach allows one to predict intrinsic alignment contamination to multiple observables quickly and consistently for a variety of cosmologies, and to marginalise over the free intrinsic alignment parameters for any estimator. The compromise is in the realism of the mocks, which are produced assuming a density-weighted NLA model. Some extensions, such as a red-blue split or additional TATT terms, would be straightforward to include.

Going further into the small scale regime, \citet{vanAlfen24} implemented an HOD model using the public software {\tt halotools}\footnote{\url{https://github.com/astropy/halotools}} to populate $N$-body simulations with central and satellite galaxies preferentially misaligned with respect of their host (sub)halo using a Dimroth-Watson distribution. They also considered a radial alignment model for satellites that overcomes missing subhalo information. In comparison to other works, \citet{vanAlfen24} explored the impact of assuming a dependence of satellite alignment with the distance to the centre of the halo. They found little difference between the cases where the alignment depends on this distance or is constant. They also allowed for anisotropy in the satellite distribution within a halo, finding that isotropic models miss a significant amount of power in the one-halo regime ($30-60\%$ depending on the statistic considered) which is needed to calibrate the model against the alignment correlation functions of TNG300. Further evidence on the significance of this contribution to the intrinsic alignment signal is provided by \citet{Rodriguez23}, who used the ACF statistic in TNG300 to show that red central galaxies show increasing alignment with the distribution of nearby galaxies as a function of mass.

\citet{Jagvaral22} benefitted from new developments in machine learning to produce intrinsic alignment mocks calibrated with a hydrodynamical simulation. Specifically, they used a deep generative model to produce intrinsic alignments mocks calibrated on the TNG100 simulation. The network was trained to produce three-dimensional shapes of haloes and galaxies, as well as projected shapes of galaxies. The resulting projected correlation function matched very well the one measured in TNG100. Even when splitting samples by mass, the predictions agreed with the simulation relatively well, with some shortcomings on the amplitude of $w_{g+}$ at low and intermediate mass. This methodology is agnostic to any specific semi-analytical model such as the ones adopted in Table \ref{tab:galaxymodels}. There is no need to model centrals and satellites specifically either. In \citet{Jagvaral24}, the authors extended the methodology to successfully predict galaxy colours and sizes, in addition to shapes and orientations. In the future, it should be tested whether it can also yield accurate calibrations with respect to other hydrodynamical simulations, or when working with a lower resolution halo catalogue.

Eventually, observables need to be constructed from each realisation of the mock catalogue. To accelerate this process, \citet{Pandya24} has recently proposed the use of deep learning for generating clustering and intrinsic alignments correlation functions given a set of HOD parameters, building on the work of \citet{vanAlfen24}. In the future, one would expect that machine learning will be able to predict the contamination of intrinsic alignments to the weak lensing observables directly.

\section{Astrophysics}\label{sec:astro}

The aim of this section is to discuss in more depth what physical processes might determine whether a galaxy is aligned or not. We know that haloes and galaxies have different alignment amplitudes and redshift evolution and for haloes, mass is a key driver of alignment amplitude. It is also clear that both can be accurately described by the LA model on large scales, but analytical estimates of the alignment amplitude (see Sect.~\ref{sec:priors}) fall short of predicting the level of alignment experienced by these objects. As a consequence, one can hypothesize that whatever process makes them align needs to have a correlation with the tidal field, but might not necessarily be rooted in tidal stretching. Alternatively, if tidal stretching is responsible for galaxy alignments, it should be primordial or at the very least build up over time sufficiently fast to explain the observed alignment amplitude. 

In Sect.~\ref{sec:lpt}, we discussed how to advect the LPT predictions of galaxy shapes to predict low redshift intrinsic alignments. \citet{Schmitz18} argued that if intrinsic alignments are imprinted at some high redshift and evolve passively until today via advection, there should be observable signatures of this process. Specifically, advection gives rise to a non-local dependence of shapes on the cosmological fields, i.e., a dependence on the history of these fields, which manifests itself in the redshift evolution of the $c_t$ coefficient in the LPT expansion (Eq. \ref{eq:lptshape}). 

In addition, there seems to be a dichotomy in the dependence of alignment amplitude with galaxy colour. We already argued in Sect.~\ref{sec:obs} that intrinsic alignments are typically observed for red samples, but not for blue galaxies. While simulations paint a different picture and often find blue galaxy alignment to be significant (see Sect.~\ref{sec:sims}), this could be a consequence of missing or incorrectly modelled sub-grid physics, or lack of sufficient resolution. From the observational evidence at hand, one would argue that whatever physical mechanism aligns red galaxies should not be the same as for the blue population. Often, blue galaxies are equated to being disc-like and less sensitive to tidal stretching \citep{Ghosh24}. Instead, their angular momentum is torqued by surrounding structures, which leads to the second order term we discussed in Sect.~\ref{sec:model}. 

One such possible mechanism, which could act through long periods of time is mergers. If mergers happen along preferential directions that are correlated with the tidal field, and if red galaxies preserve a memory of this direction, this could explain their alignment.  In this scenario, the lack of alignment of blue galaxies could be explained by less frequent mergers, or by a constant rebuilding of the gaseous disc from gas accretion \citep{Dubois16}. In \citet{Bate19,Bhowmick19,Rodriguez24}, the authors examined the alignment of the progenitors of low-redshift massive galaxies over time. \citet{Bate19} found the elliptical progenitors in Horizon-AGN to be progressively more disc-like at high redshift and transitioning from no alignment at $z=3$ to a significant radial one at $z=0$. Intriguingly, in comparison, the overall population of discs at $z=3$ did show evidence of tangential alignments in the same simulation. \citet{Bhowmick19} and \citet{Rodriguez24} also found a monotonic trend for the progenitors of massive galaxies to be less aligned at high redshift in MB-II and TNG300, but no overall tangential alignment for the disc population. This discrepancy between the findings of \citet{Bate19} and \citet{Bhowmick19} can be related to the overall discrepancy in alignment predictions made by different hydro simulations as discussed in Sect.~\ref{sec:sims}. 

A few works have looked at the interplay between sub-grid physics models and alignments. 
\citet{Tenneti17} investigated intrinsic alignments in several variations of the MB-II simulation \citep{Khandai15} where star formation and stellar and Active Galactic Nuclei (AGN) feedback recipes were altered. They found intrinsic alignments to be mostly insensitive to these variations, excepts perhaps for a weak dependence on the strength of stellar winds, which might be able to remove angular momentum from a galaxy. 

\citet{Soussana20} compared two versions of the Horizon-AGN simulation with and without AGN feedback. In contrast to \citet{Tenneti17}, they found that the intrinsic alignment signal changed in the simulations mainly as a consequence of the change in number of elliptical galaxies present. The appearance of larger simulation suites with controlled variations of baryonic feedback scenarios (e.g., \citealt{Camels,vanDaalen20,Schaye23}) will allow one to explore the sensitivity of alignments to these physical processes further in the near future.   

\subsection{Connection to angular momentum alignments}\label{sec:spin}

As described in Sect.~\ref{sec:tatt}, one of the mechanisms thought to be responsible for intrinsic alignments is tidal torquing. Tidal torquing results in a correlation between halo or galactic spins that translates into a correlation of their shapes. 
An ample literature on spin alignments exists on the formulation and validation of tidal torque theory, reviewed in \citet{Schafer09}, as well as more recent numerical \citep{Lee22,Lee24} and theoretical developments \citep{Codis15b,Lopez21,Lee23,Lopez24}. 

An interesting prediction of tidal torque theory is the fact that a mass transition scale exists above which the spin of an object orients itself perpendicular to its host filament \citep{Codis15b}. Such a transition could be related to the different signs of alignments for disc and ellipticals evidenced in several hydro-dynamical simulations \citep{Chisari16,Codis18,Kraljic20,Shi21b,Moon23}. 

Attempts to directly measure spin correlations through de-projection \citep{Pahwa16,Motloch21}, resolved kinematics \citep{Welker20,Kraljic21} or quasar polarization \citep{Hutsemekers14,Friday22} could help test these predictions directly. Finally, the kinematic lensing technique proposed by \citet{Huff13} might also serve as a probe of spin alignments \citep{Huang24}. 

\section{Cosmology}\label{sec:cosmo}

If we start from the premise that there is a connection between intrinsic alignments and the tidal field, we immediately realize that intrinsic alignments should contain cosmological information. Na\"ively, we would expect them to be sensitive to cosmology in a way similar to galaxy clustering (because there are no additional distance kernels involved). However, because intrinsic shapes are spin-2 quantities, there is some other interesting phenomenology to be harvested. Moreover, for some of these applications observational constraints are already available. We review them below and we summarize the main references in Table~\ref{tab:cosmo}.

\begin{itemize}
\item {\it Baryon acoustic oscillations}. If alignments are set by the tidal field of the large scale-structure, then it follows that on large scales they should be sensitive to the BAO feature \citep{Chisari13}. The signature appears as a trough at the baryon acoustic oscillation scale --- a consequence of the tidal field operating on a shell of mass \citep{vanDompseler23}. The presence of the BAO feature in alignments has been confirmed by \citet{Xia17,Okumura19,Okumura20,Kurita21} using $N$-body simulations. An analysis of the BOSS CMASS sample was able to extract the feature \citep{Xu23} and showed that it increases the constraints on the distance to $z=0.57$ by 10\%.

\item {\it Growth of structure.} Considering linear theory alone, one could argue that intrinsic alignments can complement geometric and dynamical constraints on dark energy \citep{Taruya20}. The clustering power spectrum is not measured in comoving space but in angular and redshift space. The consequence is that there is cosmological information in the transformation between redshifts and distances, i.e. the Alcock--Paczynski effect \citep{AP}. Furthermore, galaxy clustering is sensitive to RSD, which constrain the linear growth rate, $f$ \citep{Kaiser87}. The same is true for position-shape correlations. (In practice, there is a degeneracy with the power spectrum amplitude, such that the actual constraint is on $f\sigma_8$.) On the other hand, the intrinsic alignment auto-spectrum is insensitive to RSD, and thereby helps break the degeneracy between growth and distances. 

A Fisher forecast performed in \citet{Taruya20} suggested up to a factor of $2$ gain on the relevant cosmological parameters when BOSS data is considered (marginalizing over bias and intrinsic alignment amplitude). This proposal was put to practice in a re-analysis of intrinsic alignments and clustering in \citet{Okumura22}, which focused on constraints on the growth alone. The resulting improvement was $\sim19\%$, compared to considering clustering alone. \citet{Zwetsloot22} reported more modest gains in a Fisher forecast with a similar set-up, the reason for which is the inclusion of selection effects induced by alignments in the clustering correlations. The distribution of cosmological information in the intrinsic alignment multipoles depends on the basis of expansion \citep{Inoue24}. Further improvements on how much we might learn about the growth of structure from intrinsic alignments might be possible by adding information from velocity-shape correlations, where the velocity field is constrained from the kinetic Sunyaev--Zeldovich effect \citep{vanGemeren20,Okumura22}.

\begin{landscape}
\begin{table}
    \caption{Cosmological applications of intrinsic alignments and the corresponding references for theoretical modelling, validation in simulations and application to observations, when available. In addition, \citet{Philcox24} provides a general treatment of tensor and vector perturbation signatures in intrinsic alignments.}
    \label{tab:cosmo}
    \centering
    \begin{tabular}{p{3.5cm} p{5cm} p{4cm} p{4cm}}
    \toprule
     Application  & Theory & Simulations & Observations  \\
     \midrule
     Growth of structure  &\cite{Taruya20,Zwetsloot22,vanGemeren20,Okumura23} &  - & \cite{Okumura22}\\
     Baryon acoustic oscillations  & \cite{Chisari13,vanDompseler23}  &  \cite{Xia17,Okumura19,Okumura20,Kurita21} & \cite{Xu23} \\
     Primordial non-Gaussianity  &\cite{Schmidt15,Kogai18,Kogai21}  & \cite{Akitsu21}& \cite{Kurita23} \\
     Massive $s\neq 0$ fields  & " & - & - \\
     Primordial magnetic fields    & \cite{Schmidt15,Saga24} & - & - \\
     Gravitational wave background  &\cite{Schmidt12,Schmidt14,Chisari14,Biagetti20} & \cite{Akitsu23b} & -\\
     Parity violation & \cite{Vlah21,Biagetti20,Yin25} & -& - \\
     Isotropy & \cite{Shiraishi23} & - & - \\
     Modified gravity   & \cite{Reischke22} & \cite{LHuillier17,Chuang22} & - \\
     Relativistic effects   & \cite{Saga23} & - & - \\
     Nature of dark matter & - & \cite{Harvey21,Dome23} & -\\
    \bottomrule
    \end{tabular}
\end{table}
\end{landscape}

\item {\it Primordial non-Gaussianity.} Intrinsic alignments are sensitive to primordial non-Gaussianity in a way that is different and complementary to galaxy clustering. Alignments respond to anisotropic types of non-Gaussianity. The first theoretical derivation of this effect was presented in \citet{Schmidt15}. \citet{Akitsu21} confirmed this prediction using $N-$body numerical simulations and the first observational constraint was produced by \citet{Kurita23}. Constraints could potentially be improved by adopting a multi-tracer \citep{SeljakMulti} approach that takes advantage of the scale-dependence of galaxy shapes, as described in \citet{ChisariMulti}. This has not been realized observationally yet. The CMB bispectrum also probes this type of non-Gaussianity, but at a different physical scale. Therefore, combining both probes is interesting in order to constrain the composition of the primordial Universe.

\item {\it Massive non-zero spin fields.} Several references have proposed that galaxy shapes can be a probe of spin-2 particles in the early Universe. This is in fact one of the ways that the primordial non-Gaussianity signal might arise. 
\citet{Kogai18} extended the work done in \citet{Schmidt15} by including scale-dependence in the amplitude of the squeezed primordial non-Gaussianity, consistently with the presence of massive non-zero spin particles in the early Universe. This led to a different scale-dependence of the alignment bias (more prominent at small scales than in the case of \citealt{Schmidt15}) and distinct oscillatory features. \citet{Kogai21} generalized this work to suggest that studying higher moments of galaxy shapes could distinguish between primordial non-Gaussianity signatures generated by massive particles with different spins, as predicted for example by string theory. 

\item {\it Gravitational wave background.} When gravitational waves from inflation enter the horizon, they decay very quickly. According to \citet{Schmidt12} and \citet{Schmidt14}, they can still impart a long-lasting fossil effect on the shape of galaxies (or any other spin-2 tracers) in the Universe. This is evidenced as intrinsic alignments $B$-modes and can be searched for in auto-correlation. Cross-correlation with other probes that are sensitive to the same gravitational waves (e.g., CMB $B$-modes, \citealt{Chisari14}) could also potentially provide confirmation of the primordial origin of the signal. 

\citet{Akitsu23b} provided a numerical confirmation of the prediction that alignments are sensitive to large-scale gravitational waves via separate universe simulations. According to \citet{Biagetti20}, the detection prospects are however limited given the recent upper limits on the tensor-to-scalar ratio, $r$, from the last \textit{Planck} release \citep{Tristram22}. The same authors have argued that intrinsic alignment $B$-modes have a bigger role to play in the context of parity violation. Still, the sensitivity of different intrinsic alignments observables to gravitational waves of different frequencies remains relatively unexplored. \citet{Philcox24} provides a general formalism to predict the consequence of not only tensor but also vector perturbations (such as from cosmic strings) in the weak gravitational lensing and the intrinsic alignments of galaxies and their higher-order distortions \cite[e.g.,][]{Bacon06}. They argue for the use of both observables to distinguish between different physically interesting scenarios.

\item {\it Parity violation.} Parity is violated when the laws of physics change under the inversion of spatial coordinates. We know this to be realised within the standard model in the weak sector, but also in many beyond-standard-model scenarios. If the physics of the early Universe violated parity, this could be probed through the intrinsic alignments of galaxies. Parity-violating signatures would appear as a non-zero scalar-$B$ cross-correlation, or as an $EB$ cross-correlation \citep{Vlah20,Biagetti20,Vlah21}. There is a clear analogy to the case of CMB polarization, where up to $2.4\sigma$ evidence of $EB$ polarization has been claimed \citep{Minami20}. An explicit proposal for the use of galaxy shapes (in combination with radio polarization data) as a probe of such a cosmic ``birefringence" effect, driven by the  Chern-Simons interaction between
axions and electromagnetism, has recently been put forward by \citet{Yin25}. Observational constraints from intrinsic alignments on parity-breaking are not yet available. 

\item {\it Magnetic fields.} Magnetic fields are known the permeate the Universe. Their origin remains a mystery and it is particularly intriguing that they have been found in very low density regions of the Universe \citep{Neronov10}. This has suggested that they do not necessarily originate in galaxies, but they are instead primordial. \citet{Schmidt15} suggested that intrinsic alignments would be sensitive to anisotropic non-Gaussianities sourced by primordial magnetic fields. More generally, if magnetic fields are present in the early Universe, they can source metric perturbations (mainly vector and tensor ones) that could be detectable through different observables, including galaxy alignments \citep{Saga24}. Similarly to primordial gravitational waves, evidence of primordial magnetic fields could be found in the $BB$ correlation of intrinsic alignments on large scales, sourced by the corresponding tensor perturbations. The minimum detectable primordial magnetic field by a Stage IV survey would be of $30-300\,{\rm nG}$ according to \citet{Saga24}. However, this would depend on how efficiently galaxy shapes might respond to such a large-scale tensor perturbation.

\item {\it Isotropy.} \citet{Kogai18} proposed that intrinsic alignments can be used as a probe of the isotropy of our Universe. In many inflationary, dark matter or dark energy scenarios, cosmic isotropy is broken. This is often associated to the presence of vector fields. \citet{Kogai18} found that when isotropy is broken by the presence of vector fields: $B$-modes appear, there is a non-zero correlation between different multipoles $l,l'$ and the angular power spectra carry an azimuthal dependence. Assuming the same type of isotropy-breaking model, \citet{Shiraishi23} constructed specific intrinsic alignment observables beyond the plane-parallel approximation designed to probe this phenomenon. Their work is a generalization of isotropy-violating studies based on lower spin observables (such as clustering or peculiar velocities). For a specific mock isotropy-violating model, the authors were able to show that a signature dependent on the direction of anisotropy appears in the two-point functions of intrinsic alignments. 

\item {\it Modified gravity.} Little is known about how intrinsic alignments manifest themselves in modified gravity theories. In principle, modified gravity theories could induce a different time-dependence of the alignment signal or even introduce new operators. \citet{Reischke22} investigated the specific case of Hordenski theories. Assuming no new operators are present and all changes in the alignment signal depend on the gravitational potential, their Fisher analysis showed that intrinsic alignments can improve the constraints on the two free parameters of the theory ($\hat \alpha_B,\hat \alpha_M$) by $10$ and $30\%$, respectively, only if the alignment amplitude is known a priori. \citet{LHuillier17} found the large-scale alignment of haloes with the cosmic web to be insensitive to $f(R)$ in their simulations, while \citet{Chuang22} reported significant signatures comparable to those in clustering statistics. Possibly, the use of $w_{g+}$ in \citet{Chuang22} vs. alignment angles in \citet{LHuillier17} can explain this contradiction.

\item {\it Relativistic effects.} \citet{Saga23} have argued that galaxy position-intrinsic shape correlations can probe relativistic effects such as the gravitational redshift and RSDs (\citealt{McDonald09b}, see also Sect.~\ref{sec:model}) in an independent way from galaxy clustering and without the need to perform sample splits \citep{Saga22}. Relativistic effects leave an imprint on the position-intrinsic shape dipole that could be detectable with future surveys. This provides a test of the equivalence principle and alternative theories of gravity \citep{Bonvin18}.

\item {\it Nature of dark matter.} If dark matter deviates from the standard weakly interacting massive particle (WIMP) scenario, it might be possible to find a signature of this in the intrinsic alignments of galaxies. For now, there is no general formalism that can be used (\`a la EFT) to determine what generic signatures might arise in the alignment signal for different dark matter candidates. Moreover, because evidence of non-standard dark matter models often appears only in the non-linear regime, numerical simulations are needed. 

\citet{Harvey21} studied the impact of self-interacting dark matter (SIDM) on the intrinsic alignments of galaxies for the first time using the BAHAMAS-SIDM simulation suite, a fully hydrodynamical SIDM cosmological simulation. By measuring $w_{g+}$ for the shapes of central galaxies around other galaxies, and comparing them between SIDM and CDM simulations, they showed that SIDM produced a mass-dependent suppression of alignments, at the level of up to $50\%$. This suppression was present on scales of even tens of Mpc, which provides an advantage compared to other observables, where the impact of SIDM is restricted to the cores of galaxy clusters. Moreover, their results showed that in principle this signature can be distinguished from baryonic effects. (Note, however, that baryonic feedback parameters were not re-calibrated for SIDM compared to the CDM runs.) 

More recently, \citet{Dome23} investigated the impact of assuming a fuzzy dark matter (FDM, where dark matter is comprised of a family of ultralight bosons \citealt{Hui17}) scenario on the shapes and alignments of haloes in the cosmic web at $z\sim3-4$ and over scales of $k \sim 2-18\,h\,{\rm Mpc}^{-1}$. The motivation to study these alignments at high redshift lies on the fact that this regime would be more sensitive to differences between CDM and FDM. Examining the shape-shape and shape-position alignment angles in three-dimensions for haloes in a mass range of $10^{10}-10^{11}\,h^{-1}\,{\rm M}_\odot$, the authors found that alignment is enhanced as the mass of the FDM particle increases. They also estimated the alignment amplitude (proportional to $A_{\rm IA}$) for their halo sample and find that it can differ by up to $6.4\sigma$ from the CDM scenario at $z=4.38$. While this discrepancy is significant and could give tantalizing hints of FDM, accessing this regime of intrinsic alignments observationally remains a challenge.

\end{itemize}

One might worry about how to distinguish these cosmological signatures from systematics effects or a variation of the intrinsic alignment signal itself (e.g., intrinsic redshift evolution). Intrinsic alignment measurements rely on having a large number of good quality shapes available over large areas of the sky. Effectively, this is also a requirement for weak lensing measurements and in practice, the shapes used for intrinsic alignment studies most often come from such surveys. Therefore, care has been taken to understand at what level PSF contamination might be present. If the shapes used in an alignment study are not weak lensing shapes, then additional studies might be required to guarantee that the measured correlations are systematics-free. Ideally, an intrinsic alignment measurement also requires well-calibrated redshift distributions (ideally, even spectroscopic data). Care must also be taken to model lensing as a contaminant. 

Most of the cosmological applications of intrinsic alignments listed above can be easily distinguished from variations in the cosmology or the astrophysics of alignments. For example: parity-violation would give a distinctive $EB$ cross-correlation, and anisotropic non-Gaussianity signal yields a scale-dependent alignment bias on large-scales. But some cases are harder to distinguish or require prior knowledge on the alignment amplitude (e.g., modified gravity). Some of the predicted constraints on the scenarios above could presumably also be improved if a multi-tracer approach is considered, as proposed for non-Gaussianity \citep{ChisariMulti}. 

\section{Mitigation}\label{sec:miti}

\begin{figure*}
    \centering \includegraphics[width=0.7\textwidth]{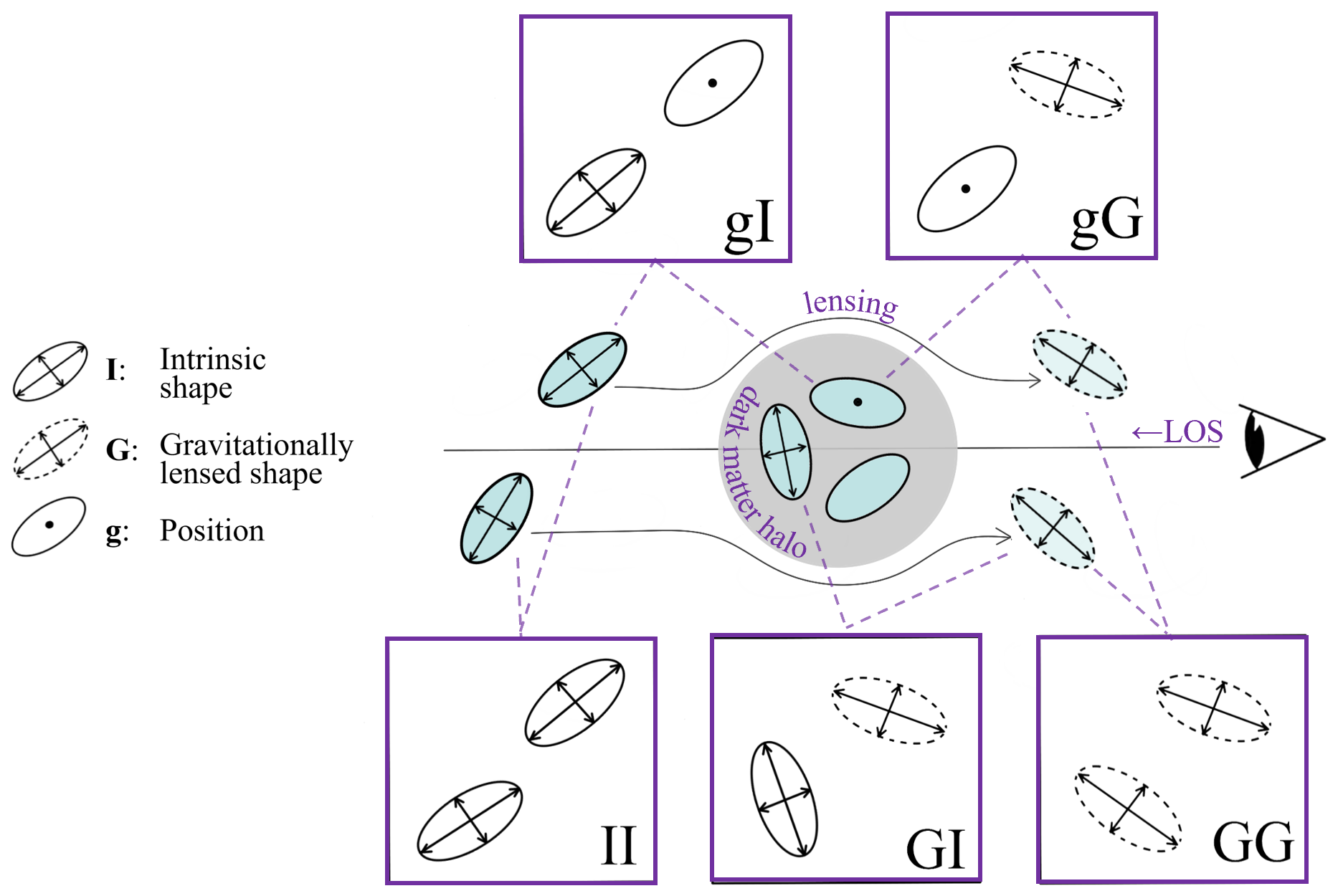}
    \caption{Different types of contributions to position-shape correlations (top) and shape-shape correlations (bottom). Galaxy shapes can be intrinsically aligned with respect to other galaxy positions ($gI$). Position tracers also lens the galaxies in the background ($gG$). At the bottom, we see the contributions to shape-shape correlations. Namely, $GG$ is the auto-correlation of lensing distortions (cosmic shear); $II$ is the auto-correlation of intrinsic shapes and $GI$ represents the lensing and intrinsic distortions sourced by the same matter field. Credit: \citet{Lamman23}.}
    \label{fig:contam}
\end{figure*}

Intrinsic alignments often act as a contaminant to other observables. The canonical example is gravitational lensing. As seen in Fig.~\ref{fig:contam}, the deflection of the path of photons along the line of sight is sourced by the clumping of matter across time. This effect is reconstructed from measuring coherent patterns in galaxy shapes. Unfortunately, such coherent patterns are similar, but usually of opposite sign, to the alignment patterns. I.e., if the alignment $E$-mode is positive, the lensing $E$-mode is negative. Without redshift information, lensed and aligned galaxies mix in projection. In addition, the shape $S_{ij}$ of each galaxy picks up a local alignment contribution as well as lensing ($G_{ij}$) by the foreground matter,
\begin{equation}
    S_{ij}({\bf x},z)=G_{ij}({\bf x},z)+I_{ij}({\bf x},z).
\end{equation}
Here, $i$ and $j$ run over the Cartesian indices of the projected shape tensor.

Separating the two signals is difficult. At low enough redshift ($z\lesssim 0.1$), one might be able to assume that lensing is negligible. At higher redshift, position-intrinsic shape correlations can be isolated using spectroscopic redshifts, but care must be taken to verify that magnification is not significant. If only photometric redshifts are available, which is often the case for lensing surveys which need to access galaxies in large numbers, separating the two signals becomes impossible and calls for joint modelling instead (but see Sect.~\ref{sec:selfcalib}). An additional complication arises from the fact that direct measurements of intrinsic alignments are performed for samples which are not representative of those observed by weak lensing surveys (i.e., because they are either spectroscopically-selected, too bright or color-selected).  

The intrinsic alignment contamination to lensing was first pointed out by \citet{Heavens00,Crittenden00,Catelan01} and developed further in multiple works, including a seminal paper by \citet{Hirata04}. To date, it is clear that alignments constitute a significant fraction of shape auto-correlations at low and intermediate redshifts \citep{Joachimi10,Kirk12,Krause16}, and even in cross-correlation with CMB lensing \citep{Hall14,Troxel14}. In principle, any study making use of galaxy shape information in auto- or cross-correlation with the large-scale structure should include intrinsic alignments.

In addition, we will see in Sect.~\ref{sec:selection} that there is yet another avenue for intrinsic alignment contamination to large-scale structure observables. If targets are selected or detected based on some quantity that is affected by their orientation, then intrinsic alignments will induce additional correlations of the targets with the large-scale structure. If significant, these would need to be modelled to avoid biases in cosmological parameter inference. The effect was identified in \citet{Hirata09} as a contaminant to redshift space distortion measurements from targeted samples for galaxy clustering.

While this section mostly focuses on how intrinsic alignments contaminate two-point statistics, other observables such as one- or three-point statistics of gravitational lensing, or three-point statistics of clustering (in the case of selection effects; \citealt{Krause11}) are generally also contaminated. A brief discussion of this issue can be found in Sect.~\ref{sec:beyond}.

\subsection{Contamination to gravitational lensing} \label{ssec:contam}

The shapes of galaxies in the Universe have a large intrinsic dispersion: there are elliptical, spirals, lenticular and irregular galaxies. Blending can have an impact on their distribution of ellipticities, as well as instrumental noise in the measurement of the shapes. On top of this intrinsic dispersion, the shape of a galaxy is generally distorted by gravitational lensing and intrinsic alignments. The gravitational lensing component correlates the shape of the galaxy with all the structure in front, while the intrinsic alignment component is thought to correlate the galaxy with all the structure in its vicinity (i.e., at the same redshift). Both effects are illustrated in Fig.~\ref{fig:contam}. 

\begin{figure}[htbp]
    \centering 
    \includegraphics[width=\textwidth]{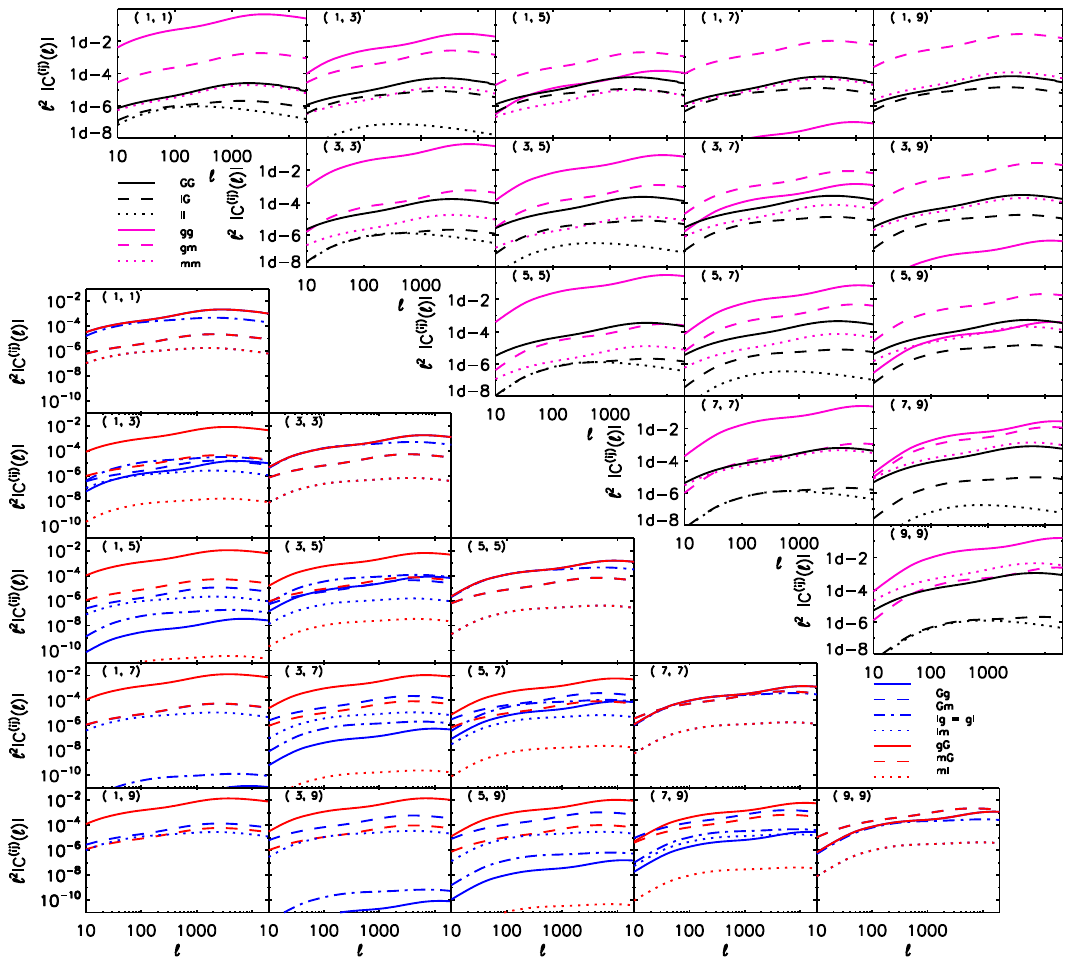}
    \caption{Angular power spectra from the NLA model for shape-shape, position-position (top) and position-shape (bottom), split by colour and line style in different contributions according to the notation of Eqs. (\ref{eq:sscontamination}) and (\ref{eq:cscontamination}). Results are presented for 9 tomographic bins for a Stage IV \textit{Euclid}-like survey. The assumed Gaussian photometric redshift error has a dispersion of $\sigma_z=0.05(1+z)$. Image reproduced with permission from \citet{Kirk12}, copyright by the author(s).}
    \label{fig:kirk12}
\end{figure}

In most situations, disentangling lensing from alignments is difficult and can only be done in specific regimes. For example, because lensing is very small at low redshift, any shape-shape or shape-position correlation between galaxies would be assumed to arise due to intrinsic alignments. Similarly, if galaxies with shapes can be reliably selected to be far behind a lens, one would assume that the lens-shape correlation will only be sourced by weak lensing. Unfortunately, in the majority of measurements of correlations involving galaxy shapes, the distinction between lensing and alignments is compromised by the fact that the uncertainty in the ensemble of galaxy redshifts is high. This prevents one from separating what is aligned from what is lensed. 

When trying to extract cosmological information from shape-shape correlations, we measure, for example, angular power spectra of the shapes of galaxies in redshift bin $i$ with the shapes of galaxies in redshift bin $j$:
\begin{equation}
    C_{ij}^{SS}(\ell) =C_{ij}^{GG}(\ell)+C_{ij}^{GI}(\ell) +C_{ij}^{IG}(\ell)+C_{ij}^{II}(\ell)
    \label{eq:sscontamination}
\end{equation}
The $C_{ij}^{GG}(\ell)$ correlation between redshift bins $i$ and $j$ indicates the cosmic shear contribution, which carries most of the cosmological information. The last term, $C_{ij}^{II}(\ell)$, is a pure intrinsic alignment auto-correlation, which is only thought to be prominent at low redshift. The cross-correlations $C_{ij}^{GI}(\ell)$ and $C_{ij}^{IG}(\ell)$ exist because the same matter field that is responsible for the alignment of a galaxy lenses another galaxy in the background. If bins $i$ and $j$ are disjoint and $I$ is the intrinsic shape of galaxies at redshifts further than bin $i$, then no correlation will exist. Aligned galaxies at higher redshifts know nothing about the matter field at lower redshifts.

In galaxy-galaxy lensing, the angular power spectrum we are interested in is that of position tracers at low redshift and galaxy shapes at high redshift, this is
\begin{equation}
C_{ij}^{pS}(\ell)=C_{ij}^{gG}(\ell)+C_{ij}^{gI}(\ell)+C_{ij}^{mG}(\ell)+C_{ij}^{mI}(\ell)
\label{eq:cscontamination}
\end{equation}
Here, the position tracer is sourcing the lensing of galaxies behind it ($g_iG_j$), it is aligning galaxies in its vicinity ($g_iI_j$) and it picks up two additional contributions from the impact of magnification. The position tracer is magnified and this magnification is generated by the same structure which distorts the shapes of galaxies, leading to the $m_iG_j$ term. The term $m_iI_j$ can only be produced if the magnified position tracers are tracing the same matter field which aligns the intrinsic galaxy shapes. Some works have proposed ways to simultaneously constrain $gI$ and $gG$ directly from the data \citep{Blazek12,Chisari14}, but these do not yet account for magnification effects. 

In a $3\times 2$pt analysis that combines galaxy positions with galaxy shapes, the dominant contaminant is normally thought to be the $GI$ term \citep{Hirata04}. Position-position correlations can also be contaminated by selection effects from intrinsic alignments, as we will see in Sect.~\ref{sec:selection}.
Note also that alignments themselves are contaminated by lensing. Especially if photometric redshifts are used, a measurement of position-shape correlations ($w_{g+}$) over a $2\Pi_{\max}$ baseline can pick up a significant $g_iG_j$ contribution that needs to be accounted for  \cite[e.g.,][]{Georgiou25}.

Finally, cross-correlations between other fields and galaxy shapes might be affected by intrinsic alignment contamination. The angular power spectrum of the cross-correlation between the lensing of the CMB and galaxy shapes in redshift bin $i$, for example, picks up an alignment contamination as follows:
\begin{equation}
C_{i}^{\kappa_{\rm CMB}S}(\ell)=C_{i}^{\kappa_{\rm CMB}G}(\ell)+C_{i}^{\kappa_{\rm CMB}I}(\ell)
\label{eq:cmbcontamination}
\end{equation}
This was first quantified in \citet{Hall14,Troxel14}, who suggested the presence of this contaminant and estimated it around $15\%$ for precursor surveys to Stage III using the NLA model. \citet{ChisariCMB} refined this estimate using different priors for red and blue galaxy alignments, and obtained a $10\%$ contamination and up to $9.5\%$ (at $2\sigma$) for each population, respectively. Additional contributions might be present given that the CMB lensing potential reconstruction picks up higher order contributions to the lensed temperature map \citep{Larsen16,Merkel17}. 
In principle, it should be straightforward to extend these calculations using perturbation theory. But this has not been explored much further, likely because such refinements are not needed given that the alignment contamination is small, and the lensing of the CMB probes mostly quasi-linear scales. 

\subsection{Selection effects from alignments}
\label{sec:selection}

Alignments create selection effects for spectroscopic surveys that can contaminate estimates of the growth history of structure. The growth factor is isolated from the impact of gravitationally-induced velocities on the redshift space statistics of galaxy clustering. This effect is called a redshift space distortion (RSD) \citep{Jackson72,Kaiser87}. In absence of any contaminant, the linear galaxy clustering power spectrum is given by 
\begin{equation}
P_g(k,\mu,z)=\left(b_{1,g}+f\mu^2\right)^2P_L(k,z).
\end{equation}
This signal is explicitly sensitive to the growth rate. To compress the information, an angular average is performed to obtain the quadrupole,  
\begin{equation}
 P_g^{(2)}(k,z)=\frac{5}{2}\int_{-1}^{1}d\mu\,P_g(k,\mu,z)\mathcal{L}_2(\mu), 
\end{equation}
with $\mathcal{L}_2$ the Legendre polynomial of order 2. The final expression for the quadrupole is
\begin{equation}
P_g^{(2)}(k,z)=\left(\frac{4}{3}b_{1,g}f+\frac{4}{7}f^2\right)P_L(k,z).
\end{equation}
and thus contains cosmological information.

\begin{figure}[ht]
\includegraphics[width=0.95\textwidth]{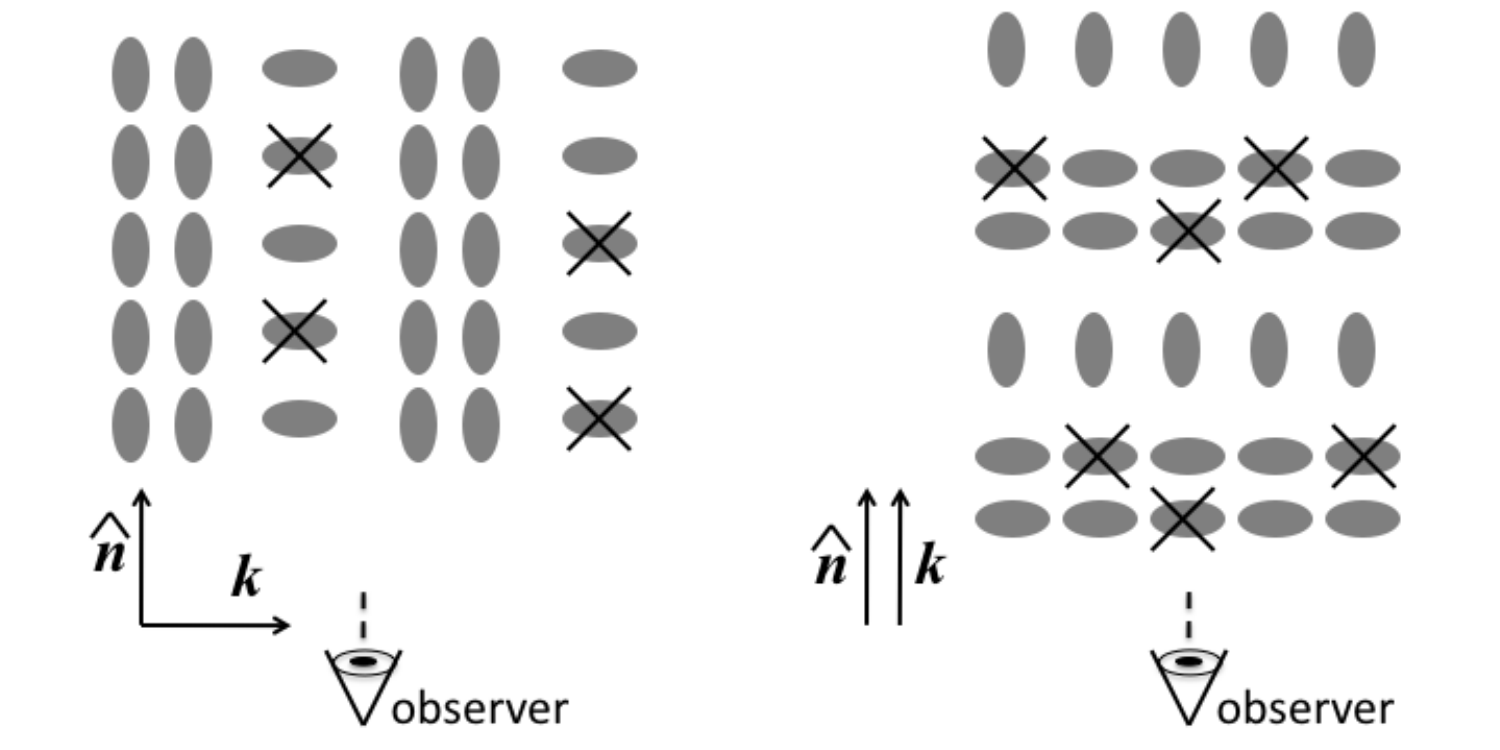}
\caption{A cartoon demonstrating how the preferential orientation of galaxies with respect to the line-of-sight, which is correlated with the large-scale structure, can induce selection effects. The left panel corresponds to a long-wavelength tidal field perpendicular to the line of sight; the right panel, to one parallel to the line of sight. Image reproduced with permission from \citet{Hirata09}, copyright by the author.}
\label{fig:selection}
\end{figure}

RSD contamination from intrinsic alignments was first proposed in \citet{Hirata09} and followed up on in multiple works: \citet{Martens18,Obuljen20,Singh21,Zwetsloot22,Lamman23,Lamman24}. The phenomenon comes about because the efficiency of selection of an object depends on the orientation, and this orientation is intrinsically correlated with the large-scale structure --- see sketch in Fig.~\ref{fig:selection}.

\citet{Hirata09} showed that the real-space galaxy density receives a correction when a selection effect that depends on the line-of-sight direction, $\epsilon(\hat{n}|{\bf x})$, is present:
\begin{equation}
1+\delta_g^{(\rm r,obs)}({\bf x},\hat{n})=[1+\delta_g^{(\rm r)}({\bf x})][1+\epsilon(\hat{n}|{\bf x})].
\end{equation}
For the specific case of selections induced by linear intrinsic alignments \citep{Catelan01}, the Fourier transform of $\epsilon(\hat{n}|{\bf x})$ takes the form
\begin{equation}
\tilde\epsilon(\hat{n}|{\bf k})=A\left[(\hat n\cdot \hat k)^2-\frac{1}{3}\right]\tilde\delta({\bf k}).
\end{equation}
The contribution to the galaxy power spectrum can then be easily obtained as
\begin{equation}
P_g(k,\mu,z)=\left[b_{1,g}-\frac{A}{3}+(f+A)\mu^2\right]^2P_L(k,z),
\end{equation}
where all power spectra are linear. $A$ is an amplitude which is correlated with the alignment amplitude $A_{\rm IA}$, but also includes a pre-factor that is sample dependent and encapsulates the efficiency with which a given target selection strategy responds to alignments with respect to the line of sight. Both monopole and quadrupole are contaminated and neither the linear bias of the sample nor the growth factor are recovered correctly if alignments are unaccounted for. (The explicit expressions can be found in the Appendix of \citealt{Zwetsloot22}.)

Detections of this effect for Stage III surveys are not entirely certain. \citet{Martens18} split BOSS galaxies (both CMASS and LRG samples, see Sect.~\ref{sec:obs}) based on their offset with respect to the fundamental plane \citep{FP}, a scaling relation between size, velocity dispersion and surface brightness of early-type galaxies. They assumed that any deviations from this plane originate in the galaxy's orientation, which is entirely given by their intrinsic alignments (plus noise). They then compared the clustering of two subsamples above and below the fundamental plane, and estimated their difference in intrinsic alignment from this comparison. This led them to claim a 2 to 3$\sigma$ detection of the impact of alignments on clustering, in line with theoretical predictions from \citet{Hirata09}. 

On the contrary, \citet{Singh21} performed an analysis of fundamental plane residuals that demonstrated they are correlated with other galaxy properties, such as luminosity, environment and colour. In addition, they found that while residual amplitudes are correlated with the intrinsic alignment amplitude, a model where this was the only source of scatter in the fundamental plane was not favoured by the data. Instead, they hypothesized that fundamental plane residuals could originate in other physical processes that are also correlated with environment. For further investigations on this topic, we refer the reader to \citet{Hearin19}, \citet{Johnston23} and \citet{Schafer24}.  

\citet{Obuljen20} examined the origin of anisotropic clustering signals in BOSS data using an entirely different methodology. They identified subsamples of galaxies with the same redshift distribution and clustering monopole, and looked for differences in the clustering quadrupole. When splitting the samples by their observed intrinsic size, they found no evidence of selection effects contaminating the clustering signal, in agreement with \citet{Singh21}. Instead, they suggested that selections relying on the stellar velocity dispersion would yield up to 5$\sigma$ evidence of contamination. Their results could therefore explain why those of \citet{Martens18} and \citet{Singh21} were in tentative conflict. 

For Stage IV surveys, the higher signal-to-noise ratio renders the selection effect from alignments potentially more problematic. \citet{Zwetsloot22} presented a forecast of how much contamination one should expect on the growth rate ($f\sigma_8$) if alignments are not modelled. The bias can be catastrophic (several $\sigma$) in this case. However, this can be mitigated if an informative alignment prior is available (e.g., from simulations) or if alignments are measured and modelled jointly. In that case, one might also gain information on the growth rate depending on the number density of the sample and the observable used for this purpose. (See also Sect.~\ref{sec:cosmo} for how alignments constrain the growth rate on their own.)

\citet{Lamman23} used mocks to predict the alignment contamination to the clustering of the LRG sample of DESI. DESI is a Stage IV spectroscopic survey covering approximately 16,000 sq. deg. of the sky and 20\% of its targets are LRGs in the redshift range from $0.4$ to $1$. Since LRGs normally show prominent intrinsic alignments, it is important to understand at what level clustering measurements from DESI might be contaminated. \citet{Lamman23} measured the alignment of this sample of galaxies directly from the data and used it to predict the contamination to the clustering quadrupole. They cross-checked their estimate with simulated mock catalogues. Their results suggested a $0.5\%$ fractional contribution from alignments to the clustering quadrupole at a typical scale between $40-80\,h^{-1}$ Mpc. They remark this is comparable to DESI's overall error budget at those scales (0.4--0.7\%) and must then be taken into account in the cosmological analysis. They also pointed out that the result is rather optimistic because at the time of the analysis, there were not enough spectroscopic redshifts to isolate physically close LRGs, and thus the alignment signal might be underestimated through a combination of dilution effects from photometric redshifts and lensing contamination. The option of performing quadrupole measurements in sub-samples less affected by selection effects was also considered, but not specifically pursued.

\subsection{Mitigation strategies}

The most commonly adopted strategy for mitigation of intrinsic alignments is to model them jointly with the observables being considered. Prior information on the alignment amplitude or how it translates to target selection are important, since totally agnostic studies might find it difficult to extract cosmological information at all \citep{Agarwal21}. Joint modelling of the alignment amplitude together with the luminosity function could also be beneficial \citep{Sarcevic25}. Self-calibration \citep{Zhang10} is a model-independent mitigation strategy to estimate alignment contamination based on exploiting the symmetry of the alignment signal around a lens. In this section, we will review both marginalization and self-calibration methods. 

In the early days, some data removal strategies were suggested to mitigate the impact of intrinsic alignments in weak lensing studies. Examples include the removal of red galaxies from the sample (known to align more strongly) \citep{Krause16}, or removing or weighting down close pairs in redshift space \citep{King02,King03,Heymans03,Heymans06}. The nulling technique \citep{Joachimi08,Joachimi09} similarly relied on the different redshift dependence of the lensing and alignment signals to achieve the mitigation. Most of these data removal strategies have been abandoned because they effectively worsen the statistical power of the surveys. An exception is the attempt to select galaxies by colour or type, as in \citet{Samuroff19,Li21,McCullough24}. Some level of marginalization is still often needed in those cases.

\subsubsection{Marginalization \& indirect constraints}
\label{sec:margin}

In weak lensing studies from Stage III and Stage IV surveys, a marginalization strategy is adopted by default. Of course, this requires a choice of model. Given the scales to be analysed and the precision of the data, the question is how many free parameters are needed and whether one has informative priors available for them. As a by-product, cosmic shear surveys indirectly set constraints on the alignment amplitude when performing their cosmological analyses. 

To illustrate the case of marginalization, we consider the example of the combination of the KiDS and DES Stage III surveys \citep{KiDS+DES}. As argued in Sect.~\ref{ssec:contam}, the cosmic shear signal picks up two contaminating alignment terms: $II$ and $GI$, cf.\ Eq.~(\ref{eq:sscontamination}). $II$ is most relevant at low redshift and in auto-correlated redshift bins. $GI$ is most relevant when the aligned galaxies are at low redshift and the sources are being lensed by the same overdensity field that drives the alignment. Due to the lack of spectroscopic redshift information, which would allow for a cleaner separation of the alignment signal, the strategy of both the KiDS and DES surveys is to model these terms jointly with the $GG$ term. 

The KiDS+DES analysis adopts as a fiducial choice the NLA model (Sect.~\ref{sec:linear}) and, when sampling over the cosmological parameter, it also samples and marginalizes over the alignment amplitude, $A_{\rm IA}$. Optionally, a power-law redshift dependence is included and the power-law index is also marginalized over. This is referred to as the NLA$-z$ model. The priors of choice for these parameters are between $[-5,5]$ or $[-6,6]$ for NLA only. One concern regarding the power-law parametrization of the redshift dependence of the alignment amplitude is that it does not allow a zero-crossing. Therefore, the alignment amplitude is assumed to preserve sign with redshift. In cosmic shear samples that mix red and blue populations, this could potentially be problematic if blue galaxies are confirmed to change alignment sign at high redshift \citep{Chisari16,Kraljic20} (see discussion in Sect.~\ref{sec:compare}). 

The final cosmic shear analysis of the KiDS survey: KiDS Legacy \citep{Wright25}, adopts several variations of the NLA model. The NLA$-k$ model adds the density-weighting term from TATT only, arguing for a similar scale dependence of other non-linear terms at the same order, which is exacerbated in projection. The NLA$-z$ model parametrizes the alignment amplitude in terms of a constant and a time-dependent term which is proportional to $(a/a_{\rm piv}-1)$, with $a_{\rm piv}$ some pivot scale factor. Finally, the NLA$-M$ model fixes the red fraction of galaxies in each tomographic bin (fixed) and assumes that $A_{\rm IA}$ has a power-law dependence with average halo mass (with the uncertainty on the mass marginalized over). 

\begin{figure}[ht]
    \centering
    \includegraphics[width=0.6\linewidth]{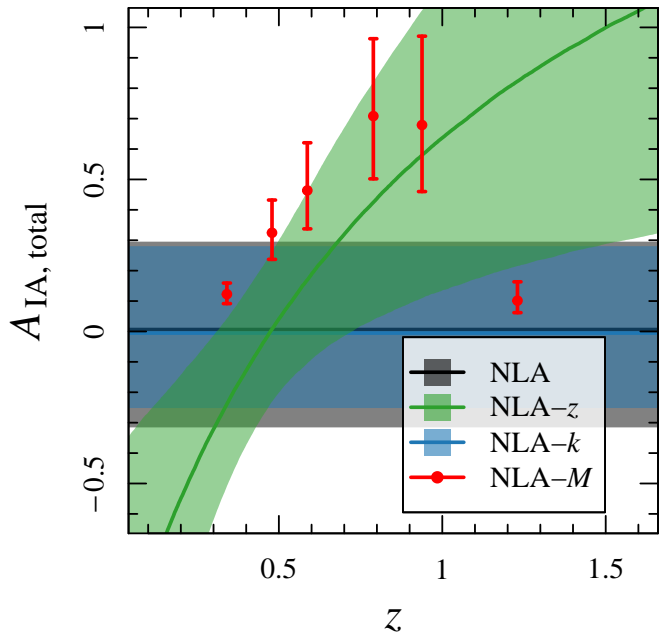}
    \caption{The intrinsic alignment amplitude as a function of redshift constrained by the KiDS Legacy survey through different versions of the NLA model. The NLA-$M$ fiducial model provides an $A_{\rm IA}$ constraint per tomographic bin where the cosmic shear signal is measured. The NLA$-z$ model frees the redshift dependence while NLA$-k$ and vanilla NLA assume the fiducial one from \citet{Hirata04}. Credit: \citet{Wright25}, reproduced with permission of the authors.}
    \label{fig:KiDSLegacy_AIA}
\end{figure}

This final model choice is adopted as the fiducial by the survey. The three model variations: NLA, NLA$-z$ and NLA$-k$ end up giving very small deviations from the fiducial, up to $0.54\sigma$. Most models yield similar constraining power in the cosmological parameters and goodness-of-fit, but NLA$-z$ loses $\sim 10\%$ power on $\Sigma_8=\sigma_8(\Omega_{\rm m}/0.3)^\alpha$, where $\alpha=\{0.58,0.6\}$ depending on the observable considered. Both NLA$-z$ and NLA$-M$ yield similar redshift-dependent $A_{\rm IA}$ posteriors, shown in Fig.~\ref{fig:KiDSLegacy_AIA}, except in the last tomographic bin ($\langle z\rangle\sim 1.2$), where there is a drop in the alignment amplitude for the fiducial model. This was attributed to the small fraction of early-type galaxies in that bin. 

Galaxy-galaxy lensing is particularly sensitive to intrinsic alignments, specially when lens-source bins have a substantial redshift overlap. Another recent significant deviation from the vanilla NLA redshift-dependence was observed in the $2\times 2$pt analysis of the DESI collaboration performed in \citet{Chen24b}. The authors modelled galaxy-galaxy lensing and galaxy clustering of bright and luminous red galaxies of DESI with background source shapes from DES Y3 using a combination of Lagrangian perturbation theory and EFT in quasi-linear scales (Sect.~\ref{sec:model}), also for intrinsic alignments. At fixed cosmology, they found the first $>5\sigma$ evidence of redshift-dependence beyond NLA. Nevertheless, when sampling over the cosmological and nuisance parameter space, they found consistent $S_8$ constraints whether they adopted a more or less flexible redshift evolution model in either NLA or EFT. Fixing the redshift dependence improved the constraints by approximately $50-60\%$. 

The KiDS survey has always opted to use the NLA model as fiducial, while DES analyses often use TATT, a slightly more complex model with five free parameters: $\{a_1,\eta_1,a_2,\eta_2,b_{\rm TA}\}$, reviewed in Sect.~\ref{sec:tatt}. But how much model complexity is needed? Ideally, the model should be as simple as possible to avoid potential biases in the cosmological parameters and retain as much constraining power. To answer this question, different authors have carried out simulated analyses where input synthetic data is contaminated by an intrinsic alignment contribution following one or another model, and tried to retrieve the cosmological parameters using a second model. A seminal example of this can be found in \citet{Krause11}. 

This strategy can also be applied in the blinded data vector. This was the procedure followed by the HSC survey in their cosmic shear analysis. \citet{Li23} decided to adopt TATT after observing no degradation of the cosmological parameter constraints from the blinded correlation functions compared to NLA. When focusing on the angular power spectra, \citet{Dalal23} considered also the cases of NLA and no alignments. None of the choices affected their $S_8$ constraints, possibly driven by the lack of detection of alignments in the survey. The NLA alignment amplitude obtained for HSC was $A_{\rm IA}=0.47\pm 0.58$.

When forecasting the impact of different alignment model choices, one can resort to sampling or opt for a simpler version via Fisher forecasting \citep{Tegmark97,Leonard24}. Fisher forecasts have the drawback of returning a Gaussian posterior and often overestimating constraining power, but are fast to perform compared to sampling the likelihood. The input data can also consist of noisy or noiseless \citep{Campos23} theoretical predictions for the data vectors or simulations (e.g., \citealt{Hoffmann22}). \citet{Campos23} presents a rigorous discussion of model comparison statistics illustrated by the case of intrinsic alignment model bias.

The procedure is similar in the case of other common probe combinations, like $3\times2$pt \citep{Leonard24}, CMB cross-correlations \citep{Robertson21} or even up to $6\times2$pt \citep{Johnston24}. While marginalization might successfully account for alignment contamination in all these cases and yield unbiased cosmological parameter constraints, caveats apply to the posterior of the intrinsic alignment parameters. It is entirely possible that these are biased even when cosmology is not, and this is mainly attributed to a degeneracy between photometric redshift uncertainties and alignments \citep{Hildebrandt17, Fischbacher23,Leonard24}. In other words, intrinsic alignments can absorb residual redshift calibration biases.

A few works have looked into whether cosmic shear parameter constraints from sub-samples of galaxies which are known to have different alignment strength are consistent with each other and with the constraints derived from the full galaxy shape sample \citep{Samuroff19,Li21,McCullough24}. The challenge lies in the need to re-calibrate redshift distributions and galaxy shapes for each sub-sample, and in the lower number density of galaxies. Generally speaking, \citet{Li21} and \citet{McCullough24} have found their cosmological constraints to be consistent with the fiducial analysis of the full sample of galaxy shapes, but with different NLA alignment amplitudes, as expected from selecting specific populations. Note that the inferred $S_8$ from the red sample can vary up to $1.4\sigma$ depending on the alignment model assumed (no alignments/NLA/TATT). This is consistent with the previous findings of \citet{Troxel18} when using TATT for the analysis of DES Y1 data. \citet{Samuroff19} suggests this is not necessarily a sign of bias of NLA, but of the model being too flexible for the data being analysed, as demonstrated by a synthetic analysis. 

Regardless,
\citet{McCullough24} quote at most a $1.3\sigma$ difference between the fiducial DES Y3 analysis and their blue sample analysis (which removes $35\%$ of the galaxies). They also find a $1.5\times$ gain in constraining power over $S_8$ in the blue sample analysis when the alignment of those galaxies is neglected. This improvement is partly due to the use of more non-linear scales in the analysis, which is warranted thanks to a more flexible feedback model. \citet{McCullough24} find that the single parameter NLA amplitude for blue galaxies is given by $A_{\rm IA} =-0.075_{-0.390}^{+0.229}$, and for TATT: $A_1 = -0.53_{-0.95}^{+0.45}$, $A_2=0.33_{-0.55}^{+1.76}$. We do not reproduce the constraints from \citet{Li23} due to significant degeneracy with the redshift calibration nuisance parameters.

Finally, \citet{Samuroff19} have extended the same procedure to a $3\times2$pt setting in the case of DES Y1, and reported similar results in terms of consistency of cosmological constraints when the shapes of early-/late-type galaxies are considered. They find intrinsic alignment amplitudes for the NLA model of $A_{\rm IA} = 2.38^{+0.32}_{-0.31}$ and $A_{\rm IA}= 0.05^{+0.10}_{-0.09}$ for early- and late-type galaxies, respectively, from DES Y1. They also quote results for the TATT model: $A_1 = 2.66 ^{+0.67}_{-0.66}$, $A_2=-2.94^{+1.94}_{-1.83}$, respectively, for early-type galaxies and $A_1 = 0.62 ^{+0.41}_{-0.41}$, $A_2 = -2.26^{+1.30}_{-1.16}$ for late-type galaxies. In the full Y1 sample which contains both galaxy types, they find $A_1 = 0.70 ^{+0.41}_{-0.38}$, $A_2 = -1.36 ^{+1.08}_{-1.41}$. No significant detection of terms beyond tidal alignment ($A_1$) was obtained. 
 
How might mitigation strategies change for Stage IV surveys? The increased precision could drive a need for more complex intrinsic alignment models unless stringent scale cuts are implemented. How much model complexity is needed and how it interplays with scale cuts is still an open question. The answer also depends on the combination of probes considered, i.e. cosmic shear vs. $3\times2$pt or other extensions, and on the precision of the redshifts of the position and shape samples. If the trend of decreasing alignment amplitude with increasing depth observed in current cosmic shear samples is monotonic, we might obtain unbiased cosmological constraints even if NLA is used. However, there is a risk that blue/disc alignments might be present at higher redshift (see Sect.~\ref{sec:hydro}), and that these alignments might not be well-described by NLA (see Sect.~\ref{sec:model}). 

\subsubsection{Self-calibration}
\label{sec:selfcalib}

Self-calibration (SC) \citep{Zhang08,Zhang10} is a model-independent mitigation technique that exploits the symmetry of the alignment signal in front and behind a lens. This symmetry is not satisfied by lensing, since galaxies in front of the lens do not get distorted by it, and this is how the two signals can be distinguished. If redshifts were known perfectly, then within a redshift bin $i$, one could select pairs of galaxies where the redshift of the galaxy with a shape $z_S$ is lower than the redshift of the galaxy whose position is being considered, $z_S<z_p$. If this selection is applied, the angular power spectrum of the position-shape correlation of such selected pairs would only pick up an alignment contribution:
\begin{equation}
C_{z_S<z_p}^{pS}(\ell)=C_{z_S<z_p}^{gI}(\ell)
\end{equation}
Unfortunately, if redshifts are not known perfectly, as in the presence of photometric redshift uncertainties, applying the cut $z_S<z_p$ would not guarantee that lensing has been removed. A quality factor is then introduced to gauge the impact of lensing, as
\begin{equation}
Q_i(\ell)\equiv \frac{C_{z_S<z_p}^{gG}(\ell)}{C_{ii}^{gG}(\ell)}.
\end{equation}
Both numerator and denominator are estimated at a fiducial cosmology and using the photometric and estimated true redshift distributions of the survey.
The $gI$ term can be isolated in a model-independent way as follows 
\begin{equation}
C_{ii}^{gI}(\ell)= \frac{C_{z_S<z_p}^{pS}(\ell)-Q_i(\ell)C_{ii}^{pS}(\ell)}{1-Q_i(\ell)}.
\end{equation}
Using this measurement to inform $GI$ contamination does require making some modelling assumptions. If we can assume the scales probed are large and the clustering of galaxies can be described linearly, then the connection can be established as
\begin{equation}
C_{ij}^{GI}(\ell)\simeq  \frac{W_{ij}(\ell)\Delta_i(\ell)}{b_g^{(i)}}C_{ii}^{gI}(\ell),
\end{equation}
where $b_g^{(i)}$ is the galaxy bias (linear or otherwise) in bin $i$ and
\begin{eqnarray}
    W_{ij}&\equiv&\int_0^\infty dz_L\int_0^\infty dz_S\,W_L(z_L,z_S)n_i(z_L)n_j(z_S), \\ 
    \Delta_{i}^{-1}&\equiv& \int_0^\infty dz\, \frac{dz}{d\chi}n_i^2(z),
\end{eqnarray}
$W_L(z_L,z_S)$ is the lensing kernel for a source at $z_S$ and a lens at $z_L$, and $n_i(z)$ is the redshift distribution of galaxies in bin $i$. 

Further improvements have been made to the SC method by adding information from II correlations and using a small-bin approximation to make $ij$ bin cross-correlation predictions more accurate \citep{Yao19}. In principle $b_{g}^{(i)}$ can also be allowed to be $\ell$-dependent by effectively considering $b_{g}^{(i)}(\ell)\simeq \int_0^\infty b_{g}(k=\ell/\chi,z)n_i(z)dz$. This avoids the assumption of linear bias but requires a theoretical model for $b_g$. Alternatively, one could envisage constraining $b_g$ from clustering measurements of the lens sample, as considered in \citet{Yao19}. 

\citet{Yao17} suggested that in comparison to marginalization techniques, SC is equally powerful in removing biases in cosmological parameters that come from poor knowledge of intrinsic alignments. An application to real data has been showcased \citep{Yao20} for KiDS450 and KV450. For KV450 data, where the photometric redshifts are of greater quality than KiDS450 due to the addition of near-infrared information, the authors found that SC reproduced the expected results from the best-fit KV450 cosmological model. In the two lowest redshift bins, the $gG$ and $gI$ signals extracted via SC were instead in tension with the model, possibly due to the presence of photometric redshift outliers. A particular challenge for applying the SC method is the lack of knowledge of the individual galaxy redshifts, due to the need of identifying sources in front of the lens. However, the calibration of photometric redshifts used in cosmic shear is normally optimized to yield the right \emph{ensemble} distribution of redshifts for galaxies in a bin.

\citet{Pedersen20} has proposed that the SC method can be used to extract the $gI$ and $GI$ signals from photometric surveys and demonstrated this in KiDS450 data. The detections of the signals achieved a $\sim 3.5\sigma$ significance but were limited to specific bins where the photometric redshift quality was sufficient for the method to be applied. In subsequent work, \citet{Yao20b} achieved a $14\sigma$ detection of intrinsic alignments in DECaLS DR3. This came with improvements to the SC method such as further scale cuts, simultaneous fitting of $A_{\rm IA}$ and galaxy bias and treatment of asymmetric redshift distributions.  

A version of the SC method also exists for some combinations of three-point statistics of galaxy shapes and galaxy positions \citep{Troxel12,Troxel12b} (Sect.~\ref{sec:beyond}). \citet{Yao23} also extended the method to include the impact of magnification on the number counts and showed an application of the SC method to break the degeneracy between the lensing and amplitudes in the cross-correlation of galaxy shapes with CMB lensing. 

\section{Beyond two-point statistics}\label{sec:beyond}

Two-point correlations are a way to compress information from two fields into a function of a single variable. When the fields are Gaussian, their two-point correlation function contains all the relevant cosmological information. The properties of non-Gaussian (e.g., non-linear) fields cannot be reconstructed from the two-point correlation alone. In the following, we discuss measurements of beyond-2pt alignment statistics and also contamination from these to beyond-2pt weak gravitational lensing statistics. In addition to the statistics described in this section, there are many other examples we have not considered \citep{Ajani23} and which could be applied to intrinsic alignments or be contaminated by them when used to measure weak gravitational lensing. 

\subsection{Measurements of beyond-2pt alignment statistics}

We give here examples of beyond-2pt measurements of intrinsic alignments which can be used for the purposes of testing intrinsic alignment models, deriving priors for mitigation or for cosmological or astrophysical applications.

\begin{itemize}
    
    \item {\it Three-point statistics.} Several possible three-point correlations measurements of intrinsic alignments are possible: position-position-shape, position-shape-shape, a triplet of shapes, and compressed versions thereof, such as ``aperture masses'' \citep{Kaiser95,Schneider96}. \citet{Linke24} used a generalization of the NLA model to describe position-position-shape aperture statistics of the LOWZ sample. This yielded a $4.7\sigma$ constraint on $A_{\rm IA}$ compatible at $1\sigma$ with the constraint from two-point statistics.
    
    Theoretical approaches consistent with the SPT/EFT formalisms and enforcing the symmetries of the spin-$2$ observable have been presented in \citet{Schmitz18} (position-position-shape tree-level only) and \citet{Vlah20,Bakx25} (tree-level, all combinations). \citet{Bakx25} shows that generalizing from two- to three-point predictions requires care. For example, they remark that all possible combinations of scalar, $E$ and $B$-mode bispectra are non-zero (no connection to parity violation). The position-position-$B$ bispectrum might even be detectable with DESI, at a signal-to-noise ratio of $5$. 
    
    \item {\it Multiplets.} An interesting new approach to consists of measuring the orientation of small groups of galaxies (from 2 to 4 members) with respect to another tracer \citep{multiplet}. Effectively, such an estimator is equivalent to a squeezed three-point function of galaxy clustering. One advantage compared to measuring a three-point function is that it is faster to compute. On the other hand, it does not contain all the possible triangle configurations of three-point statistics. Nevertheless, the application to three DESI Y1 spectroscopic samples in \citet{multiplet} has demonstrated that a very high signal-to-noise ratio is available: 29.3 for the bright galaxies in $0.1<z<0.4$ and 15 for luminous red galaxies in $0.4<z<1.1$ (C. Lamman, private communication).
    When modelled according to the LA formalism, LRG multiplets were shown to align with $A_{\rm IA}=5.7\pm 0.1$ at scales larger than $20\,h^{-1}\,{\rm Mpc}$. (This can be compared to the alignment amplitude for shapes of the same LRG sample, which the authors quote to be $A_{\rm IA}=1.96\pm 0.001$.) 
    
    Multiplets have thus been confirmed to be a biased tracer of the tidal field on linear scales. This means that they could be used to test intrinsic alignment models and for cosmological applications (see Sect.~\ref{sec:cosmo}). However, because in multiplet alignment there is no galaxy shape involved, they cannot be used to establish priors for intrinsic alignment mitigation. Another important caveat is that for this type of measurement, realistic mocks are needed to reproduce the spectroscopic selection function (e.g., the impact of fibre assignment).  
    
    \item {\it Field-level.} Field-level inference is a growing area of interest in cosmology for capturing statistical power beyond two-point estimators. In field-level inference, a forward model is compared to the field-level data to obtain constraints on cosmological parameters. In \citet{Tsaprazi22}, the authors used re-simulations of the SDSS-III BOSS volume with a gravity-only model to obtain realistic realizations of the tidal field of the Universe in that volume. In this way, they could estimate the likelihood for the observed shapes of galaxies given one realization, and then marginalize over all of them to find a posterior for the alignment amplitude (adoption a uniform prior). This resulted in a $4\sigma$ detection of the alignment amplitude of $A_{\rm IA}=2.9\pm 0.7$. This estimation was limited to scales above $15.6\,h^{-1}\,{\rm Mpc}$ due to the intrinsic grid resolution of the algorithm used for reconstructing the tidal field. If resolution could be increased, one would expect a higher gain in constraining power at the field-level in the non-linear regime, compared to two-point statistics. 
    
    \item {\it Environmental dependence of intrinsic alignments.} \citet{Chen19} has shown that LOWZ galaxies show a small but significant tendency for galaxies to align parallel to filament segments of the cosmic web. The fact that the effect is present even when nearby clusters are excluded, suggests the measurement is truly capturing the effect of matter contained in the filaments. Similar studies have been performed in cosmological hydrodynamical simulations, also predicting alignments with respect to walls \citep{Chen15,Codis18}. \citet{Codis18} notes that the alignment of spiral galaxies with filaments in Horizon-AGN at high redshift is predicted to be of different sign compared to other simulations, and that this could be the origin of the tangential alignment signal found in \citet{Chisari16}. See \citet{Wang20,Welker18} for the analogous case of the alignments of the shapes of groups of satellites with respect to filaments in both observations and numerical simulations. Theoretical predictions by \citet{Reischke19} suggest that mitigation schemes might be able to exploit this dependency.
    
\end{itemize}

\subsection{Contamination to beyond-2pt lensing statistics}

Several works have studied how three-point statistics of galaxy shapes could aid in mitigating systematics present at the two-point statistics level, including intrinsic alignments. \citet{Semboloni08} was the first to study intrinsic alignment aperture masses with this goal. They populated an $N$-body simulation with aligned elliptical (30\%) and spiral galaxies. They found that the fractional contribution of intrinsic alignments to shape-shape-shape correlations is higher than for two-point and particularly prominent in shallow surveys. 
    
\citet{Pyne21} generalized the NLA model of intrinsic alignments to predict three-point statistics and used a Fisher forecast formalism to quantify the effect of adding the bispectrum in the figure-of-merit of $(\Omega_{\rm m},\sigma_8)$ and $(w_0,w_a)$. Their results suggested that three-point statistics are particularly useful for constraining systematic effects such as intrinsic alignments. Despite some unresolved discrepancies between \citet{Semboloni08} and \citet{Pyne21} regarding the amplitude and sign of the $III$ correlation, \citet{Pyne22} was able to validate their model against numerical predictions from TNG300 and to deliver consistent $A_{\rm IA}$ amplitudes from two- and three-point statistics. 

A similar validation was performed by \citet{Burger24} in their joint analysis of two- and three-point statistics in KiDS 1000. In this case, the generalized NLA model was validated on $N$-body simulations infused with intrinsic alignments according to the procedure outlined by \citet{HD22}. \citet{Burger24} reports that the impact of intrinsic alignments on aperture mass statistics for lensing is large when low redshift bins are included in the analysis. \citet{Gomes25} seems to reach similar conclusions in a simulated analysis of the DES Y3 data set. In addition, they suggest that the inclusion of three-point statistics only improve the constraint on $A_{\rm IA}$ by 3\%. 

Another common statistic used to extract cosmological information from weak lensing is the distribution of peaks in a convergence map, which are associated to mass concentrations responsible for gravitational lensing. Consider a map of galaxy shapes in which these peaks are being identified. Na\"ively, the presence of intrinsic alignments could lead to two effects: (1) galaxies aligned radially towards overdensities could create negative convergence peaks, and (2) the signal-to-noise ratio of a positive convergence peak could be diminished from intrinsically aligned interlopers. \citet{Fan07} studied the effect of (2) specifically by simulating the impact of an intrinsic alignments as an additional ($10\%$) ellipticity dispersion in the lensing maps. The result was a significant increase in the number of false convergence peaks (doubling for signal-to-noise ratio$>3.5$, and tripling for signal-to-noise ratio$>5$). An increase in the contribution of alignments to the overall shape dispersion would worsen this estimate. In addition, the treatment of alignments as an increase in the Gaussian noise prevents the authors from addressing case (1). 

A more recent study \citep{HD22} infused intrinsic alignments using the NLA model (including density-weighting) in an $N$-body simulation and found a $10\%$ and $30\%$ decrease in the number of lensing peaks for a Stage III and IV surveys, respectively. This more realistic procedure of introducing intrinsic alignment contamination seems to indicate case (2) is dominant. \citet{Zhang22} took this one step further by using semi-analytical galaxy formation models to populate $N$-body simulations with aligned galaxies. This adds non-linear power to intrinsic alignments at small scales, compared to \citet{HD22}, because it explicitly incorporates a recipe for satellite galaxy alignment. 

For narrow redshift bins, \citet{Zhang22} find that the main contribution of alignments is to add to the shape noise. For wider redshift bins, satellite interlopers decrease the significance of weak lensing peaks, in agreement with the previous studies mentioned. An analogy can be made to $GI$ correlations being the main contaminant to cosmic shear. This justifies cosmological analyses including the effect of alignments as an effective signal-to-noise ratio change of the peaks, as in \citet{Kacprzak16}. Intriguingly, a study by \citet{Lanzieri23} on the cosmological constraining power of peak statistics in Stage IV surveys (vs. angular power spectra) suggested that peaks not only yield tighter constraints on $\sigma_8$ and $\Omega_{\rm c}$, but also on $A_{\rm IA}$. 

\section{Future data sets}\label{sec:future}

Stage III surveys such as KiDS, HSC and DES have finished or are close to finishing their observing campaigns. Their data sets have enabled a wide range of weak lensing and intrinsic alignment studies. While most direct measurements of alignments have been realized, there remain possible applications through the combination with spectroscopic data from the DESI Stage IV spectroscopic survey\footnote{\url{https://www.desi.lbl.gov/}}. We have had a glimpse of this promising avenue in \citet{Chen24b}. Photometric narrow-band surveys like PAUS and J-PAS \citep{Benitez14} are yet to deliver their final data releases. 

At the same time, Stage IV surveys are starting to collect data. ESA's 1.2-m space telescope, \textit{Euclid} \citep{Euclid}, was launched in June 2023 and will survey $14,000$ sq. deg. in the one visible (VIS) and three near-infrared (NISP) bands for the following $6$ years. The photometry used to deliver 30 galaxy shapes per arcmin$^{2}$ down to $R+I+Z=24.5$ is complemented by near-infrared low-resolution ($R=250$) spectroscopy to measure clustering. \textit{Euclid}'s first data release (Q1) already yielded an intrinsic alignment measurement of galaxies with respect to filaments \citep{Laigle25}.   

While \textit{Euclid} delivers exquisite shapes from space, the Vera C. Rubin Observatory provides photometry from the ground in $u,g,r,i,z,y$ bands. Starting towards the end of 2025, Rubin will carry out the Legacy Survey of Space and Time \cite[LSST,][]{Ivezic19}, a wide-fast-deep survey over $14,000$ sq. deg. of the sky up to $i<26$ that will provide $27$ galaxy shapes per arcmin$^2$ at full depth. Complemented with photometric clustering measurements and/or targeted spectroscopic samples from other surveys, LSST will widely improve the statistics of intrinsic alignment measurements. 

Spectra are key to isolating aligned pairs and deliver a high signal-to-noise ratio measurement of intrinsic alignments. 
In addition to DESI, the Prime Focus Spectrograph (PFS, \citealt{Takada12}) began operations on Maunakea in Hawaii in January, 2025. It is expected to deliver an [OII] emitter sample over 1,400 sq.~deg. in the redshift range of $0.8<z<2.4$, targeted from HSC spectroscopy.
In the South, 4MOST\footnote{\url{https://www.4most.eu/}}, 
on the VISTA telescope at the European Southern Observatory’s Paranal site, will complement the effort of DESI by carrying out several spectroscopic surveys which will allow us to tackle some of the open questions in this area. While the Cosmology Redshift Survey \citep{Richard19} can be expected to yield similar targets as those of DESI, WAVES \citep{Driver19} will be unique in seeking high-completeness over relatively modest volumes. WAVES-Wide ($1200$ sq. deg.) will probe $0.9$ million galaxies down to $Z\leq 21.1$ and WAVES-Deep ($70$ sq. deg.) will obtain $0.75$ million redshifts at $Z \lesssim 21.25$. Overlap with KiDS+VIKING and LSST guarantees the availability of good quality shapes. As a result, intrinsic alignment studies will be pushed to much necessary fainter magnitudes. A similar regime is expected to be probed by DESI-II in the North \citep{StageV}. 

Although these spectroscopic surveys will not reach the magnitude limit of LSST or the \textit{Euclid} satellite, cross-correlations between galaxies with shapes and galaxies with secure spectroscopic redshifts is expected to improve alignment priors. A Stage V experiment \citep{StageV} would instead directly use LSST photometric data to target 
hundreds of millions of galaxies at $z<1$ (bright or luminous red samples), at $z<0.1$ and [OII] emitters at $0.5<z<2$. Other high redshift targets might include Lyman-break galaxies or Lyman-$\alpha$ emitters, though it is unclear whether intrinsic alignments constraints will be necessary for mitigation at $z>2$. New methodologies are nevertheless starting to be proposed to measure the alignments of ELGs \cite[e.g.,][]{Shi21}.

The Roman Space Telescope \citep{Akeson19} was the highest ranked mission on the Astro2010 decadal survey. Expected to launch in 2027, Roman would achieve a depth of 26.9 and measure 27 million galaxy shapes per month (yielding a number density of $45$ arcmin$^{-2}$), deeper than any other photometric survey. In addition to imaging, it will be equipped to perform slitless spectroscopy. The high latitude wide area survey (HLWAS) will deliver $\sim 10$ million H$\alpha$ redshifts in $0.5<z<1.9$ and $\sim 3$ million [OIII] redshifts in $1<z<2.8$.

These photometric and spectroscopic data sets will probe intrinsic alignments more robustly, in fainter regimes and further in redshift, allowing us to test the model and simulation predictions we have described in this review, and to deliver cosmological and astrophysical constraints from alignments. The same experiments might open the door for implementing mitigation techniques based on multi-shape measurements \citep{Leonard18,MacMahon24}, either from one or multiple surveys \citep{EuclidDDP}. New samples of galaxy clusters will provide further constraints on the mass-dependence and redshift evolution of galaxy alignments and possibly serve for cosmological applications \citep{Vedder20}.

\section{Open problems}\label{sec:discuss}

In Sect.~\ref{sec:model} we introduced the existing models for shape alignments. Even though these cover everything from the linear to the non-linear regime, several open problems remain. Perturbative models can describe the quasilinear scales, but are not guaranteed to mitigate intrinsic alignments in the non-linear regime. Here, the halo model is the tool of choice, but how to seamlessly bridge the quasilinear with the non-linear scales is not yet clear. Mitigation strategies that are model-dependent might face this issue in coming years. This will depend on how much model complexity is needed, which is tied to the specific observables and constraining power of Stage IV surveys. 

Direct priors for the model parameters cannot be predicted from first principles, although we reviewed some interesting possibilities in Sect.~\ref{sec:priors}. Except for a few selected cases, priors available from observations are limited to bright samples at low redshift. Several works have attempted to derive this missing information from cosmological hydrodynamical simulations. While such simulations have made significant advances in the modelling of galaxy alignments, they do not fully agree on the behaviour of blue/disc galaxies at high redshift. It is difficult to envisage that this discrepancy will be resolved in the short term, as it requires a concerted simulation effort to compare different techniques over a sufficiently large and well-resolved volume. On the other hand, this implies that intrinsic alignments might also constrain different astrophysical processes during the formation and evolution of galaxies, and might inform sub-grid implementations. To achieve fully realistic simulated predictions, effort will also be needed in going from stellar particle distributions to observed shapes, depending on each experiment.

Current data sets have been underused for intrinsic alignment model testing. The direct observational studies of intrinsic alignments presented in Sect.~\ref{sec:obs} are often restricted to fitting the NLA model and variations thereof with different redshift dependence. In comparison, cosmic shear analyses have adopted more sophisticated intrinsic alignment models to assess whether a bias in the cosmological parameters is incurred. There is still a significant amount of information at small scales that could be harnessed from spectroscopic or colour-selected samples, but degeneracies with non-linear clustering complicate extracting information on such scales.

In Sect.~\ref{sec:cosmo}, we discussed the cosmological applications of the intrinsic alignment signal. There, we argued that intrinsic alignments are sensitive to cosmology even on linear scales and they can potentially constrain models out of the reach for two-point clustering statistics. Some works have already provided cosmological constraints from galaxy alignments which could be improved by the use of a multi-tracer strategy. But how to optimally split samples or perform shape measurements to enable the multi-tracer technique is an open question.

Beyond two-point statistics offer an avenue to extract more information from galaxy shapes, whether from intrinsic alignments or weak lensing effects. How much more information can be extracted compared to the two-point case remains to be properly quantified. In addition, the role of intrinsic alignments as a contaminant is not fully explored for some of these statistics.

\section{Conclusions}\label{sec:conclu}

Intrinsic alignments are a spin-$2$ observable that contaminates weak lensing when attempting to measure it from the auto- and cross-correlation of galaxy shapes, and galaxy clustering via selection effects. Their presence is well-established observationally and in numerical simulations of structure formation over a wide range of halo and galaxy. A dichotomy arises between red and blue galaxies, where the first shows a significant alignment trend that seems to depend on halo mass and the second, no detectable alignment. 

Observational constraints remain consistent with predictions from cosmological hydrodynamical simulations. While simulations predict some level of blue (spiral) alignment, albeit with contradicting signs and redshift evolution, the observational error bars remain too large to distinguish between these options, which could be connected to key astrophysical processes during the evolution of a galaxy. 
Nevertheless, within the level of alignment predicted by existing hydrodynamical simulations, both red and blue populations remain a significant source of contamination for large-scale structure observables that must be mitigated over or self-calibrated, even if Stage III surveys are finding ever decreasing alignment amplitudes at higher apparent magnitude limits. 

The development of more accurate models that can alleviate the impact of alignments in cosmological parameter extraction is an active area of research. Since the turn of the millennium, we have gone from linear and quadratic models to SPT, EFT or (hybrid) LPT formulations that push the modelling to quasilinear scales. On fully non-linear scales, we must resort to the halo model or semi-analytic prescriptions for creating mocks, which rely on calibrating the connection between the shapes and alignments of central and satellite galaxies with respect to their host haloes and subhaloes using either observations or hydrodynamical simulations. 

Spectroscopy from Stage IV and possibly Stage V surveys will enable a range of cosmological applications of intrinsic alignments, some of them providing unique information thanks to the spin-$2$ nature of galaxy shapes. Observables beyond two-point might also play a role, whether through their sensitivity to cosmology, by breaking degeneracies with cosmological parameters, or simply providing observational constraints that can be translated into alignment priors. In just over a quarter of a century, the possibilities that have emerged for intrinsic alignments to play a role in cosmology and galaxy evolution are numerous, and it is likely some are yet to be revealed.

\bmhead{Acknowledgments}

I acknowledge support from the project ``A rising tide: Galaxy intrinsic alignments as a new probe of cosmology and galaxy evolution'' (with project number VI.Vidi.203.011) of the Talent programme Vidi which is (partly) financed by the Dutch Research Council (NWO). 

I am thankful to Rachel Mandelbaum and Michael Strauss for introducing me into the topic of intrinsic alignments, and to Fabian Schmidt and Cora Dvorkin for their theoretical insights and enthusiasm for the topic. I am indebted to several colleagues and collaborators for their valuable feedback: Henk Hoekstra, Christos Georgiou, Ji Yao, David Navarro Giron\'es, Elizabeth Gonz\'alez, Dennis Neumann, Jonathan Blazek, Toshiki Kurita, Casper Vedder, Thomas Bakx and Marloes van Heukelum. I thank Jamie McCullough, Claire Lamman, Angus Wright, Marloes van Heukelum, David Navarro Giron\'es, Christos Georgiou and Roohi Dalal for sharing numbers and figures for this review, Maria Cristina Fortuna for the illustrations, and the anonymous referee for their feedback on the manuscript. 

I would like to thank Instituto de F\'isica de Altas Energ\'ias (IFAE) for their hospitality while this work was in progress. I also gratefully acknowledge support from the Institute for Advanced Study and Princeton University. 

The recent echoIA community workshops\footnote{\url{https://github.com/echo-IA}} on intrinsic alignments have been instrumental for writing this review. I thank Jonathan Blazek and Benjamin Joachimi for pushing this initiative and specifically acknowledge the workshops held at the Lorentz centre in 2023 and Harvard University in 2024.

This version of the article has been accepted for publication, after peer review (when applicable) but is
not the Version of Record and does not reflect post-acceptance improvements, or any corrections. The Version of Record is
available online at: \url{http://dx.doi.org/10.1007/s00159-025-00161-8}.

\phantomsection
\addcontentsline{toc}{section}{References}
\bibliography{main}

@ARTICLE{Mandelbaum14,
       author = {{Mandelbaum}, Rachel and {Rowe}, Barnaby and {Bosch}, James and {Chang}, Chihway and {Courbin}, Frederic and {Gill}, Mandeep and {Jarvis}, Mike and {Kannawadi}, Arun and {Kacprzak}, Tomasz and {Lackner}, Claire and {Leauthaud}, Alexie and {Miyatake}, Hironao and {Nakajima}, Reiko and {Rhodes}, Jason and {Simet}, Melanie and {Zuntz}, Joe and {Armstrong}, Bob and {Bridle}, Sarah and {Coupon}, Jean and {Dietrich}, J{\"o}rg P. and {Gentile}, Marc and {Heymans}, Catherine and {Jurling}, Alden S. and {Kent}, Stephen M. and {Kirkby}, David and {Margala}, Daniel and {Massey}, Richard and {Melchior}, Peter and {Peterson}, John and {Roodman}, Aaron and {Schrabback}, Tim},
        title = "{The Third Gravitational Lensing Accuracy Testing (GREAT3) Challenge Handbook}",
      journal = {\apjs},
         year = 2014,
        month = may,
       volume = {212},
       number = {1},
          eid = {5},
        pages = {5},
          doi = {10.1088/0067-0049/212/1/5},
archivePrefix = {arXiv},
       eprint = {1308.4982},
 primaryClass = {astro-ph.CO},
       adsurl = {https://ui.adsabs.harvard.edu/abs/2014ApJS..212....5M}
}

@ARTICLE{CMBPol,
       author = {{Hu}, Wayne and {White}, Martin},
        title = "{A CMB polarization primer}",
      journal = {\na},
         year = 1997,
        month = oct,
       volume = {2},
       number = {4},
        pages = {323-344},
          doi = {10.1016/S1384-1076(97)00022-5},
archivePrefix = {arXiv},
       eprint = {astro-ph/9706147},
 primaryClass = {astro-ph},
       adsurl = {https://ui.adsabs.harvard.edu/abs/1997NewA....2..323H}
}

@ARTICLE{Bartelmann,
       author = {{Bartelmann}, M. and {Schneider}, P.},
        title = "{Weak gravitational lensing}",
      journal = {\physrep},
         year = 2001,
        month = jan,
       volume = {340},
       number = {4-5},
        pages = {291-472},
          doi = {10.1016/S0370-1573(00)00082-X},
archivePrefix = {arXiv},
       eprint = {astro-ph/9912508},
 primaryClass = {astro-ph},
       adsurl = {https://ui.adsabs.harvard.edu/abs/2001PhR...340..291B}
}

@ARTICLE{Kilbinger,
       author = {{Kilbinger}, Martin},
        title = "{Cosmology with cosmic shear observations: a review}",
      journal = {Reports on Progress in Physics},
         year = 2015,
        month = jul,
       volume = {78},
       number = {8},
          eid = {086901},
        pages = {086901},
          doi = {10.1088/0034-4885/78/8/086901},
archivePrefix = {arXiv},
       eprint = {1411.0115},
 primaryClass = {astro-ph.CO},
       adsurl = {https://ui.adsabs.harvard.edu/abs/2015RPPh...78h6901K}
}

@ARTICLE{Mandelbaum,
       author = {{Mandelbaum}, Rachel},
        title = "{Weak Lensing for Precision Cosmology}",
      journal = {\araa},
         year = 2018,
        month = sep,
       volume = {56},
        pages = {393-433},
          doi = {10.1146/annurev-astro-081817-051928},
archivePrefix = {arXiv},
       eprint = {1710.03235},
 primaryClass = {astro-ph.CO},
       adsurl = {https://ui.adsabs.harvard.edu/abs/2018ARA&A..56..393M}
}

@ARTICLE{Desjacques18,
       author = {{Desjacques}, Vincent and {Jeong}, Donghui and {Schmidt}, Fabian},
        title = "{Large-scale galaxy bias}",
      journal = {\physrep},
         year = 2018,
        month = feb,
       volume = {733},
        pages = {1-193},
          doi = {10.1016/j.physrep.2017.12.002},
archivePrefix = {arXiv},
       eprint = {1611.09787},
 primaryClass = {astro-ph.CO},
       adsurl = {https://ui.adsabs.harvard.edu/abs/2018PhR...733....1D}
}

@ARTICLE{Schafer09,
       author = {{Sch{\"a}fer}, Bj{\"o}rn Malte},
        title = "{Galactic Angular Momenta and Angular Momentum Correlations in the Cosmological Large-Scale Structure}",
      journal = {International Journal of Modern Physics D},
         year = 2009,
        month = jan,
       volume = {18},
       number = {2},
        pages = {173-222},
          doi = {10.1142/S0218271809014388},
archivePrefix = {arXiv},
       eprint = {0808.0203},
 primaryClass = {astro-ph},
       adsurl = {https://ui.adsabs.harvard.edu/abs/2009IJMPD..18..173S}
}

@ARTICLE{Kirk15,
       author = {{Kirk}, Donnacha and {Brown}, Michael L. and {Hoekstra}, Henk and {Joachimi}, Benjamin and {Kitching}, Thomas D. and {Mandelbaum}, Rachel and {Sif{\'o}n}, Crist{\'o}bal and {Cacciato}, Marcello and {Choi}, Ami and {Kiessling}, Alina and {Leonard}, Adrienne and {Rassat}, Anais and {Sch{\"a}fer}, Bj{\"o}rn Malte},
        title = "{Galaxy Alignments: Observations and Impact on Cosmology}",
      journal = {\ssr},
         year = 2015,
        month = nov,
       volume = {193},
       number = {1-4},
        pages = {139-211},
          doi = {10.1007/s11214-015-0213-4},
archivePrefix = {arXiv},
       eprint = {1504.05465},
 primaryClass = {astro-ph.GA},
       adsurl = {https://ui.adsabs.harvard.edu/abs/2015SSRv..193..139K}
}

@ARTICLE{Kiessling15,
       author = {{Kiessling}, Alina and {Cacciato}, Marcello and {Joachimi}, Benjamin and {Kirk}, Donnacha and {Kitching}, Thomas D. and {Leonard}, Adrienne and {Mandelbaum}, Rachel and {Sch{\"a}fer}, Bj{\"o}rn Malte and {Sif{\'o}n}, Crist{\'o}bal and {Brown}, Michael L. and {Rassat}, Anais},
        title = "{Galaxy Alignments: Theory, Modelling \& Simulations}",
      journal = {\ssr},
         year = 2015,
        month = nov,
       volume = {193},
       number = {1-4},
        pages = {67-136},
          doi = {10.1007/s11214-015-0203-6},
archivePrefix = {arXiv},
       eprint = {1504.05546},
 primaryClass = {astro-ph.GA},
       adsurl = {https://ui.adsabs.harvard.edu/abs/2015SSRv..193...67K}
}

@ARTICLE{IAGuide,
       author = {{Lamman}, Claire and {Tsaprazi}, Eleni and {Shi}, Jingjing and {{\v{S}}ar{\v{c}}evi{\'c}}, Nikolina Niko and {Pyne}, Susan and {Legnani}, Elisa and {Ferreira}, Tassia},
        title = "{The IA Guide: A Breakdown of Intrinsic Alignment Formalisms}",
      journal = {The Open Journal of Astrophysics},
         year = 2024,
        month = feb,
       volume = {7},
          eid = {14},
        pages = {14},
          doi = {10.21105/astro.2309.08605},
archivePrefix = {arXiv},
       eprint = {2309.08605},
 primaryClass = {astro-ph.CO},
       adsurl = {https://ui.adsabs.harvard.edu/abs/2024OJAp....7E..14L}
}

@ARTICLE{Joachimi15,
   author = {{Joachimi}, B. and {Cacciato}, M. and {Kitching}, T.~D. and 
	{Leonard}, A. and {Mandelbaum}, R. and {Sch{\"a}fer}, B.~M. and 
	{Sif{\'o}n}, C. and {Hoekstra}, H. and {Kiessling}, A. and {Kirk}, D. and 
	{Rassat}, A.},
    title = "{Galaxy Alignments: An Overview}",
  journal = {\ssr},
archivePrefix = "arXiv",
   eprint = {1504.05456},
     year = 2015,
    month = nov,
   volume = 193,
    pages = {1-65},
      doi = {10.1007/s11214-015-0177-4},
   adsurl = {http://adsabs.harvard.edu/abs/2015SSRv..193....1J}
}

@ARTICLE{Troxel15,
       author = {{Troxel}, M.~A. and {Ishak}, Mustapha},
        title = "{The intrinsic alignment of galaxies and its impact on weak gravitational
        lensing in an era of precision cosmology}",
      journal = {\physrep},
         year = 2015,
        month = Feb,
       volume = {558},
        pages = {1-59},
          doi = {10.1016/j.physrep.2014.11.001},
archivePrefix = {arXiv},
       eprint = {1407.6990},
 primaryClass = {astro-ph.CO},
       adsurl = {https://ui.adsabs.harvard.edu/#abs/2015PhR...558....1T}
}

@ARTICLE{Brown02,
   author = {{Brown}, M.~L. and {Taylor}, A.~N. and {Hambly}, N.~C. and {Dye}, S.
	},
    title = "{Measurement of intrinsic alignments in galaxy ellipticities}",
  journal = {\mnras},
   eprint = {astro-ph/0009499},
     year = 2002,
    month = jul,
   volume = 333,
    pages = {501-509},
      doi = {10.1046/j.1365-8711.2002.05354.x},
   adsurl = {http://adsabs.harvard.edu/abs/2002MNRAS.333..501B}
}

@ARTICLE{Binggeli82,
       author = {{Binggeli}, B.},
        title = "{The shape and orientation of clusters of galaxies.}",
      journal = {\aap},
         year = 1982,
        month = mar,
       volume = {107},
        pages = {338-349},
       adsurl = {https://ui.adsabs.harvard.edu/abs/1982A&A...107..338B}
}

@ARTICLE{Kaiser92,
       author = {{Kaiser}, Nick},
        title = "{Weak Gravitational Lensing of Distant Galaxies}",
      journal = {\apj},
         year = 1992,
        month = apr,
       volume = {388},
        pages = {272},
          doi = {10.1086/171151},
       adsurl = {https://ui.adsabs.harvard.edu/abs/1992ApJ...388..272K}
}

@ARTICLE{Weinberg13,
       author = {{Weinberg}, David H. and {Mortonson}, Michael J. and {Eisenstein}, Daniel J. and {Hirata}, Christopher and {Riess}, Adam G. and {Rozo}, Eduardo},
        title = "{Observational probes of cosmic acceleration}",
      journal = {\physrep},
         year = 2013,
        month = sep,
       volume = {530},
       number = {2},
        pages = {87-255},
          doi = {10.1016/j.physrep.2013.05.001},
archivePrefix = {arXiv},
       eprint = {1201.2434},
 primaryClass = {astro-ph.CO},
       adsurl = {https://ui.adsabs.harvard.edu/abs/2013PhR...530...87W}
}

@ARTICLE{Vedder20,
       author = {{Vedder}, C.~J.~G. and {Chisari}, N.~E.},
        title = "{Galaxy clusters as intrinsic alignment tracers: present and future}",
      journal = {\mnras},
         year = 2021,
        month = jan,
       volume = {500},
       number = {4},
        pages = {5561-5569},
          doi = {10.1093/mnras/staa3633},
archivePrefix = {arXiv},
       eprint = {2011.06904},
 primaryClass = {astro-ph.CO},
       adsurl = {https://ui.adsabs.harvard.edu/abs/2021MNRAS.500.5561V}
}

@ARTICLE{Aihara18,
       author = {{Aihara}, Hiroaki and {Arimoto}, Nobuo and {Armstrong}, Robert and {Arnouts}, St{\'e}phane and {Bahcall}, Neta A. and {Bickerton}, Steven and {Bosch}, James and {Bundy}, Kevin and {Capak}, Peter L. and {Chan}, James H.~H. and {Chiba}, Masashi and {Coupon}, Jean and {Egami}, Eiichi and {Enoki}, Motohiro and {Finet}, Francois and {Fujimori}, Hiroki and {Fujimoto}, Seiji and {Furusawa}, Hisanori and {Furusawa}, Junko and {Goto}, Tomotsugu and {Goulding}, Andy and {Greco}, Johnny P. and {Greene}, Jenny E. and {Gunn}, James E. and {Hamana}, Takashi and {Harikane}, Yuichi and {Hashimoto}, Yasuhiro and {Hattori}, Takashi and {Hayashi}, Masao and {Hayashi}, Yusuke and {He{\l}miniak}, Krzysztof G. and {Higuchi}, Ryo and {Hikage}, Chiaki and {Ho}, Paul T.~P. and {Hsieh}, Bau-Ching and {Huang}, Kuiyun and {Huang}, Song and {Ikeda}, Hiroyuki and {Imanishi}, Masatoshi and {Inoue}, Akio K. and {Iwasawa}, Kazushi and {Iwata}, Ikuru and {Jaelani}, Anton T. and {Jian}, Hung-Yu and {Kamata}, Yukiko and {Karoji}, Hiroshi and {Kashikawa}, Nobunari and {Katayama}, Nobuhiko and {Kawanomoto}, Satoshi and {Kayo}, Issha and {Koda}, Jin and {Koike}, Michitaro and {Kojima}, Takashi and {Komiyama}, Yutaka and {Konno}, Akira and {Koshida}, Shintaro and {Koyama}, Yusei and {Kusakabe}, Haruka and {Leauthaud}, Alexie and {Lee}, Chien-Hsiu and {Lin}, Lihwai and {Lin}, Yen-Ting and {Lupton}, Robert H. and {Mandelbaum}, Rachel and {Matsuoka}, Yoshiki and {Medezinski}, Elinor and {Mineo}, Sogo and {Miyama}, Shoken and {Miyatake}, Hironao and {Miyazaki}, Satoshi and {Momose}, Rieko and {More}, Anupreeta and {More}, Surhud and {Moritani}, Yuki and {Moriya}, Takashi J. and {Morokuma}, Tomoki and {Mukae}, Shiro and {Murata}, Ryoma and {Murayama}, Hitoshi and {Nagao}, Tohru and {Nakata}, Fumiaki and {Niida}, Mana and {Niikura}, Hiroko and {Nishizawa}, Atsushi J. and {Obuchi}, Yoshiyuki and {Oguri}, Masamune and {Oishi}, Yukie and {Okabe}, Nobuhiro and {Okamoto}, Sakurako and {Okura}, Yuki and {Ono}, Yoshiaki and {Onodera}, Masato and {Onoue}, Masafusa and {Osato}, Ken and {Ouchi}, Masami and {Price}, Paul A. and {Pyo}, Tae-Soo and {Sako}, Masao and {Sawicki}, Marcin and {Shibuya}, Takatoshi and {Shimasaku}, Kazuhiro and {Shimono}, Atsushi and {Shirasaki}, Masato and {Silverman}, John D. and {Simet}, Melanie and {Speagle}, Joshua and {Spergel}, David N. and {Strauss}, Michael A. and {Sugahara}, Yuma and {Sugiyama}, Naoshi and {Suto}, Yasushi and {Suyu}, Sherry H. and {Suzuki}, Nao and {Tait}, Philip J. and {Takada}, Masahiro and {Takata}, Tadafumi and {Tamura}, Naoyuki and {Tanaka}, Manobu M. and {Tanaka}, Masaomi and {Tanaka}, Masayuki and {Tanaka}, Yoko and {Terai}, Tsuyoshi and {Terashima}, Yuichi and {Toba}, Yoshiki and {Tominaga}, Nozomu and {Toshikawa}, Jun and {Turner}, Edwin L. and {Uchida}, Tomohisa and {Uchiyama}, Hisakazu and {Umetsu}, Keiichi and {Uraguchi}, Fumihiro and {Urata}, Yuji and {Usuda}, Tomonori and {Utsumi}, Yousuke and {Wang}, Shiang-Yu and {Wang}, Wei-Hao and {Wong}, Kenneth C. and {Yabe}, Kiyoto and {Yamada}, Yoshihiko and {Yamanoi}, Hitomi and {Yasuda}, Naoki and {Yeh}, Sherry and {Yonehara}, Atsunori and {Yuma}, Suraphong},
        title = "{The Hyper Suprime-Cam SSP Survey: Overview and survey design}",
      journal = {\pasj},
         year = 2018,
        month = jan,
       volume = {70},
          eid = {S4},
        pages = {S4},
          doi = {10.1093/pasj/psx066},
archivePrefix = {arXiv},
       eprint = {1704.05858},
 primaryClass = {astro-ph.IM},
       adsurl = {https://ui.adsabs.harvard.edu/abs/2018PASJ...70S...4A}
}

@ARTICLE{DES,
       author = {{Abbott}, T.~M.~C. and {Aguena}, M. and {Alarcon}, A. and {Allam}, S. and {Alves}, O. and {Amon}, A. and {Andrade-Oliveira}, F. and {Annis}, J. and {Avila}, S. and {Bacon}, D. and {Baxter}, E. and {Bechtol}, K. and {Becker}, M.~R. and {Bernstein}, G.~M. and {Bhargava}, S. and {Birrer}, S. and {Blazek}, J. and {Brandao-Souza}, A. and {Bridle}, S.~L. and {Brooks}, D. and {Buckley-Geer}, E. and {Burke}, D.~L. and {Camacho}, H. and {Campos}, A. and {Carnero Rosell}, A. and {Carrasco Kind}, M. and {Carretero}, J. and {Castander}, F.~J. and {Cawthon}, R. and {Chang}, C. and {Chen}, A. and {Chen}, R. and {Choi}, A. and {Conselice}, C. and {Cordero}, J. and {Costanzi}, M. and {Crocce}, M. and {da Costa}, L.~N. and {da Silva Pereira}, M.~E. and {Davis}, C. and {Davis}, T.~M. and {De Vicente}, J. and {DeRose}, J. and {Desai}, S. and {Di Valentino}, E. and {Diehl}, H.~T. and {Dietrich}, J.~P. and {Dodelson}, S. and {Doel}, P. and {Doux}, C. and {Drlica-Wagner}, A. and {Eckert}, K. and {Eifler}, T.~F. and {Elsner}, F. and {Elvin-Poole}, J. and {Everett}, S. and {Evrard}, A.~E. and {Fang}, X. and {Farahi}, A. and {Fernandez}, E. and {Ferrero}, I. and {Fert{\'e}}, A. and {Fosalba}, P. and {Friedrich}, O. and {Frieman}, J. and {Garc{\'\i}a-Bellido}, J. and {Gatti}, M. and {Gaztanaga}, E. and {Gerdes}, D.~W. and {Giannantonio}, T. and {Giannini}, G. and {Gruen}, D. and {Gruendl}, R.~A. and {Gschwend}, J. and {Gutierrez}, G. and {Harrison}, I. and {Hartley}, W.~G. and {Herner}, K. and {Hinton}, S.~R. and {Hollowood}, D.~L. and {Honscheid}, K. and {Hoyle}, B. and {Huff}, E.~M. and {Huterer}, D. and {Jain}, B. and {James}, D.~J. and {Jarvis}, M. and {Jeffrey}, N. and {Jeltema}, T. and {Kovacs}, A. and {Krause}, E. and {Kron}, R. and {Kuehn}, K. and {Kuropatkin}, N. and {Lahav}, O. and {Leget}, P. -F. and {Lemos}, P. and {Liddle}, A.~R. and {Lidman}, C. and {Lima}, M. and {Lin}, H. and {MacCrann}, N. and {Maia}, M.~A.~G. and {Marshall}, J.~L. and {Martini}, P. and {McCullough}, J. and {Melchior}, P. and {Mena-Fern{\'a}ndez}, J. and {Menanteau}, F. and {Miquel}, R. and {Mohr}, J.~J. and {Morgan}, R. and {Muir}, J. and {Myles}, J. and {Nadathur}, S. and {Navarro-Alsina}, A. and {Nichol}, R.~C. and {Ogando}, R.~L.~C. and {Omori}, Y. and {Palmese}, A. and {Pandey}, S. and {Park}, Y. and {Paz-Chinch{\'o}n}, F. and {Petravick}, D. and {Pieres}, A. and {Plazas Malag{\'o}n}, A.~A. and {Porredon}, A. and {Prat}, J. and {Raveri}, M. and {Rodriguez-Monroy}, M. and {Rollins}, R.~P. and {Romer}, A.~K. and {Roodman}, A. and {Rosenfeld}, R. and {Ross}, A.~J. and {Rykoff}, E.~S. and {Samuroff}, S. and {S{\'a}nchez}, C. and {Sanchez}, E. and {Sanchez}, J. and {Sanchez Cid}, D. and {Scarpine}, V. and {Schubnell}, M. and {Scolnic}, D. and {Secco}, L.~F. and {Serrano}, S. and {Sevilla-Noarbe}, I. and {Sheldon}, E. and {Shin}, T. and {Smith}, M. and {Soares-Santos}, M. and {Suchyta}, E. and {Swanson}, M.~E.~C. and {Tabbutt}, M. and {Tarle}, G. and {Thomas}, D. and {To}, C. and {Troja}, A. and {Troxel}, M.~A. and {Tucker}, D.~L. and {Tutusaus}, I. and {Varga}, T.~N. and {Walker}, A.~R. and {Weaverdyck}, N. and {Wechsler}, R. and {Weller}, J. and {Yanny}, B. and {Yin}, B. and {Zhang}, Y. and {Zuntz}, J. and {DES Collaboration}},
        title = "{Dark Energy Survey Year 3 results: Cosmological constraints from galaxy clustering and weak lensing}",
      journal = {\prd},
         year = 2022,
        month = jan,
       volume = {105},
       number = {2},
          eid = {023520},
        pages = {023520},
          doi = {10.1103/PhysRevD.105.023520},
archivePrefix = {arXiv},
       eprint = {2105.13549},
 primaryClass = {astro-ph.CO},
       adsurl = {https://ui.adsabs.harvard.edu/abs/2022PhRvD.105b3520A}
}

@ARTICLE{Sanchez21,
       author = {{Sanchez}, Javier and {Mendoza}, Ismael and {Kirkby}, David P. and {Burchat}, Patricia R. and {LSST Dark Energy Science Collaboration}},
        title = "{Effects of overlapping sources on cosmic shear estimation: Statistical sensitivity and pixel-noise bias}",
      journal = {\jcap},
         year = 2021,
        month = jul,
       volume = {2021},
       number = {7},
          eid = {043},
        pages = {043},
          doi = {10.1088/1475-7516/2021/07/043},
archivePrefix = {arXiv},
       eprint = {2103.02078},
 primaryClass = {astro-ph.CO},
       adsurl = {https://ui.adsabs.harvard.edu/abs/2021JCAP...07..043S}
}

@ARTICLE{HeymansCFHT,
       author = {{Heymans}, Catherine and {Van Waerbeke}, Ludovic and {Miller}, Lance and {Erben}, Thomas and {Hildebrandt}, Hendrik and {Hoekstra}, Henk and {Kitching}, Thomas D. and {Mellier}, Yannick and {Simon}, Patrick and {Bonnett}, Christopher and {Coupon}, Jean and {Fu}, Liping and {Harnois D{\'e}raps}, Joachim and {Hudson}, Michael J. and {Kilbinger}, Martin and {Kuijken}, Koenraad and {Rowe}, Barnaby and {Schrabback}, Tim and {Semboloni}, Elisabetta and {van Uitert}, Edo and {Vafaei}, Sanaz and {Velander}, Malin},
        title = "{CFHTLenS: the Canada-France-Hawaii Telescope Lensing Survey}",
      journal = {\mnras},
         year = 2012,
        month = nov,
       volume = {427},
       number = {1},
        pages = {146-166},
          doi = {10.1111/j.1365-2966.2012.21952.x},
archivePrefix = {arXiv},
       eprint = {1210.0032},
 primaryClass = {astro-ph.CO},
       adsurl = {https://ui.adsabs.harvard.edu/abs/2012MNRAS.427..146H}
}

@ARTICLE{Navarro25,
       author = {{Navarro-Giron{\'e}s}, D. and {Crocce}, M. and {Gazta{\~n}aga}, E. and {Wittje}, A. and {Siudek}, M. and {Hoekstra}, H. and {Hildebrandt}, H. and {Joachimi}, B. and {Paviot}, R. and {Baugh}, C.~M. and {Carretero}, J. and {Casas}, R. and {Castander}, F.~J. and {Eriksen}, M. and {Fernandez}, E. and {Fosalba}, P. and {Garc{\'\i}a-Bellido}, J. and {Miquel}, R. and {Padilla}, C. and {Renard}, P. and {S{\'a}nchez}, E. and {Serrano}, S. and {Sevilla-Noarbe}, I. and {Tallada-Cresp{\'\i}}, P.},
        title = "{The PAU Survey: Measuring intrinsic galaxy alignments in deep wide fields as a function of colour, luminosity, stellar mass and redshift}",
      journal = {arXiv e-prints},
         year = 2025,
        month = may,
          eid = {arXiv:2505.15470},
        pages = {arXiv:2505.15470},
archivePrefix = {arXiv},
       eprint = {2505.15470},
 primaryClass = {astro-ph.CO},
       adsurl = {https://ui.adsabs.harvard.edu/abs/2025arXiv250515470N}
}

@ARTICLE{Navarro24,
       author = {{Navarro-Giron{\'e}s}, D. and {Gazta{\~n}aga}, E. and {Crocce}, M. and {Wittje}, A. and {Hildebrandt}, H. and {Wright}, A.~H. and {Siudek}, M. and {Eriksen}, M. and {Serrano}, S. and {Renard}, P. and {Gonzalez}, E.~J. and {Baugh}, C.~M. and {Cabayol}, L. and {Carretero}, J. and {Casas}, R. and {Castander}, F.~J. and {Daza-Perilla}, I.~V. and {De Vicente}, J. and {Fernandez}, E. and {Garc{\'\i}a-Bellido}, J. and {Hoekstra}, H. and {Manzoni}, G. and {Miquel}, R. and {Padilla}, C. and {S{\'a}nchez}, E. and {Sevilla-Noarbe}, I. and {Tallada-Cresp{\'\i}}, P.},
        title = "{The PAU survey: photometric redshift estimation in deep wide fields}",
      journal = {\mnras},
         year = 2024,
        month = oct,
       volume = {534},
       number = {2},
        pages = {1504-1527},
          doi = {10.1093/mnras/stae1686},
archivePrefix = {arXiv},
       eprint = {2312.07581},
 primaryClass = {astro-ph.CO},
       adsurl = {https://ui.adsabs.harvard.edu/abs/2024MNRAS.534.1504N}
}

@ARTICLE{Cannon06,
       author = {{Cannon}, Russell and {Drinkwater}, Michael and {Edge}, Alastair and {Eisenstein}, Daniel and {Nichol}, Robert and {Outram}, Phillip and {Pimbblet}, Kevin and {de Propris}, Roberto and {Roseboom}, Isaac and {Wake}, David and {Allen}, Paul and {Bland-Hawthorn}, Joss and {Bridges}, Terry and {Carson}, Daniel and {Chiu}, Kuenley and {Colless}, Matthew and {Couch}, Warrick and {Croom}, Scott and {Driver}, Simon and {Fine}, Stephen and {Hewett}, Paul and {Loveday}, Jon and {Ross}, Nicholas and {Sadler}, Elaine M. and {Shanks}, Tom and {Sharp}, Robert and {Smith}, J. Allyn and {Stoughton}, Chris and {Weilbacher}, Peter and {Brunner}, Robert J. and {Meiksin}, Avery and {Schneider}, Donald P.},
        title = "{The 2dF-SDSS LRG and QSO (2SLAQ) Luminous Red Galaxy Survey}",
      journal = {\mnras},
         year = 2006,
        month = oct,
       volume = {372},
       number = {1},
        pages = {425-442},
          doi = {10.1111/j.1365-2966.2006.10875.x},
archivePrefix = {arXiv},
       eprint = {astro-ph/0607631},
 primaryClass = {astro-ph},
       adsurl = {https://ui.adsabs.harvard.edu/abs/2006MNRAS.372..425C}
}

@ARTICLE{Benitez14,
       author = {{Benitez}, N. and {Dupke}, R. and {Moles}, M. and {Sodre}, L. and {Cenarro}, J. and {Marin-Franch}, A. and {Taylor}, K. and {Cristobal}, D. and {Fernandez-Soto}, A. and {Mendes de Oliveira}, C. and {Cepa-Nogue}, J. and {Abramo}, L.~R. and {Alcaniz}, J.~S. and {Overzier}, R. and {Hernandez-Monteagudo}, C. and {Alfaro}, E.~J. and {Kanaan}, A. and {Carvano}, J.~M. and {Reis}, R.~R.~R. and {Martinez Gonzalez}, E. and {Ascaso}, B. and {Ballesteros}, F. and {Xavier}, H.~S. and {Varela}, J. and {Ederoclite}, A. and {Vazquez Ramio}, H. and {Broadhurst}, T. and {Cypriano}, E. and {Angulo}, R. and {Diego}, J.~M. and {Zandivarez}, A. and {Diaz}, E. and {Melchior}, P. and {Umetsu}, K. and {Spinelli}, P.~F. and {Zitrin}, A. and {Coe}, D. and {Yepes}, G. and {Vielva}, P. and {Sahni}, V. and {Marcos-Caballero}, A. and {Kitaura}, F. -S. and {Maroto}, A.~L. and {Masip}, M. and {Tsujikawa}, S. and {Carneiro}, S. and {Gonzalez Nuevo}, J. and {Carvalho}, G.~C. and {Reboucas}, M.~J. and {Carvalho}, J.~C. and {Abdalla}, E. and {Bernui}, A. and {Pigozzo}, C. and {Ferreira}, E.~G.~M. and {Chandrachani Devi}, N. and {Bengaly}, Jr., C.~A.~P. and {Campista}, M. and {Amorim}, A. and {Asari}, N.~V. and {Bongiovanni}, A. and {Bonoli}, S. and {Bruzual}, G. and {Cardiel}, N. and {Cava}, A. and {Cid Fernandes}, R. and {Coelho}, P. and {Cortesi}, A. and {Delgado}, R.~G. and {Diaz Garcia}, L. and {Espinosa}, J.~M.~R. and {Galliano}, E. and {Gonzalez-Serrano}, J.~I. and {Falcon-Barroso}, J. and {Fritz}, J. and {Fernandes}, C. and {Gorgas}, J. and {Hoyos}, C. and {Jimenez-Teja}, Y. and {Lopez-Aguerri}, J.~A. and {Lopez-San Juan}, C. and {Mateus}, A. and {Molino}, A. and {Novais}, P. and {OMill}, A. and {Oteo}, I. and {Perez-Gonzalez}, P.~G. and {Poggianti}, B. and {Proctor}, R. and {Ricciardelli}, E. and {Sanchez-Blazquez}, P. and {Storchi-Bergmann}, T. and {Telles}, E. and {Schoennell}, W. and {Trujillo}, N. and {Vazdekis}, A. and {Viironen}, K. and {Daflon}, S. and {Aparicio-Villegas}, T. and {Rocha}, D. and {Ribeiro}, T. and {Borges}, M. and {Martins}, S.~L. and {Marcolino}, W. and {Martinez-Delgado}, D. and {Perez-Torres}, M.~A. and {Siffert}, B.~B. and {Calvao}, M.~O. and {Sako}, M. and {Kessler}, R. and {Alvarez-Candal}, A. and {De Pra}, M. and {Roig}, F. and {Lazzaro}, D. and {Gorosabel}, J. and {Lopes de Oliveira}, R. and {Lima-Neto}, G.~B. and {Irwin}, J. and {Liu}, J.~F. and {Alvarez}, E. and {Balmes}, I. and {Chueca}, S. and {Costa-Duarte}, M.~V. and {da Costa}, A.~A. and {Dantas}, M.~L.~L. and {Diaz}, A.~Y. and {Fabregat}, J. and {Ferrari}, F. and {Gavela}, B. and {Gracia}, S.~G. and {Gruel}, N. and {Gutierrez}, J.~L.~L. and {Guzman}, R. and {Hernandez-Fernandez}, J.~D. and {Herranz}, D. and {Hurtado-Gil}, L. and {Jablonsky}, F. and {Laporte}, R. and {Le Tiran}, L.~L. and {Licandro}, J and {Lima}, M. and {Martin}, E. and {Martinez}, V. and {Montero}, J.~J.~C. and {Penteado}, P. and {Pereira}, C.~B. and {Peris}, V. and {Quilis}, V. and {Sanchez-Portal}, M. and {Soja}, A.~C. and {Solano}, E. and {Torra}, J. and {Valdivielso}, L.},
        title = "{J-PAS: The Javalambre-Physics of the Accelerated Universe Astrophysical Survey}",
      journal = {arXiv e-prints},
         year = 2014,
        month = mar,
          eid = {arXiv:1403.5237},
        pages = {arXiv:1403.5237},
          doi = {10.48550/arXiv.1403.5237},
archivePrefix = {arXiv},
       eprint = {1403.5237},
 primaryClass = {astro-ph.CO},
       adsurl = {https://ui.adsabs.harvard.edu/abs/2014arXiv1403.5237B}
}

@ARTICLE{Rozo16,
       author = {{Rozo}, E. and {Rykoff}, E.~S. and {Abate}, A. and {Bonnett}, C. and {Crocce}, M. and {Davis}, C. and {Hoyle}, B. and {Leistedt}, B. and {Peiris}, H.~V. and {Wechsler}, R.~H. and {Abbott}, T. and {Abdalla}, F.~B. and {Banerji}, M. and {Bauer}, A.~H. and {Benoit-L{\'e}vy}, A. and {Bernstein}, G.~M. and {Bertin}, E. and {Brooks}, D. and {Buckley-Geer}, E. and {Burke}, D.~L. and {Capozzi}, D. and {Rosell}, A. Carnero and {Carollo}, D. and {Kind}, M. Carrasco and {Carretero}, J. and {Castander}, F.~J. and {Childress}, M.~J. and {Cunha}, C.~E. and {D'Andrea}, C.~B. and {Davis}, T. and {DePoy}, D.~L. and {Desai}, S. and {Diehl}, H.~T. and {Dietrich}, J.~P. and {Doel}, P. and {Eifler}, T.~F. and {Evrard}, A.~E. and {Neto}, A. Fausti and {Flaugher}, B. and {Fosalba}, P. and {Frieman}, J. and {Gaztanaga}, E. and {Gerdes}, D.~W. and {Glazebrook}, K. and {Gruen}, D. and {Gruendl}, R.~A. and {Honscheid}, K. and {James}, D.~J. and {Jarvis}, M. and {Kim}, A.~G. and {Kuehn}, K. and {Kuropatkin}, N. and {Lahav}, O. and {Lidman}, C. and {Lima}, M. and {Maia}, M.~A.~G. and {March}, M. and {Martini}, P. and {Melchior}, P. and {Miller}, C.~J. and {Miquel}, R. and {Mohr}, J.~J. and {Nichol}, R.~C. and {Nord}, B. and {O'Neill}, C.~R. and {Ogando}, R. and {Plazas}, A.~A. and {Romer}, A.~K. and {Roodman}, A. and {Sako}, M. and {Sanchez}, E. and {Santiago}, B. and {Schubnell}, M. and {Sevilla-Noarbe}, I. and {Smith}, R.~C. and {Soares-Santos}, M. and {Sobreira}, F. and {Suchyta}, E. and {Swanson}, M.~E.~C. and {Thaler}, J. and {Thomas}, D. and {Uddin}, S. and {Vikram}, V. and {Walker}, A.~R. and {Wester}, W. and {Zhang}, Y. and {da Costa}, L.~N.},
        title = "{redMaGiC: selecting luminous red galaxies from the DES Science Verification data}",
      journal = {\mnras},
         year = 2016,
        month = sep,
       volume = {461},
       number = {2},
        pages = {1431-1450},
          doi = {10.1093/mnras/stw1281},
archivePrefix = {arXiv},
       eprint = {1507.05460},
 primaryClass = {astro-ph.IM},
       adsurl = {https://ui.adsabs.harvard.edu/abs/2016MNRAS.461.1431R}
}

@ARTICLE{lensfit,
       author = {{Miller}, L. and {Heymans}, C. and {Kitching}, T.~D. and {van Waerbeke}, L. and {Erben}, T. and {Hildebrandt}, H. and {Hoekstra}, H. and {Mellier}, Y. and {Rowe}, B.~T.~P. and {Coupon}, J. and {Dietrich}, J.~P. and {Fu}, L. and {Harnois-D{\'e}raps}, J. and {Hudson}, M.~J. and {Kilbinger}, M. and {Kuijken}, K. and {Schrabback}, T. and {Semboloni}, E. and {Vafaei}, S. and {Velander}, M.},
        title = "{Bayesian galaxy shape measurement for weak lensing surveys - III. Application to the Canada-France-Hawaii Telescope Lensing Survey}",
      journal = {\mnras},
         year = 2013,
        month = mar,
       volume = {429},
       number = {4},
        pages = {2858-2880},
          doi = {10.1093/mnras/sts454},
archivePrefix = {arXiv},
       eprint = {1210.8201},
 primaryClass = {astro-ph.CO},
       adsurl = {https://ui.adsabs.harvard.edu/abs/2013MNRAS.429.2858M}
}

@ARTICLE{Blandford91,
       author = {{Blandford}, R.~D. and {Saust}, A.~B. and {Brainerd}, T.~G. and {Villumsen}, J.~V.},
        title = "{The distortion of distant galaxy images by large-scale structure.}",
      journal = {\mnras},
         year = 1991,
        month = aug,
       volume = {251},
        pages = {600},
          doi = {10.1093/mnras/251.4.600},
       adsurl = {https://ui.adsabs.harvard.edu/abs/1991MNRAS.251..600B}
}

@ARTICLE{Ivezic19,
       author = {{Ivezi{\'c}}, {\v{Z}}eljko and {Kahn}, Steven M. and {Tyson}, J. Anthony and {Abel}, Bob and {Acosta}, Emily and {Allsman}, Robyn and {Alonso}, David and {AlSayyad}, Yusra and {Anderson}, Scott F. and {Andrew}, John and {Angel}, James Roger P. and {Angeli}, George Z. and {Ansari}, Reza and {Antilogus}, Pierre and {Araujo}, Constanza and {Armstrong}, Robert and {Arndt}, Kirk T. and {Astier}, Pierre and {Aubourg}, {\'E}ric and {Auza}, Nicole and {Axelrod}, Tim S. and {Bard}, Deborah J. and {Barr}, Jeff D. and {Barrau}, Aurelian and {Bartlett}, James G. and {Bauer}, Amanda E. and {Bauman}, Brian J. and {Baumont}, Sylvain and {Bechtol}, Ellen and {Bechtol}, Keith and {Becker}, Andrew C. and {Becla}, Jacek and {Beldica}, Cristina and {Bellavia}, Steve and {Bianco}, Federica B. and {Biswas}, Rahul and {Blanc}, Guillaume and {Blazek}, Jonathan and {Blandford}, Roger D. and {Bloom}, Josh S. and {Bogart}, Joanne and {Bond}, Tim W. and {Booth}, Michael T. and {Borgland}, Anders W. and {Borne}, Kirk and {Bosch}, James F. and {Boutigny}, Dominique and {Brackett}, Craig A. and {Bradshaw}, Andrew and {Brandt}, William Nielsen and {Brown}, Michael E. and {Bullock}, James S. and {Burchat}, Patricia and {Burke}, David L. and {Cagnoli}, Gianpietro and {Calabrese}, Daniel and {Callahan}, Shawn and {Callen}, Alice L. and {Carlin}, Jeffrey L. and {Carlson}, Erin L. and {Chandrasekharan}, Srinivasan and {Charles-Emerson}, Glenaver and {Chesley}, Steve and {Cheu}, Elliott C. and {Chiang}, Hsin-Fang and {Chiang}, James and {Chirino}, Carol and {Chow}, Derek and {Ciardi}, David R. and {Claver}, Charles F. and {Cohen-Tanugi}, Johann and {Cockrum}, Joseph J. and {Coles}, Rebecca and {Connolly}, Andrew J. and {Cook}, Kem H. and {Cooray}, Asantha and {Covey}, Kevin R. and {Cribbs}, Chris and {Cui}, Wei and {Cutri}, Roc and {Daly}, Philip N. and {Daniel}, Scott F. and {Daruich}, Felipe and {Daubard}, Guillaume and {Daues}, Greg and {Dawson}, William and {Delgado}, Francisco and {Dellapenna}, Alfred and {de Peyster}, Robert and {de Val-Borro}, Miguel and {Digel}, Seth W. and {Doherty}, Peter and {Dubois}, Richard and {Dubois-Felsmann}, Gregory P. and {Durech}, Josef and {Economou}, Frossie and {Eifler}, Tim and {Eracleous}, Michael and {Emmons}, Benjamin L. and {Fausti Neto}, Angelo and {Ferguson}, Henry and {Figueroa}, Enrique and {Fisher-Levine}, Merlin and {Focke}, Warren and {Foss}, Michael D. and {Frank}, James and {Freemon}, Michael D. and {Gangler}, Emmanuel and {Gawiser}, Eric and {Geary}, John C. and {Gee}, Perry and {Geha}, Marla and {Gessner}, Charles J.~B. and {Gibson}, Robert R. and {Gilmore}, D. Kirk and {Glanzman}, Thomas and {Glick}, William and {Goldina}, Tatiana and {Goldstein}, Daniel A. and {Goodenow}, Iain and {Graham}, Melissa L. and {Gressler}, William J. and {Gris}, Philippe and {Guy}, Leanne P. and {Guyonnet}, Augustin and {Haller}, Gunther and {Harris}, Ron and {Hascall}, Patrick A. and {Haupt}, Justine and {Hernandez}, Fabio and {Herrmann}, Sven and {Hileman}, Edward and {Hoblitt}, Joshua and {Hodgson}, John A. and {Hogan}, Craig and {Howard}, James D. and {Huang}, Dajun and {Huffer}, Michael E. and {Ingraham}, Patrick and {Innes}, Walter R. and {Jacoby}, Suzanne H. and {Jain}, Bhuvnesh and {Jammes}, Fabrice and {Jee}, M. James and {Jenness}, Tim and {Jernigan}, Garrett and {Jevremovi{\'c}}, Darko and {Johns}, Kenneth and {Johnson}, Anthony S. and {Johnson}, Margaret W.~G. and {Jones}, R. Lynne and {Juramy-Gilles}, Claire and {Juri{\'c}}, Mario and {Kalirai}, Jason S. and {Kallivayalil}, Nitya J. and {Kalmbach}, Bryce and {Kantor}, Jeffrey P. and {Karst}, Pierre and {Kasliwal}, Mansi M. and {Kelly}, Heather and {Kessler}, Richard and {Kinnison}, Veronica and {Kirkby}, David and {Knox}, Lloyd and {Kotov}, Ivan V. and {Krabbendam}, Victor L. and {Krughoff}, K. Simon and {Kub{\'a}nek}, Petr and {Kuczewski}, John and {Kulkarni}, Shri and {Ku}, John and {Kurita}, Nadine R. and {Lage}, Craig S. and {Lambert}, Ron and {Lange}, Travis and {Langton}, J. Brian and {Le Guillou}, Laurent and {Levine}, Deborah and {Liang}, Ming and {Lim}, Kian-Tat and {Lintott}, Chris J. and {Long}, Kevin E. and {Lopez}, Margaux and {Lotz}, Paul J. and {Lupton}, Robert H. and {Lust}, Nate B. and {MacArthur}, Lauren A. and {Mahabal}, Ashish and {Mandelbaum}, Rachel and {Markiewicz}, Thomas W. and {Marsh}, Darren S. and {Marshall}, Philip J. and {Marshall}, Stuart and {May}, Morgan and {McKercher}, Robert and {McQueen}, Michelle and {Meyers}, Joshua and {Migliore}, Myriam and {Miller}, Michelle and {Mills}, David J.},
        title = "{LSST: From Science Drivers to Reference Design and Anticipated Data Products}",
      journal = {\apj},
         year = 2019,
        month = mar,
       volume = {873},
       number = {2},
          eid = {111},
        pages = {111},
          doi = {10.3847/1538-4357/ab042c},
archivePrefix = {arXiv},
       eprint = {0805.2366},
 primaryClass = {astro-ph},
       adsurl = {https://ui.adsabs.harvard.edu/abs/2019ApJ...873..111I}
}

@ARTICLE{Euclid,
       author = {{Laureijs}, R. and {Amiaux}, J. and {Arduini}, S. and {Augu{\`e}res}, J. -L. and {Brinchmann}, J. and {Cole}, R. and {Cropper}, M. and {Dabin}, C. and {Duvet}, L. and {Ealet}, A. and {Garilli}, B. and {Gondoin}, P. and {Guzzo}, L. and {Hoar}, J. and {Hoekstra}, H. and {Holmes}, R. and {Kitching}, T. and {Maciaszek}, T. and {Mellier}, Y. and {Pasian}, F. and {Percival}, W. and {Rhodes}, J. and {Saavedra Criado}, G. and {Sauvage}, M. and {Scaramella}, R. and {Valenziano}, L. and {Warren}, S. and {Bender}, R. and {Castander}, F. and {Cimatti}, A. and {Le F{\`e}vre}, O. and {Kurki-Suonio}, H. and {Levi}, M. and {Lilje}, P. and {Meylan}, G. and {Nichol}, R. and {Pedersen}, K. and {Popa}, V. and {Rebolo Lopez}, R. and {Rix}, H. -W. and {Rottgering}, H. and {Zeilinger}, W. and {Grupp}, F. and {Hudelot}, P. and {Massey}, R. and {Meneghetti}, M. and {Miller}, L. and {Paltani}, S. and {Paulin-Henriksson}, S. and {Pires}, S. and {Saxton}, C. and {Schrabback}, T. and {Seidel}, G. and {Walsh}, J. and {Aghanim}, N. and {Amendola}, L. and {Bartlett}, J. and {Baccigalupi}, C. and {Beaulieu}, J. -P. and {Benabed}, K. and {Cuby}, J. -G. and {Elbaz}, D. and {Fosalba}, P. and {Gavazzi}, G. and {Helmi}, A. and {Hook}, I. and {Irwin}, M. and {Kneib}, J. -P. and {Kunz}, M. and {Mannucci}, F. and {Moscardini}, L. and {Tao}, C. and {Teyssier}, R. and {Weller}, J. and {Zamorani}, G. and {Zapatero Osorio}, M.~R. and {Boulade}, O. and {Foumond}, J.~J. and {Di Giorgio}, A. and {Guttridge}, P. and {James}, A. and {Kemp}, M. and {Martignac}, J. and {Spencer}, A. and {Walton}, D. and {Bl{\"u}mchen}, T. and {Bonoli}, C. and {Bortoletto}, F. and {Cerna}, C. and {Corcione}, L. and {Fabron}, C. and {Jahnke}, K. and {Ligori}, S. and {Madrid}, F. and {Martin}, L. and {Morgante}, G. and {Pamplona}, T. and {Prieto}, E. and {Riva}, M. and {Toledo}, R. and {Trifoglio}, M. and {Zerbi}, F. and {Abdalla}, F. and {Douspis}, M. and {Grenet}, C. and {Borgani}, S. and {Bouwens}, R. and {Courbin}, F. and {Delouis}, J. -M. and {Dubath}, P. and {Fontana}, A. and {Frailis}, M. and {Grazian}, A. and {Koppenh{\"o}fer}, J. and {Mansutti}, O. and {Melchior}, M. and {Mignoli}, M. and {Mohr}, J. and {Neissner}, C. and {Noddle}, K. and {Poncet}, M. and {Scodeggio}, M. and {Serrano}, S. and {Shane}, N. and {Starck}, J. -L. and {Surace}, C. and {Taylor}, A. and {Verdoes-Kleijn}, G. and {Vuerli}, C. and {Williams}, O.~R. and {Zacchei}, A. and {Altieri}, B. and {Escudero Sanz}, I. and {Kohley}, R. and {Oosterbroek}, T. and {Astier}, P. and {Bacon}, D. and {Bardelli}, S. and {Baugh}, C. and {Bellagamba}, F. and {Benoist}, C. and {Bianchi}, D. and {Biviano}, A. and {Branchini}, E. and {Carbone}, C. and {Cardone}, V. and {Clements}, D. and {Colombi}, S. and {Conselice}, C. and {Cresci}, G. and {Deacon}, N. and {Dunlop}, J. and {Fedeli}, C. and {Fontanot}, F. and {Franzetti}, P. and {Giocoli}, C. and {Garcia-Bellido}, J. and {Gow}, J. and {Heavens}, A. and {Hewett}, P. and {Heymans}, C. and {Holland}, A. and {Huang}, Z. and {Ilbert}, O. and {Joachimi}, B. and {Jennins}, E. and {Kerins}, E. and {Kiessling}, A. and {Kirk}, D. and {Kotak}, R. and {Krause}, O. and {Lahav}, O. and {van Leeuwen}, F. and {Lesgourgues}, J. and {Lombardi}, M. and {Magliocchetti}, M. and {Maguire}, K. and {Majerotto}, E. and {Maoli}, R. and {Marulli}, F. and {Maurogordato}, S. and {McCracken}, H. and {McLure}, R. and {Melchiorri}, A. and {Merson}, A. and {Moresco}, M. and {Nonino}, M. and {Norberg}, P. and {Peacock}, J. and {Pello}, R. and {Penny}, M. and {Pettorino}, V. and {Di Porto}, C. and {Pozzetti}, L. and {Quercellini}, C. and {Radovich}, M. and {Rassat}, A. and {Roche}, N. and {Ronayette}, S. and {Rossetti}, E.},
        title = "{Euclid Definition Study Report}",
      journal = {arXiv e-prints},
         year = 2011,
        month = oct,
          eid = {arXiv:1110.3193},
        pages = {arXiv:1110.3193},
          doi = {10.48550/arXiv.1110.3193},
archivePrefix = {arXiv},
       eprint = {1110.3193},
 primaryClass = {astro-ph.CO},
       adsurl = {https://ui.adsabs.harvard.edu/abs/2011arXiv1110.3193L}
}

@ARTICLE{Laigle25,
       author = {{Euclid Collaboration} and {Laigle}, C. and {Gouin}, C. and {Sarron}, F. and {Quilley}, L. and {Pichon}, C. and {Kraljic}, K. and {Durret}, F. and {Chisari}, N.~E. and {Kuchner}, U. and {Malavasi}, N. and {Magliocchetti}, M. and {McCracken}, H.~J. and {Sorce}, J.~G. and {Kang}, Y. and {McPartland}, C.~J.~R. and {Toft}, S. and {Aghanim}, N. and {Altieri}, B. and {Amara}, A. and {Andreon}, S. and {Auricchio}, N. and {Aussel}, H. and {Baccigalupi}, C. and {Baldi}, M. and {Balestra}, A. and {Bardelli}, S. and {Basset}, A. and {Battaglia}, P. and {Bernardeau}, F. and {Biviano}, A. and {Bonchi}, A. and {Branchini}, E. and {Brescia}, M. and {Brinchmann}, J. and {Camera}, S. and {Ca{\~n}as-Herrera}, G. and {Capobianco}, V. and {Carbone}, C. and {Carretero}, J. and {Casas}, S. and {Castellano}, M. and {Castignani}, G. and {Cavuoti}, S. and {Chambers}, K.~C. and {Cimatti}, A. and {Colodro-Conde}, C. and {Congedo}, G. and {Conselice}, C.~J. and {Conversi}, L. and {Copin}, Y. and {Courbin}, F. and {Courtois}, H.~M. and {Cropper}, M. and {Da Silva}, A. and {Degaudenzi}, H. and {de la Torre}, S. and {De Lucia}, G. and {Di Giorgio}, A.~M. and {Dolding}, C. and {Dole}, H. and {Dubath}, F. and {Duncan}, C.~A.~J. and {Dupac}, X. and {Ealet}, A. and {Escoffier}, S. and {Farina}, M. and {Farinelli}, R. and {Faustini}, F. and {Ferriol}, S. and {Finelli}, F. and {Fotopoulou}, S. and {Frailis}, M. and {Franceschi}, E. and {Galeotta}, S. and {George}, K. and {Gillard}, W. and {Gillis}, B. and {Giocoli}, C. and {G{\'o}mez-Alvarez}, P. and {Gracia-Carpio}, J. and {Granett}, B.~R. and {Grazian}, A. and {Grupp}, F. and {Gwyn}, S. and {Haugan}, S.~V.~H. and {Hoekstra}, H. and {Holmes}, W. and {Hook}, I.~M. and {Hormuth}, F. and {Hornstrup}, A. and {Hudelot}, P. and {Jahnke}, K. and {Jhabvala}, M. and {Joachimi}, B. and {Keih{\"a}nen}, E. and {Kermiche}, S. and {Kiessling}, A. and {Kilbinger}, M. and {Kubik}, B. and {Kuijken}, K. and {K{\"u}mmel}, M. and {Kunz}, M. and {Kurki-Suonio}, H. and {Le Boulc'h}, Q. and {Le Brun}, A.~M.~C. and {Le Mignant}, D. and {Liebing}, P. and {Ligori}, S. and {Lilje}, P.~B. and {Lindholm}, V. and {Lloro}, I. and {Mainetti}, G. and {Maino}, D. and {Maiorano}, E. and {Mansutti}, O. and {Marcin}, S. and {Marggraf}, O. and {Martinelli}, M. and {Martinet}, N. and {Marulli}, F. and {Massey}, R. and {Maurogordato}, S. and {Medinaceli}, E. and {Mei}, S. and {Melchior}, M. and {Mellier}, Y. and {Meneghetti}, M. and {Merlin}, E. and {Meylan}, G. and {Mora}, A. and {Moresco}, M. and {Moscardini}, L. and {Nakajima}, R. and {Neissner}, C. and {Niemi}, S. -M. and {Nightingale}, J.~W. and {Padilla}, C. and {Paltani}, S. and {Pasian}, F. and {Pedersen}, K. and {Percival}, W.~J. and {Pettorino}, V. and {Pires}, S. and {Polenta}, G. and {Poncet}, M. and {Popa}, L.~A. and {Pozzetti}, L. and {Raison}, F. and {Rebolo}, R. and {Renzi}, A. and {Rhodes}, J. and {Riccio}, G. and {Romelli}, E. and {Roncarelli}, M. and {Rusholme}, B. and {Saglia}, R. and {Sakr}, Z. and {S{\'a}nchez}, A.~G. and {Sapone}, D. and {Sartoris}, B. and {Schewtschenko}, J.~A. and {Schirmer}, M. and {Schneider}, P. and {Schrabback}, T. and {Scodeggio}, M. and {Secroun}, A. and {Seidel}, G. and {Seiffert}, M. and {Serrano}, S. and {Simon}, P. and {Sirignano}, C. and {Sirri}, G. and {Skottfelt}, J. and {Stanco}, L. and {Steinwagner}, J. and {Tallada-Cresp{\'\i}}, P. and {Taylor}, A.~N. and {Teplitz}, H.~I. and {Tereno}, I. and {Tessore}, N. and {Toledo-Moreo}, R. and {Torradeflot}, F. and {Tutusaus}, I. and {Valenziano}, L. and {Valiviita}, J. and {Vassallo}, T. and {Verdoes Kleijn}, G. and {Veropalumbo}, A. and {Wang}, Y. and {Weller}, J. and {Zacchei}, A. and {Zamorani}, G. and {Zerbi}, F.~M. and {Zinchenko}, I.~A. and {Zucca}, E. and {Allevato}, V. and {Ballardini}, M. and {Bolzonella}, M. and {Bozzo}, E.},
        title = "{Euclid Quick Data Release (Q1). Galaxy shapes and alignments in the cosmic web}",
      journal = {arXiv e-prints},
         year = 2025,
        month = mar,
          eid = {arXiv:2503.15333},
        pages = {arXiv:2503.15333},
          doi = {10.48550/arXiv.2503.15333},
archivePrefix = {arXiv},
       eprint = {2503.15333},
 primaryClass = {astro-ph.GA},
       adsurl = {https://ui.adsabs.harvard.edu/abs/2025arXiv250315333E}
}

@INPROCEEDINGS{EuclidDDP,
       author = {{Guy}, Leanne P. and {Cuillandre}, Jean-Charles and {Bachelet}, Etienne and {Banerji}, Manda and {Bauer}, Franz E. and {Collett}, Thomas and {Conselice}, Christopher J. and {Eggl}, Siegfried and {Ferguson}, Annette and {Fontana}, Adriano and {Heymans}, Catherine and {Hook}, Isobel M. and {Aubourg}, {\'E}ric and {Aussel}, Herv{\'e} and {Bosch}, James and {Carry}, Benoit and {Hoekstra}, Henk and {Kuijken}, Konrad and {Lanusse}, Francois and {Melchior}, Peter and {Mohr}, Joseph and {Moresco}, Michele and {Nakajima}, Reiko and {Paltani}, St{\'e}phane and {Troxel}, Michael and {Allevato}, Viola and {Amara}, Adam and {Andreon}, Stefano and {Anguita}, Timo and {Bardelli}, Sandro and {Bechtol}, Keith and {Birrer}, Simon and {Bisigello}, Laura and {Bolzonella}, Micol and {Botticella}, Maria Teresa and {Bouy}, Herv{\'e} and {Brinchmann}, Jarle and {Brough}, Sarah and {Camera}, Stefano and {Cantiello}, Michele and {Cappellaro}, Enrico and {Carlin}, Jeffrey L. and {Castander}, Francisco J. and {Castellano}, Marco and {Chari}, Ranga Ram and {Chisari}, Nora Elisa and {Collins}, Christopher and {Courbin}, Fr{\'e}d{\'e}ric and {Cuby}, Jean-Gabriel and {Cucciati}, Olga and {Daylan}, Tansu and {Diego}, Jose M. and {Duc}, Pierre-Alain and {Fotopoulou}, Sotiria and {Fouchez}, Dominique and {Gavazzi}, Rapha{\"e}l and {Gruen}, Daniel and {Hatfield}, Peter and {Hildebrandt}, Hendrik and {Landt}, Hermine and {Hunt}, Leslie K. and {Ibata}, Rodrigo and {Ilbert}, Olivier and {Jasche}, Jens and {Joachimi}, Benjamin and {Joseph}, R{\'e}my and {Knight}, Matthew and {Kotak}, Rubina and {Laigle}, Clotilde and {Lan{\c{c}}on}, Ariane and {Larsen}, S{\o}ren S. and {Lavaux}, Guilhem and {Leclercq}, Florent and {Leonard}, C. Danielle and {von der Linden}, Anja and {Liu}, Xin and {Longo}, Giuseppe and {Magliocchetti}, Manuela and {Maraston}, Claudia and {Marshall}, Phil and {Mart{\'\i}n}, Eduardo L. and {Mattila}, Seppo and {Maturi}, Matteo and {McCracken}, Henry Joy and {Metcalf}, R. Benton and {Montes}, Mireia and {Mortlock}, Daniel and {Moscardini}, Lauro and {Narayan}, Gautham and {Paolillo}, Maurizio and {Papaderos}, Polychronis and {Pello}, Roser and {Pozzetti}, Lucia and {Radovich}, Mario and {Rejkuba}, Marina and {Rom{\'a}n}, Javier and {S{\'a}nchez-Janssen}, Rub{\'e}n and {Sarpa}, Elena and {Sartoris}, Barbara and {Schrabback}, Tim and {Sluse}, Dominique and {Smartt}, Stephen J. and {Smith}, Graham P. and {Snodgrass}, Colin and {Talia}, Margherita and {Tao}, Charling and {Toft}, Sune and {Tortora}, Crescenzo and {Tutusaus}, Isaac and {Usher}, Christopher and {van Velzen}, Sjoert and {Verma}, Aprajita and {Vernardos}, Georgios and {Voggel}, Karina and {Wandelt}, Benjamin and {Watkins}, Aaron E. and {Weller}, Jochen and {Wright}, Angus H. and {Yoachim}, Peter and {Yoon}, Ilsang and {Zucca}, Elena},
        title = "{Rubin-Euclid Derived Data Products: Initial Recommendations}",
    booktitle = {Zenodo id. 5836022},
         year = 2022,
       volume = {58},
        month = jan,
          eid = {5836022},
        pages = {5836022},
          doi = {10.5281/zenodo.5836022},
archivePrefix = {arXiv},
       eprint = {2201.03862},
 primaryClass = {astro-ph.IM},
       adsurl = {https://ui.adsabs.harvard.edu/abs/2022zndo...5836022G}
}

@ARTICLE{multiplet,
       author = {{Lamman}, Claire and {Eisenstein}, Daniel and {Forero-Romero}, Jaime E. and {Aguilar}, Jessica Nicole and {Ahlen}, Steven and {Bailey}, Stephen and {Bianchi}, Davide and {Brooks}, David and {Claybaugh}, Todd and {de la Macorra}, Axel and {Doel}, Peter and {Ferraro}, Simone and {Font-Ribera}, Andreu and {Gazta{\~n}aga}, Enrique and {Gontcho A Gontcho}, Satya and {Gutierrez}, Gaston and {Honscheid}, Klaus and {Howlett}, Cullan and {Kremin}, Anthony and {Lambert}, Andrew and {Landriau}, Martin and {Le Guillou}, Laurent and {Levi}, Michael E. and {Meisner}, Aaron and {Miquel}, Ramon and {Moustakas}, John and {Newman}, Jeffrey A. and {Niz}, Gustavo and {Prada}, Francisco and {P{\'e}rez-R{\`a}fols}, Ignasi and {Ross}, Ashley J. and {Rossi}, Graziano and {Sanchez}, Eusebio and {Schubnell}, Michael and {Sprayberry}, David and {Tarl{\'e}}, Gregory and {Vargas-Maga{\~n}a}, Mariana and {Weaver}, Benjamin Alan and {Zou}, Hu},
        title = "{Detection of the large-scale tidal field with galaxy multiplet alignment in the DESI Y1 spectroscopic survey}",
      journal = {\mnras},
         year = 2024,
        month = nov,
       volume = {534},
       number = {4},
        pages = {3540-3551},
          doi = {10.1093/mnras/stae2290},
archivePrefix = {arXiv},
       eprint = {2408.11056},
 primaryClass = {astro-ph.CO},
       adsurl = {https://ui.adsabs.harvard.edu/abs/2024MNRAS.534.3540L}
}

@ARTICLE{MacMahon24,
       author = {{MacMahon-Gell{\'e}r}, Charlie and {Leonard}, C. Danielle},
        title = "{Intrinsic alignment from multiple shear estimates: a first application to data and forecasts for stage IV}",
      journal = {\mnras},
         year = 2024,
        month = feb,
       volume = {528},
       number = {2},
        pages = {2980-2999},
          doi = {10.1093/mnras/stae054},
archivePrefix = {arXiv},
       eprint = {2306.11428},
 primaryClass = {astro-ph.CO},
       adsurl = {https://ui.adsabs.harvard.edu/abs/2024MNRAS.528.2980M}
}

@ARTICLE{Takada12,
       author = {{Takada}, Masahiro and {Ellis}, Richard S. and {Chiba}, Masashi and {Greene}, Jenny E. and {Aihara}, Hiroaki and {Arimoto}, Nobuo and {Bundy}, Kevin and {Cohen}, Judith and {Dor{\'e}}, Olivier and {Graves}, Genevieve and {Gunn}, James E. and {Heckman}, Timothy and {Hirata}, Christopher M. and {Ho}, Paul and {Kneib}, Jean-Paul and {Le F{\`e}vre}, Olivier and {Lin}, Lihwai and {More}, Surhud and {Murayama}, Hitoshi and {Nagao}, Tohru and {Ouchi}, Masami and {Seiffert}, Michael and {Silverman}, John D. and {Sodr{\'e}}, Laerte and {Spergel}, David N. and {Strauss}, Michael A. and {Sugai}, Hajime and {Suto}, Yasushi and {Takami}, Hideki and {Wyse}, Rosemary},
        title = "{Extragalactic science, cosmology, and Galactic archaeology with the Subaru Prime Focus Spectrograph}",
      journal = {\pasj},
         year = 2014,
        month = feb,
       volume = {66},
       number = {1},
          eid = {R1},
        pages = {R1},
          doi = {10.1093/pasj/pst019},
archivePrefix = {arXiv},
       eprint = {1206.0737},
 primaryClass = {astro-ph.CO},
       adsurl = {https://ui.adsabs.harvard.edu/abs/2014PASJ...66R...1T}
}

@ARTICLE{Akeson19,
       author = {{Akeson}, Rachel and {Armus}, Lee and {Bachelet}, Etienne and {Bailey}, Vanessa and {Bartusek}, Lisa and {Bellini}, Andrea and {Benford}, Dominic and {Bennett}, David and {Bhattacharya}, Aparna and {Bohlin}, Ralph and {Boyer}, Martha and {Bozza}, Valerio and {Bryden}, Geoffrey and {Calchi Novati}, Sebastiano and {Carpenter}, Kenneth and {Casertano}, Stefano and {Choi}, Ami and {Content}, David and {Dayal}, Pratika and {Dressler}, Alan and {Dor{\'e}}, Olivier and {Fall}, S. Michael and {Fan}, Xiaohui and {Fang}, Xiao and {Filippenko}, Alexei and {Finkelstein}, Steven and {Foley}, Ryan and {Furlanetto}, Steven and {Kalirai}, Jason and {Gaudi}, B. Scott and {Gilbert}, Karoline and {Girard}, Julien and {Grady}, Kevin and {Greene}, Jenny and {Guhathakurta}, Puragra and {Heinrich}, Chen and {Hemmati}, Shoubaneh and {Hendel}, David and {Henderson}, Calen and {Henning}, Thomas and {Hirata}, Christopher and {Ho}, Shirley and {Huff}, Eric and {Hutter}, Anne and {Jansen}, Rolf and {Jha}, Saurabh and {Johnson}, Samson and {Jones}, David and {Kasdin}, Jeremy and {Kelly}, Patrick and {Kirshner}, Robert and {Koekemoer}, Anton and {Kruk}, Jeffrey and {Lewis}, Nikole and {Macintosh}, Bruce and {Madau}, Piero and {Malhotra}, Sangeeta and {Mandel}, Kaisey and {Massara}, Elena and {Masters}, Daniel and {McEnery}, Julie and {McQuinn}, Kristen and {Melchior}, Peter and {Melton}, Mark and {Mennesson}, Bertrand and {Peeples}, Molly and {Penny}, Matthew and {Perlmutter}, Saul and {Pisani}, Alice and {Plazas}, Andr{\'e}s and {Poleski}, Radek and {Postman}, Marc and {Ranc}, Cl{\'e}ment and {Rauscher}, Bernard and {Rest}, Armin and {Roberge}, Aki and {Robertson}, Brant and {Rodney}, Steven and {Rhoads}, James and {Rhodes}, Jason and {Ryan}, Jr., Russell and {Sahu}, Kailash and {Sand}, David and {Scolnic}, Dan and {Seth}, Anil and {Shvartzvald}, Yossi and {Siellez}, Karelle and {Smith}, Arfon and {Spergel}, David and {Stassun}, Keivan and {Street}, Rachel and {Strolger}, Louis-Gregory and {Szalay}, Alexander and {Trauger}, John and {Troxel}, M.~A. and {Turnbull}, Margaret and {van der Marel}, Roeland and {von der Linden}, Anja and {Wang}, Yun and {Weinberg}, David and {Williams}, Benjamin and {Windhorst}, Rogier and {Wollack}, Edward and {Wu}, Hao-Yi and {Yee}, Jennifer and {Zimmerman}, Neil},
        title = "{The Wide Field Infrared Survey Telescope: 100 Hubbles for the 2020s}",
      journal = {arXiv e-prints},
         year = 2019,
        month = feb,
          eid = {arXiv:1902.05569},
        pages = {arXiv:1902.05569},
          doi = {10.48550/arXiv.1902.05569},
archivePrefix = {arXiv},
       eprint = {1902.05569},
 primaryClass = {astro-ph.IM},
       adsurl = {https://ui.adsabs.harvard.edu/abs/2019arXiv190205569A}
}

@ARTICLE{StageV,
       author = {{Schlegel}, David J. and {Ferraro}, Simone and {Aldering}, Greg and {Baltay}, Charles and {BenZvi}, Segev and {Besuner}, Robert and {Blanc}, Guillermo A. and {Bolton}, Adam S. and {Bonaca}, Ana and {Brooks}, David and {Buckley-Geer}, Elizabeth and {Cai}, Zheng and {DeRose}, Joseph and {Dey}, Arjun and {Doel}, Peter and {Drlica-Wagner}, Alex and {Fan}, Xiaohui and {Gutierrez}, Gaston and {Green}, Daniel and {Guy}, Julien and {Huterer}, Dragan and {Infante}, Leopoldo and {Jelinsky}, Patrick and {Karagiannis}, Dionysios and {Kent}, Stephen M. and {Kim}, Alex G. and {Kneib}, Jean-Paul and {Kollmeier}, Juna A. and {Kremin}, Anthony and {Lahav}, Ofer and {Landriau}, Martin and {Lang}, Dustin and {Leauthaud}, Alexie and {Levi}, Michael E. and {Linder}, Eric V. and {Magneville}, Christophe and {Martini}, Paul and {McDonald}, Patrick and {Miller}, Christopher J. and {Myers}, Adam D. and {Newman}, Jeffrey A. and {Nugent}, Peter E. and {Palanque-Delabrouille}, Nathalie and {Padmanabhan}, Nikhil and {Palmese}, Antonella and {Poppett}, Claire and {Prochaska}, Jason X. and {Raichoor}, Anand and {Ramirez}, Solange and {Sailer}, Noah and {Schaan}, Emmanuel and {Schubnell}, Michael and {Seljak}, Uros and {Seo}, Hee-Jong and {Silber}, Joseph and {Simon}, Joshua D. and {Slepian}, Zachary and {Soares-Santos}, Marcelle and {Tarle}, Greg and {Valluri}, Monica and {Weaverdyck}, Noah J. and {Wechsler}, Risa H. and {White}, Martin and {Yeche}, Christophe and {Zhou}, Rongpu},
        title = "{A Spectroscopic Road Map for Cosmic Frontier: DESI, DESI-II, Stage-5}",
      journal = {arXiv e-prints},
         year = 2022,
        month = sep,
          eid = {arXiv:2209.03585},
        pages = {arXiv:2209.03585},
          doi = {10.48550/arXiv.2209.03585},
archivePrefix = {arXiv},
       eprint = {2209.03585},
 primaryClass = {astro-ph.CO},
       adsurl = {https://ui.adsabs.harvard.edu/abs/2022arXiv220903585S}
}

@ARTICLE{Richard19,
       author = {{Richard}, J. and {Kneib}, J. -P. and {Blake}, C. and {Raichoor}, A. and {Comparat}, J. and {Shanks}, T. and {Sorce}, J. and {Sahl{\'e}n}, M. and {Howlett}, C. and {Tempel}, E. and {McMahon}, R. and {Bilicki}, M. and {Roukema}, B. and {Loveday}, J. and {Pryer}, D. and {Buchert}, T. and {Zhao}, C. and {CRS Team}},
        title = "{4MOST Consortium Survey 8: Cosmology Redshift Survey (CRS)}",
      journal = {The Messenger},
         year = 2019,
        month = mar,
       volume = {175},
        pages = {50-53},
          doi = {10.18727/0722-6691/5127},
archivePrefix = {arXiv},
       eprint = {1903.02474},
 primaryClass = {astro-ph.CO},
       adsurl = {https://ui.adsabs.harvard.edu/abs/2019Msngr.175...50R}
}

@ARTICLE{Driver19,
       author = {{Driver}, S.~P. and {Liske}, J. and {Davies}, L.~J.~M. and {Robotham}, A.~S.~G. and {Baldry}, I.~K. and {Brown}, M.~J.~I. and {Cluver}, M. and {Kuijken}, K. and {Loveday}, J. and {McMahon}, R. and {Meyer}, M.~J. and {Norberg}, P. and {Owers}, M. and {Power}, C. and {Taylor}, E.~N. and {WAVES Team}},
        title = "{4MOST Consortium Survey 7: Wide-Area VISTA Extragalactic Survey (WAVES)}",
      journal = {The Messenger},
         year = 2019,
        month = mar,
       volume = {175},
        pages = {46-49},
          doi = {10.18727/0722-6691/5126},
archivePrefix = {arXiv},
       eprint = {1903.02473},
 primaryClass = {astro-ph.GA},
       adsurl = {https://ui.adsabs.harvard.edu/abs/2019Msngr.175...46D}
}

@ARTICLE{Friday22,
       author = {{Friday}, Tracey and {Clowes}, Roger G. and {Williger}, Gerard M.},
        title = "{Correlated orientations of the axes of large quasar groups on Gpc scales}",
      journal = {\mnras},
         year = 2022,
        month = apr,
       volume = {511},
       number = {3},
        pages = {4159-4178},
          doi = {10.1093/mnras/stac269},
archivePrefix = {arXiv},
       eprint = {2201.11474},
 primaryClass = {astro-ph.CO},
       adsurl = {https://ui.adsabs.harvard.edu/abs/2022MNRAS.511.4159F}
}

@ARTICLE{Hutsemekers14,
       author = {{Hutsem{\'e}kers}, D. and {Braibant}, L. and {Pelgrims}, V. and {Sluse}, D.},
        title = "{Alignment of quasar polarizations with large-scale structures}",
      journal = {\aap},
         year = 2014,
        month = dec,
       volume = {572},
          eid = {A18},
        pages = {A18},
          doi = {10.1051/0004-6361/201424631},
       adsurl = {https://ui.adsabs.harvard.edu/abs/2014A&A...572A..18H}
}

@ARTICLE{Kraljic21,
       author = {{Kraljic}, Katarina and {Duckworth}, Christopher and {Tojeiro}, Rita and {Alam}, Shadab and {Bizyaev}, Dmitry and {Weijmans}, Anne-Marie and {Boardman}, Nicholas Fraser and {Lane}, Richard R.},
        title = "{SDSS-IV MaNGA: 3D spin alignment of spiral and S0 galaxies}",
      journal = {\mnras},
         year = 2021,
        month = jul,
       volume = {504},
       number = {3},
        pages = {4626-4633},
          doi = {10.1093/mnras/stab1109},
archivePrefix = {arXiv},
       eprint = {2104.08275},
 primaryClass = {astro-ph.GA},
       adsurl = {https://ui.adsabs.harvard.edu/abs/2021MNRAS.504.4626K}
}

@ARTICLE{Welker20,
       author = {{Welker}, C. and {Bland-Hawthorn}, J. and {van de Sande}, J. and {Lagos}, C. and {Elahi}, P. and {Obreschkow}, D. and {Bryant}, J. and {Pichon}, C. and {Cortese}, L. and {Richards}, S.~N. and {Croom}, S.~M. and {Goodwin}, M. and {Lawrence}, J.~S. and {Sweet}, S. and {Lopez-Sanchez}, A. and {Medling}, A. and {Owers}, M.~S. and {Dubois}, Y. and {Devriendt}, J.},
        title = "{The SAMI Galaxy Survey: first detection of a transition in spin orientation with respect to cosmic filaments in the stellar kinematics of galaxies}",
      journal = {\mnras},
         year = 2020,
        month = jan,
       volume = {491},
       number = {2},
        pages = {2864-2884},
          doi = {10.1093/mnras/stz2860},
archivePrefix = {arXiv},
       eprint = {1909.12371},
 primaryClass = {astro-ph.GA},
       adsurl = {https://ui.adsabs.harvard.edu/abs/2020MNRAS.491.2864W}
}

@ARTICLE{Niederste10,
       author = {{Niederste-Ostholt}, Martin and {Strauss}, Michael A. and {Dong}, Feng and {Koester}, Benjamin P. and {McKay}, Timothy A.},
        title = "{Alignment of brightest cluster galaxies with their host clusters}",
      journal = {\mnras},
         year = 2010,
        month = jul,
       volume = {405},
       number = {3},
        pages = {2023-2036},
          doi = {10.1111/j.1365-2966.2010.16597.x},
archivePrefix = {arXiv},
       eprint = {1003.0322},
 primaryClass = {astro-ph.CO},
       adsurl = {https://ui.adsabs.harvard.edu/abs/2010MNRAS.405.2023N}
}

@ARTICLE{Schneider13,
       author = {{Schneider}, Michael D. and {Cole}, Shaun and {Frenk}, Carlos S. and {Kelvin}, Lee and {Mandelbaum}, Rachel and {Norberg}, Peder and {Bland-Hawthorn}, Joss and {Brough}, Sarah and {Driver}, Simon and {Hopkins}, Andrew and {Liske}, Jochen and {Loveday}, Jon and {Robotham}, Aaron},
        title = "{Galaxy And Mass Assembly (GAMA): galaxy radial alignments in GAMA groups}",
      journal = {\mnras},
         year = 2013,
        month = aug,
       volume = {433},
       number = {4},
        pages = {2727-2738},
          doi = {10.1093/mnras/stt855},
archivePrefix = {arXiv},
       eprint = {1306.4963},
 primaryClass = {astro-ph.CO},
       adsurl = {https://ui.adsabs.harvard.edu/abs/2013MNRAS.433.2727S}
}

@ARTICLE{Scoville07,
       author = {{Scoville}, N. and {Aussel}, H. and {Brusa}, M. and {Capak}, P. and {Carollo}, C.~M. and {Elvis}, M. and {Giavalisco}, M. and {Guzzo}, L. and {Hasinger}, G. and {Impey}, C. and {Kneib}, J. -P. and {LeFevre}, O. and {Lilly}, S.~J. and {Mobasher}, B. and {Renzini}, A. and {Rich}, R.~M. and {Sanders}, D.~B. and {Schinnerer}, E. and {Schminovich}, D. and {Shopbell}, P. and {Taniguchi}, Y. and {Tyson}, N.~D.},
        title = "{The Cosmic Evolution Survey (COSMOS): Overview}",
      journal = {\apjs},
         year = 2007,
        month = sep,
       volume = {172},
       number = {1},
        pages = {1-8},
          doi = {10.1086/516585},
archivePrefix = {arXiv},
       eprint = {astro-ph/0612305},
 primaryClass = {astro-ph},
       adsurl = {https://ui.adsabs.harvard.edu/abs/2007ApJS..172....1S}
}

@BOOK{CIC,
       author = {{Hockney}, R.~W. and {Eastwood}, J.~W.},
        title = "{Computer Simulation Using Particles}",
      address = {New York},
    publisher = {McGraw-Hill},
         year = 1981,
       adsurl = {https://ui.adsabs.harvard.edu/abs/1981csup.book.....H}
}

@ARTICLE{FKP,
       author = {{Feldman}, Hume A. and {Kaiser}, Nick and {Peacock}, John A.},
        title = "{Power-Spectrum Analysis of Three-dimensional Redshift Surveys}",
      journal = {\apj},
         year = 1994,
        month = may,
       volume = {426},
        pages = {23},
          doi = {10.1086/174036},
archivePrefix = {arXiv},
       eprint = {astro-ph/9304022},
 primaryClass = {astro-ph},
       adsurl = {https://ui.adsabs.harvard.edu/abs/1994ApJ...426...23F}
}

@ARTICLE{Li23,
       author = {{Li}, Xiangchong and {Zhang}, Tianqing and {Sugiyama}, Sunao and {Dalal}, Roohi and {Terasawa}, Ryo and {Rau}, Markus M. and {Mandelbaum}, Rachel and {Takada}, Masahiro and {More}, Surhud and {Strauss}, Michael A. and {Miyatake}, Hironao and {Shirasaki}, Masato and {Hamana}, Takashi and {Oguri}, Masamune and {Luo}, Wentao and {Nishizawa}, Atsushi J. and {Takahashi}, Ryuichi and {Nicola}, Andrina and {Osato}, Ken and {Kannawadi}, Arun and {Sunayama}, Tomomi and {Armstrong}, Robert and {Bosch}, James and {Komiyama}, Yutaka and {Lupton}, Robert H. and {Lust}, Nate B. and {MacArthur}, Lauren A. and {Miyazaki}, Satoshi and {Murayama}, Hitoshi and {Nishimichi}, Takahiro and {Okura}, Yuki and {Price}, Paul A. and {Tait}, Philip J. and {Tanaka}, Masayuki and {Wang}, Shiang-Yu},
        title = "{Hyper Suprime-Cam Year 3 results: Cosmology from cosmic shear two-point correlation functions}",
      journal = {\prd},
         year = 2023,
        month = dec,
       volume = {108},
       number = {12},
          eid = {123518},
        pages = {123518},
          doi = {10.1103/PhysRevD.108.123518},
archivePrefix = {arXiv},
       eprint = {2304.00702},
 primaryClass = {astro-ph.CO},
       adsurl = {https://ui.adsabs.harvard.edu/abs/2023PhRvD.108l3518L}
}

@ARTICLE{Dalal23,
       author = {{Dalal}, Roohi and {Li}, Xiangchong and {Nicola}, Andrina and {Zuntz}, Joe and {Strauss}, Michael A. and {Sugiyama}, Sunao and {Zhang}, Tianqing and {Rau}, Markus M. and {Mandelbaum}, Rachel and {Takada}, Masahiro and {More}, Surhud and {Miyatake}, Hironao and {Kannawadi}, Arun and {Shirasaki}, Masato and {Taniguchi}, Takanori and {Takahashi}, Ryuichi and {Osato}, Ken and {Hamana}, Takashi and {Oguri}, Masamune and {Nishizawa}, Atsushi J. and {Malag{\'o}n}, Andr{\'e}s A. Plazas and {Sunayama}, Tomomi and {Alonso}, David and {Slosar}, An{\v{z}}e and {Luo}, Wentao and {Armstrong}, Robert and {Bosch}, James and {Hsieh}, Bau-Ching and {Komiyama}, Yutaka and {Lupton}, Robert H. and {Lust}, Nate B. and {MacArthur}, Lauren A. and {Miyazaki}, Satoshi and {Murayama}, Hitoshi and {Nishimichi}, Takahiro and {Okura}, Yuki and {Price}, Paul A. and {Tait}, Philip J. and {Tanaka}, Masayuki and {Wang}, Shiang-Yu},
        title = "{Hyper Suprime-Cam Year 3 results: Cosmology from cosmic shear power spectra}",
      journal = {\prd},
         year = 2023,
        month = dec,
       volume = {108},
       number = {12},
          eid = {123519},
        pages = {123519},
          doi = {10.1103/PhysRevD.108.123519},
archivePrefix = {arXiv},
       eprint = {2304.00701},
 primaryClass = {astro-ph.CO},
       adsurl = {https://ui.adsabs.harvard.edu/abs/2023PhRvD.108l3519D}
}

@ARTICLE{Bilicki21,
       author = {{Bilicki}, M. and {Dvornik}, A. and {Hoekstra}, H. and {Wright}, A.~H. and {Chisari}, N.~E. and {Vakili}, M. and {Asgari}, M. and {Giblin}, B. and {Heymans}, C. and {Hildebrandt}, H. and {Holwerda}, B.~W. and {Hopkins}, A. and {Johnston}, H. and {Kannawadi}, A. and {Kuijken}, K. and {Nakoneczny}, S.~J. and {Shan}, H.~Y. and {Sonnenfeld}, A. and {Valentijn}, E.},
        title = "{Bright galaxy sample in the Kilo-Degree Survey Data Release 4. Selection, photometric redshifts, and physical properties}",
      journal = {\aap},
         year = 2021,
        month = sep,
       volume = {653},
          eid = {A82},
        pages = {A82},
          doi = {10.1051/0004-6361/202140352},
archivePrefix = {arXiv},
       eprint = {2101.06010},
 primaryClass = {astro-ph.GA},
       adsurl = {https://ui.adsabs.harvard.edu/abs/2021A&A...653A..82B}
}

@ARTICLE{Leauthaud07,
       author = {{Leauthaud}, Alexie and {Massey}, Richard and {Kneib}, Jean-Paul and {Rhodes}, Jason and {Johnston}, David E. and {Capak}, Peter and {Heymans}, Catherine and {Ellis}, Richard S. and {Koekemoer}, Anton M. and {Le F{\`e}vre}, Oliver and {Mellier}, Yannick and {R{\'e}fr{\'e}gier}, Alexandre and {Robin}, Annie C. and {Scoville}, Nick and {Tasca}, Lidia and {Taylor}, James E. and {Van Waerbeke}, Ludovic},
        title = "{Weak Gravitational Lensing with COSMOS: Galaxy Selection and Shape Measurements}",
      journal = {\apjs},
         year = 2007,
        month = sep,
       volume = {172},
       number = {1},
        pages = {219-238},
          doi = {10.1086/516598},
archivePrefix = {arXiv},
       eprint = {astro-ph/0702359},
 primaryClass = {astro-ph},
       adsurl = {https://ui.adsabs.harvard.edu/abs/2007ApJS..172..219L}
}

@ARTICLE{Troxel18,
       author = {{Troxel}, M.~A. and {MacCrann}, N. and {Zuntz}, J. and {Eifler}, T.~F. and {Krause}, E. and {Dodelson}, S. and {Gruen}, D. and {Blazek}, J. and {Friedrich}, O. and {Samuroff}, S. and {Prat}, J. and {Secco}, L.~F. and {Davis}, C. and {Fert{\'e}}, A. and {DeRose}, J. and {Alarcon}, A. and {Amara}, A. and {Baxter}, E. and {Becker}, M.~R. and {Bernstein}, G.~M. and {Bridle}, S.~L. and {Cawthon}, R. and {Chang}, C. and {Choi}, A. and {De Vicente}, J. and {Drlica-Wagner}, A. and {Elvin-Poole}, J. and {Frieman}, J. and {Gatti}, M. and {Hartley}, W.~G. and {Honscheid}, K. and {Hoyle}, B. and {Huff}, E.~M. and {Huterer}, D. and {Jain}, B. and {Jarvis}, M. and {Kacprzak}, T. and {Kirk}, D. and {Kokron}, N. and {Krawiec}, C. and {Lahav}, O. and {Liddle}, A.~R. and {Peacock}, J. and {Rau}, M.~M. and {Refregier}, A. and {Rollins}, R.~P. and {Rozo}, E. and {Rykoff}, E.~S. and {S{\'a}nchez}, C. and {Sevilla-Noarbe}, I. and {Sheldon}, E. and {Stebbins}, A. and {Varga}, T.~N. and {Vielzeuf}, P. and {Wang}, M. and {Wechsler}, R.~H. and {Yanny}, B. and {Abbott}, T.~M.~C. and {Abdalla}, F.~B. and {Allam}, S. and {Annis}, J. and {Bechtol}, K. and {Benoit-L{\'e}vy}, A. and {Bertin}, E. and {Brooks}, D. and {Buckley-Geer}, E. and {Burke}, D.~L. and {Carnero Rosell}, A. and {Carrasco Kind}, M. and {Carretero}, J. and {Castander}, F.~J. and {Crocce}, M. and {Cunha}, C.~E. and {D'Andrea}, C.~B. and {da Costa}, L.~N. and {DePoy}, D.~L. and {Desai}, S. and {Diehl}, H.~T. and {Dietrich}, J.~P. and {Doel}, P. and {Fernandez}, E. and {Flaugher}, B. and {Fosalba}, P. and {Garc{\'\i}a-Bellido}, J. and {Gaztanaga}, E. and {Gerdes}, D.~W. and {Giannantonio}, T. and {Goldstein}, D.~A. and {Gruendl}, R.~A. and {Gschwend}, J. and {Gutierrez}, G. and {James}, D.~J. and {Jeltema}, T. and {Johnson}, M.~W.~G. and {Johnson}, M.~D. and {Kent}, S. and {Kuehn}, K. and {Kuhlmann}, S. and {Kuropatkin}, N. and {Li}, T.~S. and {Lima}, M. and {Lin}, H. and {Maia}, M.~A.~G. and {March}, M. and {Marshall}, J.~L. and {Martini}, P. and {Melchior}, P. and {Menanteau}, F. and {Miquel}, R. and {Mohr}, J.~J. and {Neilsen}, E. and {Nichol}, R.~C. and {Nord}, B. and {Petravick}, D. and {Plazas}, A.~A. and {Romer}, A.~K. and {Roodman}, A. and {Sako}, M. and {Sanchez}, E. and {Scarpine}, V. and {Schindler}, R. and {Schubnell}, M. and {Smith}, M. and {Smith}, R.~C. and {Soares-Santos}, M. and {Sobreira}, F. and {Suchyta}, E. and {Swanson}, M.~E.~C. and {Tarle}, G. and {Thomas}, D. and {Tucker}, D.~L. and {Vikram}, V. and {Walker}, A.~R. and {Weller}, J. and {Zhang}, Y. and {DES Collaboration}},
        title = "{Dark Energy Survey Year 1 results: Cosmological constraints from cosmic shear}",
      journal = {\prd},
         year = 2018,
        month = aug,
       volume = {98},
       number = {4},
          eid = {043528},
        pages = {043528},
          doi = {10.1103/PhysRevD.98.043528},
archivePrefix = {arXiv},
       eprint = {1708.01538},
 primaryClass = {astro-ph.CO},
       adsurl = {https://ui.adsabs.harvard.edu/abs/2018PhRvD..98d3528T}
}

@ARTICLE{Hildebrandt17,
       author = {{Hildebrandt}, H. and {Viola}, M. and {Heymans}, C. and {Joudaki}, S. and {Kuijken}, K. and {Blake}, C. and {Erben}, T. and {Joachimi}, B. and {Klaes}, D. and {Miller}, L. and {Morrison}, C.~B. and {Nakajima}, R. and {Verdoes Kleijn}, G. and {Amon}, A. and {Choi}, A. and {Covone}, G. and {de Jong}, J.~T.~A. and {Dvornik}, A. and {Fenech Conti}, I. and {Grado}, A. and {Harnois-D{\'e}raps}, J. and {Herbonnet}, R. and {Hoekstra}, H. and {K{\"o}hlinger}, F. and {McFarland}, J. and {Mead}, A. and {Merten}, J. and {Napolitano}, N. and {Peacock}, J.~A. and {Radovich}, M. and {Schneider}, P. and {Simon}, P. and {Valentijn}, E.~A. and {van den Busch}, J.~L. and {van Uitert}, E. and {Van Waerbeke}, L.},
        title = "{KiDS-450: cosmological parameter constraints from tomographic weak gravitational lensing}",
      journal = {\mnras},
         year = 2017,
        month = feb,
       volume = {465},
       number = {2},
        pages = {1454-1498},
          doi = {10.1093/mnras/stw2805},
archivePrefix = {arXiv},
       eprint = {1606.05338},
 primaryClass = {astro-ph.CO},
       adsurl = {https://ui.adsabs.harvard.edu/abs/2017MNRAS.465.1454H}
}

@ARTICLE{Singh17,
       author = {{Singh}, Sukhdeep and {Mandelbaum}, Rachel and {Seljak}, Uro{\v{s}} and {Slosar}, An{\v{z}}e and {Vazquez Gonzalez}, Jose},
        title = "{Galaxy-galaxy lensing estimators and their covariance properties}",
      journal = {\mnras},
         year = 2017,
        month = nov,
       volume = {471},
       number = {4},
        pages = {3827-3844},
          doi = {10.1093/mnras/stx1828},
archivePrefix = {arXiv},
       eprint = {1611.00752},
 primaryClass = {astro-ph.CO},
       adsurl = {https://ui.adsabs.harvard.edu/abs/2017MNRAS.471.3827S}
}

@ARTICLE{Secco22,
       author = {{Secco}, L.~F. and {Samuroff}, S. and {Krause}, E. and {Jain}, B. and {Blazek}, J. and {Raveri}, M. and {Campos}, A. and {Amon}, A. and {Chen}, A. and {Doux}, C. and {Choi}, A. and {Gruen}, D. and {Bernstein}, G.~M. and {Chang}, C. and {DeRose}, J. and {Myles}, J. and {Fert{\'e}}, A. and {Lemos}, P. and {Huterer}, D. and {Prat}, J. and {Troxel}, M.~A. and {MacCrann}, N. and {Liddle}, A.~R. and {Kacprzak}, T. and {Fang}, X. and {S{\'a}nchez}, C. and {Pandey}, S. and {Dodelson}, S. and {Chintalapati}, P. and {Hoffmann}, K. and {Alarcon}, A. and {Alves}, O. and {Andrade-Oliveira}, F. and {Baxter}, E.~J. and {Bechtol}, K. and {Becker}, M.~R. and {Brandao-Souza}, A. and {Camacho}, H. and {Carnero Rosell}, A. and {Carrasco Kind}, M. and {Cawthon}, R. and {Cordero}, J.~P. and {Crocce}, M. and {Davis}, C. and {Di Valentino}, E. and {Drlica-Wagner}, A. and {Eckert}, K. and {Eifler}, T.~F. and {Elidaiana}, M. and {Elsner}, F. and {Elvin-Poole}, J. and {Everett}, S. and {Fosalba}, P. and {Friedrich}, O. and {Gatti}, M. and {Giannini}, G. and {Gruendl}, R.~A. and {Harrison}, I. and {Hartley}, W.~G. and {Herner}, K. and {Huang}, H. and {Huff}, E.~M. and {Jarvis}, M. and {Jeffrey}, N. and {Kuropatkin}, N. and {Leget}, P. -F. and {Muir}, J. and {Mccullough}, J. and {Navarro Alsina}, A. and {Omori}, Y. and {Park}, Y. and {Porredon}, A. and {Rollins}, R. and {Roodman}, A. and {Rosenfeld}, R. and {Ross}, A.~J. and {Rykoff}, E.~S. and {Sanchez}, J. and {Sevilla-Noarbe}, I. and {Sheldon}, E.~S. and {Shin}, T. and {Troja}, A. and {Tutusaus}, I. and {Varga}, T.~N. and {Weaverdyck}, N. and {Wechsler}, R.~H. and {Yanny}, B. and {Yin}, B. and {Zhang}, Y. and {Zuntz}, J. and {Abbott}, T.~M.~C. and {Aguena}, M. and {Allam}, S. and {Annis}, J. and {Bacon}, D. and {Bertin}, E. and {Bhargava}, S. and {Bridle}, S.~L. and {Brooks}, D. and {Buckley-Geer}, E. and {Burke}, D.~L. and {Carretero}, J. and {Costanzi}, M. and {da Costa}, L.~N. and {De Vicente}, J. and {Diehl}, H.~T. and {Dietrich}, J.~P. and {Doel}, P. and {Ferrero}, I. and {Flaugher}, B. and {Frieman}, J. and {Garc{\'\i}a-Bellido}, J. and {Gaztanaga}, E. and {Gerdes}, D.~W. and {Giannantonio}, T. and {Gschwend}, J. and {Gutierrez}, G. and {Hinton}, S.~R. and {Hollowood}, D.~L. and {Honscheid}, K. and {Hoyle}, B. and {James}, D.~J. and {Jeltema}, T. and {Kuehn}, K. and {Lahav}, O. and {Lima}, M. and {Lin}, H. and {Maia}, M.~A.~G. and {Marshall}, J.~L. and {Martini}, P. and {Melchior}, P. and {Menanteau}, F. and {Miquel}, R. and {Mohr}, J.~J. and {Morgan}, R. and {Ogando}, R.~L.~C. and {Palmese}, A. and {Paz-Chinch{\'o}n}, F. and {Petravick}, D. and {Pieres}, A. and {Plazas Malag{\'o}n}, A.~A. and {Rodriguez-Monroy}, M. and {Romer}, A.~K. and {Sanchez}, E. and {Scarpine}, V. and {Schubnell}, M. and {Scolnic}, D. and {Serrano}, S. and {Smith}, M. and {Soares-Santos}, M. and {Suchyta}, E. and {Swanson}, M.~E.~C. and {Tarle}, G. and {Thomas}, D. and {To}, C. and {DES Collaboration}},
        title = "{Dark Energy Survey Year 3 results: Cosmology from cosmic shear and robustness to modeling uncertainty}",
      journal = {\prd},
         year = 2022,
        month = jan,
       volume = {105},
       number = {2},
          eid = {023515},
        pages = {023515},
          doi = {10.1103/PhysRevD.105.023515},
archivePrefix = {arXiv},
       eprint = {2105.13544},
 primaryClass = {astro-ph.CO},
       adsurl = {https://ui.adsabs.harvard.edu/abs/2022PhRvD.105b3515S}
}

@ARTICLE{Wang09,
       author = {{Wang}, Yougang and {Park}, Changbom and {Yang}, Xiaohu and {Choi}, Yun-Young and {Chen}, Xuelei},
        title = "{Alignments of Group Galaxies with Neighboring Groups}",
      journal = {\apj},
         year = 2009,
        month = sep,
       volume = {703},
       number = {1},
        pages = {951-963},
          doi = {10.1088/0004-637X/703/1/951},
archivePrefix = {arXiv},
       eprint = {0810.3359},
 primaryClass = {astro-ph},
       adsurl = {https://ui.adsabs.harvard.edu/abs/2009ApJ...703..951W}
}

@ARTICLE{Paz11,
       author = {{Paz}, Dante J. and {Sgr{\'o}}, Mario A. and {Merch{\'a}n}, Manuel and {Padilla}, Nelson},
        title = "{Alignments of galaxy group shapes with large-scale structure}",
      journal = {\mnras},
         year = 2011,
        month = jul,
       volume = {414},
       number = {3},
        pages = {2029-2039},
          doi = {10.1111/j.1365-2966.2011.18518.x},
archivePrefix = {arXiv},
       eprint = {1102.2229},
 primaryClass = {astro-ph.CO},
       adsurl = {https://ui.adsabs.harvard.edu/abs/2011MNRAS.414.2029P}
}

@ARTICLE{Smargon12,
       author = {{Smargon}, A. and {Mandelbaum}, R. and {Bahcall}, N. and {Niederste-Ostholt}, M.},
        title = "{Detection of intrinsic cluster alignments to 100 h$^{-1}$ Mpc in the Sloan Digital Sky Survey}",
      journal = {\mnras},
         year = 2012,
        month = jun,
       volume = {423},
       number = {1},
        pages = {856-861},
          doi = {10.1111/j.1365-2966.2012.20923.x},
archivePrefix = {arXiv},
       eprint = {1109.6020},
 primaryClass = {astro-ph.CO},
       adsurl = {https://ui.adsabs.harvard.edu/abs/2012MNRAS.423..856S}
}

@ARTICLE{Simet17,
       author = {{Simet}, Melanie and {McClintock}, Tom and {Mandelbaum}, Rachel and {Rozo}, Eduardo and {Rykoff}, Eli and {Sheldon}, Erin and {Wechsler}, Risa H.},
        title = "{Weak lensing measurement of the mass-richness relation of SDSS redMaPPer clusters}",
      journal = {\mnras},
         year = 2017,
        month = apr,
       volume = {466},
       number = {3},
        pages = {3103-3118},
          doi = {10.1093/mnras/stw3250},
archivePrefix = {arXiv},
       eprint = {1603.06953},
 primaryClass = {astro-ph.CO},
       adsurl = {https://ui.adsabs.harvard.edu/abs/2017MNRAS.466.3103S}
}

@ARTICLE{redmapper,
       author = {{Rykoff}, E.~S. and {Rozo}, E. and {Busha}, M.~T. and {Cunha}, C.~E. and {Finoguenov}, A. and {Evrard}, A. and {Hao}, J. and {Koester}, B.~P. and {Leauthaud}, A. and {Nord}, B. and {Pierre}, M. and {Reddick}, R. and {Sadibekova}, T. and {Sheldon}, E.~S. and {Wechsler}, R.~H.},
        title = "{redMaPPer. I. Algorithm and SDSS DR8 Catalog}",
      journal = {\apj},
         year = 2014,
        month = apr,
       volume = {785},
       number = {2},
          eid = {104},
        pages = {104},
          doi = {10.1088/0004-637X/785/2/104},
archivePrefix = {arXiv},
       eprint = {1303.3562},
 primaryClass = {astro-ph.CO},
       adsurl = {https://ui.adsabs.harvard.edu/abs/2014ApJ...785..104R}
}

@ARTICLE{Slosar09,
       author = {{Slosar}, An{\v{z}}e and {White}, Martin},
        title = "{Alignment of galaxy spins in the vicinity of voids}",
      journal = {\jcap},
         year = 2009,
        month = jun,
       volume = {2009},
       number = {6},
          eid = {009},
        pages = {009},
          doi = {10.1088/1475-7516/2009/06/009},
archivePrefix = {arXiv},
       eprint = {0811.3216},
 primaryClass = {astro-ph},
       adsurl = {https://ui.adsabs.harvard.edu/abs/2009JCAP...06..009S}
}

@ARTICLE{Varela12,
       author = {{Varela}, Jes{\'u}s and {Betancort-Rijo}, Juan and {Trujillo}, Ignacio and {Ricciardelli}, Elena},
        title = "{The Orientation of Disk Galaxies around Large Cosmic Voids}",
      journal = {\apj},
         year = 2012,
        month = jan,
       volume = {744},
       number = {2},
          eid = {82},
        pages = {82},
          doi = {10.1088/0004-637X/744/2/82},
archivePrefix = {arXiv},
       eprint = {1109.2056},
 primaryClass = {astro-ph.CO},
       adsurl = {https://ui.adsabs.harvard.edu/abs/2012ApJ...744...82V}
}

@ARTICLE{Trujillo06,
       author = {{Trujillo}, Ignacio and {Carretero}, Conrado and {Patiri}, Santiago G.},
        title = "{Detection of the Effect of Cosmological Large-Scale Structure on the Orientation of Galaxies}",
      journal = {\apjl},
         year = 2006,
        month = apr,
       volume = {640},
       number = {2},
        pages = {L111-L114},
          doi = {10.1086/503548},
archivePrefix = {arXiv},
       eprint = {astro-ph/0511680},
 primaryClass = {astro-ph},
       adsurl = {https://ui.adsabs.harvard.edu/abs/2006ApJ...640L.111T}
}

@ARTICLE{Melchior11,
       author = {{Melchior}, P. and {Viola}, M. and {Sch{\"a}fer}, B.~M. and {Bartelmann}, M.},
        title = "{Weak gravitational lensing with DEIMOS}",
      journal = {\mnras},
         year = 2011,
        month = apr,
       volume = {412},
       number = {3},
        pages = {1552-1558},
          doi = {10.1111/j.1365-2966.2010.17875.x},
archivePrefix = {arXiv},
       eprint = {1008.1076},
 primaryClass = {astro-ph.IM},
       adsurl = {https://ui.adsabs.harvard.edu/abs/2011MNRAS.412.1552M}
}

@ARTICLE{Georgiou25,
       author = {{Georgiou}, Christos and {Chisari}, Nora Elisa and {Bilicki}, Maciej and {La Barbera}, Francesco and {Napolitano}, Nicola R. and {Roy}, Nivya and {Tortora}, Crescenzo},
        title = "{Intrinsic galaxy alignments in the KiDS-1000 bright sample: dependence on colour, luminosity, morphology and galaxy scale}",
      journal = {arXiv e-prints},
         year = 2025,
        month = feb,
          eid = {arXiv:2502.09452},
        pages = {arXiv:2502.09452},
          doi = {10.48550/arXiv.2502.09452},
archivePrefix = {arXiv},
       eprint = {2502.09452},
 primaryClass = {astro-ph.CO},
       adsurl = {https://ui.adsabs.harvard.edu/abs/2025arXiv250209452G}
}

@ARTICLE{Reid16,
       author = {{Reid}, Beth and {Ho}, Shirley and {Padmanabhan}, Nikhil and {Percival}, Will J. and {Tinker}, Jeremy and {Tojeiro}, Rita and {White}, Martin and {Eisenstein}, Daniel J. and {Maraston}, Claudia and {Ross}, Ashley J. and {S{\'a}nchez}, Ariel G. and {Schlegel}, David and {Sheldon}, Erin and {Strauss}, Michael A. and {Thomas}, Daniel and {Wake}, David and {Beutler}, Florian and {Bizyaev}, Dmitry and {Bolton}, Adam S. and {Brownstein}, Joel R. and {Chuang}, Chia-Hsun and {Dawson}, Kyle and {Harding}, Paul and {Kitaura}, Francisco-Shu and {Leauthaud}, Alexie and {Masters}, Karen and {McBride}, Cameron K. and {More}, Surhud and {Olmstead}, Matthew D. and {Oravetz}, Daniel and {Nuza}, Sebasti{\'a}n E. and {Pan}, Kaike and {Parejko}, John and {Pforr}, Janine and {Prada}, Francisco and {Rodr{\'\i}guez-Torres}, Sergio and {Salazar-Albornoz}, Salvador and {Samushia}, Lado and {Schneider}, Donald P. and {Sc{\'o}ccola}, Claudia G. and {Simmons}, Audrey and {Vargas-Magana}, Mariana},
        title = "{SDSS-III Baryon Oscillation Spectroscopic Survey Data Release 12: galaxy target selection and large-scale structure catalogues}",
      journal = {\mnras},
         year = 2016,
        month = jan,
       volume = {455},
       number = {2},
        pages = {1553-1573},
          doi = {10.1093/mnras/stv2382},
archivePrefix = {arXiv},
       eprint = {1509.06529},
 primaryClass = {astro-ph.CO},
       adsurl = {https://ui.adsabs.harvard.edu/abs/2016MNRAS.455.1553R}
}

@ARTICLE{Raichoor17,
       author = {{Raichoor}, A. and {Comparat}, J. and {Delubac}, T. and {Kneib}, J. -P. and {Y{\`e}che}, Ch and {Dawson}, K.~S. and {Percival}, W.~J. and {Dey}, A. and {Lang}, D. and {Schlegel}, D.~J. and {Gorgoni}, C. and {Bautista}, J. and {Brownstein}, J.~R. and {Mariappan}, V. and {Seo}, H. -J. and {Tinker}, J.~L. and {Ross}, A.~J. and {Wang}, Y. and {Zhao}, G. -B. and {Moustakas}, J. and {Palanque-Delabrouille}, N. and {Jullo}, E. and {Newmann}, J.~A. and {Prada}, F. and {Zhu}, G.~B.},
        title = "{The SDSS-IV extended Baryon Oscillation Spectroscopic Survey: final emission line galaxy target selection}",
      journal = {\mnras},
         year = 2017,
        month = nov,
       volume = {471},
       number = {4},
        pages = {3955-3973},
          doi = {10.1093/mnras/stx1790},
archivePrefix = {arXiv},
       eprint = {1704.00338},
 primaryClass = {astro-ph.CO},
       adsurl = {https://ui.adsabs.harvard.edu/abs/2017MNRAS.471.3955R}
}

@ARTICLE{Motloch21,
       author = {{Motloch}, Pavel and {Yu}, Hao-Ran and {Pen}, Ue-Li and {Xie}, Yuanbo},
        title = "{An observed correlation between galaxy spins and initial conditions}",
      journal = {Nature Astronomy},
         year = 2021,
        month = jan,
       volume = {5},
        pages = {283-288},
          doi = {10.1038/s41550-020-01262-3},
archivePrefix = {arXiv},
       eprint = {2003.04800},
 primaryClass = {astro-ph.CO},
       adsurl = {https://ui.adsabs.harvard.edu/abs/2021NatAs...5..283M}
}

@ARTICLE{Xu23,
       author = {{Xu}, Kun and {Jing}, Y.~P. and {Zhao}, Gong-Bo and {Cuesta}, Antonio J.},
        title = "{Evidence for baryon acoustic oscillations from galaxy-ellipticity correlations.}",
      journal = {Nature Astronomy},
         year = 2023,
        month = oct,
       volume = {7},
        pages = {1259-1264},
          doi = {10.1038/s41550-023-02035-4},
archivePrefix = {arXiv},
       eprint = {2306.09407},
 primaryClass = {astro-ph.CO},
       adsurl = {https://ui.adsabs.harvard.edu/abs/2023NatAs...7.1259X}
}

@ARTICLE{Gatti21,
       author = {{Gatti}, M. and {Sheldon}, E. and {Amon}, A. and {Becker}, M. and {Troxel}, M. and {Choi}, A. and {Doux}, C. and {MacCrann}, N. and {Navarro-Alsina}, A. and {Harrison}, I. and {Gruen}, D. and {Bernstein}, G. and {Jarvis}, M. and {Secco}, L.~F. and {Fert{\'e}}, A. and {Shin}, T. and {McCullough}, J. and {Rollins}, R.~P. and {Chen}, R. and {Chang}, C. and {Pandey}, S. and {Tutusaus}, I. and {Prat}, J. and {Elvin-Poole}, J. and {Sanchez}, C. and {Plazas}, A.~A. and {Roodman}, A. and {Zuntz}, J. and {Abbott}, T.~M.~C. and {Aguena}, M. and {Allam}, S. and {Annis}, J. and {Avila}, S. and {Bacon}, D. and {Bertin}, E. and {Bhargava}, S. and {Brooks}, D. and {Burke}, D.~L. and {Carnero Rosell}, A. and {Carrasco Kind}, M. and {Carretero}, J. and {Castander}, F.~J. and {Conselice}, C. and {Costanzi}, M. and {Crocce}, M. and {da Costa}, L.~N. and {Davis}, T.~M. and {De Vicente}, J. and {Desai}, S. and {Diehl}, H.~T. and {Dietrich}, J.~P. and {Doel}, P. and {Drlica-Wagner}, A. and {Eckert}, K. and {Everett}, S. and {Ferrero}, I. and {Frieman}, J. and {Garc{\'\i}a-Bellido}, J. and {Gerdes}, D.~W. and {Giannantonio}, T. and {Gruendl}, R.~A. and {Gschwend}, J. and {Gutierrez}, G. and {Hartley}, W.~G. and {Hinton}, S.~R. and {Hollowood}, D.~L. and {Honscheid}, K. and {Hoyle}, B. and {Huff}, E.~M. and {Huterer}, D. and {Jain}, B. and {James}, D.~J. and {Jeltema}, T. and {Krause}, E. and {Kron}, R. and {Kuropatkin}, N. and {Lima}, M. and {Maia}, M.~A.~G. and {Marshall}, J.~L. and {Miquel}, R. and {Morgan}, R. and {Myles}, J. and {Palmese}, A. and {Paz-Chinch{\'o}n}, F. and {Rykoff}, E.~S. and {Samuroff}, S. and {Sanchez}, E. and {Scarpine}, V. and {Schubnell}, M. and {Serrano}, S. and {Sevilla-Noarbe}, I. and {Smith}, M. and {Soares-Santos}, M. and {Suchyta}, E. and {Swanson}, M.~E.~C. and {Tarle}, G. and {Thomas}, D. and {To}, C. and {Tucker}, D.~L. and {Varga}, T.~N. and {Wechsler}, R.~H. and {Weller}, J. and {Wester}, W. and {Wilkinson}, R.~D.},
        title = "{Dark energy survey year 3 results: weak lensing shape catalogue}",
      journal = {\mnras},
         year = 2021,
        month = jul,
       volume = {504},
       number = {3},
        pages = {4312-4336},
          doi = {10.1093/mnras/stab918},
archivePrefix = {arXiv},
       eprint = {2011.03408},
 primaryClass = {astro-ph.CO},
       adsurl = {https://ui.adsabs.harvard.edu/abs/2021MNRAS.504.4312G}
}

@ARTICLE{HervasPeters24,
       author = {{Hervas Peters}, Fabian and {Kilbinger}, Martin and {Paviot}, Romain and {Baumont}, Lucie and {Russier}, Elisa and {Zhang}, Ziwen and {Murray}, Calum and {Pettorino}, Valeria and {de Boer}, Thomas and {Fabbro}, S{\'e}bastien and {Guerrini}, Sacha and {Hildebrandt}, Hendrik and {Hudson}, Mike and {Van Waerbeke}, Ludovic and {Wittje}, Anna},
        title = "{UNIONS: a direct measurement of intrinsic alignment with BOSS/eBOSS spectroscopy}",
      journal = {arXiv e-prints},
         year = 2024,
        month = dec,
          eid = {arXiv:2412.01790},
        pages = {arXiv:2412.01790},
          doi = {10.48550/arXiv.2412.01790},
archivePrefix = {arXiv},
       eprint = {2412.01790},
 primaryClass = {astro-ph.CO},
       adsurl = {https://ui.adsabs.harvard.edu/abs/2024arXiv241201790H}
}

@ARTICLE{McCullough24,
       author = {{McCullough}, J. and {Amon}, A. and {Legnani}, E. and {Gruen}, D. and {Roodman}, A. and {Friedrich}, O. and {MacCrann}, N. and {Becker}, M.~R. and {Myles}, J. and {Dodelson}, S. and {Samuroff}, S. and {Blazek}, J. and {Prat}, J. and {Honscheid}, K. and {Pieres}, A. and {Fert{\'e}}, A. and {Alarcon}, A. and {Drlica-Wagner}, A. and {Choi}, A. and {Navarro-Alsina}, A. and {Campos}, A. and {Plazas Malag{\'o}n}, A.~A. and {Porredon}, A. and {Farahi}, A. and {Ross}, A.~J. and {Carnero Rosell}, A. and {Yin}, B. and {Flaugher}, B. and {Yanny}, B. and {S{\'a}nchez}, C. and {Chang}, C. and {Davis}, C. and {To}, C. and {Doux}, C. and {Brooks}, D. and {James}, D.~J. and {Sanchez Cid}, D. and {Hollowood}, D.~L. and {Huterer}, D. and {Rykoff}, E.~S. and {Gaztanaga}, E. and {Huff}, E.~M. and {Suchyta}, E. and {Sheldon}, E. and {Sanchez}, E. and {Tarsitano}, F. and {Andrade-Oliveira}, F. and {Castander}, F.~J. and {Bernstein}, G.~M. and {Gutierrez}, G. and {Giannini}, G. and {Tarle}, G. and {Diehl}, H.~T. and {Huang}, H. and {Harrison}, I. and {Sevilla-Noarbe}, I. and {Tutusaus}, I. and {Ferrero}, I. and {Elvin-Poole}, J. and {Marshall}, J.~L. and {Muir}, J. and {Weller}, J. and {Zuntz}, J. and {Carretero}, J. and {DeRose}, J. and {Frieman}, J. and {Cordero}, J. and {De Vicente}, J. and {Garc{\'\i}a-Bellido}, J. and {Mena-Fern{\'a}ndez}, J. and {Eckert}, K. and {Romer}, A.~K. and {Bechtol}, K. and {Herner}, K. and {Kuehn}, K. and {Secco}, L.~F. and {da Costa}, L.~N. and {Paterno}, M. and {Soares-Santos}, 21 M. and {Gatti}, M. and {Raveri}, M. and {Yamamoto}, M. and {Smith}, M. and {Carrasco Kind}, M. and {Troxel}, M.~A. and {Aguena}, M. and {Jarvis}, M. and {Swanson}, M.~E.~C. and {Weaverdyck}, N. and {Lahav}, O. and {Doel}, P. and {Wiseman}, P. and {Miquel}, R. and {Gruendl}, R.~A. and {Cawthon}, R. and {Allam}, S. and {Hinton}, S.~R. and {Bridle}, S.~L. and {Bocquet}, S. and {Desai}, S. and {Pandey}, S. and {Everett}, S. and {Lee}, S. and {Shin}, T. and {Palmese}, A. and {Conselice}, C. and {Burke}, D.~L. and {Buckley-Geer}, E. and {Lima}, M. and {Vincenzi}, M. and {Pereira}, M.~E.~S. and {Crocce}, M. and {Schubnell}, M. and {Jeffrey}, N. and {Alves}, O. and {Vikram}, V. and {Zhang}, Y. and {DES Collaboration}},
        title = "{Dark Energy Survey Year 3: Blue Shear}",
      journal = {arXiv e-prints},
         year = 2024,
        month = oct,
          eid = {arXiv:2410.22272},
        pages = {arXiv:2410.22272},
          doi = {10.48550/arXiv.2410.22272},
archivePrefix = {arXiv},
       eprint = {2410.22272},
 primaryClass = {astro-ph.CO},
       adsurl = {https://ui.adsabs.harvard.edu/abs/2024arXiv241022272M}
}

@ARTICLE{Li21,
       author = {{Li}, Shun-Sheng and {Kuijken}, Konrad and {Hoekstra}, Henk and {Hildebrandt}, Hendrik and {Joachimi}, Benjamin and {Kannawadi}, Arun},
        title = "{KiDS+VIKING-450: An internal-consistency test for cosmic shear tomography with a colour-based split of source galaxies}",
      journal = {\aap},
         year = 2021,
        month = feb,
       volume = {646},
          eid = {A175},
        pages = {A175},
          doi = {10.1051/0004-6361/202039254},
archivePrefix = {arXiv},
       eprint = {2009.00367},
 primaryClass = {astro-ph.CO},
       adsurl = {https://ui.adsabs.harvard.edu/abs/2021A&A...646A.175L}
}

@ARTICLE{Strauss02,
       author = {{Strauss}, Michael A. and {Weinberg}, David H. and {Lupton}, Robert H. and {Narayanan}, Vijay K. and {Annis}, James and {Bernardi}, Mariangela and {Blanton}, Michael and {Burles}, Scott and {Connolly}, A.~J. and {Dalcanton}, Julianne and {Doi}, Mamoru and {Eisenstein}, Daniel and {Frieman}, Joshua A. and {Fukugita}, Masataka and {Gunn}, James E. and {Ivezi{\'c}}, {\v{Z}}eljko and {Kent}, Stephen and {Kim}, Rita S.~J. and {Knapp}, G.~R. and {Kron}, Richard G. and {Munn}, Jeffrey A. and {Newberg}, Heidi Jo and {Nichol}, R.~C. and {Okamura}, Sadanori and {Quinn}, Thomas R. and {Richmond}, Michael W. and {Schlegel}, David J. and {Shimasaku}, Kazuhiro and {SubbaRao}, Mark and {Szalay}, Alexander S. and {Vanden Berk}, Dan and {Vogeley}, Michael S. and {Yanny}, Brian and {Yasuda}, Naoki and {York}, Donald G. and {Zehavi}, Idit},
        title = "{Spectroscopic Target Selection in the Sloan Digital Sky Survey: The Main Galaxy Sample}",
      journal = {\aj},
         year = 2002,
        month = sep,
       volume = {124},
       number = {3},
        pages = {1810-1824},
          doi = {10.1086/342343},
archivePrefix = {arXiv},
       eprint = {astro-ph/0206225},
 primaryClass = {astro-ph},
       adsurl = {https://ui.adsabs.harvard.edu/abs/2002AJ....124.1810S}
}

@ARTICLE{York00,
       author = {{York}, Donald G. and {Adelman}, J. and {Anderson}, Jr., John E. and {Anderson}, Scott F. and {Annis}, James and {Bahcall}, Neta A. and {Bakken}, J.~A. and {Barkhouser}, Robert and {Bastian}, Steven and {Berman}, Eileen and {Boroski}, William N. and {Bracker}, Steve and {Briegel}, Charlie and {Briggs}, John W. and {Brinkmann}, J. and {Brunner}, Robert and {Burles}, Scott and {Carey}, Larry and {Carr}, Michael A. and {Castander}, Francisco J. and {Chen}, Bing and {Colestock}, Patrick L. and {Connolly}, A.~J. and {Crocker}, J.~H. and {Csabai}, Istv{\'a}n and {Czarapata}, Paul C. and {Davis}, John Eric and {Doi}, Mamoru and {Dombeck}, Tom and {Eisenstein}, Daniel and {Ellman}, Nancy and {Elms}, Brian R. and {Evans}, Michael L. and {Fan}, Xiaohui and {Federwitz}, Glenn R. and {Fiscelli}, Larry and {Friedman}, Scott and {Frieman}, Joshua A. and {Fukugita}, Masataka and {Gillespie}, Bruce and {Gunn}, James E. and {Gurbani}, Vijay K. and {de Haas}, Ernst and {Haldeman}, Merle and {Harris}, Frederick H. and {Hayes}, J. and {Heckman}, Timothy M. and {Hennessy}, G.~S. and {Hindsley}, Robert B. and {Holm}, Scott and {Holmgren}, Donald J. and {Huang}, Chi-hao and {Hull}, Charles and {Husby}, Don and {Ichikawa}, Shin-Ichi and {Ichikawa}, Takashi and {Ivezi{\'c}}, {\v{Z}}eljko and {Kent}, Stephen and {Kim}, Rita S.~J. and {Kinney}, E. and {Klaene}, Mark and {Kleinman}, A.~N. and {Kleinman}, S. and {Knapp}, G.~R. and {Korienek}, John and {Kron}, Richard G. and {Kunszt}, Peter Z. and {Lamb}, D.~Q. and {Lee}, B. and {Leger}, R. French and {Limmongkol}, Siriluk and {Lindenmeyer}, Carl and {Long}, Daniel C. and {Loomis}, Craig and {Loveday}, Jon and {Lucinio}, Rich and {Lupton}, Robert H. and {MacKinnon}, Bryan and {Mannery}, Edward J. and {Mantsch}, P.~M. and {Margon}, Bruce and {McGehee}, Peregrine and {McKay}, Timothy A. and {Meiksin}, Avery and {Merelli}, Aronne and {Monet}, David G. and {Munn}, Jeffrey A. and {Narayanan}, Vijay K. and {Nash}, Thomas and {Neilsen}, Eric and {Neswold}, Rich and {Newberg}, Heidi Jo and {Nichol}, R.~C. and {Nicinski}, Tom and {Nonino}, Mario and {Okada}, Norio and {Okamura}, Sadanori and {Ostriker}, Jeremiah P. and {Owen}, Russell and {Pauls}, A. George and {Peoples}, John and {Peterson}, R.~L. and {Petravick}, Donald and {Pier}, Jeffrey R. and {Pope}, Adrian and {Pordes}, Ruth and {Prosapio}, Angela and {Rechenmacher}, Ron and {Quinn}, Thomas R. and {Richards}, Gordon T. and {Richmond}, Michael W. and {Rivetta}, Claudio H. and {Rockosi}, Constance M. and {Ruthmansdorfer}, Kurt and {Sandford}, Dale and {Schlegel}, David J. and {Schneider}, Donald P. and {Sekiguchi}, Maki and {Sergey}, Gary and {Shimasaku}, Kazuhiro and {Siegmund}, Walter A. and {Smee}, Stephen and {Smith}, J. Allyn and {Snedden}, S. and {Stone}, R. and {Stoughton}, Chris and {Strauss}, Michael A. and {Stubbs}, Christopher and {SubbaRao}, Mark and {Szalay}, Alexander S. and {Szapudi}, Istvan and {Szokoly}, Gyula P. and {Thakar}, Anirudda R. and {Tremonti}, Christy and {Tucker}, Douglas L. and {Uomoto}, Alan and {Vanden Berk}, Dan and {Vogeley}, Michael S. and {Waddell}, Patrick and {Wang}, Shu-i. and {Watanabe}, Masaru and {Weinberg}, David H. and {Yanny}, Brian and {Yasuda}, Naoki and {SDSS Collaboration}},
        title = "{The Sloan Digital Sky Survey: Technical Summary}",
      journal = {\aj},
         year = 2000,
        month = sep,
       volume = {120},
       number = {3},
        pages = {1579-1587},
          doi = {10.1086/301513},
archivePrefix = {arXiv},
       eprint = {astro-ph/0006396},
 primaryClass = {astro-ph},
       adsurl = {https://ui.adsabs.harvard.edu/abs/2000AJ....120.1579Y}
}

@ARTICLE{Liske15,
       author = {{Liske}, J. and {Baldry}, I.~K. and {Driver}, S.~P. and {Tuffs}, R.~J. and {Alpaslan}, M. and {Andrae}, E. and {Brough}, S. and {Cluver}, M.~E. and {Grootes}, M.~W. and {Gunawardhana}, M.~L.~P. and {Kelvin}, L.~S. and {Loveday}, J. and {Robotham}, A.~S.~G. and {Taylor}, E.~N. and {Bamford}, S.~P. and {Bland-Hawthorn}, J. and {Brown}, M.~J.~I. and {Drinkwater}, M.~J. and {Hopkins}, A.~M. and {Meyer}, M.~J. and {Norberg}, P. and {Peacock}, J.~A. and {Agius}, N.~K. and {Andrews}, S.~K. and {Bauer}, A.~E. and {Ching}, J.~H.~Y. and {Colless}, M. and {Conselice}, C.~J. and {Croom}, S.~M. and {Davies}, L.~J.~M. and {De Propris}, R. and {Dunne}, L. and {Eardley}, E.~M. and {Ellis}, S. and {Foster}, C. and {Frenk}, C.~S. and {H{\"a}u{\ss}ler}, B. and {Holwerda}, B.~W. and {Howlett}, C. and {Ibarra}, H. and {Jarvis}, M.~J. and {Jones}, D.~H. and {Kafle}, P.~R. and {Lacey}, C.~G. and {Lange}, R. and {Lara-L{\'o}pez}, M.~A. and {L{\'o}pez-S{\'a}nchez}, {\'A}. R. and {Maddox}, S. and {Madore}, B.~F. and {McNaught-Roberts}, T. and {Moffett}, A.~J. and {Nichol}, R.~C. and {Owers}, M.~S. and {Palamara}, D. and {Penny}, S.~J. and {Phillipps}, S. and {Pimbblet}, K.~A. and {Popescu}, C.~C. and {Prescott}, M. and {Proctor}, R. and {Sadler}, E.~M. and {Sansom}, A.~E. and {Seibert}, M. and {Sharp}, R. and {Sutherland}, W. and {V{\'a}zquez-Mata}, J.~A. and {van Kampen}, E. and {Wilkins}, S.~M. and {Williams}, R. and {Wright}, A.~H.},
        title = "{Galaxy And Mass Assembly (GAMA): end of survey report and data release 2}",
      journal = {\mnras},
         year = 2015,
        month = sep,
       volume = {452},
       number = {2},
        pages = {2087-2126},
          doi = {10.1093/mnras/stv1436},
archivePrefix = {arXiv},
       eprint = {1506.08222},
 primaryClass = {astro-ph.GA},
       adsurl = {https://ui.adsabs.harvard.edu/abs/2015MNRAS.452.2087L}
}

@ARTICLE{deJong13,
       author = {{de Jong}, Jelte T.~A. and {Verdoes Kleijn}, Gijs A. and {Kuijken}, Konrad H. and {Valentijn}, Edwin A.},
        title = "{The Kilo-Degree Survey}",
      journal = {Experimental Astronomy},
         year = 2013,
        month = jan,
       volume = {35},
       number = {1-2},
        pages = {25-44},
          doi = {10.1007/s10686-012-9306-1},
archivePrefix = {arXiv},
       eprint = {1206.1254},
 primaryClass = {astro-ph.CO},
       adsurl = {https://ui.adsabs.harvard.edu/abs/2013ExA....35...25D}
}

@ARTICLE{HirataJK,
       author = {{Hirata}, Christopher M. and {Mandelbaum}, Rachel and {Seljak}, Uro{\v{s}} and {Guzik}, Jacek and {Padmanabhan}, Nikhil and {Blake}, Cullen and {Brinkmann}, Jonathan and {Bud{\'a}vari}, Tamas and {Connolly}, Andrew and {Csabai}, Istvan and {Scranton}, Ryan and {Szalay}, Alexander S.},
        title = "{Galaxy-galaxy weak lensing in the Sloan Digital Sky Survey: intrinsic alignments and shear calibration errors}",
      journal = {\mnras},
         year = 2004,
        month = sep,
       volume = {353},
       number = {2},
        pages = {529-549},
          doi = {10.1111/j.1365-2966.2004.08090.x},
archivePrefix = {arXiv},
       eprint = {astro-ph/0403255},
 primaryClass = {astro-ph},
       adsurl = {https://ui.adsabs.harvard.edu/abs/2004MNRAS.353..529H}
}

@ARTICLE{Hirata03,
       author = {{Hirata}, Christopher and {Seljak}, Uro{\v{s}}},
        title = "{Shear calibration biases in weak-lensing surveys}",
      journal = {\mnras},
         year = 2003,
        month = aug,
       volume = {343},
       number = {2},
        pages = {459-480},
          doi = {10.1046/j.1365-8711.2003.06683.x},
archivePrefix = {arXiv},
       eprint = {astro-ph/0301054},
 primaryClass = {astro-ph},
       adsurl = {https://ui.adsabs.harvard.edu/abs/2003MNRAS.343..459H}
}

@ARTICLE{Mandelbaum05a,
       author = {{Mandelbaum}, Rachel and {Hirata}, Christopher M. and {Seljak}, Uro{\v{s}} and {Guzik}, Jacek and {Padmanabhan}, Nikhil and {Blake}, Cullen and {Blanton}, Michael R. and {Lupton}, Robert and {Brinkmann}, Jonathan},
        title = "{Systematic errors in weak lensing: application to SDSS galaxy-galaxy weak lensing}",
      journal = {\mnras},
         year = 2005,
        month = aug,
       volume = {361},
       number = {4},
        pages = {1287-1322},
          doi = {10.1111/j.1365-2966.2005.09282.x},
archivePrefix = {arXiv},
       eprint = {astro-ph/0501201},
 primaryClass = {astro-ph},
       adsurl = {https://ui.adsabs.harvard.edu/abs/2005MNRAS.361.1287M}
}

@ARTICLE{Heymans04,
       author = {{Heymans}, Catherine and {Brown}, Michael and {Heavens}, Alan and {Meisenheimer}, Klaus and {Taylor}, Andy and {Wolf}, Christian},
        title = "{Weak lensing with COMBO-17: estimation and removal of intrinsic alignments}",
      journal = {\mnras},
         year = 2004,
        month = jan,
       volume = {347},
       number = {3},
        pages = {895-908},
          doi = {10.1111/j.1365-2966.2004.07264.x},
archivePrefix = {arXiv},
       eprint = {astro-ph/0310174},
 primaryClass = {astro-ph},
       adsurl = {https://ui.adsabs.harvard.edu/abs/2004MNRAS.347..895H}
}

@ARTICLE{Johnston21,
       author = {{Johnston}, Harry and {Joachimi}, Benjamin and {Norberg}, Peder and {Hoekstra}, Henk and {Eriksen}, Martin and {Fortuna}, Maria Cristina and {Manzoni}, Giorgio and {Serrano}, Santiago and {Siudek}, Malgorzata and {Tortorelli}, Luca and {Asorey}, Jacobo and {Cabayol}, Laura and {Carretero}, Jorge and {Casas}, Ricard and {Castander}, Francisco and {Crocce}, Martin and {Fernandez}, Enrique and {Garc{\'\i}a-Bellido}, Juan and {Gaztanaga}, Enrique and {Hildebrandt}, Hendrik and {Miquel}, Ramon and {Navarro-Girones}, David and {Padilla}, Cristobal and {Sanchez}, Eusebio and {Sevilla-Noarbe}, Ignacio and {Tallada-Cresp{\'\i}}, Pau},
        title = "{The PAU Survey: Intrinsic alignments and clustering of narrow-band photometric galaxies}",
      journal = {\aap},
         year = 2021,
        month = feb,
       volume = {646},
          eid = {A147},
        pages = {A147},
          doi = {10.1051/0004-6361/202039682},
archivePrefix = {arXiv},
       eprint = {2010.09696},
 primaryClass = {astro-ph.GA},
       adsurl = {https://ui.adsabs.harvard.edu/abs/2021A&A...646A.147J}
}

@ARTICLE{Chen24b,
       author = {{Chen}, S. and {DeRose}, J. and {Zhou}, R. and {White}, M. and {Ferraro}, S. and {Blake}, C. and {Lange}, J.~U. and {Wechsler}, R.~H. and {Aguilar}, J. and {Ahlen}, S. and {Brooks}, D. and {Claybaugh}, T. and {Dawson}, K. and {de la Macorra}, A. and {Doel}, P. and {Font-Ribera}, A. and {Gazta{\~n}aga}, E. and {Gontcho}, S. Gontcho A and {Gutierrez}, G. and {Honscheid}, K. and {Howlett}, C. and {Kehoe}, R. and {Kirkby}, D. and {Kisner}, T. and {Kremin}, A. and {Landriau}, M. and {Le Guillou}, L. and {Manera}, M. and {Meisner}, A. and {Miquel}, R. and {Newman}, J.~A. and {Niz}, G. and {Palanque-Delabrouille}, N. and {Percival}, W.~J. and {Prada}, F. and {Rossi}, G. and {Sanchez}, E. and {Schlegel}, D. and {Schubnell}, M. and {Sprayberry}, D. and {Tarl{\'e}}, G. and {Weaver}, B.~A.},
        title = "{Not all lensing is low: An analysis of DESI$\times$DES using the Lagrangian Effective Theory of LSS}",
      journal = {arXiv e-prints},
         year = 2024,
        month = jul,
          eid = {arXiv:2407.04795},
        pages = {arXiv:2407.04795},
          doi = {10.48550/arXiv.2407.04795},
archivePrefix = {arXiv},
       eprint = {2407.04795},
 primaryClass = {astro-ph.CO},
       adsurl = {https://ui.adsabs.harvard.edu/abs/2024arXiv240704795C}
}

@article{dAssignies21,
   title={Intrinsic alignments of galaxies around cosmic voids},
   volume={509},
   ISSN={1365-2966},
   DOI={10.1093/mnras/stab2986},
   number={2},
   journal={Monthly Notices of the Royal Astronomical Society},
   publisher={Oxford University Press (OUP)},
   author={d’Assignies D., William and Chisari, Nora Elisa and Hamaus, Nico and Singh, Sukhdeep},
   year={2021},
   month=oct, pages={1985–1994} }

@ARTICLE{Huang18,
       author = {{Huang}, Hung-Jin and {Mandelbaum}, Rachel and {Freeman}, Peter E. and {Chen}, Yen-Chi and {Rozo}, Eduardo and {Rykoff}, Eli},
        title = "{Intrinsic alignment in redMaPPer clusters - II. Radial alignment of satellites towards cluster centres}",
      journal = {\mnras},
         year = 2018,
        month = mar,
       volume = {474},
       number = {4},
        pages = {4772-4794},
          doi = {10.1093/mnras/stx2995},
archivePrefix = {arXiv},
       eprint = {1704.06273},
 primaryClass = {astro-ph.GA},
       adsurl = {https://ui.adsabs.harvard.edu/abs/2018MNRAS.474.4772H}
}

@ARTICLE{Georgiou19b,
       author = {{Georgiou}, Christos and {Johnston}, Harry and {Hoekstra}, Henk and {Viola}, Massimo and {Kuijken}, Konrad and {Joachimi}, Benjamin and {Chisari}, Nora Elisa and {Farrow}, Daniel J. and {Hildebrandt}, Hendrik and {Holwerda}, Benne W. and {Kannawadi}, Arun},
        title = "{The dependence of intrinsic alignment of galaxies on wavelength using KiDS and GAMA}",
      journal = {\aap},
         year = 2019,
        month = feb,
       volume = {622},
          eid = {A90},
        pages = {A90},
          doi = {10.1051/0004-6361/201834219},
archivePrefix = {arXiv},
       eprint = {1809.03602},
 primaryClass = {astro-ph.CO},
       adsurl = {https://ui.adsabs.harvard.edu/abs/2019A&A...622A..90G}
}

@ARTICLE{Tsaprazi22,
       author = {{Tsaprazi}, Eleni and {Nguyen}, Nhat-Minh and {Jasche}, Jens and {Schmidt}, Fabian and {Lavaux}, Guilhem},
        title = "{Field-level inference of galaxy intrinsic alignment from the SDSS-III BOSS survey}",
      journal = {\jcap},
         year = 2022,
        month = aug,
       volume = {2022},
       number = {8},
          eid = {003},
        pages = {003},
          doi = {10.1088/1475-7516/2022/08/003},
archivePrefix = {arXiv},
       eprint = {2112.04484},
 primaryClass = {astro-ph.CO},
       adsurl = {https://ui.adsabs.harvard.edu/abs/2022JCAP...08..003T}
}

@ARTICLE{Robertson21,
       author = {{Robertson}, Naomi Clare and {Alonso}, David and {Harnois-D{\'e}raps}, Joachim and {Darwish}, Omar and {Kannawadi}, Arun and {Amon}, Alexandra and {Asgari}, Marika and {Bilicki}, Maciej and {Calabrese}, Erminia and {Choi}, Steve K. and {Devlin}, Mark J. and {Dunkley}, Jo and {Dvornik}, Andrej and {Erben}, Thomas and {Ferraro}, Simone and {Fortuna}, Maria Cristina and {Giblin}, Benjamin and {Han}, Dongwon and {Heymans}, Catherine and {Hildebrandt}, Hendrik and {Hill}, J. Colin and {Hilton}, Matt and {Ho}, Shuay-Pwu P. and {Hoekstra}, Henk and {Hubmayr}, Johannes and {Hughes}, John P. and {Joachimi}, Benjamin and {Joudaki}, Shahab and {Knowles}, Kenda and {Kuijken}, Konrad and {Madhavacheril}, Mathew S. and {Moodley}, Kavilan and {Miller}, Lance and {Namikawa}, Toshiya and {Nati}, Federico and {Niemack}, Michael D. and {Page}, Lyman A. and {Partridge}, Bruce and {Schaan}, Emmanuel and {Schillaci}, Alessandro and {Schneider}, Peter and {Sehgal}, Neelima and {Sherwin}, Blake D. and {Sif{\'o}n}, Crist{\'o}bal and {Staggs}, Suzanne T. and {Tr{\"o}ster}, Tilman and {van Engelen}, Alexander and {Valentijn}, Edwin and {Wollack}, Edward J. and {Wright}, Angus H. and {Xu}, Zhilei},
        title = "{Strong detection of the CMB lensing and galaxy weak lensing cross-correlation from ACT-DR4, Planck Legacy, and KiDS-1000}",
      journal = {\aap},
         year = 2021,
        month = may,
       volume = {649},
          eid = {A146},
        pages = {A146},
          doi = {10.1051/0004-6361/202039975},
archivePrefix = {arXiv},
       eprint = {2011.11613},
 primaryClass = {astro-ph.CO},
       adsurl = {https://ui.adsabs.harvard.edu/abs/2021A&A...649A.146R}
}

@article{Tonegawa24,
       author = {{Tonegawa}, Motonari and {Okumura}, Teppei and {Hayashi}, Masao},
        title = "{Intrinsic alignments and spin correlations of [O II] emitters at $z = 1.2$ and $z = 1.5$ from the HSC narrow-band survey}",
      journal = {\pasj},
         year = 2025,
        month = apr,
       volume = {77},
       number = {2},
        pages = {389-402},
          doi = {10.1093/pasj/psaf006},
archivePrefix = {arXiv},
       eprint = {2408.11462},
 primaryClass = {astro-ph.CO},
       adsurl = {https://ui.adsabs.harvard.edu/abs/2025PASJ...77..389T}
}

@ARTICLE{Tonegawa21,
       author = {{Tonegawa}, Motonari and {Okumura}, Teppei},
        title = "{First Evidence of Intrinsic Alignments of Red Galaxies at z > 1: Cross Correlation between CFHTLenS and FastSound Samples}",
      journal = {\apjl},
         year = 2022,
        month = jan,
       volume = {924},
       number = {1},
          eid = {L3},
        pages = {L3},
          doi = {10.3847/2041-8213/ac4246},
archivePrefix = {arXiv},
       eprint = {2109.14297},
 primaryClass = {astro-ph.CO},
       adsurl = {https://ui.adsabs.harvard.edu/abs/2022ApJ...924L...3T}
}

@misc{FortunaMass,
      title={KiDS-1000: weak lensing and intrinsic alignment around luminous red galaxies}, 
      author={Maria Cristina Fortuna and Andrej Dvornik and Henk Hoekstra and Nora Elisa Chisari and Marika Asgari and Maciej Bilicki and Catherine Heymans and Hendrik Hildebrandt and Koen Kuijken and Angus H. Wright and Ji Yao},
      year={2024},
      eprint={2409.15416},
      archivePrefix={arXiv},
      primaryClass={astro-ph.CO},
   }

@article{vanUitert17,
    author = {van Uitert, Edo and Joachimi, Benjamin},
    title = "{Intrinsic alignment of redMaPPer clusters: cluster shape–matter density correlation}",
    journal = {\mnras},
    volume = {468},
    number = {4},
    pages = {4502-4512},
    year = {2017},
    month = {04},
    issn = {0035-8711},
    doi = {10.1093/mnras/stx756}
}

@article{Samuroff19,
    author = "Samuroff, S. and others",
    collaboration = "DES",
    title = "{Dark Energy Survey Year 1 Results: Constraints on Intrinsic Alignments and their Colour Dependence from Galaxy Clustering and Weak Lensing}",
    eprint = "1811.06989",
    archivePrefix = "arXiv",
    primaryClass = "astro-ph.CO",
    reportNumber = "FERMILAB-PUB-18-622-AE",
    doi = "10.1093/mnras/stz2197",
    journal = "MNRAS",
    volume = "489",
    number = "4",
    pages = "5453--5482",
    year = "2019"
}

@ARTICLE{Georgiou19,
       author = {{Georgiou}, Christos and {Chisari}, Nora Elisa and
         {Fortuna}, Maria Cristina and {Hoekstra}, Henk and {Kuijken}, Konrad and
         {Joachimi}, Benjamin and {Vakili}, Mohammadjavad and {Bilicki}, Maciej and
         {Dvornik}, Andrej and {Erben}, Thomas and {Giblin}, Benjamin and
         {Heymans}, Catherine and {Napolitano}, Nicola R. and {Shan}, HuanYuan},
        title = "{GAMA+KiDS: Alignment of galaxies in galaxy groups and its dependence on galaxy scale}",
      journal = {\aap},
         year = "2019",
        month = "Aug",
       volume = {628},
          eid = {A31},
        pages = {A31},
          doi = {10.1051/0004-6361/201935810},
archivePrefix = {arXiv},
       eprint = {1905.00370},
 primaryClass = {astro-ph.CO},
       adsurl = {https://ui.adsabs.harvard.edu/abs/2019A&A...628A..31G}
}

@ARTICLE{MegaZ,
   author = {{Joachimi}, B. and {Mandelbaum}, R. and {Abdalla}, F.~B. and 
	{Bridle}, S.~L.},
    title = "{Constraints on intrinsic alignment contamination of weak lensing surveys using the MegaZ-LRG sample}",
  journal = {\aap},
archivePrefix = "arXiv",
   eprint = {1008.3491},
 primaryClass = "astro-ph.CO",
     year = 2011,
    month = mar,
   volume = 527,
      eid = {A26},
    pages = {A26},
      doi = {10.1051/0004-6361/201015621},
   adsurl = {http://adsabs.harvard.edu/abs/2011A%26A...527A..26J}
}

@ARTICLE{Hirata07,
   author = {{Hirata}, C.~M. and {Mandelbaum}, R. and {Ishak}, M. and {Seljak}, U. and 
	{Nichol}, R. and {Pimbblet}, K.~A. and {Ross}, N.~P. and {Wake}, D.
	},
    title = "{Intrinsic galaxy alignments from the 2SLAQ and SDSS surveys: luminosity and redshift scalings and implications for weak lensing surveys}",
  journal = {\mnras},
   eprint = {astro-ph/0701671},
     year = 2007,
    month = nov,
   volume = 381,
    pages = {1197-1218},
      doi = {10.1111/j.1365-2966.2007.12312.x},
   adsurl = {http://adsabs.harvard.edu/abs/2007MNRAS.381.1197H}
}

@ARTICLE{Schechter76,
       author = {{Schechter}, P.},
        title = "{An analytic expression for the luminosity function for galaxies.}",
      journal = {\apj},
         year = 1976,
        month = jan,
       volume = {203},
        pages = {297-306},
          doi = {10.1086/154079},
       adsurl = {https://ui.adsabs.harvard.edu/abs/1976ApJ...203..297S}
}

@ARTICLE{Mandelbaum06,
   author = {{Mandelbaum}, R. and {Hirata}, C.~M. and {Ishak}, M. and {Seljak}, U. and 
	{Brinkmann}, J.},
    title = "{Detection of large-scale intrinsic ellipticity-density correlation from the Sloan Digital Sky Survey and implications for weak lensing surveys}",
  journal = {\mnras},
   eprint = {astro-ph/0509026},
     year = 2006,
    month = apr,
   volume = 367,
    pages = {611-626},
      doi = {10.1111/j.1365-2966.2005.09946.x},
   adsurl = {http://adsabs.harvard.edu/abs/2006MNRAS.367..611M}
}

@ARTICLE{Chen19,
       author = {{Chen}, Yen-Chi and {Ho}, Shirley and {Blazek}, Jonathan and {He}, Siyu and
         {Mandelbaum}, Rachel and {Melchior}, Peter and {Singh}, Sukhdeep},
        title = "{Detecting galaxy-filament alignments in the Sloan Digital Sky Survey III}",
      journal = {\mnras},
         year = "2019",
        month = "May",
       volume = {485},
       number = {2},
        pages = {2492-2504},
          doi = {10.1093/mnras/stz539},
archivePrefix = {arXiv},
       eprint = {1805.00159},
 primaryClass = {astro-ph.CO},
       adsurl = {https://ui.adsabs.harvard.edu/abs/2019MNRAS.485.2492C}
}

@ARTICLE{JohnstonKiDSGAMA,
       author = {{Johnston}, Harry and {Georgiou}, Christos and {Joachimi}, Benjamin and {Hoekstra}, Henk and {Chisari}, Nora Elisa and {Farrow}, Daniel and {Fortuna}, Maria Cristina and {Heymans}, Catherine and {Joudaki}, Shahab and {Kuijken}, Konrad and {Wright}, Angus},
        title = "{KiDS+GAMA: Intrinsic alignment model constraints for current and future weak lensing cosmology}",
      journal = {\aap},
         year = 2019,
        month = apr,
       volume = {624},
          eid = {A30},
        pages = {A30},
          doi = {10.1051/0004-6361/201834714},
archivePrefix = {arXiv},
       eprint = {1811.09598},
 primaryClass = {astro-ph.CO},
       adsurl = {https://ui.adsabs.harvard.edu/abs/2019A&A...624A..30J}
}

@article{Shi21,
   title={An Optimal Estimator of Intrinsic Alignments for Star-forming Galaxies in IllustrisTNG Simulation},
   volume={917},
   ISSN={1538-4357},
   DOI={10.3847/1538-4357/ac0cfa},
   number={2},
   journal={The Astrophysical Journal},
   publisher={American Astronomical Society},
   author={Shi, Jingjing and Osato, Ken and Kurita, Toshiki and Takada, Masahiro},
   year={2021},
   month=aug,
   pages={109}
}

@article{Shi24,
       author = {{Shi}, Jingjing and {Sunayama}, Tomomi and {Kurita}, Toshiki and {Takada}, Masahiro and {Sugiyama}, Sunao and {Mandelbaum}, Rachel and {Miyatake}, Hironao and {More}, Surhud and {Nishimichi}, Takahiro and {Johnston}, Harry},
        title = "{The intrinsic alignment of galaxy clusters and impact of projection effects}",
      journal = {\mnras},
         year = 2024,
        month = feb,
       volume = {528},
       number = {2},
        pages = {1487-1499},
          doi = {10.1093/mnras/stae064},
archivePrefix = {arXiv},
       eprint = {2306.09661},
 primaryClass = {astro-ph.CO},
       adsurl = {https://ui.adsabs.harvard.edu/abs/2024MNRAS.528.1487S}
}

@ARTICLE{SinghShapes,
       author = {{Singh}, Sukhdeep and {Mandelbaum}, Rachel},
        title = "{Intrinsic alignments of BOSS LOWZ galaxies - II. Impact of shape measurement methods}",
      journal = {"MNRAS"},
         year = 2016,
        month = apr,
       volume = {457},
       number = {3},
        pages = {2301-2317},
          doi = {10.1093/mnras/stw144},
archivePrefix = {arXiv},
       eprint = {1510.06752},
 primaryClass = {astro-ph.CO},
       adsurl = {https://ui.adsabs.harvard.edu/abs/2016MNRAS.457.2301S}
}

@ARTICLE{Hoekstra05,
       author = {{Hoekstra}, H. and {Hsieh}, B.~C. and {Yee}, H.~K.~C. and {Lin}, H. and {Gladders}, M.~D.},
        title = "{Virial Masses and the Baryon Fraction in Galaxies}",
      journal = {\apj},
         year = 2005,
        month = dec,
       volume = {635},
       number = {1},
        pages = {73-85},
          doi = {10.1086/496913},
archivePrefix = {arXiv},
       eprint = {astro-ph/0510097},
 primaryClass = {astro-ph},
       adsurl = {https://ui.adsabs.harvard.edu/abs/2005ApJ...635...73H}
}

@ARTICLE{Behroozi13,
       author = {{Behroozi}, Peter S. and {Wechsler}, Risa H. and {Conroy}, Charlie},
        title = "{The Average Star Formation Histories of Galaxies in Dark Matter Halos from z = 0-8}",
      journal = {\apj},
         year = 2013,
        month = jun,
       volume = {770},
       number = {1},
          eid = {57},
        pages = {57},
          doi = {10.1088/0004-637X/770/1/57},
archivePrefix = {arXiv},
       eprint = {1207.6105},
 primaryClass = {astro-ph.CO},
       adsurl = {https://ui.adsabs.harvard.edu/abs/2013ApJ...770...57B}
}

@ARTICLE{Vakili19,
       author = {{Vakili}, Mohammadjavad and {Bilicki}, Maciej and {Hoekstra}, Henk and {Chisari}, Nora Elisa and {Brown}, Michael J.~I. and {Georgiou}, Christos and {Kannawadi}, Arun and {Kuijken}, Konrad and {Wright}, Angus H.},
        title = "{Luminous red galaxies in the Kilo-Degree Survey: selection with broad-band photometry and weak lensing measurements}",
      journal = {\mnras},
         year = 2019,
        month = aug,
       volume = {487},
       number = {3},
        pages = {3715-3733},
          doi = {10.1093/mnras/stz1249},
archivePrefix = {arXiv},
       eprint = {1811.02518},
 primaryClass = {astro-ph.CO},
       adsurl = {https://ui.adsabs.harvard.edu/abs/2019MNRAS.487.3715V}
}

@ARTICLE{Fortuna24,
       author = {{Fortuna}, Maria Cristina and {Dvornik}, Andrej and {Hoekstra}, Henk and {Chisari}, Nora Elisa and {Asgari}, Marika and {Bilicki}, Maciej and {Heymans}, Catherine and {Hildebrandt}, Hendrik and {Kuijken}, Koen and {Wright}, Angus H. and {Yao}, Ji},
        title = "{KiDS-1000: weak lensing and intrinsic alignment around luminous red galaxies}",
      journal = {arXiv e-prints},
         year = 2024,
        month = sep,
          eid = {arXiv:2409.15416},
        pages = {arXiv:2409.15416},
          doi = {10.48550/arXiv.2409.15416},
archivePrefix = {arXiv},
       eprint = {2409.15416},
 primaryClass = {astro-ph.CO},
       adsurl = {https://ui.adsabs.harvard.edu/abs/2024arXiv240915416F}
}

@ARTICLE{BJ02,
       author = {{Bernstein}, G.~M. and {Jarvis}, M.},
        title = "{Shapes and Shears, Stars and Smears: Optimal Measurements for Weak Lensing}",
      journal = {\aj},
         year = 2002,
        month = feb,
       volume = {123},
       number = {2},
        pages = {583-618},
          doi = {10.1086/338085},
archivePrefix = {arXiv},
       eprint = {astro-ph/0107431},
 primaryClass = {astro-ph},
       adsurl = {https://ui.adsabs.harvard.edu/abs/2002AJ....123..583B}
}

@ARTICLE{LS93,
       author = {{Landy}, Stephen D. and {Szalay}, Alexander S.},
        title = "{Bias and Variance of Angular Correlation Functions}",
      journal = {\apj},
         year = 1993,
        month = jul,
       volume = {412},
        pages = {64},
          doi = {10.1086/172900},
       adsurl = {https://ui.adsabs.harvard.edu/abs/1993ApJ...412...64L}
}

@ARTICLE{SinghBOSS,
       author = {{Singh}, Sukhdeep and {Mandelbaum}, Rachel and {More}, Surhud},
        title = "{Intrinsic alignments of SDSS-III BOSS LOWZ sample galaxies}",
      journal = {\mnras},
         year = 2015,
        month = jun,
       volume = {450},
       number = {2},
        pages = {2195-2216},
          doi = {10.1093/mnras/stv778},
archivePrefix = {arXiv},
       eprint = {1411.1755},
 primaryClass = {astro-ph.CO},
       adsurl = {https://ui.adsabs.harvard.edu/abs/2015MNRAS.450.2195S}
}

@ARTICLE{Baldauf10,
       author = {{Baldauf}, Tobias and {Smith}, Robert E. and {Seljak}, Uro{\v{s}} and {Mandelbaum}, Rachel},
        title = "{Algorithm for the direct reconstruction of the dark matter correlation function from weak lensing and galaxy clustering}",
      journal = {\prd},
         year = 2010,
        month = mar,
       volume = {81},
       number = {6},
          eid = {063531},
        pages = {063531},
          doi = {10.1103/PhysRevD.81.063531},
archivePrefix = {arXiv},
       eprint = {0911.4973},
 primaryClass = {astro-ph.CO},
       adsurl = {https://ui.adsabs.harvard.edu/abs/2010PhRvD..81f3531B}
}

@ARTICLE{Kurita22,
       author = {{Kurita}, Toshiki and {Takada}, Masahiro},
        title = "{Analysis method for 3D power spectrum of projected tensor fields with fast estimator and window convolution modeling: An application to intrinsic alignments}",
      journal = {\prd},
         year = 2022,
        month = jun,
       volume = {105},
       number = {12},
          eid = {123501},
        pages = {123501},
          doi = {10.1103/PhysRevD.105.123501},
archivePrefix = {arXiv},
       eprint = {2202.11839},
 primaryClass = {astro-ph.CO},
       adsurl = {https://ui.adsabs.harvard.edu/abs/2022PhRvD.105l3501K}
}

@ARTICLE{Kurita23,
       author = {{Kurita}, Toshiki and {Takada}, Masahiro},
        title = "{Constraints on anisotropic primordial non-Gaussianity from intrinsic alignments of SDSS-III BOSS galaxies}",
      journal = {\prd},
         year = 2023,
        month = oct,
       volume = {108},
       number = {8},
          eid = {083533},
        pages = {083533},
          doi = {10.1103/PhysRevD.108.083533},
archivePrefix = {arXiv},
       eprint = {2302.02925},
 primaryClass = {astro-ph.CO},
       adsurl = {https://ui.adsabs.harvard.edu/abs/2023PhRvD.108h3533K}
}

@ARTICLE{SamuroffEBOSS,
       author = {{Samuroff}, S. and {Mandelbaum}, R. and {Blazek}, J. and {Campos}, A. and {MacCrann}, N. and {Zacharegkas}, G. and {Amon}, A. and {Prat}, J. and {Singh}, S. and {Elvin-Poole}, J. and {Ross}, A.~J. and {Alarcon}, A. and {Baxter}, E. and {Bechtol}, K. and {Becker}, M.~R. and {Bernstein}, G.~M. and {Rosell}, A. Carnero and {Kind}, M. Carrasco and {Cawthon}, R. and {Chang}, C. and {Chen}, R. and {Choi}, A. and {Crocce}, M. and {Davis}, C. and {DeRose}, J. and {Dodelson}, S. and {Doux}, C. and {Drlica-Wagner}, A. and {Eckert}, K. and {Everett}, S. and {Fert{\'e}}, A. and {Gatti}, M. and {Giannini}, G. and {Gruen}, D. and {Gruendl}, R.~A. and {Harrison}, I. and {Herner}, K. and {Huff}, E.~M. and {Jarvis}, M. and {Kuropatkin}, N. and {Leget}, P. -F. and {Lemos}, P. and {McCullough}, J. and {Myles}, J. and {Navarro-Alsina}, A. and {Pandey}, S. and {Porredon}, A. and {Raveri}, M. and {Rodriguez-Monroy}, M. and {Rollins}, R.~P. and {Roodman}, A. and {Rossi}, G. and {Rykoff}, E.~S. and {S{\'a}nchez}, C. and {Secco}, L.~F. and {Sevilla-Noarbe}, I. and {Sheldon}, E. and {Shin}, T. and {Troxel}, M.~A. and {Tutusaus}, I. and {Weaverdyck}, N. and {Yanny}, B. and {Yin}, B. and {Zhang}, Y. and {Zuntz}, J. and {Aguena}, M. and {Alves}, O. and {Annis}, J. and {Bacon}, D. and {Bertin}, E. and {Bocquet}, S. and {Brooks}, D. and {Burke}, D.~L. and {Carretero}, J. and {Costanzi}, M. and {da Costa}, L.~N. and {Pereira}, M.~E.~S. and {De Vicente}, J. and {Desai}, S. and {Diehl}, H.~T. and {Dietrich}, J.~P. and {Doel}, P. and {Ferrero}, I. and {Flaugher}, B. and {Frieman}, J. and {Garc{\'\i}a-Bellido}, J. and {Hinton}, S.~R. and {Hollowood}, D.~L. and {Honscheid}, K. and {James}, D.~J. and {Kuehn}, K. and {Lahav}, O. and {Marshall}, J.~L. and {Melchior}, P. and {Mena-Fern{\'a}ndez}, J. and {Menanteau}, F. and {Miquel}, R. and {Newman}, J. and {Palmese}, A. and {Pieres}, A. and {Malag{\'o}n}, A.~A. Plazas and {Sanchez}, E. and {Scarpine}, V. and {Smith}, M. and {Suchyta}, E. and {Swanson}, M.~E.~C. and {Tarle}, G. and {To}, C. and {DES Collaboration}},
        title = "{The Dark Energy Survey Year 3 and eBOSS: constraining galaxy intrinsic alignments across luminosity and colour space}",
      journal = {\mnras},
         year = 2023,
        month = sep,
       volume = {524},
       number = {2},
        pages = {2195-2223},
          doi = {10.1093/mnras/stad2013},
archivePrefix = {arXiv},
       eprint = {2212.11319},
 primaryClass = {astro-ph.CO},
       adsurl = {https://ui.adsabs.harvard.edu/abs/2023MNRAS.524.2195S}
}

@ARTICLE{SinghMultipoles,
       author = {{Singh}, Sukhdeep and {Shakir}, Ali and {Jagvaral}, Yesukhei and {Mandelbaum}, Rachel},
        title = "{Increasing the power of survey data with multipole-based intrinsic alignment estimators}",
      journal = {arXiv e-prints},
         year = 2023,
        month = jul,
          eid = {arXiv:2307.02545},
        pages = {arXiv:2307.02545},
          doi = {10.48550/arXiv.2307.02545},
archivePrefix = {arXiv},
       eprint = {2307.02545},
 primaryClass = {astro-ph.CO},
       adsurl = {https://ui.adsabs.harvard.edu/abs/2023arXiv230702545S}
}

@ARTICLE{Mandelbaum11,
       author = {{Mandelbaum}, Rachel and {Blake}, Chris and {Bridle}, Sarah and {Abdalla}, Filipe B. and {Brough}, Sarah and {Colless}, Matthew and {Couch}, Warrick and {Croom}, Scott and {Davis}, Tamara and {Drinkwater}, Michael J. and {Forster}, Karl and {Glazebrook}, Karl and {Jelliffe}, Ben and {Jurek}, Russell J. and {Li}, I. -Hui and {Madore}, Barry and {Martin}, Chris and {Pimbblet}, Kevin and {Poole}, Gregory B. and {Pracy}, Michael and {Sharp}, Rob and {Wisnioski}, Emily and {Woods}, David and {Wyder}, Ted},
        title = "{The WiggleZ Dark Energy Survey: direct constraints on blue galaxy intrinsic alignments at intermediate redshifts}",
      journal = {\mnras},
         year = 2011,
        month = jan,
       volume = {410},
       number = {2},
        pages = {844-859},
          doi = {10.1111/j.1365-2966.2010.17485.x},
archivePrefix = {arXiv},
       eprint = {0911.5347},
 primaryClass = {astro-ph.CO},
       adsurl = {https://ui.adsabs.harvard.edu/abs/2011MNRAS.410..844M}
}

@ARTICLE{Burger24,
       author = {{Burger}, Pierre A. and {Porth}, Lucas and {Heydenreich}, Sven and {Linke}, Laila and {Wielders}, Niek and {Schneider}, Peter and {Asgari}, Marika and {Castro}, Tiago and {Dolag}, Klaus and {Harnois-D{\'e}raps}, Joachim and {Hildebrandt}, Hendrik and {Kuijken}, Konrad and {Martinet}, Nicolas},
        title = "{KiDS-1000 cosmology: Combined second- and third-order shear statistics}",
      journal = {\aap},
         year = 2024,
        month = mar,
       volume = {683},
          eid = {A103},
        pages = {A103},
          doi = {10.1051/0004-6361/202347986},
archivePrefix = {arXiv},
       eprint = {2309.08602},
 primaryClass = {astro-ph.CO},
       adsurl = {https://ui.adsabs.harvard.edu/abs/2024A&A...683A.103B}
}

@ARTICLE{Gomes25,
       author = {{Gomes}, R.~C.~H. and {Sugiyama}, S. and {Jain}, B. and {Jarvis}, M. and {Anbajagane}, D. and {Gatti}, M. and {Gebauer}, D. and {Gong}, Z. and {Halder}, A. and {Marques}, G.~A. and {Pandey}, S. and {Marshall}, J.~L. and {Allam}, S. and {Alves}, O. and {Andrade-Oliveira}, F. and {Bacon}, D. and {Blazek}, J. and {Bocquet}, S. and {Brooks}, D. and {Carnero Rosell}, A. and {Carretero}, J. and {da Costa}, L.~N. and {Doel}, P. and {Doux}, C. and {Everett}, S. and {Flaugher}, B. and {Frieman}, J. and {Garc{\'\i}a-Bellido}, J. and {Gaztanaga}, E. and {Gruen}, D. and {Gruendl}, R.~A. and {Gutierrez}, G. and {Herner}, K. and {Hinton}, S.~R. and {Hollowood}, D.~L. and {Honscheid}, K. and {Huterer}, D. and {James}, D.~J. and {Jeffrey}, N. and {Mena-Fern{\'a}ndez}, J. and {Miquel}, R. and {Muir}, J. and {Ogando}, R.~L.~C. and {Pereira}, M.~E.~S. and {Pieres}, A. and {Plazas Malag{\'o}n}, A.~A. and {Samuroff}, S. and {Sanchez}, E. and {Sanchez Cid}, D. and {Santiago}, B. and {Sevilla-Noarbe}, I. and {Smith}, M. and {Suchyta}, E. and {Swanson}, M.~E.~C. and {Tarle}, G. and {To}, C. and {Vikram}, V. and {Weaverdyck}, N. and {Weller}, J.},
        title = "{Cosmology with second and third-order shear statistics for the Dark Energy Survey: Methods and simulated analysis}",
      journal = {arXiv e-prints},
         year = 2025,
        month = mar,
          eid = {arXiv:2503.03964},
        pages = {arXiv:2503.03964},
          doi = {10.48550/arXiv.2503.03964},
archivePrefix = {arXiv},
       eprint = {2503.03964},
 primaryClass = {astro-ph.CO},
       adsurl = {https://ui.adsabs.harvard.edu/abs/2025arXiv250303964G}
}

@ARTICLE{Ajani23,
       author = {{Euclid Collaboration} and {Ajani}, V. and {Baldi}, M. and {Barthelemy}, A. and {Boyle}, A. and {Burger}, P. and {Cardone}, V.~F. and {Cheng}, S. and {Codis}, S. and {Giocoli}, C. and {Harnois-D{\'e}raps}, J. and {Heydenreich}, S. and {Kansal}, V. and {Kilbinger}, M. and {Linke}, L. and {Llinares}, C. and {Martinet}, N. and {Parroni}, C. and {Peel}, A. and {Pires}, S. and {Porth}, L. and {Tereno}, I. and {Uhlemann}, C. and {Vicinanza}, M. and {Vinciguerra}, S. and {Aghanim}, N. and {Auricchio}, N. and {Bonino}, D. and {Branchini}, E. and {Brescia}, M. and {Brinchmann}, J. and {Camera}, S. and {Capobianco}, V. and {Carbone}, C. and {Carretero}, J. and {Castander}, F.~J. and {Castellano}, M. and {Cavuoti}, S. and {Cimatti}, A. and {Cledassou}, R. and {Congedo}, G. and {Conselice}, C.~J. and {Conversi}, L. and {Corcione}, L. and {Courbin}, F. and {Cropper}, M. and {Da Silva}, A. and {Degaudenzi}, H. and {Di Giorgio}, A.~M. and {Dinis}, J. and {Douspis}, M. and {Dubath}, F. and {Dupac}, X. and {Farrens}, S. and {Ferriol}, S. and {Fosalba}, P. and {Frailis}, M. and {Franceschi}, E. and {Galeotta}, S. and {Garilli}, B. and {Gillis}, B. and {Grazian}, A. and {Grupp}, F. and {Hoekstra}, H. and {Holmes}, W. and {Hornstrup}, A. and {Hudelot}, P. and {Jahnke}, K. and {Jhabvala}, M. and {K{\"u}mmel}, M. and {Kitching}, T. and {Kunz}, M. and {Kurki-Suonio}, H. and {Lilje}, P.~B. and {Lloro}, I. and {Maiorano}, E. and {Mansutti}, O. and {Marggraf}, O. and {Markovic}, K. and {Marulli}, F. and {Massey}, R. and {Mei}, S. and {Mellier}, Y. and {Meneghetti}, M. and {Moresco}, M. and {Moscardini}, L. and {Niemi}, S. -M. and {Nightingale}, J. and {Nutma}, T. and {Padilla}, C. and {Paltani}, S. and {Pedersen}, K. and {Pettorino}, V. and {Polenta}, G. and {Poncet}, M. and {Popa}, L.~A. and {Raison}, F. and {Renzi}, A. and {Rhodes}, J. and {Riccio}, G. and {Romelli}, E. and {Roncarelli}, M. and {Rossetti}, E. and {Saglia}, R. and {Sapone}, D. and {Sartoris}, B. and {Schneider}, P. and {Schrabback}, T. and {Secroun}, A. and {Seidel}, G. and {Serrano}, S. and {Sirignano}, C. and {Stanco}, L. and {Starck}, J. -L. and {Tallada-Cresp{\'\i}}, P. and {Taylor}, A.~N. and {Toledo-Moreo}, R. and {Torradeflot}, F. and {Tutusaus}, I. and {Valentijn}, E.~A. and {Valenziano}, L. and {Vassallo}, T. and {Wang}, Y. and {Weller}, J. and {Zamorani}, G. and {Zoubian}, J. and {Andreon}, S. and {Bardelli}, S. and {Boucaud}, A. and {Bozzo}, E. and {Colodro-Conde}, C. and {Di Ferdinando}, D. and {Fabbian}, G. and {Farina}, M. and {Graci{\'a}-Carpio}, J. and {Keih{\"a}nen}, E. and {Lindholm}, V. and {Maino}, D. and {Mauri}, N. and {Neissner}, C. and {Schirmer}, M. and {Scottez}, V. and {Zucca}, E. and {Akrami}, Y. and {Baccigalupi}, C. and {Balaguera-Antol{\'\i}nez}, A. and {Ballardini}, M. and {Bernardeau}, F. and {Biviano}, A. and {Blanchard}, A. and {Borgani}, S. and {Borlaff}, A.~S. and {Burigana}, C. and {Cabanac}, R. and {Cappi}, A. and {Carvalho}, C.~S. and {Casas}, S. and {Castignani}, G. and {Castro}, T. and {Chambers}, K.~C. and {Cooray}, A.~R. and {Coupon}, J. and {Courtois}, H.~M. and {Davini}, S. and {de la Torre}, S. and {De Lucia}, G. and {Desprez}, G. and {Dole}, H. and {Escartin}, J.~A. and {Escoffier}, S. and {Ferrero}, I. and {Finelli}, F. and {Ganga}, K. and {Garcia-Bellido}, J. and {George}, K. and {Giacomini}, F. and {Gozaliasl}, G. and {Hildebrandt}, H. and {Jimenez Mu{\~n}oz}, A. and {Joachimi}, B. and {Kajava}, J.~J.~E. and {Kirkpatrick}, C.~C. and {Legrand}, L. and {Loureiro}, A. and {Magliocchetti}, M. and {Maoli}, R. and {Marcin}, S. and {Martinelli}, M. and {Martins}, C.~J.~A.~P. and {Matthew}, S. and {Maurin}, L. and {Metcalf}, R.~B. and {Monaco}, P. and {Morgante}, G. and {Nadathur}, S. and {Nucita}, A.~A. and {Popa}, V. and {Potter}, D. and {Pourtsidou}, A. and {P{\"o}ntinen}, M.},
        title = "{Euclid preparation. XXVIII. Forecasts for ten different higher-order weak lensing statistics}",
      journal = {\aap},
         year = 2023,
        month = jul,
       volume = {675},
          eid = {A120},
        pages = {A120},
          doi = {10.1051/0004-6361/202346017},
archivePrefix = {arXiv},
       eprint = {2301.12890},
 primaryClass = {astro-ph.CO},
       adsurl = {https://ui.adsabs.harvard.edu/abs/2023A&A...675A.120E}
}

@ARTICLE{FJ12,
       author = {{Schmidt}, Fabian and {Jeong}, Donghui},
        title = "{Cosmic rulers}",
      journal = {\prd},
         year = 2012,
        month = oct,
       volume = {86},
       number = {8},
          eid = {083527},
        pages = {083527},
          doi = {10.1103/PhysRevD.86.083527},
archivePrefix = {arXiv},
       eprint = {1204.3625},
 primaryClass = {astro-ph.CO},
       adsurl = {https://ui.adsabs.harvard.edu/abs/2012PhRvD..86h3527S}
}

@ARTICLE{Huff13,
       author = {{Huff}, Eric M. and {Krause}, Elisabeth and {Eifler}, Tim and {Fang}, Xiao and {George}, Matthew R. and {Schlegel}, David},
        title = "{Cosmic shear without shape noise}",
      journal = {arXiv e-prints},
         year = 2013,
        month = nov,
          eid = {arXiv:1311.1489},
        pages = {arXiv:1311.1489},
          doi = {10.48550/arXiv.1311.1489},
archivePrefix = {arXiv},
       eprint = {1311.1489},
 primaryClass = {astro-ph.CO},
       adsurl = {https://ui.adsabs.harvard.edu/abs/2013arXiv1311.1489H}
}

@ARTICLE{Huang24,
       author = {{Huang}, Yu-Hsiu and {Krause}, Elisabeth and {Xu}, Jiachuan and {Eifler}, Tim and {R.~S.}, Pranjal and {Huff}, Eric},
        title = "{Astrophysical systematics in kinematic lensing: Quantifying an intrinsic alignment analog}",
      journal = {\prd},
         year = 2024,
        month = aug,
       volume = {110},
       number = {4},
          eid = {043509},
        pages = {043509},
          doi = {10.1103/PhysRevD.110.043509},
archivePrefix = {arXiv},
       eprint = {2404.00197},
 primaryClass = {astro-ph.CO},
       adsurl = {https://ui.adsabs.harvard.edu/abs/2024PhRvD.110d3509H}
}

@ARTICLE{Lopez21,
       author = {{L{\'o}pez}, Pablo and {Cautun}, Marius and {Paz}, Dante and {Merch{\'a}n}, Manuel and {van de Weygaert}, Rien},
        title = "{Deviations from tidal torque theory: Evolution of the halo spin-filament alignment}",
      journal = {\mnras},
         year = 2021,
        month = apr,
       volume = {502},
       number = {4},
        pages = {5528-5545},
          doi = {10.1093/mnras/stab451},
archivePrefix = {arXiv},
       eprint = {2012.01638},
 primaryClass = {astro-ph.GA},
       adsurl = {https://ui.adsabs.harvard.edu/abs/2021MNRAS.502.5528L}
}

@ARTICLE{Lopez24,
       author = {{L{\'o}pez}, Pablo},
        title = "{Early Evolution of Spin Direction in Dark Matter Halos and the Effect of the Surrounding Large-scale Tidal Field}",
      journal = {\pasp},
         year = 2024,
        month = mar,
       volume = {136},
       number = {3},
          eid = {037001},
        pages = {037001},
          doi = {10.1088/1538-3873/ad31c9},
archivePrefix = {arXiv},
       eprint = {2404.04223},
 primaryClass = {astro-ph.CO},
       adsurl = {https://ui.adsabs.harvard.edu/abs/2024PASP..136c7001L}
}

@ARTICLE{Codis15b,
       author = {{Codis}, Sandrine and {Pichon}, Christophe and {Pogosyan}, Dmitry},
        title = "{Spin alignments within the cosmic web: a theory of constrained tidal torques near filaments}",
      journal = {\mnras},
         year = 2015,
        month = oct,
       volume = {452},
       number = {4},
        pages = {3369-3393},
          doi = {10.1093/mnras/stv1570},
archivePrefix = {arXiv},
       eprint = {1504.06073},
 primaryClass = {astro-ph.CO},
       adsurl = {https://ui.adsabs.harvard.edu/abs/2015MNRAS.452.3369C}
}

@ARTICLE{Smith05,
       author = {{Smith}, R.~E. and {Watts}, P.~I.~R.},
        title = "{Triaxial haloes, intrinsic alignments and the dark matter power spectrum}",
      journal = {\mnras},
         year = 2005,
        month = jun,
       volume = {360},
       number = {1},
        pages = {203-215},
          doi = {10.1111/j.1365-2966.2005.09053.x},
archivePrefix = {arXiv},
       eprint = {astro-ph/0412441},
 primaryClass = {astro-ph},
       adsurl = {https://ui.adsabs.harvard.edu/abs/2005MNRAS.360..203S}
}

@ARTICLE{Okumura20b,
       author = {{Okumura}, Teppei and {Taruya}, Atsushi},
        title = "{Anisotropies of galaxy ellipticity correlations in real and redshift space: angular dependence in linear tidal alignment model}",
      journal = {\mnras},
         year = 2020,
        month = mar,
       volume = {493},
       number = {1},
        pages = {L124-L128},
          doi = {10.1093/mnrasl/slaa024},
archivePrefix = {arXiv},
       eprint = {1912.04118},
 primaryClass = {astro-ph.CO},
       adsurl = {https://ui.adsabs.harvard.edu/abs/2020MNRAS.493L.124O}
}

@ARTICLE{Bakx25,
       author = {{Bakx}, Thomas and {Kurita}, Toshiki and {Eggemeier}, Alexander and {Chisari}, Nora Elisa and {Vlah}, Zvonimir},
        title = "{The Bispectrum of Intrinsic Alignments: Theory Modelling and Forecasts for Stage IV Galaxy Surveys}",
      journal = {arXiv e-prints},
         year = 2025,
        month = apr,
          eid = {arXiv:2504.10009},
        pages = {arXiv:2504.10009},
          doi = {10.48550/arXiv.2504.10009},
archivePrefix = {arXiv},
       eprint = {2504.10009},
 primaryClass = {astro-ph.CO},
       adsurl = {https://ui.adsabs.harvard.edu/abs/2025arXiv250410009B}
}

@ARTICLE{Hui08,
       author = {{Hui}, Lam and {Zhang}, Jun},
        title = "{Density-Ellipticity Correlations, Galaxy-Galaxy Lensing, and the Importance of Non-Gaussianity in Intrinsic Alignment}",
      journal = {\apj},
         year = 2008,
        month = dec,
       volume = {688},
       number = {2},
        pages = {742-756},
          doi = {10.1086/589872},
archivePrefix = {arXiv},
       eprint = {astro-ph/0205512},
 primaryClass = {astro-ph},
       adsurl = {https://ui.adsabs.harvard.edu/abs/2008ApJ...688..742H}
}

@ARTICLE{Scoccimarro04,
       author = {{Scoccimarro}, Rom{\'a}n},
        title = "{Redshift-space distortions, pairwise velocities, and nonlinearities}",
      journal = {\prd},
         year = 2004,
        month = oct,
       volume = {70},
       number = {8},
          eid = {083007},
        pages = {083007},
          doi = {10.1103/PhysRevD.70.083007},
archivePrefix = {arXiv},
       eprint = {astro-ph/0407214},
 primaryClass = {astro-ph},
       adsurl = {https://ui.adsabs.harvard.edu/abs/2004PhRvD..70h3007S}
}

@ARTICLE{Hamilton92,
       author = {{Hamilton}, A.~J.~S.},
        title = "{Measuring Omega and the Real Correlation Function from the Redshift Correlation Function}",
      journal = {\apjl},
         year = 1992,
        month = jan,
       volume = {385},
        pages = {L5},
          doi = {10.1086/186264},
       adsurl = {https://ui.adsabs.harvard.edu/abs/1992ApJ...385L...5H}
}

@ARTICLE{Crittenden00,
       author = {{Crittenden}, Robert G. and {Natarajan}, Priyamvada and {Pen}, Ue-Li and {Theuns}, Tom},
        title = "{Spin-induced Galaxy Alignments and Their Implications for Weak-Lensing Measurements}",
      journal = {\apj},
         year = 2001,
        month = oct,
       volume = {559},
       number = {2},
        pages = {552-571},
          doi = {10.1086/322370},
archivePrefix = {arXiv},
       eprint = {astro-ph/0009052},
 primaryClass = {astro-ph},
       adsurl = {https://ui.adsabs.harvard.edu/abs/2001ApJ...559..552C}
}

@ARTICLE{vandenBosch02,
       author = {{van den Bosch}, Frank C. and {Abel}, Tom and {Croft}, Rupert A.~C. and {Hernquist}, Lars and {White}, Simon D.~M.},
        title = "{The Angular Momentum of Gas in Protogalaxies. I. Implications for the Formation of Disk Galaxies}",
      journal = {\apj},
         year = 2002,
        month = sep,
       volume = {576},
       number = {1},
        pages = {21-35},
          doi = {10.1086/341619},
archivePrefix = {arXiv},
       eprint = {astro-ph/0201095},
 primaryClass = {astro-ph},
       adsurl = {https://ui.adsabs.harvard.edu/abs/2002ApJ...576...21V}
}

@ARTICLE{Heavens88,
       author = {{Heavens}, Alan and {Peacock}, John},
        title = "{Tidal torques and local density maxima}",
      journal = {\mnras},
         year = 1988,
        month = may,
       volume = {232},
        pages = {339-360},
          doi = {10.1093/mnras/232.2.339},
       adsurl = {https://ui.adsabs.harvard.edu/abs/1988MNRAS.232..339H}
}

@ARTICLE{Croft00,
       author = {{Croft}, Rupert A.~C. and {Metzler}, Christopher A.},
        title = "{Weak-Lensing Surveys and the Intrinsic Correlation of Galaxy Ellipticities}",
      journal = {\apj},
         year = 2000,
        month = dec,
       volume = {545},
       number = {2},
        pages = {561-571},
          doi = {10.1086/317856},
archivePrefix = {arXiv},
       eprint = {astro-ph/0005384},
 primaryClass = {astro-ph},
       adsurl = {https://ui.adsabs.harvard.edu/abs/2000ApJ...545..561C}
}

@ARTICLE{Fischbacher23,
       author = {{Fischbacher}, Silvan and {Kacprzak}, Tomasz and {Blazek}, Jonathan and {Refregier}, Alexandre},
        title = "{Redshift requirements for cosmic shear with intrinsic alignment}",
      journal = {\jcap},
         year = 2023,
        month = jan,
       volume = {2023},
       number = {1},
          eid = {033},
        pages = {033},
          doi = {10.1088/1475-7516/2023/01/033},
archivePrefix = {arXiv},
       eprint = {2207.01627},
 primaryClass = {astro-ph.CO},
       adsurl = {https://ui.adsabs.harvard.edu/abs/2023JCAP...01..033F}
}

@ARTICLE{Chisari19,
       author = {{Chisari}, Nora Elisa and {Alonso}, David and {Krause}, Elisabeth and {Leonard}, C. Danielle and {Bull}, Philip and {Neveu}, J{\'e}r{\'e}my and {Villarreal}, Antonia Sierra and {Singh}, Sukhdeep and {McClintock}, Thomas and {Ellison}, John and {Du}, Zilong and {Zuntz}, Joe and {Mead}, Alexander and {Joudaki}, Shahab and {Lorenz}, Christiane S. and {Tr{\"o}ster}, Tilman and {Sanchez}, Javier and {Lanusse}, Francois and {Ishak}, Mustapha and {Hlozek}, Ren{\'e}e and {Blazek}, Jonathan and {Campagne}, Jean-Eric and {Almoubayyed}, Husni and {Eifler}, Tim and {Kirby}, Matthew and {Kirkby}, David and {Plaszczynski}, St{\'e}phane and {Slosar}, An{\v{z}}e and {Vrastil}, Michal and {Wagoner}, Erika L. and {LSST Dark Energy Science Collaboration}},
        title = "{Core Cosmology Library: Precision Cosmological Predictions for LSST}",
      journal = {\apjs},
         year = 2019,
        month = may,
       volume = {242},
       number = {1},
          eid = {2},
        pages = {2},
          doi = {10.3847/1538-4365/ab1658},
archivePrefix = {arXiv},
       eprint = {1812.05995},
 primaryClass = {astro-ph.CO},
       adsurl = {https://ui.adsabs.harvard.edu/abs/2019ApJS..242....2C}
}

@ARTICLE{Asgari23,
       author = {{Asgari}, Marika and {Mead}, Alexander J. and {Heymans}, Catherine},
        title = "{The halo model for cosmology: a pedagogical review}",
      journal = {The Open Journal of Astrophysics},
         year = 2023,
        month = nov,
       volume = {6},
          eid = {39},
        pages = {39},
          doi = {10.21105/astro.2303.08752},
archivePrefix = {arXiv},
       eprint = {2303.08752},
 primaryClass = {astro-ph.CO},
       adsurl = {https://ui.adsabs.harvard.edu/abs/2023OJAp....6E..39A}
}

@ARTICLE{Cooray02,
       author = {{Cooray}, Asantha and {Sheth}, Ravi},
        title = "{Halo models of large scale structure}",
      journal = {\physrep},
         year = 2002,
        month = dec,
       volume = {372},
       number = {1},
        pages = {1-129},
          doi = {10.1016/S0370-1573(02)00276-4},
archivePrefix = {arXiv},
       eprint = {astro-ph/0206508},
 primaryClass = {astro-ph},
       adsurl = {https://ui.adsabs.harvard.edu/abs/2002PhR...372....1C}
}

@ARTICLE{Coles07,
       author = {{Coles}, Peter and {Erdogdu}, Pirin},
        title = "{Scale dependent galaxy bias}",
      journal = {\jcap},
         year = 2007,
        month = oct,
       volume = {2007},
       number = {10},
          eid = {007},
        pages = {007},
          doi = {10.1088/1475-7516/2007/10/007},
archivePrefix = {arXiv},
       eprint = {0706.0412},
 primaryClass = {astro-ph},
       adsurl = {https://ui.adsabs.harvard.edu/abs/2007JCAP...10..007C}
}

@ARTICLE{Matsubara99,
       author = {{Matsubara}, Takahiko},
        title = "{Stochasticity of Bias and Nonlocality of Galaxy Formation: Linear Scales}",
      journal = {\apj},
         year = 1999,
        month = nov,
       volume = {525},
       number = {2},
        pages = {543-553},
          doi = {10.1086/307931},
archivePrefix = {arXiv},
       eprint = {astro-ph/9906029},
 primaryClass = {astro-ph},
       adsurl = {https://ui.adsabs.harvard.edu/abs/1999ApJ...525..543M}
}

@ARTICLE{Merkel17,
       author = {{Merkel}, Philipp M. and {Sch{\"a}fer}, Bj{\"o}rn Malte},
        title = "{Imitating intrinsic alignments: a bias to the CMB lensing-galaxy shape cross-correlation power spectrum induced by the large-scale structure bispectrum}",
      journal = {\mnras},
         year = 2017,
        month = oct,
       volume = {471},
       number = {2},
        pages = {2431-2437},
          doi = {10.1093/mnras/stx1664},
archivePrefix = {arXiv},
       eprint = {1709.04444},
 primaryClass = {astro-ph.CO},
       adsurl = {https://ui.adsabs.harvard.edu/abs/2017MNRAS.471.2431M}
}

@ARTICLE{Fang17,
       author = {{Fang}, Xiao and {Blazek}, Jonathan A. and {McEwen}, Joseph E. and {Hirata}, Christopher M.},
        title = "{FAST-PT II: an algorithm to calculate convolution integrals of general tensor quantities in cosmological perturbation theory}",
      journal = {\jcap},
         year = 2017,
        month = feb,
       volume = {2017},
       number = {2},
          eid = {030},
        pages = {030},
          doi = {10.1088/1475-7516/2017/02/030},
archivePrefix = {arXiv},
       eprint = {1609.05978},
 primaryClass = {astro-ph.CO},
       adsurl = {https://ui.adsabs.harvard.edu/abs/2017JCAP...02..030F}
}

@ARTICLE{McEwen16,
       author = {{McEwen}, Joseph E. and {Fang}, Xiao and {Hirata}, Christopher M. and {Blazek}, Jonathan A.},
        title = "{FAST-PT: a novel algorithm to calculate convolution integrals in cosmological perturbation theory}",
      journal = {\jcap},
         year = 2016,
        month = sep,
       volume = {2016},
       number = {9},
          eid = {015},
        pages = {015},
          doi = {10.1088/1475-7516/2016/09/015},
archivePrefix = {arXiv},
       eprint = {1603.04826},
 primaryClass = {astro-ph.CO},
       adsurl = {https://ui.adsabs.harvard.edu/abs/2016JCAP...09..015M}
}

@ARTICLE{Georgiou24,
       author = {{Georgiou}, Christos and {Bakx}, Thomas and {van Donkersgoed}, Juliard and {Chisari}, Nora Elisa},
        title = "{B-modes from galaxy cluster alignments in future surveys}",
      journal = {The Open Journal of Astrophysics},
         year = 2024,
        month = may,
       volume = {7},
          eid = {36},
        pages = {36},
          doi = {10.33232/001c.117495},
archivePrefix = {arXiv},
       eprint = {2309.03841},
 primaryClass = {astro-ph.CO},
       adsurl = {https://ui.adsabs.harvard.edu/abs/2024OJAp....7E..36G}
}

@article{Vedder21,
   title={Galaxy clusters as intrinsic alignment tracers: present and future},
   volume={500},
   ISSN={1365-2966},
   DOI={10.1093/mnras/staa3633},
   number={4},
   journal={Monthly Notices of the Royal Astronomical Society},
   publisher={Oxford University Press (OUP)},
   author={Vedder, C J G and Chisari, N E},
   year={2020},
   month=nov, pages={5561–5569} }

@ARTICLE{Bett12,
       author = {{Bett}, Philip},
        title = "{Halo shapes from weak lensing: the impact of galaxy-halo misalignment}",
      journal = {\mnras},
         year = 2012,
        month = mar,
       volume = {420},
       number = {4},
        pages = {3303-3323},
          doi = {10.1111/j.1365-2966.2011.20258.x},
archivePrefix = {arXiv},
       eprint = {1108.3717},
 primaryClass = {astro-ph.CO},
       adsurl = {https://ui.adsabs.harvard.edu/abs/2012MNRAS.420.3303B}
}

@ARTICLE{Joachimi13b,
       author = {{Joachimi}, B. and {Semboloni}, E. and {Hilbert}, S. and {Bett}, P.~E. and {Hartlap}, J. and {Hoekstra}, H. and {Schneider}, P.},
        title = "{Intrinsic galaxy shapes and alignments - II. Modelling the intrinsic alignment contamination of weak lensing surveys}",
      journal = {\mnras},
         year = 2013,
        month = nov,
       volume = {436},
       number = {1},
        pages = {819-838},
          doi = {10.1093/mnras/stt1618},
archivePrefix = {arXiv},
       eprint = {1305.5791},
 primaryClass = {astro-ph.CO},
       adsurl = {https://ui.adsabs.harvard.edu/abs/2013MNRAS.436..819J}
}

@ARTICLE{Reischke19,
       author = {{Reischke}, Robert and {Sch{\"a}fer}, Bj{\"o}rn Malte},
        title = "{Environmental dependence of ellipticity correlation functions of intrinsic alignments}",
      journal = {\jcap},
         year = 2019,
        month = apr,
       volume = {2019},
       number = {4},
          eid = {031},
        pages = {031},
          doi = {10.1088/1475-7516/2019/04/031},
archivePrefix = {arXiv},
       eprint = {1812.06918},
 primaryClass = {astro-ph.CO},
       adsurl = {https://ui.adsabs.harvard.edu/abs/2019JCAP...04..031R}
}

@ARTICLE{Ghosh24,
       author = {{Ghosh}, Basundhara and {Nussbaumer}, Kai and {Giesel}, Eileen Sophie and {Sch{\"a}fer}, Bj{\"o}rn Malte},
        title = "{A unified linear intrinsic alignment model for elliptical and disc galaxies and the resulting ellipticity spectra}",
      journal = {The Open Journal of Astrophysics},
         year = 2024,
        month = may,
       volume = {7},
          eid = {41},
        pages = {41},
          doi = {10.33232/001c.117965},
archivePrefix = {arXiv},
       eprint = {2312.03353},
 primaryClass = {astro-ph.CO},
       adsurl = {https://ui.adsabs.harvard.edu/abs/2024OJAp....7E..41G}
}

@ARTICLE{Tugendhat18,
       author = {{Tugendhat}, Tim M. and {Sch{\"a}fer}, Bj{\"o}rn Malte},
        title = "{Angular ellipticity correlations in a composite alignment model for elliptical and spiral galaxies and inference from weak lensing}",
      journal = {\mnras},
         year = 2018,
        month = may,
       volume = {476},
       number = {3},
        pages = {3460-3477},
          doi = {10.1093/mnras/sty323},
archivePrefix = {arXiv},
       eprint = {1709.02630},
 primaryClass = {astro-ph.CO},
       adsurl = {https://ui.adsabs.harvard.edu/abs/2018MNRAS.476.3460T}
}

@ARTICLE{SuperSample,
       author = {{Taruya}, Atsushi and {Akitsu}, Kazuyuki},
        title = "{Lagrangian approach to super-sample effects on biased tracers at field level: galaxy density fields and intrinsic alignments}",
      journal = {\jcap},
         year = 2021,
        month = nov,
       volume = {2021},
       number = {11},
          eid = {061},
        pages = {061},
          doi = {10.1088/1475-7516/2021/11/061},
archivePrefix = {arXiv},
       eprint = {2106.04789},
 primaryClass = {astro-ph.CO},
       adsurl = {https://ui.adsabs.harvard.edu/abs/2021JCAP...11..061T}
}

@ARTICLE{SuperSample2,
       author = {{Ansarifard}, Saeed and {Movahed}, S.~M.~S.},
        title = "{Cosmological consequences of intrinsic alignments supersample covariance}",
      journal = {\mnras},
         year = 2020,
        month = dec,
       volume = {499},
       number = {4},
        pages = {6094-6104},
          doi = {10.1093/mnras/staa3214},
archivePrefix = {arXiv},
       eprint = {2011.01551},
 primaryClass = {astro-ph.CO},
       adsurl = {https://ui.adsabs.harvard.edu/abs/2020MNRAS.499.6094A}
}

@ARTICLE{LHuillier17,
       author = {{L'Huillier}, Benjamin and {Winther}, Hans A. and {Mota}, David F. and {Park}, Changbom and {Kim}, Juhan},
        title = "{Dark matter haloes in modified gravity and dark energy: interaction rate, small- and large-scale alignment}",
      journal = {\mnras},
         year = 2017,
        month = jul,
       volume = {468},
       number = {3},
        pages = {3174-3183},
          doi = {10.1093/mnras/stx700},
archivePrefix = {arXiv},
       eprint = {1703.07357},
 primaryClass = {astro-ph.CO},
       adsurl = {https://ui.adsabs.harvard.edu/abs/2017MNRAS.468.3174L}
}

@ARTICLE{Chuang22,
       author = {{Chuang}, Yao-Tsung and {Okumura}, Teppei and {Shirasaki}, Masato},
        title = "{Distinguishing between {\ensuremath{\Lambda}}CDM and f(R) gravity models using halo ellipticity correlations in simulations}",
      journal = {\mnras},
         year = 2022,
        month = sep,
       volume = {515},
       number = {3},
        pages = {4464-4470},
          doi = {10.1093/mnras/stac2029},
archivePrefix = {arXiv},
       eprint = {2111.01417},
 primaryClass = {astro-ph.CO},
       adsurl = {https://ui.adsabs.harvard.edu/abs/2022MNRAS.515.4464C}
}

@ARTICLE{Inoue24,
       author = {{Inoue}, Takuya and {Okumura}, Teppei and {Saga}, Shohei and {Taruya}, Atsushi},
        title = "{Information content in anisotropic cosmological fields: Impact of different multipole expansion scheme for galaxy density and ellipticity correlations}",
      journal = {arXiv e-prints},
         year = 2024,
        month = jun,
          eid = {arXiv:2406.19669},
        pages = {arXiv:2406.19669},
          doi = {10.48550/arXiv.2406.19669},
archivePrefix = {arXiv},
       eprint = {2406.19669},
 primaryClass = {astro-ph.CO},
       adsurl = {https://ui.adsabs.harvard.edu/abs/2024arXiv240619669I}
}

@ARTICLE{Minami20,
       author = {{Minami}, Yuto and {Komatsu}, Eiichiro},
        title = "{New Extraction of the Cosmic Birefringence from the Planck 2018 Polarization Data}",
      journal = {\prl},
         year = 2020,
        month = nov,
       volume = {125},
       number = {22},
          eid = {221301},
        pages = {221301},
          doi = {10.1103/PhysRevLett.125.221301},
archivePrefix = {arXiv},
       eprint = {2011.11254},
 primaryClass = {astro-ph.CO},
       adsurl = {https://ui.adsabs.harvard.edu/abs/2020PhRvL.125v1301M}
}

@ARTICLE{Okumura24,
       author = {{Okumura}, Teppei and {Taruya}, Atsushi and {Kurita}, Toshiki and {Nishimichi}, Takahiro},
        title = "{Nonlinear redshift space distortion in halo ellipticity correlations: Analytical model and N -body simulations}",
      journal = {\prd},
         year = 2024,
        month = may,
       volume = {109},
       number = {10},
          eid = {103501},
        pages = {103501},
          doi = {10.1103/PhysRevD.109.103501},
archivePrefix = {arXiv},
       eprint = {2310.07384},
 primaryClass = {astro-ph.CO},
       adsurl = {https://ui.adsabs.harvard.edu/abs/2024PhRvD.109j3501O}
}

@ARTICLE{Taruya24,
       author = {{Taruya}, Atsushi and {Kurita}, Toshiki and {Okumura}, Teppei},
        title = "{Improving redshift-space power spectra of halo intrinsic alignments from perturbation theory}",
      journal = {arXiv e-prints},
         year = 2024,
        month = sep,
          eid = {arXiv:2409.06616},
        pages = {arXiv:2409.06616},
          doi = {10.48550/arXiv.2409.06616},
archivePrefix = {arXiv},
       eprint = {2409.06616},
 primaryClass = {astro-ph.CO},
       adsurl = {https://ui.adsabs.harvard.edu/abs/2024arXiv240906616T}
}

@ARTICLE{Saga23,
       author = {{Saga}, Shohei and {Okumura}, Teppei and {Taruya}, Atsushi and {Inoue}, Takuya},
        title = "{Relativistic distortions in galaxy density-ellipticity correlations: gravitational redshift and peculiar velocity effects}",
      journal = {\mnras},
         year = 2023,
        month = feb,
       volume = {518},
       number = {4},
        pages = {4976-4990},
          doi = {10.1093/mnras/stac3462},
archivePrefix = {arXiv},
       eprint = {2207.03454},
 primaryClass = {astro-ph.CO},
       adsurl = {https://ui.adsabs.harvard.edu/abs/2023MNRAS.518.4976S}
}

@ARTICLE{Blazek15,
       author = {{Blazek}, Jonathan and {Vlah}, Zvonimir and {Seljak}, Uro{\v{s}}},
        title = "{Tidal alignment of galaxies}",
      journal = {\jcap},
         year = 2015,
        month = aug,
       volume = {2015},
       number = {8},
        pages = {015-015},
          doi = {10.1088/1475-7516/2015/08/015},
archivePrefix = {arXiv},
       eprint = {1504.02510},
 primaryClass = {astro-ph.CO},
       adsurl = {https://ui.adsabs.harvard.edu/abs/2015JCAP...08..015B}
}

@ARTICLE{Schmitz18,
       author = {{Schmitz}, D.~M. and {Hirata}, C.~M. and {Blazek}, J. and {Krause}, E.},
        title = "{Time evolution of intrinsic alignments of galaxies}",
      journal = {\jcap},
         year = 2018,
        month = jul,
       volume = {2018},
       number = {7},
          eid = {030},
        pages = {030},
          doi = {10.1088/1475-7516/2018/07/030},
archivePrefix = {arXiv},
       eprint = {1805.02649},
 primaryClass = {astro-ph.CO},
       adsurl = {https://ui.adsabs.harvard.edu/abs/2018JCAP...07..030S}
}

@ARTICLE{Carrasco12,
       author = {{Carrasco}, John Joseph M. and {Hertzberg}, Mark P. and {Senatore}, Leonardo},
        title = "{The effective field theory of cosmological large scale structures}",
      journal = {Journal of High Energy Physics},
         year = 2012,
        month = sep,
       volume = {2012},
          eid = {82},
        pages = {82},
          doi = {10.1007/JHEP09(2012)082},
archivePrefix = {arXiv},
       eprint = {1206.2926},
 primaryClass = {astro-ph.CO},
       adsurl = {https://ui.adsabs.harvard.edu/abs/2012JHEP...09..082C}
}

@ARTICLE{Mcdonald09,
       author = {{McDonald}, Patrick and {Roy}, Arabindo},
        title = "{Clustering of dark matter tracers: generalizing bias for the coming era of precision LSS}",
      journal = {\jcap},
         year = 2009,
        month = aug,
       volume = {2009},
       number = {8},
          eid = {020},
        pages = {020},
          doi = {10.1088/1475-7516/2009/08/020},
archivePrefix = {arXiv},
       eprint = {0902.0991},
 primaryClass = {astro-ph.CO},
       adsurl = {https://ui.adsabs.harvard.edu/abs/2009JCAP...08..020M}
}

@ARTICLE{BaumannEFT,
       author = {{Baumann}, Daniel and {Nicolis}, Alberto and {Senatore}, Leonardo and {Zaldarriaga}, Matias},
        title = "{Cosmological non-linearities as an effective fluid}",
      journal = {\jcap},
         year = 2012,
        month = jul,
       volume = {2012},
       number = {7},
          eid = {051},
        pages = {051},
          doi = {10.1088/1475-7516/2012/07/051},
archivePrefix = {arXiv},
       eprint = {1004.2488},
 primaryClass = {astro-ph.CO},
       adsurl = {https://ui.adsabs.harvard.edu/abs/2012JCAP...07..051B}
}

@ARTICLE{FortunaHM,
       author = {{Fortuna}, Maria Cristina and {Hoekstra}, Henk and {Joachimi}, Benjamin and {Johnston}, Harry and {Chisari}, Nora Elisa and {Georgiou}, Christos and {Mahony}, Constance},
        title = "{The halo model as a versatile tool to predict intrinsic alignments}",
      journal = {\mnras},
         year = 2021,
        month = feb,
       volume = {501},
       number = {2},
        pages = {2983-3002},
          doi = {10.1093/mnras/staa3802},
archivePrefix = {arXiv},
       eprint = {2003.02700},
 primaryClass = {astro-ph.CO},
       adsurl = {https://ui.adsabs.harvard.edu/abs/2021MNRAS.501.2983F}
}

@ARTICLE{Catelan01,
       author = {{Catelan}, Paolo and {Kamionkowski}, Marc and {Blandford}, Roger D.},
        title = "{Intrinsic and extrinsic galaxy alignment}",
      journal = {\mnras},
         year = 2001,
        month = jan,
       volume = {320},
       number = {1},
        pages = {L7-L13},
          doi = {10.1046/j.1365-8711.2001.04105.x},
archivePrefix = {arXiv},
       eprint = {astro-ph/0005470},
 primaryClass = {astro-ph},
       adsurl = {https://ui.adsabs.harvard.edu/abs/2001MNRAS.320L...7C}
}

@ARTICLE{Bridle07,
       author = {{Bridle}, Sarah and {King}, Lindsay},
        title = "{Dark energy constraints from cosmic shear power spectra: impact of intrinsic alignments on photometric redshift requirements}",
      journal = {New Journal of Physics},
         year = 2007,
        month = dec,
       volume = {9},
       number = {12},
        pages = {444},
          doi = {10.1088/1367-2630/9/12/444},
archivePrefix = {arXiv},
       eprint = {0705.0166},
 primaryClass = {astro-ph},
       adsurl = {https://ui.adsabs.harvard.edu/abs/2007NJPh....9..444B}
}

@ARTICLE{Blazek11,
       author = {{Blazek}, Jonathan and {McQuinn}, Matthew and {Seljak}, Uro{\v{s}}},
        title = "{Testing the tidal alignment model of galaxy intrinsic alignment}",
      journal = {\jcap},
         year = 2011,
        month = may,
       volume = {2011},
       number = {5},
          eid = {010},
        pages = {010},
          doi = {10.1088/1475-7516/2011/05/010},
archivePrefix = {arXiv},
       eprint = {1101.4017},
 primaryClass = {astro-ph.CO},
       adsurl = {https://ui.adsabs.harvard.edu/abs/2011JCAP...05..010B}
}

@ARTICLE{Blazek19,
       author = {{Blazek}, Jonathan A. and {MacCrann}, Niall and {Troxel}, M.~A. and {Fang}, Xiao},
        title = "{Beyond linear galaxy alignments}",
      journal = {\prd},
         year = 2019,
        month = nov,
       volume = {100},
       number = {10},
          eid = {103506},
        pages = {103506},
          doi = {10.1103/PhysRevD.100.103506},
archivePrefix = {arXiv},
       eprint = {1708.09247},
 primaryClass = {astro-ph.CO},
       adsurl = {https://ui.adsabs.harvard.edu/abs/2019PhRvD.100j3506B}
}

@ARTICLE{Vlah20,
       author = {{Vlah}, Zvonimir and {Chisari}, Nora Elisa and {Schmidt}, Fabian},
        title = "{An EFT description of galaxy intrinsic alignments}",
      journal = {\jcap},
         year = 2020,
        month = jan,
       volume = {2020},
       number = {1},
          eid = {025},
        pages = {025},
          doi = {10.1088/1475-7516/2020/01/025},
archivePrefix = {arXiv},
       eprint = {1910.08085},
 primaryClass = {astro-ph.CO},
       adsurl = {https://ui.adsabs.harvard.edu/abs/2020JCAP...01..025V}
}

@ARTICLE{Vlah21,
       author = {{Vlah}, Zvonimir and {Chisari}, Nora Elisa and {Schmidt}, Fabian},
        title = "{Galaxy shape statistics in the effective field theory}",
      journal = {\jcap},
         year = 2021,
        month = may,
       volume = {2021},
       number = {5},
          eid = {061},
        pages = {061},
          doi = {10.1088/1475-7516/2021/05/061},
archivePrefix = {arXiv},
       eprint = {2012.04114},
 primaryClass = {astro-ph.CO},
       adsurl = {https://ui.adsabs.harvard.edu/abs/2021JCAP...05..061V}
}

@ARTICLE{Stucker21,
       author = {{St{\"u}cker}, Jens and {Schmidt}, Andreas S. and {White}, Simon D.~M. and {Schmidt}, Fabian and {Hahn}, Oliver},
        title = "{Measuring the tidal response of structure formation: anisotropic separate universe simulations using TREEPM}",
      journal = {\mnras},
         year = 2021,
        month = may,
       volume = {503},
       number = {1},
        pages = {1473-1489},
          doi = {10.1093/mnras/stab473},
archivePrefix = {arXiv},
       eprint = {2003.06427},
 primaryClass = {astro-ph.CO},
       adsurl = {https://ui.adsabs.harvard.edu/abs/2021MNRAS.503.1473S}
}

@ARTICLE{Bakx23,
       author = {{Bakx}, Thomas and {Kurita}, Toshiki and {Elisa Chisari}, Nora and {Vlah}, Zvonimir and {Schmidt}, Fabian},
        title = "{Effective field theory of intrinsic alignments at one loop order: a comparison to dark matter simulations}",
      journal = {\jcap},
         year = 2023,
        month = oct,
       volume = {2023},
       number = {10},
          eid = {005},
        pages = {005},
          doi = {10.1088/1475-7516/2023/10/005},
archivePrefix = {arXiv},
       eprint = {2303.15565},
 primaryClass = {astro-ph.CO},
       adsurl = {https://ui.adsabs.harvard.edu/abs/2023JCAP...10..005B}
}

@ARTICLE{Chen24,
       author = {{Chen}, Shi-Fan and {Kokron}, Nickolas},
        title = "{A Lagrangian theory for galaxy shape statistics}",
      journal = {\jcap},
         year = 2024,
        month = jan,
       volume = {2024},
       number = {1},
          eid = {027},
        pages = {027},
          doi = {10.1088/1475-7516/2024/01/027},
archivePrefix = {arXiv},
       eprint = {2309.16761},
 primaryClass = {astro-ph.CO},
       adsurl = {https://ui.adsabs.harvard.edu/abs/2024JCAP...01..027C}
}

@ARTICLE{Maion24,
       author = {{Maion}, Francisco and {Angulo}, Raul E. and {Bakx}, Thomas and {Chisari}, Nora Elisa and {Kurita}, Toshiki and {Pellejero-Ib{\'a}{\~n}ez}, Marcos},
        title = "{HYMALAIA: a hybrid lagrangian model for intrinsic alignments}",
      journal = {\mnras},
         year = 2024,
        month = jun,
       volume = {531},
       number = {2},
        pages = {2684-2700},
          doi = {10.1093/mnras/stae1331},
archivePrefix = {arXiv},
       eprint = {2307.13754},
 primaryClass = {astro-ph.CO},
       adsurl = {https://ui.adsabs.harvard.edu/abs/2024MNRAS.531.2684M}
}

@ARTICLE{Maion24b,
       author = {{Maion}, Francisco and {St{\"u}cker}, Jens and {Angulo}, Raul E.},
        title = "{Probabilistic Estimators of Lagrangian Shape Biases: Universal Relations and Physical Insights}",
      journal = {arXiv e-prints},
         year = 2024,
        month = sep,
          eid = {arXiv:2409.14995},
        pages = {arXiv:2409.14995},
          doi = {10.48550/arXiv.2409.14995},
archivePrefix = {arXiv},
       eprint = {2409.14995},
 primaryClass = {astro-ph.CO},
       adsurl = {https://ui.adsabs.harvard.edu/abs/2024arXiv240914995M}
}

@ARTICLE{Schneider10,
       author = {{Schneider}, Michael D. and {Bridle}, Sarah},
        title = "{A halo model for intrinsic alignments of galaxy ellipticities}",
      journal = {\mnras},
         year = 2010,
        month = mar,
       volume = {402},
       number = {4},
        pages = {2127-2139},
          doi = {10.1111/j.1365-2966.2009.15956.x},
archivePrefix = {arXiv},
       eprint = {0903.3870},
 primaryClass = {astro-ph.CO},
       adsurl = {https://ui.adsabs.harvard.edu/abs/2010MNRAS.402.2127S}
}

@ARTICLE{Wright25,
       author = {{Wright}, Angus H. and {St{\"o}lzner}, Benjamin and {Asgari}, Marika and {Bilicki}, Maciej and {Giblin}, Benjamin and {Heymans}, Catherine and {Hildebrandt}, Hendrik and {Hoekstra}, Henk and {Joachimi}, Benjamin and {Kuijken}, Konrad and {Li}, Shun-Sheng and {Reischke}, Robert and {von Wietersheim-Kramsta}, Maximilian and {Yoon}, Mijin and {Burger}, Pierre and {Chisari}, Nora Elisa and {de Jong}, Jelte and {Dvornik}, Andrej and {Georgiou}, Christos and {Harnois-D{\'e}raps}, Joachim and {Jalan}, Priyanka and {William}, Anjitha John and {Joudaki}, Shahab and {Lesci}, Giorgio Francesco and {Linke}, Laila and {Loureiro}, Arthur and {Mahony}, Constance and {Maturi}, Matteo and {Miller}, Lance and {Moscardini}, Lauro and {Napolitano}, Nicola R. and {Porth}, Lucas and {Radovich}, Mario and {Schneider}, Peter and {Tr{\"o}ster}, Tilman and {Wittje}, Anna and {Yan}, Ziang and {Zhang}, Yun-Hao},
        title = "{KiDS-Legacy: Cosmological constraints from cosmic shear with the complete Kilo-Degree Survey}",
      journal = {arXiv e-prints},
         year = 2025,
        month = mar,
          eid = {arXiv:2503.19441},
        pages = {arXiv:2503.19441},
          doi = {10.48550/arXiv.2503.19441},
archivePrefix = {arXiv},
       eprint = {2503.19441},
 primaryClass = {astro-ph.CO},
       adsurl = {https://ui.adsabs.harvard.edu/abs/2025arXiv250319441W}
}

@ARTICLE{Fortuna21,
       author = {{Fortuna}, Maria Cristina and {Hoekstra}, Henk and {Johnston}, Harry and {Vakili}, Mohammadjavad and {Kannawadi}, Arun and {Georgiou}, Christos and {Joachimi}, Benjamin and {Wright}, Angus H. and {Asgari}, Marika and {Bilicki}, Maciej and {Heymans}, Catherine and {Hildebrandt}, Hendrik and {Kuijken}, Konrad and {Von Wietersheim-Kramsta}, Maximilian},
        title = "{KiDS-1000: Constraints on the intrinsic alignment of luminous red galaxies}",
      journal = {\aap},
         year = 2021,
        month = oct,
       volume = {654},
          eid = {A76},
        pages = {A76},
          doi = {10.1051/0004-6361/202140706},
archivePrefix = {arXiv},
       eprint = {2109.02556},
 primaryClass = {astro-ph.CO},
       adsurl = {https://ui.adsabs.harvard.edu/abs/2021A&A...654A..76F}
}

@ARTICLE{Okumura19,
       author = {{Okumura}, Teppei and {Taruya}, Atsushi and {Nishimichi}, Takahiro},
        title = "{Intrinsic alignment statistics of density and velocity fields at large scales: Formulation, modeling, and baryon acoustic oscillation features}",
      journal = {\prd},
         year = 2019,
        month = nov,
       volume = {100},
       number = {10},
          eid = {103507},
        pages = {103507},
          doi = {10.1103/PhysRevD.100.103507},
archivePrefix = {arXiv},
       eprint = {1907.00750},
 primaryClass = {astro-ph.CO},
       adsurl = {https://ui.adsabs.harvard.edu/abs/2019PhRvD.100j3507O}
}

@ARTICLE{Chisari13,
       author = {{Chisari}, Nora Elisa and {Dvorkin}, Cora},
        title = "{Cosmological information in the intrinsic alignments of luminous red galaxies}",
      journal = {\jcap},
         year = 2013,
        month = dec,
       volume = {2013},
       number = {12},
          eid = {029},
        pages = {029},
          doi = {10.1088/1475-7516/2013/12/029},
archivePrefix = {arXiv},
       eprint = {1308.5972},
 primaryClass = {astro-ph.CO},
       adsurl = {https://ui.adsabs.harvard.edu/abs/2013JCAP...12..029C}
}

@ARTICLE{vanDompseler23,
       author = {{van Dompseler}, Dennis and {Georgiou}, Christos and {Chisari}, Nora Elisa},
        title = "{The alignment of galaxies at the Baryon Acoustic Oscillation scale}",
      journal = {The Open Journal of Astrophysics},
         year = 2023,
        month = jun,
       volume = {6},
          eid = {19},
        pages = {19},
          doi = {10.21105/astro.2301.04649},
archivePrefix = {arXiv},
       eprint = {2301.04649},
 primaryClass = {astro-ph.CO},
       adsurl = {https://ui.adsabs.harvard.edu/abs/2023OJAp....6E..19V}
}

@ARTICLE{Leonard18,
       author = {{Leonard}, C. Danielle and {Mandelbaum}, Rachel and {LSST Dark Energy Science Collaboration}},
        title = "{Measuring the scale dependence of intrinsic alignments usingmultiple shear estimates}",
      journal = {\mnras},
         year = 2018,
        month = sep,
       volume = {479},
       number = {1},
        pages = {1412-1426},
          doi = {10.1093/mnras/sty1444},
archivePrefix = {arXiv},
       eprint = {1802.08263},
 primaryClass = {astro-ph.CO},
       adsurl = {https://ui.adsabs.harvard.edu/abs/2018MNRAS.479.1412L}
}

@ARTICLE{Sarcevic25,
       author = {{{\v{S}}ar{\v{c}}evi{\'c}}, Nikolina and {Leonard}, C. Danielle and {Rau}, Markus M. and {the LSST Dark Energy Science Collaboration}},
        title = "{Joint modelling of astrophysical systematics for cosmology with LSST cosmic shear}",
      journal = {\mnras},
         year = 2025,
        month = feb,
       volume = {537},
       number = {2},
        pages = {1924-1948},
          doi = {10.1093/mnras/staf156},
archivePrefix = {arXiv},
       eprint = {2406.03352},
 primaryClass = {astro-ph.CO},
       adsurl = {https://ui.adsabs.harvard.edu/abs/2025MNRAS.537.1924S}
}

@ARTICLE{Blazek12,
       author = {{Blazek}, Jonathan and {Mandelbaum}, Rachel and {Seljak}, Uro{\v{s}} and {Nakajima}, Reiko},
        title = "{Separating intrinsic alignment and galaxy-galaxy lensing}",
      journal = {\jcap},
         year = 2012,
        month = may,
       volume = {2012},
       number = {5},
          eid = {041},
        pages = {041},
          doi = {10.1088/1475-7516/2012/05/041},
archivePrefix = {arXiv},
       eprint = {1204.2264},
 primaryClass = {astro-ph.CO},
       adsurl = {https://ui.adsabs.harvard.edu/abs/2012JCAP...05..041B}
}

@ARTICLE{Yao23,
       author = {{Yao}, Ji and {Shan}, Huanyuan and {Zhang}, Pengjie and {Liu}, Xiangkun and {Heymans}, Catherine and {Joachimi}, Benjamin and {Asgari}, Marika and {Bilicki}, Maciej and {Hildebrandt}, Hendrik and {Kuijken}, Konrad and {Tr{\"o}ster}, Tilman and {van den Busch}, Jan Luca and {Wright}, Angus and {Yan}, Ziang},
        title = "{KiDS-1000: Cross-correlation with Planck cosmic microwave background lensing and intrinsic alignment removal with self-calibration}",
      journal = {\aap},
         year = 2023,
        month = may,
       volume = {673},
          eid = {A111},
        pages = {A111},
          doi = {10.1051/0004-6361/202346020},
archivePrefix = {arXiv},
       eprint = {2301.13437},
 primaryClass = {astro-ph.CO},
       adsurl = {https://ui.adsabs.harvard.edu/abs/2023A&A...673A.111Y}
}

@ARTICLE{Heymans03,
       author = {{Heymans}, Catherine and {Heavens}, Alan},
        title = "{Weak gravitational lensing: reducing the contamination by intrinsic alignments}",
      journal = {\mnras},
         year = 2003,
        month = mar,
       volume = {339},
       number = {3},
        pages = {711-720},
          doi = {10.1046/j.1365-8711.2003.06213.x},
archivePrefix = {arXiv},
       eprint = {astro-ph/0208220},
 primaryClass = {astro-ph},
       adsurl = {https://ui.adsabs.harvard.edu/abs/2003MNRAS.339..711H}
}

@ARTICLE{Heymans06,
       author = {{Heymans}, Catherine and {White}, Martin and {Heavens}, Alan and {Vale}, Chris and {van Waerbeke}, Ludovic},
        title = "{Potential sources of contamination to weak lensing measurements: constraints from N-body simulations}",
      journal = {\mnras},
         year = 2006,
        month = sep,
       volume = {371},
       number = {2},
        pages = {750-760},
          doi = {10.1111/j.1365-2966.2006.10705.x},
archivePrefix = {arXiv},
       eprint = {astro-ph/0604001},
 primaryClass = {astro-ph},
       adsurl = {https://ui.adsabs.harvard.edu/abs/2006MNRAS.371..750H}
}

@ARTICLE{Johnston24,
       author = {{Johnston}, Harry and {Chisari}, Nora Elisa and {Joudaki}, Shahab and {Reischke}, Robert and {St{\"o}lzner}, Benjamin and {Loureiro}, Arthur and {Mahony}, Constance and {Unruh}, Sandra and {Wright}, Angus H. and {Asgari}, Marika and {Bilicki}, Maciej and {Burger}, Pierre and {Dvornik}, Andrej and {Georgiou}, Christos and {Giblin}, Benjamin and {Heymans}, Catherine and {Hildebrandt}, Hendrik and {Joachimi}, Benjamin and {Kuijken}, Konrad and {Li}, Shun-Sheng and {Linke}, Laila and {Porth}, Lucas and {Shan}, HuanYuan and {Tr{\"o}ster}, Tilman and {van den Busch}, Jan Luca and {von Wietersheim-Kramsta}, Maximilian and {Yan}, Ziang and {Zhang}, Yun-Hao},
        title = "{6x2pt: Forecasting gains from joint weak lensing and galaxy clustering analyses with spectroscopic-photometric galaxy cross-correlations}",
      journal = {arXiv e-prints},
         year = 2024,
        month = sep,
          eid = {arXiv:2409.17377},
        pages = {arXiv:2409.17377},
          doi = {10.48550/arXiv.2409.17377},
archivePrefix = {arXiv},
       eprint = {2409.17377},
 primaryClass = {astro-ph.CO},
       adsurl = {https://ui.adsabs.harvard.edu/abs/2024arXiv240917377J}
}

@ARTICLE{Tegmark97,
       author = {{Tegmark}, Max and {Taylor}, Andy N. and {Heavens}, Alan F.},
        title = "{Karhunen-Lo{\`e}ve Eigenvalue Problems in Cosmology: How Should We Tackle Large Data Sets?}",
      journal = {\apj},
         year = 1997,
        month = may,
       volume = {480},
       number = {1},
        pages = {22-35},
          doi = {10.1086/303939},
archivePrefix = {arXiv},
       eprint = {astro-ph/9603021},
 primaryClass = {astro-ph},
       adsurl = {https://ui.adsabs.harvard.edu/abs/1997ApJ...480...22T}
}

@ARTICLE{Campos23,
       author = {{Campos}, A. and {Samuroff}, S. and {Mandelbaum}, R.},
        title = "{An empirical approach to model selection: weak lensing and intrinsic alignments}",
      journal = {\mnras},
         year = 2023,
        month = oct,
       volume = {525},
       number = {2},
        pages = {1885-1901},
          doi = {10.1093/mnras/stad2213},
archivePrefix = {arXiv},
       eprint = {2211.02800},
 primaryClass = {astro-ph.CO},
       adsurl = {https://ui.adsabs.harvard.edu/abs/2023MNRAS.525.1885C}
}

@ARTICLE{Leonard24,
       author = {{Leonard}, C. Danielle and {Rau}, Markus Michael and {Mandelbaum}, Rachel},
        title = "{Photometric redshifts and intrinsic alignments: Degeneracies and biases in the 3 {\texttimes}2 pt analysis}",
      journal = {\prd},
         year = 2024,
        month = apr,
       volume = {109},
       number = {8},
          eid = {083528},
        pages = {083528},
          doi = {10.1103/PhysRevD.109.083528},
archivePrefix = {arXiv},
       eprint = {2401.06060},
 primaryClass = {astro-ph.CO},
       adsurl = {https://ui.adsabs.harvard.edu/abs/2024PhRvD.109h3528L}
}

@ARTICLE{KiDS+DES,
       author = {{Dark Energy Survey and Kilo-Degree Survey Collaboration} and {Abbott}, T.~M.~C. and {Aguena}, M. and {Alarcon}, A. and {Alves}, O. and {Amon}, A. and {Andrade-Oliveira}, F. and {Asgari}, M. and {Avila}, S. and {Bacon}, D. and {Bechtol}, K. and {Becker}, M.~R. and {Bernstein}, G.~M. and {Bertin}, E. and {Bilicki}, M. and {Blazek}, J. and {Bocquet}, S. and {Brooks}, D. and {Burger}, P. and {Burke}, D.~L. and {Camacho}, H. and {Campos}, A. and {Carnero Rosell}, A. and {Carrasco Kind}, M. and {Carretero}, J. and {Castander}, F.~J. and {Cawthon}, R. and {Chang}, C. and {Chen}, R. and {Choi}, A. and {Conselice}, C. and {Cordero}, J. and {Crocce}, M. and {da Costa}, L.~N. and {da Silva Pereira}, M.~E. and {Dalal}, R. and {Davis}, C. and {de Jong}, J.~T.~A. and {DeRose}, J. and {Desai}, S. and {Diehl}, H.~T. and {Dodelson}, S. and {Doel}, P. and {Doux}, C. and {Drlica-Wagner}, A. and {Dvornik}, A. and {Eckert}, K. and {Eifler}, T.~F. and {Elvin-Poole}, J. and {Everett}, S. and {Fang}, X. and {Ferrero}, I. and {Fert{\'e}}, A. and {Flaugher}, B. and {Friedrich}, O. and {Frieman}, J. and {Garc{\'\i}a-Bellido}, J. and {Gatti}, M. and {Giannini}, G. and {Giblin}, B. and {Gruen}, D. and {Gruendl}, R.~A. and {Gutierrez}, G. and {Harrison}, I. and {Hartley}, W.~G. and {Herner}, K. and {Heymans}, C. and {Hildebrandt}, H. and {Hinton}, S.~R. and {Hoekstra}, H. and {Hollowood}, D.~L. and {Honscheid}, K. and {Huang}, H. and {Huff}, E.~M. and {Huterer}, D. and {James}, D.~J. and {Jarvis}, M. and {Jeffrey}, N. and {Jeltema}, T. and {Joachimi}, B. and {Joudaki}, S. and {Kannawadi}, A. and {Krause}, E. and {Kuehn}, K. and {Kuijken}, K. and {Kuropatkin}, N. and {Lahav}, O. and {Leget}, P. -F. and {Lemos}, P. and {Li}, S. -S. and {Li}, X. and {Liddle}, A.~R. and {Lima}, M. and {Lin}, C. -A. and {Lin}, H. and {MacCrann}, N. and {Mahony}, C. and {Marshall}, J.~L. and {McCullough}, J. and {Mena-Fern{\'a}ndez}, J. and {Menanteau}, F. and {Miquel}, R. and {Mohr}, J.~J. and {Muir}, J. and {Myles}, J. and {Napolitano}, N. and {Navarro-Alsina}, A. and {Ogando}, R.~L.~C. and {Palmese}, A. and {Pandey}, S. and {Park}, Y. and {Paterno}, M. and {Peacock}, J.~A. and {Petravick}, D. and {Pieres}, A. and {Plazas Malag{\'o}n}, A.~A. and {Porredon}, A. and {Prat}, J. and {Radovich}, M. and {Raveri}, M. and {Reischke}, R. and {Robertson}, N.~C. and {Rollins}, R.~P. and {Romer}, A.~K. and {Roodman}, A. and {Rykoff}, E.~S. and {Samuroff}, S. and {S{\'a}nchez}, C. and {Sanchez}, E. and {Sanchez}, J. and {Schneider}, P. and {Secco}, L.~F. and {Sevilla-Noarbe}, I. and {Shan}, H. -Y. and {Sheldon}, E. and {Shin}, T. and {Sif{\'o}n}, C. and {Smith}, M. and {Soares-Santos}, M. and {St{\"o}lzner}, B. and {Suchyta}, E. and {Swanson}, M.~E.~C. and {Tarle}, G. and {Thomas}, D. and {To}, C. and {Troxel}, M.~A. and {Tr{\"o}ster}, T. and {Tutusaus}, I. and {van den Busch}, J.~L. and {Varga}, T.~N. and {Walker}, A.~R. and {Weaverdyck}, N. and {Wechsler}, R.~H. and {Weller}, J. and {Wiseman}, P. and {Wright}, A.~H. and {Yanny}, B. and {Yin}, B. and {Yoon}, M. and {Zhang}, Y. and {Zuntz}, J.},
        title = "{DES Y3 + KiDS-1000: Consistent cosmology combining cosmic shear surveys}",
      journal = {The Open Journal of Astrophysics},
         year = 2023,
        month = oct,
       volume = {6},
          eid = {36},
        pages = {36},
          doi = {10.21105/astro.2305.17173},
archivePrefix = {arXiv},
       eprint = {2305.17173},
 primaryClass = {astro-ph.CO},
       adsurl = {https://ui.adsabs.harvard.edu/abs/2023OJAp....6E..36D}
}

@ARTICLE{Larsen16,
       author = {{Larsen}, Patricia and {Challinor}, Anthony},
        title = "{Intrinsic alignment contamination to CMB lensing-galaxy weak lensing correlations from tidal torquing}",
      journal = {\mnras},
         year = 2016,
        month = oct,
       volume = {461},
       number = {4},
        pages = {4343-4352},
          doi = {10.1093/mnras/stw1645},
archivePrefix = {arXiv},
       eprint = {1510.02617},
 primaryClass = {astro-ph.CO},
       adsurl = {https://ui.adsabs.harvard.edu/abs/2016MNRAS.461.4343L}
}

@ARTICLE{Yao17,
       author = {{Yao}, Ji and {Ishak}, Mustapha and {Lin}, Weikang and {Troxel}, Michael},
        title = "{Effects of self-calibration of intrinsic alignment on cosmological parameter constraints from future cosmic shear surveys}",
      journal = {\jcap},
         year = 2017,
        month = oct,
       volume = {2017},
       number = {10},
          eid = {056},
        pages = {056},
          doi = {10.1088/1475-7516/2017/10/056},
archivePrefix = {arXiv},
       eprint = {1707.01072},
 primaryClass = {astro-ph.CO},
       adsurl = {https://ui.adsabs.harvard.edu/abs/2017JCAP...10..056Y}
}

@ARTICLE{Yao19,
       author = {{Yao}, Ji and {Ishak}, Mustapha and {Troxel}, M.~A. and {LSST Dark Energy Science Collaboration}},
        title = "{Self-calibration method for II and GI types of intrinsic alignments of galaxies}",
      journal = {\mnras},
         year = 2019,
        month = feb,
       volume = {483},
       number = {1},
        pages = {276-288},
          doi = {10.1093/mnras/sty3188},
archivePrefix = {arXiv},
       eprint = {1809.07273},
 primaryClass = {astro-ph.CO},
       adsurl = {https://ui.adsabs.harvard.edu/abs/2019MNRAS.483..276Y}
}

@ARTICLE{Yao20b,
       author = {{Yao}, Ji and {Shan}, Huanyuan and {Zhang}, Pengjie and {Kneib}, Jean-Paul and {Jullo}, Eric},
        title = "{Unveiling the Intrinsic Alignment of Galaxies with Self-calibration and DECaLS DR3 Data}",
      journal = {\apj},
         year = 2020,
        month = dec,
       volume = {904},
       number = {2},
          eid = {135},
        pages = {135},
          doi = {10.3847/1538-4357/abc175},
archivePrefix = {arXiv},
       eprint = {2002.09826},
 primaryClass = {astro-ph.CO},
       adsurl = {https://ui.adsabs.harvard.edu/abs/2020ApJ...904..135Y}
}

@ARTICLE{Yao20,
       author = {{Yao}, Ji and {Pedersen}, Eske M. and {Ishak}, Mustapha and {Zhang}, Pengjie and {Agashe}, Anish and {Xu}, Haojie and {Shan}, Huanyuan},
        title = "{Separating the intrinsic alignment signal and the lensing signal using self-calibration in photo-z surveys with KiDS450 and KV450 Data}",
      journal = {\mnras},
         year = 2020,
        month = jul,
       volume = {495},
       number = {4},
        pages = {3900-3919},
          doi = {10.1093/mnras/staa1354},
archivePrefix = {arXiv},
       eprint = {1911.01582},
 primaryClass = {astro-ph.CO},
       adsurl = {https://ui.adsabs.harvard.edu/abs/2020MNRAS.495.3900Y}
}

@ARTICLE{Troxel12b,
       author = {{Troxel}, M.~A. and {Ishak}, M.},
        title = "{Self-calibrating the gravitational shear-intrinsic ellipticity-intrinsic ellipticity cross-correlation}",
      journal = {\mnras},
         year = 2012,
        month = nov,
       volume = {427},
       number = {1},
        pages = {442-457},
          doi = {10.1111/j.1365-2966.2012.21912.x},
archivePrefix = {arXiv},
       eprint = {1205.1547},
 primaryClass = {astro-ph.CO},
       adsurl = {https://ui.adsabs.harvard.edu/abs/2012MNRAS.427..442T}
}

@ARTICLE{Troxel12,
       author = {{Troxel}, M.~A. and {Ishak}, M.},
        title = "{Self-calibration technique for three-point intrinsic alignment correlations in weak lensing surveys}",
      journal = {\mnras},
         year = 2012,
        month = jan,
       volume = {419},
       number = {2},
        pages = {1804-1823},
          doi = {10.1111/j.1365-2966.2011.20205.x},
archivePrefix = {arXiv},
       eprint = {1109.4896},
 primaryClass = {astro-ph.CO},
       adsurl = {https://ui.adsabs.harvard.edu/abs/2012MNRAS.419.1804T}
}

@ARTICLE{Pedersen20,
       author = {{Pedersen}, Eske M. and {Yao}, Ji and {Ishak}, Mustapha and {Zhang}, Pengjie},
        title = "{First Detection of the GI-type of Intrinsic Alignments of Galaxies Using the Self-calibration Method in a Photometric Galaxy Survey}",
      journal = {\apjl},
         year = 2020,
        month = aug,
       volume = {899},
       number = {1},
          eid = {L5},
        pages = {L5},
          doi = {10.3847/2041-8213/aba51b},
archivePrefix = {arXiv},
       eprint = {1911.01614},
 primaryClass = {astro-ph.CO},
       adsurl = {https://ui.adsabs.harvard.edu/abs/2020ApJ...899L...5P}
}

@ARTICLE{Agarwal21,
       author = {{Agarwal}, Nishant and {Desjacques}, Vincent and {Jeong}, Donghui and {Schmidt}, Fabian},
        title = "{Information content in the redshift-space galaxy power spectrum and bispectrum}",
      journal = {\jcap},
         year = 2021,
        month = mar,
       volume = {2021},
       number = {3},
          eid = {021},
        pages = {021},
          doi = {10.1088/1475-7516/2021/03/021},
archivePrefix = {arXiv},
       eprint = {2007.04340},
 primaryClass = {astro-ph.CO},
       adsurl = {https://ui.adsabs.harvard.edu/abs/2021JCAP...03..021A}
}

@ARTICLE{King03,
       author = {{King}, L.~J. and {Schneider}, P.},
        title = "{Separating cosmic shear from intrinsic galaxy alignments: Correlation function tomography}",
      journal = {\aap},
         year = 2003,
        month = jan,
       volume = {398},
        pages = {23-30},
          doi = {10.1051/0004-6361:20021614},
archivePrefix = {arXiv},
       eprint = {astro-ph/0209474},
 primaryClass = {astro-ph},
       adsurl = {https://ui.adsabs.harvard.edu/abs/2003A&A...398...23K}
}

@ARTICLE{King02,
       author = {{King}, L. and {Schneider}, P.},
        title = "{Suppressing the contribution of intrinsic galaxy alignments to the shear two-point correlation function}",
      journal = {\aap},
         year = 2002,
        month = dec,
       volume = {396},
        pages = {411-418},
          doi = {10.1051/0004-6361:20021372},
archivePrefix = {arXiv},
       eprint = {astro-ph/0208256},
 primaryClass = {astro-ph},
       adsurl = {https://ui.adsabs.harvard.edu/abs/2002A&A...396..411K}
}

@ARTICLE{Pandya24,
       author = {{Pandya}, Sneh and {Yang}, Yuanyuan and {Van Alfen}, Nicholas and {Blazek}, Jonathan and {Walters}, Robin},
        title = "{Learning Galaxy Intrinsic Alignment Correlations}",
      journal = {arXiv e-prints},
         year = 2024,
        month = apr,
          eid = {arXiv:2404.13702},
        pages = {arXiv:2404.13702},
          doi = {10.48550/arXiv.2404.13702},
archivePrefix = {arXiv},
       eprint = {2404.13702},
 primaryClass = {astro-ph.CO},
       adsurl = {https://ui.adsabs.harvard.edu/abs/2024arXiv240413702P}
}

@ARTICLE{Krause11,
       author = {{Krause}, Elisabeth and {Hirata}, Christopher M.},
        title = "{Tidal alignments as a contaminant of the galaxy bispectrum}",
      journal = {\mnras},
         year = 2011,
        month = feb,
       volume = {410},
       number = {4},
        pages = {2730-2740},
          doi = {10.1111/j.1365-2966.2010.17638.x},
archivePrefix = {arXiv},
       eprint = {1004.3611},
 primaryClass = {astro-ph.CO},
       adsurl = {https://ui.adsabs.harvard.edu/abs/2011MNRAS.410.2730K}
}

@ARTICLE{Joachimi09,
       author = {{Joachimi}, B. and {Schneider}, P.},
        title = "{The removal of shear-ellipticity correlations from the cosmic shear signal. Influence of photometric redshift errors on the nulling technique}",
      journal = {\aap},
         year = 2009,
        month = nov,
       volume = {507},
       number = {1},
        pages = {105-129},
          doi = {10.1051/0004-6361/200912420},
archivePrefix = {arXiv},
       eprint = {0905.0393},
 primaryClass = {astro-ph.CO},
       adsurl = {https://ui.adsabs.harvard.edu/abs/2009A&A...507..105J}
}

@ARTICLE{Joachimi08,
       author = {{Joachimi}, B. and {Schneider}, P.},
        title = "{The removal of shear-ellipticity correlations from the cosmic shear signal via nulling techniques}",
      journal = {\aap},
         year = 2008,
        month = sep,
       volume = {488},
       number = {3},
        pages = {829-843},
          doi = {10.1051/0004-6361:200809971},
archivePrefix = {arXiv},
       eprint = {0804.2292},
 primaryClass = {astro-ph},
       adsurl = {https://ui.adsabs.harvard.edu/abs/2008A&A...488..829J}
}

@ARTICLE{Troxel14,
       author = {{Troxel}, M.~A. and {Ishak}, Mustapha},
        title = "{Cross-correlation between cosmic microwave background lensing and galaxy intrinsic alignment as a contaminant to gravitational lensing cross-correlated probes of the Universe}",
      journal = {\prd},
         year = 2014,
        month = mar,
       volume = {89},
       number = {6},
          eid = {063528},
        pages = {063528},
          doi = {10.1103/PhysRevD.89.063528},
archivePrefix = {arXiv},
       eprint = {1401.7051},
 primaryClass = {astro-ph.CO},
       adsurl = {https://ui.adsabs.harvard.edu/abs/2014PhRvD..89f3528T}
}

@ARTICLE{Hall14,
       author = {{Hall}, A. and {Taylor}, A.},
        title = "{Intrinsic alignments in the cross-correlation of cosmic shear and cosmic microwave background weak lensing.}",
      journal = {\mnras},
         year = 2014,
        month = sep,
       volume = {443},
        pages = {L119-L123},
          doi = {10.1093/mnrasl/slu094},
archivePrefix = {arXiv},
       eprint = {1401.6018},
 primaryClass = {astro-ph.CO},
       adsurl = {https://ui.adsabs.harvard.edu/abs/2014MNRAS.443L.119H}
}

@ARTICLE{Zhang08,
       author = {{Zhang}, Pengjie},
        title = "{Self-calibration of Gravitational Shear-Galaxy Intrinsic Ellipticity Correlation in Weak Lensing Surveys}",
      journal = {\apj},
         year = 2010,
        month = sep,
       volume = {720},
       number = {2},
        pages = {1090-1101},
          doi = {10.1088/0004-637X/720/2/1090},
archivePrefix = {arXiv},
       eprint = {0811.0613},
 primaryClass = {astro-ph},
       adsurl = {https://ui.adsabs.harvard.edu/abs/2010ApJ...720.1090Z}
}

@ARTICLE{Zhang10,
       author = {{Zhang}, Pengjie},
        title = "{A proposal on the galaxy intrinsic alignment self-calibration in weak lensing surveys}",
      journal = {\mnras},
         year = 2010,
        month = jul,
       volume = {406},
       number = {1},
        pages = {L95-L99},
          doi = {10.1111/j.1745-3933.2010.00893.x},
archivePrefix = {arXiv},
       eprint = {1003.5219},
 primaryClass = {astro-ph.CO},
       adsurl = {https://ui.adsabs.harvard.edu/abs/2010MNRAS.406L..95Z}
}

@ARTICLE{ChisariCMB,
       author = {{Chisari}, Nora Elisa and {Dunkley}, Joanna and {Miller}, Lance and {Allison}, Rupert},
        title = "{Contamination of early-type galaxy alignments to galaxy lensing-CMB lensing cross-correlation}",
      journal = {\mnras},
         year = 2015,
        month = oct,
       volume = {453},
       number = {1},
        pages = {682-689},
          doi = {10.1093/mnras/stv1655},
archivePrefix = {arXiv},
       eprint = {1507.03906},
 primaryClass = {astro-ph.CO},
       adsurl = {https://ui.adsabs.harvard.edu/abs/2015MNRAS.453..682C}
}

@ARTICLE{Joachimi10,
       author = {{Joachimi}, B. and {Bridle}, S.~L.},
        title = "{Simultaneous measurement of cosmology and intrinsic alignments using joint cosmic shear and galaxy number density correlations}",
      journal = {\aap},
         year = 2010,
        month = nov,
       volume = {523},
          eid = {A1},
        pages = {A1},
          doi = {10.1051/0004-6361/200913657},
archivePrefix = {arXiv},
       eprint = {0911.2454},
 primaryClass = {astro-ph.CO},
       adsurl = {https://ui.adsabs.harvard.edu/abs/2010A&A...523A...1J}
}

@ARTICLE{Hirata04,
       author = {{Hirata}, Christopher M. and {Seljak}, Uro{\v{s}}},
        title = "{Intrinsic alignment-lensing interference as a contaminant of cosmic shear}",
      journal = {\prd},
         year = 2004,
        month = sep,
       volume = {70},
       number = {6},
          eid = {063526},
        pages = {063526},
          doi = {10.1103/PhysRevD.70.063526},
archivePrefix = {arXiv},
       eprint = {astro-ph/0406275},
 primaryClass = {astro-ph},
       adsurl = {https://ui.adsabs.harvard.edu/abs/2004PhRvD..70f3526H}
}

@ARTICLE{Kirk12,
       author = {{Kirk}, Donnacha and {Rassat}, Anais and {Host}, Ole and {Bridle}, Sarah},
        title = "{The cosmological impact of intrinsic alignment model choice for cosmic shear}",
      journal = {\mnras},
         year = 2012,
        month = aug,
       volume = {424},
       number = {3},
        pages = {1647-1657},
          doi = {10.1111/j.1365-2966.2012.21099.x},
archivePrefix = {arXiv},
       eprint = {1112.4752},
 primaryClass = {astro-ph.CO},
       adsurl = {https://ui.adsabs.harvard.edu/abs/2012MNRAS.424.1647K}
}

@ARTICLE{Krause16,
       author = {{Krause}, Elisabeth and {Eifler}, Tim and {Blazek}, Jonathan},
        title = "{The impact of intrinsic alignment on current and future cosmic shear surveys}",
      journal = {\mnras},
         year = 2016,
        month = feb,
       volume = {456},
       number = {1},
        pages = {207-222},
          doi = {10.1093/mnras/stv2615},
archivePrefix = {arXiv},
       eprint = {1506.08730},
 primaryClass = {astro-ph.CO},
       adsurl = {https://ui.adsabs.harvard.edu/abs/2016MNRAS.456..207K}
}

@ARTICLE{Hirata09,
       author = {{Hirata}, Christopher M.},
        title = "{Tidal alignments as a contaminant of redshift space distortions}",
      journal = {\mnras},
         year = 2009,
        month = oct,
       volume = {399},
       number = {2},
        pages = {1074-1087},
          doi = {10.1111/j.1365-2966.2009.15353.x},
archivePrefix = {arXiv},
       eprint = {0903.4929},
 primaryClass = {astro-ph.CO},
       adsurl = {https://ui.adsabs.harvard.edu/abs/2009MNRAS.399.1074H}
}

@ARTICLE{Martens18,
       author = {{Martens}, Daniel and {Hirata}, Christopher M. and {Ross}, Ashley J. and {Fang}, Xiao},
        title = "{A radial measurement of the galaxy tidal alignment magnitude with BOSS data}",
      journal = {\mnras},
         year = 2018,
        month = jul,
       volume = {478},
       number = {1},
        pages = {711-732},
          doi = {10.1093/mnras/sty1100},
archivePrefix = {arXiv},
       eprint = {1802.07708},
 primaryClass = {astro-ph.CO},
       adsurl = {https://ui.adsabs.harvard.edu/abs/2018MNRAS.478..711M}
}

@ARTICLE{Lamman25,
       author = {{Lamman}, Claire and {Blazek}, Jonathan and {Eisenstein}, Daniel J.},
        title = "{Optimal intrinsic alignment estimators in the presence of redshift-space distortions}",
      journal = {arXiv e-prints},
         year = 2025,
        month = apr,
          eid = {arXiv:2504.16076},
        pages = {arXiv:2504.16076},
          doi = {10.48550/arXiv.2504.16076},
archivePrefix = {arXiv},
       eprint = {2504.16076},
 primaryClass = {astro-ph.CO},
       adsurl = {https://ui.adsabs.harvard.edu/abs/2025arXiv250416076L}
}

@ARTICLE{Lamman24,
       author = {{Lamman}, Claire and {Eisenstein}, Daniel and {Aguilar}, Jessica Nicole and {Ahlen}, Steven and {Brooks}, David and {Claybaugh}, Todd and {de la Macorra}, Axel and {Dey}, Arjun and {Dey}, Biprateep and {Doel}, Peter and {Ferraro}, Simone and {Font-Ribera}, Andreu and {Forero-Romero}, Jaime E. and {Gontcho}, Satya Gontcho A. and {Guy}, Julien and {Kehoe}, Robert and {Kremin}, Anthony and {Le Guillou}, Laurent and {Levi}, Michael and {Manera}, Marc and {Miquel}, Ramon and {Newman}, Jeffrey A. and {Nie}, Jundan and {Palanque-Delabrouille}, Nathalie and {Prada}, Francisco and {Rezaie}, Mehdi and {Rossi}, Graziano and {Sanchez}, Eusebio and {Schubnell}, Michael and {Hee-Jong}, Seo and {Tarl{\'e}}, Gregory and {Weaver}, Benjamin Alan and {Zhou}, Zhimin},
        title = "{Redshift-dependent RSD bias from intrinsic alignment with DESI Year 1 spectra}",
      journal = {\mnras},
         year = 2024,
        month = mar,
       volume = {528},
       number = {4},
        pages = {6559-6567},
          doi = {10.1093/mnras/stae317},
archivePrefix = {arXiv},
       eprint = {2312.04518},
 primaryClass = {astro-ph.CO},
       adsurl = {https://ui.adsabs.harvard.edu/abs/2024MNRAS.528.6559L}
}

@ARTICLE{Lamman23,
       author = {{Lamman}, Claire and {Eisenstein}, Daniel and {Aguilar}, Jessica Nicole and {Brooks}, David and {de la Macorra}, Axel and {Doel}, Peter and {Font-Ribera}, Andreu and {Gontcho A Gontcho}, Satya and {Honscheid}, Klaus and {Kehoe}, Robert and {Kisner}, Theodore and {Kremin}, Anthony and {Landriau}, Martin and {Levi}, Michael and {Miquel}, Ramon and {Moustakas}, John and {Palanque-Delabrouille}, Nathalie and {Poppett}, Claire and {Schubnell}, Michael and {Tarl{\'e}}, Gregory},
        title = "{Intrinsic alignment as an RSD contaminant in the DESI survey}",
      journal = {\mnras},
         year = 2023,
        month = jun,
       volume = {522},
       number = {1},
        pages = {117-129},
          doi = {10.1093/mnras/stad950},
archivePrefix = {arXiv},
       eprint = {2209.03949},
 primaryClass = {astro-ph.CO},
       adsurl = {https://ui.adsabs.harvard.edu/abs/2023MNRAS.522..117L}
}

@ARTICLE{Zwetsloot22,
       author = {{Zwetsloot}, Karel and {Chisari}, Nora Elisa},
        title = "{Impact of intrinsic alignments on clustering constraints of the growth rate}",
      journal = {\mnras},
         year = 2022,
        month = oct,
       volume = {516},
       number = {1},
        pages = {787-793},
          doi = {10.1093/mnras/stac2283},
archivePrefix = {arXiv},
       eprint = {2208.07062},
 primaryClass = {astro-ph.CO},
       adsurl = {https://ui.adsabs.harvard.edu/abs/2022MNRAS.516..787Z}
}

@ARTICLE{Obuljen20,
       author = {{Obuljen}, Andrej and {Percival}, Will J. and {Dalal}, Neal},
        title = "{Detection of anisotropic galaxy assembly bias in BOSS DR12}",
      journal = {\jcap},
         year = 2020,
        month = oct,
       volume = {2020},
       number = {10},
          eid = {058},
        pages = {058},
          doi = {10.1088/1475-7516/2020/10/058},
archivePrefix = {arXiv},
       eprint = {2004.07240},
 primaryClass = {astro-ph.CO},
       adsurl = {https://ui.adsabs.harvard.edu/abs/2020JCAP...10..058O}
}

@ARTICLE{Singh21,
       author = {{Singh}, Sukhdeep and {Yu}, Byeonghee and {Seljak}, Uro{\v{s}}},
        title = "{Fundamental Plane of BOSS galaxies: correlations with galaxy properties, density field, and impact on RSD measurements}",
      journal = {\mnras},
         year = 2021,
        month = mar,
       volume = {501},
       number = {3},
        pages = {4167-4183},
          doi = {10.1093/mnras/staa3263},
archivePrefix = {arXiv},
       eprint = {2001.07700},
 primaryClass = {astro-ph.CO},
       adsurl = {https://ui.adsabs.harvard.edu/abs/2021MNRAS.501.4167S}
}

@ARTICLE{HD22,
       author = {{Harnois-D{\'e}raps}, Joachim and {Martinet}, Nicolas and {Reischke}, Robert},
        title = "{Cosmic shear beyond 2-point statistics: Accounting for galaxy intrinsic alignment with projected tidal fields}",
      journal = {\mnras},
         year = 2022,
        month = jan,
       volume = {509},
       number = {3},
        pages = {3868-3888},
          doi = {10.1093/mnras/stab3222},
archivePrefix = {arXiv},
       eprint = {2107.08041},
 primaryClass = {astro-ph.CO},
       adsurl = {https://ui.adsabs.harvard.edu/abs/2022MNRAS.509.3868H}
}

@ARTICLE{Heavens00,
       author = {{Heavens}, Alan and {Refregier}, Alexandre and {Heymans}, Catherine},
        title = "{Intrinsic correlation of galaxy shapes: implications for weak lensing measurements}",
      journal = {\mnras},
         year = 2000,
        month = dec,
       volume = {319},
       number = {2},
        pages = {649-656},
          doi = {10.1046/j.1365-8711.2000.03907.x},
archivePrefix = {arXiv},
       eprint = {astro-ph/0005269},
 primaryClass = {astro-ph},
       adsurl = {https://ui.adsabs.harvard.edu/abs/2000MNRAS.319..649H}
}

\restoregeometry

\end{document}